\documentclass[prd,aps,nofootinbib,floatfix,11pt]{revtex4}

\usepackage{amsmath,graphicx,epsfig,amssymb,dsfont,mathtools,}
\usepackage[usenames]{color}
\usepackage{ulem} 
\usepackage{bigstrut}
\usepackage{slashed}
\usepackage{multirow}
\usepackage{makecell}
\usepackage{verbatim}
\usepackage{amsmath,graphicx,color,epsfig}
  \allowdisplaybreaks[1]
\renewcommand{\baselinestretch}{1.3}

\allowdisplaybreaks
\renewcommand{\thesubsection}{\Roman{subsection}}

\begin{document}
\title{Weak Decays of Doubly Heavy Baryons: Multi-body Decay Channels  }
\author{Yu-Ji Shi~\footnote{Email:shiyuji@sjtu.edu.cn}, Wei Wang~\footnote{Email:wei.wang@sjtu.edu.cn}, Ye Xing~\footnote{Email:xingye\_guang@sjtu.edu.cn}, and Ji Xu~\footnote{Email:xuji1991@sjtu.edu.cn}}
\affiliation{ INPAC, Shanghai Key Laboratory for Particle Physics and Cosmology,  MOE Key Laboratory for Particle Physics, Astrophysics and Cosmology,  \\ School of Physics and Astronomy, Shanghai Jiao Tong University, Shanghai  200240,   China }

\begin{abstract}
The newly-discovered  $\Xi_{cc}^{++}$  decays into the  $ \Lambda_{c}^+ K^-\pi^+\pi^+$, but    the experimental data has indicated that this decay is not saturated by  any two-body intermediate state.  In this work,
we analyze the multi-body weak decays of doubly heavy baryons
$\Xi_{cc}$,  $\Omega_{cc}$, $\Xi_{bc}$, $\Omega_{bc}$, $\Xi_{bb}$ and $\Omega_{bb}$, in particular the three-body nonleptonic decays and four-body semileptonic decays. We classify various decay modes according to the quark-level transitions and present an estimate of the typical  branching fractions for a few golden decay channels. Decay amplitudes are then parametrized in terms of a few   SU(3) irreducible amplitudes. With these amplitudes,   we find  a number of    relations for decay widths, which can be   examined in future.
\end{abstract}

\maketitle

\section{Introduction}

Nowadays  Lattice QCD is the sole approach that can study nonperturbative strong interactions from   first principle.
Despite the fact that there have been  great progresses on   Lattice QCD,  hadron structures  are still often encoded by phenomenological approaches like  quark models or QCD sum rules.  The quark model can be used to classify the hadrons,  in which a baryon is assigned as a three-quark system. Among various baryonic states,    doubly heavy baryons  are of particular interest  since they provide a platform to study the nonperturbative dynamics in the presence of heavy quarks. These states   have been searched for a long time~\cite{Mattson:2002vu,Ocherashvili:2004hi,Kato:2013ynr,Aaij:2013voa,Aubert:2006qw,Ratti:2003ez}, and  in 2017  the LHCb collaboration has announced an   observation of the  $\Xi_{cc}^{++}(ccu)$ with the mass $
m_{\Xi_{cc}^{++}} = (3621.40\pm 0.72\pm 0.27\pm 0.14) {\rm MeV}$~\cite{1707.01621}.  This analysis is based on the 1.7 $fb^{-1}$ data accumulated at 13 TeV, and confirmed in the additional sample of data collected at 8 TeV.

By all means the observation of  $\Xi_{cc}^{++}(ccu)$ is a milestone in hadron physics on both theoretical and experimental sides. One would  anticipate  that more experimental  data on  production and decays of  doubly heavy baryons  will be released  based on the larger data sample to be collected by LHCb in future~\cite{Bediaga:2012py}.   On the other side, to reveal the internal structures of   doubly heavy baryons,  more detailed  theoretical efforts  are  needed~\cite{Chen:2017sbg,Yu:2017zst,Wang:2017mqp,Li:2017cfz,Meng:2017udf,Wang:2017azm,Karliner:2017qjm,Gutsche:2017hux,Li:2017pxa,Guo:2017vcf,Lu:2017meb,Xiao:2017udy,Sharma:2017txj,Ma:2017nik,Meng:2017dni,Li:2017ndo,Wang:2017qvg}.

To handle weak decays of heavy mesons,  factorization approach is widely adopted in order to separate high-energy and low-energy degrees of freedoms. High-energy contributions are calculable using the  ordinary  perturbation theory.  The low-energy degrees, or equivalently the long-distance contributions, are usually parameterized as low energy inputs such as   light-cone distribution amplitudes. In terms of heavy baryon decays,  neither the low-energy inputs nor the short-distance coefficients are available in the literature. Only recently  the ``decay constants" were    studied  in QCD sum rules~\cite{Hu:2017dzi}.

This work is an extension of a series of previous works~\cite{Li:2017ndo,Yu:2017zst,Wang:2017mqp,Wang:2017azm,Hu:2017dzi}. In Ref.~\cite{Wang:2017azm},  instead of  factorization,   we have adopted the flavor SU(3) symmetry and classified various decays of doubly heavy baryons. In that work, however, we have limited ourselves to   two-body nonleptonic decay modes.  The $\Xi_{cc}^{++}$ baryon  has been  firstly observed in the   mode $\Xi_{cc}^{++} \to \Lambda_{c}^+ K^-\pi^+\pi^+$~\cite{1707.01621}, and  experimental data has indicated that this mode is not saturated  by  two-body intermediate state. This   motivates us to study the multi-body decays.  The main objective of this work is to do so, and we will focus on the cases where the final states contain one additional light meson, namely three-body nonleptonic decay and four-body semileptonic decays.

The rest of this paper is organized as follows. In Sec.~\ref{sec:particle_multiplet}, we will collect   representations for the particle multiplets in the SU(3) symmetry.  In Sec.~\ref{sec:golden_channels}, we will give a list of golden channels that can be used to reconstruct the doubly heavy baryons, and we present an estimate of their branching fractions.  In Sec.~\ref{sec:semileptonic}, we will analyze the  semileptonic decays of the doubly-heavy baryons, in which the final state contains two hadrons.  The three-body nonleptonic decays of doubly-charmed baryons, doubly-bottom baryons and the baryons with $b,c$ quarks are investigated in Sec.~\ref{sec:ccq_nonleptonic},~\ref{sec:bbq_nonleptonic} and Sec.~\ref{sec:bcq_nonleptonic}, respectively. The last section contains a brief summary.

\section{Particle Multiplets}
\label{sec:particle_multiplet}

In this section, we  start with the representations for the multiplets of the flavor SU(3) group.
Quantum numbers of the doubly heavy baryons are derived from the quark model.
These  baryons can form an SU(3) triplet:
\begin{eqnarray}
 T_{cc}  = \left(\begin{array}{c}  \Xi^{++}_{cc}(ccu)  \\  \Xi^+_{cc}(ccd)  \\  \Omega^+_{cc}(ccs)
\end{array}\right)\,,\;\;
  T_{bc} = \left(\begin{array}{c}  \Xi^+_{bc}(bcu)  \\  \Xi^0_{bc}(bcd)  \\  \Omega^0_{bc}(bcs)
\end{array}\right)\,,\;\;
  T_{bb} = \left(\begin{array}{c}  \Xi^0_{bb}(bbu)  \\  \Xi^-_{bb}(bbd)  \\  \Omega^-_{bb}(bbs)
\end{array}\right).
\end{eqnarray}

The light baryons form  an SU(3) octet and a decuplet. The octet has the expression:
\begin{eqnarray}
T_8= \left(\begin{array}{ccc} \frac{1}{\sqrt{2}}\Sigma^0+\frac{1}{\sqrt{6}}\Lambda^0 & \Sigma^+  &  p  \\ \Sigma^-  &  -\frac{1}{\sqrt{2}}\Sigma^0+\frac{1}{\sqrt{6}}\Lambda^0 & n \\ \Xi^-   & \Xi^0  & -\sqrt{\frac{2}{3}}\Lambda^0
  \end{array} \right),
\end{eqnarray}
and the light   decuplet is given as
\begin{eqnarray}
(T_{10})^{111} &=&  \Delta^{++},\;\;\; (T_{10})^{112}= (T_{10})^{121}=(T_{10})^{211}= \frac{1}{\sqrt3} \Delta^+,\nonumber\\
(T_{10})^{222} &=&  \Delta^{-},\;\;\; (T_{10})^{122}= (T_{10})^{212}=(T_{10})^{221}= \frac{1}{\sqrt3} \Delta^0, \nonumber\\
(T_{10})^{113} &=& (T_{10})^{131}=(T_{10})^{311}= \frac{1}{\sqrt3} \Sigma^{\prime+},\;\;(T_{10})^{223} = (T_{10})^{232}=(T_{10})^{322}= \frac{1}{\sqrt3} \Sigma^{\prime-},\nonumber\\
(T_{10})^{123} &=& (T_{10})^{132}=(T_{10})^{213}=(T_{10})^{231}=(T_{10})^{312}=(T_{10})^{321}= \frac{1}{\sqrt6} \Sigma^{\prime0},\nonumber\\
(T_{10})^{133} &=& (T_{10})^{313}=(T_{10})^{331}= \frac{1}{\sqrt3} \Xi^{\prime0},\;\;(T_{10})^{233} = (T_{10})^{323}=(T_{10})^{332}= \frac{1}{\sqrt3}  \Xi^{\prime-}, \nonumber\\
(T_{10})^{333}&=& \Omega^-.
\end{eqnarray}
In the meson sector,
the light pseudo-scalar meson is an octet, which can be represented  as:
\begin{eqnarray}
 M_{8}=\begin{pmatrix}
 \frac{\pi^0}{\sqrt{2}}+\frac{\eta}{\sqrt{6}}  &\pi^+ & K^+\\
 \pi^-&-\frac{\pi^0}{\sqrt{2}}+\frac{\eta}{\sqrt{6}}&{K^0}\\
 K^-&\bar K^0 &-2\frac{\eta}{\sqrt{6}}
 \end{pmatrix},
\end{eqnarray}
and  we shall not consider the flavor singlet $\eta_1$ in this work. This is also applicable to the vector meson octet and other light  mesons.

Charmed baryons  form an anti-triplet or sextet:
\begin{eqnarray}
 T_{\bf{c\bar 3}}= \left(\begin{array}{ccc} 0 & \Lambda_c^+  &  \Xi_c^+  \\ -\Lambda_c^+ & 0 & \Xi_c^0 \\ -\Xi_c^+   &  -\Xi_c^0  & 0
  \end{array} \right), \;\;
 T_{\bf{c6}} = \left(\begin{array}{ccc} \Sigma_c^{++} &  \frac{1}{\sqrt{2}}\Sigma_c^+   & \frac{1}{\sqrt{2}} \Xi_c^{\prime+}\\
  \frac{1}{\sqrt{2}}\Sigma_c^+& \Sigma_c^{0} & \frac{1}{\sqrt{2}} \Xi_c^{\prime0} \\
  \frac{1}{\sqrt{2}} \Xi_c^{\prime+}   &  \frac{1}{\sqrt{2}} \Xi_c^{\prime0}  & \Omega_c^0
  \end{array} \right)\,.
\end{eqnarray}
Charmed mesons forms an SU(3) anti-triplet:
\begin{eqnarray}
D_i=\left(\begin{array}{ccc} D^0, & D^+, & D^+_s  \end{array} \right), \;\;\;
\overline D^i=\left(\begin{array}{ccc}\overline D^0, & D^-, & D^-_s  \end{array} \right).
\end{eqnarray}
The above classification is also applicable to   bottom mesons.

\section{Golden Decay Channels}
\label{sec:golden_channels}

Before presenting the decay amplitudes for various channels,
 we will make a list of the golden channels and give an estimate of the decay branching fractions in this section~\cite{Yu:2017zst}.
In the following list we give, a hadron  is generic and can be replaced by the states with the same quark structure, for instance  one can replace  $\overline K^0$ by $\overline K^{*0}$ which decays into  $K^-\pi^+$.  Since a neutral meson is very difficult to reconstruct at LHC, we have removed the modes involving the  $\pi^0, \eta, \rho^+, \omega$.

\subsection{$\Xi_{cc}$ and $\Omega_{cc}$}

\begin{figure}
\includegraphics[width=0.8\columnwidth]{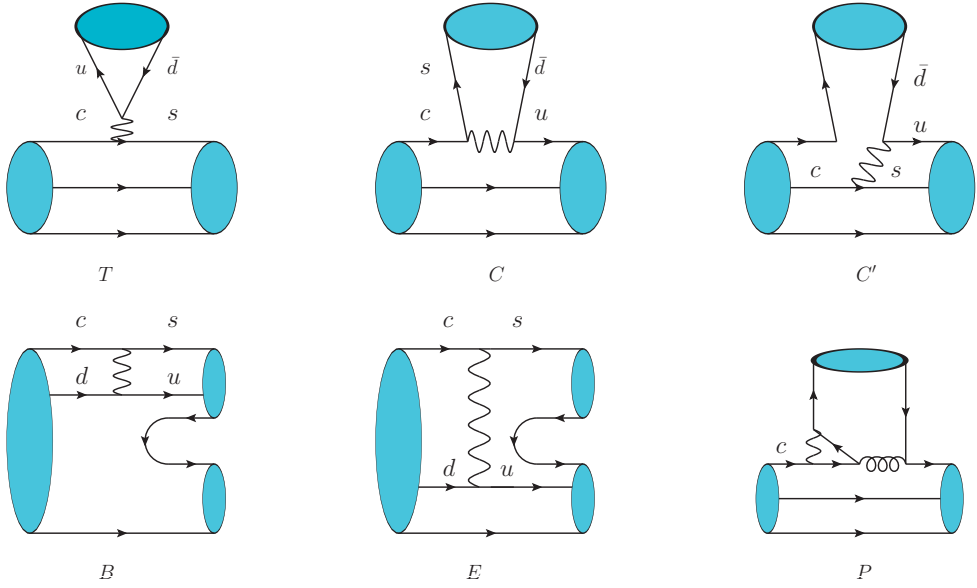}
\caption{Feynman diagrams for two-body decays of doubly charmed baryons. We only show one type of penguin diagrams. The $C,C', B,E$ diagrams are suppressed by $1/N_c$.    }
\label{fig:two_body}
\end{figure}

The Feynman diagrams for the Cabibbo-allowed decays are given in Fig.~\ref{fig:two_body}.  We only show one type of penguin diagrams. The $C,C', B,E$ diagrams are suppressed by $1/N_c$ compared to the tree amplitude $T$.
For the $\Xi_{cc}$ and $\Omega_{cc}$ decays, we collect  Cabibbo allowed decays in Tab.~\ref{tab:ccq_golden}.  From the $D$ and $\Lambda_c$ decay data, we infer that these Cabibbo allowed decay channels have typical branching fractions at a few percent level.

\begin{table}
 \caption{Cabibbo allowed $ccq$ decays whose typical branching ratios are   at  a few percents level.  }\label{tab:ccq_golden}\begin{tabular}{c  c   c  c|c|c|c|c|c}\hline\hline
$\Xi_{cc}^{++}\to  \Xi_c^+  \pi^+   $ & \qquad $\Xi_{cc}^{+}\to (\Lambda_c^+  \overline K^0 , \Lambda_c^+  \overline K^{*0})  $ &\qquad $\Omega_{cc}^{+}\to ( \Xi_c^+  \overline K^0 , \Xi_c^+  \overline K^{*0})  $   \\
 &\qquad $\Xi_{cc}^{+}\to  \Xi_c^+  \rho^0  $ &\qquad  \\
 &\qquad$\Xi_{cc}^{+}\to  \Xi_c^0  \pi^+   $ &\qquad  \\\hline
\hline
$\Xi_{cc}^{++}\to (\Sigma_{c}^{++}  \overline K^0 , \Sigma_{c}^{++}  \overline K^{*0})  $ & \qquad $\Xi_{cc}^{+}\to  \Sigma_{c}^{++}  K^-   $ & \qquad $\Omega_{cc}^{+}\to (\Xi_{c}^{\prime+}  \overline K^0 , \Xi_{c}^{\prime+}  \overline K^{*0})  $  \\
$\Xi_{cc}^{++}\to  \Xi_{c}^{\prime+}  \pi^+   $ &\qquad  $\Xi_{cc}^{+}\to (\Sigma_{c}^{+}  \overline K^0 , \Sigma_{c}^{+}  \overline K^{*0})  $ &\qquad $\Omega_{cc}^{+}\to  \Omega_{c}^{0}  \pi^+  $ \\
 &\qquad$\Xi_{cc}^{+}\to  (\Xi_{c}^{\prime+}  \rho^0,\Xi_{c}^{\prime0}  \pi^+) $ &\qquad  \\
 &\qquad$\Xi_{cc}^{+}\to  \Omega_{c}^{0}  K^+    $ &\qquad  \\\hline
\hline
$\Xi_{cc}^{++}\to (\Sigma^+  D^+,\Sigma^{\prime+}  D^+) $ &\qquad $\Xi_{cc}^{+}\to (\Lambda^0  D^+, \Sigma^0  D^+,\Sigma^{\prime0}  D^+) $ &\qquad $\Omega_{cc}^{+}\to (\Xi^0  D^+,\Xi^{\prime0}  D^+) $ \\
&\qquad$\Xi_{cc}^{+}\to (\Sigma^+  D^0,\Sigma^{\prime+}  D^0) $ &\qquad   \\
&\qquad$\Xi_{cc}^{+}\to (\Xi^0  D^+_s,\Xi^{\prime0}  D^+_s ) $ &\qquad  \\\hline
\end{tabular}
\end{table}

\subsection{$\Xi_{bc}$ and $\Omega_{bc}$}


\begin{table}
 \caption{Decay modes of the $bcq$  baryons.  Typical branching ratios for charm quark decays might be  a few percents, while   BR for decays induced by b quark may reach the order $10^{-3}$.  }\label{tab:bcq_golden}\begin{tabular}{c  c   c  cc|c|c|c|c}\hline\hline
Charm: $\Xi_{bc}^+$ &$\Xi_{bc}^0$ &  $\Omega_{bc}^0$   \\\hline
$\Xi_{bc}^{+}\to \Xi_b^0  \pi^+   $ &\qquad $\Xi_{bc}^{0}\to (\Lambda_b^0  \overline K^0 , \Lambda_b^0  \overline K^{*0})  $ &\qquad $\Omega_{bc}^{0}\to (\Xi_b^0  \overline K^0 , \Xi_b^0  \overline K^{*0})  $   \\
&\qquad$\Xi_{bc}^{0}\to  (\Xi_b^0  \rho^0,\Xi_b^-  \pi^+ )    $    \\\hline
\hline
$\Xi_{bc}^{+}\to (\Sigma_{b}^{+}  \overline K^0 , \Sigma_{b}^{+}  \overline K^{*0})  $ &\qquad $\Xi_{bc}^{0}\to \Sigma_{b}^{+}  K^-   $ &\qquad $\Omega_{bc}^{0}\to (\Xi_{b}^{\prime0} \overline K^0 , \Xi_{b}^{\prime0} \overline K^{*0})  $  \\
$\Xi_{bc}^{+}\to  \Xi_{b}^{\prime0}  \pi^+  $ &\qquad $\Xi_{bc}^{0}\to (\Sigma_{b}^{0}  \overline K^0 , \Sigma_{b}^{0}  \overline K^{*0})  $ &\qquad $\Omega_{bc}^{0}\to  \Omega_{b}^{-}  \pi^+   $  \\
&\qquad $\Xi_{bc}^{0}\to (\Xi_{b}^{\prime0}  \rho^0,\Xi_{b}^{\prime-}  \pi^+) $ &\qquad  \\
&\qquad $\Xi_{bc}^{0}\to  \Omega_{b}^{-}  K^+    $ &\qquad  \\\hline
\hline
$\Xi_{bc}^{+}\to (\Sigma^+  B^0,\Sigma^{\prime+}  B^0) $ &\qquad $\Xi_{bc}^{0}\to (\Sigma^+  B^-,\Sigma^{\prime+}  B^-) $ &\qquad  $\Omega_{cc}^{0}\to (\Xi^0  B^0,\Xi^{\prime0}  B^0) $ \\
&\qquad $\Xi_{bc}^{0}\to (\Sigma^0   B^0,\Sigma^{\prime0}  B^0,\Lambda^0 B^0) $ &\qquad   \\
&\qquad $\Xi_{bc}^{0}\to (\Xi^0  B_s^0, \Xi^{\prime0}  B_s^0) $ &\qquad  \\ \hline
\hline
Bottom: $\Xi_{bc}^+$ &$\Xi_{bc}^0$ &  $\Omega_{bc}^0$   \\\hline
%
$\Xi_{bc}^{+}\to (\Xi_c^+J/\psi, \Xi_{c}^{\prime+}J/\psi) $ &\qquad $\Xi_{bc}^{0}\to (\Xi_c^0J/\psi,\Xi_{c}^{\prime0}J/\psi) $ &\qquad $\Omega_{bc}^{0}\to \Omega_{c}^{0}J/\psi $  \\\hline
\hline
$\Xi_{bc}^{+}\to \Xi_{cc}^{++}  D^-_s $ &\qquad $\Xi_{bc}^{0}\to \Xi_{cc}^{+}  D^-_s $ &\qquad $\Omega_{bc}^{0}\to \Omega_{cc}^{+}  D^-_s $\\
$\Xi_{bc}^{+}\to \Omega_{cc}^{+}  \overline D^0 $ &\qquad $\Xi_{bc}^{0}\to \Omega_{cc}^{+}  D^- $ &\qquad  \\\hline
\hline
$\Xi_{bc}^{+}\to (\Xi_{cc}^{++}  \pi^-,\Xi_{cc}^{+}  \rho^0)   $ &\qquad  $\Xi_{bc}^{0}\to  \Xi_{cc}^{+}  \pi^-  $ &\qquad $\Omega_{bc}^{0}\to  \Xi_{cc}^{+}  K^-    $  \\
$\Xi_{bc}^{+}\to (\Omega_{cc}^{+}  K^0 , \Omega_{cc}^{+}  K^{*0})   $ &\qquad  &\qquad $\Omega_{bc}^{0}\to  \Omega_{cc}^{+}  \pi^-    $   \\\hline
\hline
$\Xi_{bc}^{+}\to (\Lambda_c^+  D^0,\Sigma_{c}^{+}  D^0,\Sigma_{c}^{0}  D^+) $ &\qquad $\Xi_{bc}^{0}\to \Sigma_{c}^{0}  D^0 $ &\qquad $\Omega_{bc}^{0}\to (\Xi_c^0  D^0,\Xi_{c}^{\prime0}  D^0 ) $  \\
 $\Xi_{bc}^{+}\to (\Xi_c^0  D^+_s, \Xi_{c}^{\prime0}  D^+_s) $ &\qquad &\qquad   \\\hline
\hline
\end{tabular}
\end{table}

A list of possible modes to reconstruct the $bcq$ baryons is given in Tab.~\ref{tab:bcq_golden}.
For the charm quark decay, the typical branching fractions might be  a few percents. The final state contains either a bottom meson, or a bottom baryon, whose decay branching fraction is then at the order  $10^{-3}$. So the branching fraction to reconstruct the $\Xi_{bc}$ and $\Omega_{bc}$ is very likely at the order  $10^{-5}$.

If the bottom quark decay first in the $bcq$ baryons, the branching fraction might be even smaller than $10^{-3}$, since the total width of $\Xi_{bc}$ and $\Omega_{bc}$ is dominated by charm quark decay. In this case, the branching fraction to reconstruct   $\Xi_{bc}$ and $\Omega_{bc}$ might be even smaller than $10^{-5}$.

\subsection{$\Xi_{bb}$ and $\Omega_{bb}$}

 \begin{table}
\caption{Doubly bottom baryon decays with branching fractions at the order of $10^{-3}$.    }\label{tab:bbq_golden}\begin{tabular}{ c  c   c   c|c|c|c|c|c}\hline\hline
$\Xi_{bb}^{0}\to (\Xi_b^0J/\psi,\Xi_b^{\prime0}J/\psi) $ &\qquad $\Xi_{bb}^{-}\to( \Xi_b^-J/\psi, \Xi_b^{\prime-}J/\psi) $ &\qquad $\Omega_{bb}^{-}\to \Omega_{b}^{-}J/\psi $  \\\hline
\hline
$\Xi_{bb}^{0}\to \Xi_{bc}^{+}  D^-_s $ &\qquad $\Xi_{bb}^{-}\to \Xi_{bc}^{0}  D^-_s $ &\qquad $\Omega_{bb}^{-}\to \Omega_{bc}^{0}  D^-_s $\\
$\Xi_{bb}^{0}\to \Omega_{bc}^{0}  \overline D^0 $ &\qquad $\Xi_{bb}^{-}\to \Omega_{bc}^{0}  D^- $ &\qquad \\\hline
\hline
$\Xi_{bb}^{0}\to  (\Xi_{bc}^{+}  \pi^-, \Xi_{bc}^{0}  \rho^0)  $ &\qquad  $\Xi_{bb}^{-}\to  \Xi_{bc}^{0}  \pi^-  $ &\qquad $\Omega_{bb}^{-}\to  \Xi_{bc}^{0}  K^- $   \\
$\Xi_{bb}^{0}\to (\Omega_{bc}^{0}  K^0 , \Omega_{bc}^{0}  K^{*0})  $ &\qquad  &\qquad $\Omega_{bb}^{-}\to  \Omega_{bc}^{0}  \pi^-  $ \\ \hline
\hline
$\Xi_{bb}^{0}\to (\Lambda_b^0  D^0,\Sigma_{b}^{0}  D^0) $ &\qquad $\Xi_{bb}^{-}\to \Sigma_{b}^{-}  D^0 $ &\qquad $\Omega_{bb}^{-}\to (\Xi_b^-  D^0, \Xi_{b}^{\prime-}  D^0 ) $ \\
$\Xi_{bb}^{0}\to \Sigma_{b}^{-}  D^+ $ &\qquad &\qquad\\
$\Xi_{bb}^{0}\to (\Xi_b^-  D^+_s,\Xi_{b}^{\prime-}  D^+_s ) $ &\qquad &\qquad\\\hline
\hline
\end{tabular}
\end{table}

The channels that can be used to reconstruct the $\Xi_{bb}$ and $\Omega_{bb}$ are collected in Tab.~\ref{tab:bbq_golden}.   Their typical branching fractions are at the order $10^{-3}$.  However in order to reconstruct the bottom meson and bottom baryons in the final state, the price to pay is another factor of $10^{-3}$. Including the fraction for $J/\psi$ or $D$ or charmed baryons, we have the largest decay branching fraction for $\Xi_{bb}$ and $\Omega_{bb}$ at the order of $10^{-8}$.

\section{Semi-Leptonic decays}
\label{sec:semileptonic}

\subsection{$\Xi_{cc}$ and $\Omega_{cc}$ decays}

  \begin{table}
\caption{Semileptonic   $\Xi_{cc}$ and $\Omega_{cc}$ decays into a singly charmed baryon (anti-triplet or sextet) and a light meson. }\label{tab:ccq_bar3cqqand8mson}\begin{tabular}{cc|cc|ccc}
\hline
channel  & amplitude  & channel  & amplitude & channel  & amplitude \tabularnewline
\hline
$\Xi_{cc}^{++}\to\Lambda_{c}^{+}\pi^{0}\ell^{+}\nu_{\ell}$  & $\frac{\left(a_{2}-a_{1}\right)V_{cd}^*}{\sqrt{2}}$ & $\Xi_{cc}^{+}\to\Xi_{c}^{+}\pi^{-}\ell^{+}\nu_{\ell}$  & $a_{2}V_{cs}^*$ &
$\Xi_{cc}^{++}\to\Lambda_{c}^{+}\overline{K}^{0}\ell^{+}\nu_{\ell}$  & $a_{1}V_{cs}^*$  \tabularnewline
\hline  $\Xi_{cc}^{+}\to\Xi_{c}^{0}\pi^{0}\ell^{+}\nu_{\ell}$  & $-\frac{a_{2}V_{cs}^*}{\sqrt{2}}$&
$\Xi_{cc}^{++}\to\Lambda_{c}^{+}\eta \ell^{+}\nu_{\ell}$  & $\frac{\left(a_{1}+a_{2}\right)V_{cd}^*}{\sqrt{6}}$ & $\Xi_{cc}^{+}\to\Xi_{c}^{0}K^{0}\ell^{+}\nu_{\ell}$  & $\left(a_{1}-a_{2}\right)V_{cd}^*$\tabularnewline
\hline
$\Xi_{cc}^{++}\to\Xi_{c}^{+}\pi^{0}\ell^{+}\nu_{\ell}$  & $\frac{a_{2}V_{cs}^*}{\sqrt{2}}$ & $\Xi_{cc}^{+}\to\Xi_{c}^{0}\eta \ell^{+}\nu_{\ell}$  & $\frac{\left(a_{2}-2a_{1}\right)V_{cs}^*}{\sqrt{6}}$&
$\Xi_{cc}^{++}\to\Xi_{c}^{+}K^{0}\ell^{+}\nu_{\ell}$  & $a_{1}V_{cd}^*$   \tabularnewline
\hline $\Omega_{cc}^{+}\to\Lambda_{c}^{+}K^{-}\ell^{+}\nu_{\ell}$  & $a_{2}V_{cd}^*$&
$\Xi_{cc}^{++}\to\Xi_{c}^{+}\eta \ell^{+}\nu_{\ell}$  & $\frac{\left(a_{2}-2a_{1}\right)V_{cs}^*}{\sqrt{6}}$ & $\Omega_{cc}^{+}\to\Xi_{c}^{+}\pi^{-}\ell^{+}\nu_{\ell}$  & $-a_{1}V_{cd}^*$\tabularnewline
\hline
$\Xi_{cc}^{++}\to\Xi_{c}^{0}\pi^{+}\ell^{+}\nu_{\ell}$  & $a_{2}V_{cs}^*$ & $\Omega_{cc}^{+}\to\Xi_{c}^{+}K^{-}\ell^{+}\nu_{\ell}$  & $\left(a_{2}-a_{1}\right)V_{cs}^*$&
$\Xi_{cc}^{++}\to\Xi_{c}^{0}K^{+}\ell^{+}\nu_{\ell}$  & $-a_{2}V_{cd}^*$ & \tabularnewline
\hline$\Omega_{cc}^{+}\to\Xi_{c}^{0}\pi^{0}\ell^{+}\nu_{\ell}$  & $\frac{a_{1}V_{cd}^*}{\sqrt{2}}$&
$\Xi_{cc}^{+}\to\Lambda_{c}^{+}\pi^{-}\ell^{+}\nu_{\ell}$  & $\left(a_{2}-a_{1}\right)V_{cd}^*$ & $\Omega_{cc}^{+}\to\Xi_{c}^{0}\overline{K}^{0}\ell^{+}\nu_{\ell}$  & $\left(a_{2}-a_{1}\right)V_{cs}^*$\tabularnewline
\hline
$\Xi_{cc}^{+}\to\Lambda_{c}^{+}K^{-}\ell^{+}\nu_{\ell}$  & $-a_{1}V_{cs}^*$ & $\Omega_{cc}^{+}\to\Xi_{c}^{0}\eta \ell^{+}\nu_{\ell}$  & $-\frac{\left(a_{1}-2a_{2}\right)V_{cd}^*}{\sqrt{6}}$\tabularnewline
\hline
\hline
$\Xi_{cc}^{++}\to\Sigma_{c}^{++}\pi^{-}\ell^{+}\nu_{\ell}$  & $a_{3}V_{cd}^*$ & $\Xi_{cc}^{+}\to\Xi_{c}^{\prime0}\pi^{0}\ell^{+}\nu_{\ell}$  & $-\frac{1}{2}a_{4}V_{cs}^*$&
$\Xi_{cc}^{++}\to\Sigma_{c}^{++}K^{-}\ell^{+}\nu_{\ell}$  & $a_{3}V_{cs}^*$  \tabularnewline
\hline  $\Xi_{cc}^{+}\to\Xi_{c}^{\prime0}K^{0}\ell^{+}\nu_{\ell}$  & $\frac{\left(a_{3}+a_{4}\right)V_{cd}^*}{\sqrt{2}}$& $\Xi_{cc}^{++}\to\Sigma_{c}^{+}\pi^{0}\ell^{+}\nu_{\ell}$  & $\frac{a_{4}-a_{3}}{2} V_{cd}^*$ & $\Xi_{cc}^{+}\to\Xi_{c}^{\prime0}\eta \ell^{+}\nu_{\ell}$  & $\frac{\left(a_{4}-2a_{3}\right)V_{cs}^*}{2\sqrt{3}}$\tabularnewline
\hline
$\Xi_{cc}^{++}\to\Sigma_{c}^{+}\overline{K}^{0}\ell^{+}\nu_{\ell}$  & $\frac{a_{3}V_{cs}^*}{\sqrt{2}}$ & $\Xi_{cc}^{+}\to\Omega_{c}^{0}K^{0}\ell^{+}\nu_{\ell}$  & $a_{4}V_{cs}^*$&
$\Xi_{cc}^{++}\to\Sigma_{c}^{+}\eta \ell^{+}\nu_{\ell}$  & $\frac{\left(a_{3}+a_{4}\right)V_{cd}^*}{2\sqrt{3}}$ &\tabularnewline
\hline $\Omega_{cc}^{+}\to\Sigma_{c}^{+}K^{-}\ell^{+}\nu_{\ell}$  & $\frac{a_{4}V_{cd}^*}{\sqrt{2}}$&
$\Xi_{cc}^{++}\to\Sigma_{c}^{0}\pi^{+}\ell^{+}\nu_{\ell}$  & $a_{4}V_{cd}^*$ & $\Omega_{cc}^{+}\to\Sigma_{c}^{0}\overline{K}^{0}\ell^{+}\nu_{\ell}$  & $a_{4}V_{cd}^*$\tabularnewline
\hline
$\Xi_{cc}^{++}\to\Xi_{c}^{\prime+}\pi^{0}\ell^{+}\nu_{\ell}$  & $\frac{1}{2}a_{4}V_{cs}^*$ & $\Omega_{cc}^{+}\to\Xi_{c}^{\prime+}\pi^{-}\ell^{+}\nu_{\ell}$  & $\frac{a_{3}V_{cd}^*}{\sqrt{2}}$& $\Xi_{cc}^{++}\to\Xi_{c}^{\prime+}K^{0}\ell^{+}\nu_{\ell}$  & $\frac{a_{3}V_{cd}^*}{\sqrt{2}}$ &\tabularnewline
\hline
 $\Omega_{cc}^{+}\to\Xi_{c}^{\prime+}K^{-}\ell^{+}\nu_{\ell}$  & $\frac{\left(a_{3}+a_{4}\right)V_{cs}^*}{\sqrt{2}}$&
$\Xi_{cc}^{++}\to\Xi_{c}^{\prime+}\eta \ell^{+}\nu_{\ell}$  & $\frac{\left(a_{4}-2a_{3}\right)V_{cs}^*}{2\sqrt{3}}$ & $\Omega_{cc}^{+}\to\Xi_{c}^{\prime0}\pi^{0}\ell^{+}\nu_{\ell}$  & $-\frac{1}{2}a_{3}V_{cd}^*$\tabularnewline
\hline
$\Xi_{cc}^{++}\to\Xi_{c}^{\prime0}\pi^{+}\ell^{+}\nu_{\ell}$  & $\frac{a_{4}V_{cs}^*}{\sqrt{2}}$ & $\Omega_{cc}^{+}\to\Xi_{c}^{\prime0}\overline{K}^{0}\ell^{+}\nu_{\ell}$  & $\frac{\left(a_{3}+a_{4}\right)V_{cs}^*}{\sqrt{2}}$&
$\Xi_{cc}^{++}\to\Xi_{c}^{\prime0}K^{+}\ell^{+}\nu_{\ell}$  & $\frac{a_{4}V_{cd}^*}{\sqrt{2}}$ &\tabularnewline
\hline $\Omega_{cc}^{+}\to\Xi_{c}^{\prime0}\eta \ell^{+}\nu_{\ell}$  & $\frac{\left(a_{3}-2a_{4}\right)V_{cd}^*}{2\sqrt{3}}$& $\Xi_{cc}^{++}\to\Omega_{c}^{0}K^{+}\ell^{+}\nu_{\ell}$  & $a_{4}V_{cs}^*$ & $\Omega_{cc}^{+}\to\Omega_{c}^{0}K^{0}\ell^{+}\nu_{\ell}$  & $a_{3}V_{cd}^*$\tabularnewline
\hline
$\Xi_{cc}^{+}\to\Sigma_{c}^{+}\pi^{-}\ell^{+}\nu_{\ell}$  & $\frac{\left(a_{3}+a_{4}\right)V_{cd}^*}{\sqrt{2}}$ & $\Omega_{cc}^{+}\to\Omega_{c}^{0}\eta \ell^{+}\nu_{\ell}$  & $-\sqrt{\frac{2}{3}}\left(a_{3}+a_{4}\right)V_{cs}^*$&
$\Xi_{cc}^{+}\to\Sigma_{c}^{+}K^{-}\ell^{+}\nu_{\ell}$  & $\frac{a_{3}V_{cs}^*}{\sqrt{2}}$ &\tabularnewline
\hline $\Xi_{cc}^{+}\to\Sigma_{c}^{0}\eta \ell^{+}\nu_{\ell}$  & $\frac{\left(a_{3}+a_{4}\right)V_{cd}^*}{\sqrt{6}}$&
$\Xi_{cc}^{+}\to\Sigma_{c}^{0}\pi^{0}\ell^{+}\nu_{\ell}$  & $-\frac{\left(a_{3}+a_{4}\right)V_{cd}^*}{\sqrt{2}}$ & $\Xi_{cc}^{+}\to\Xi_{c}^{\prime+}\pi^{-}\ell^{+}\nu_{\ell}$  & $\frac{a_{4}V_{cs}^*}{\sqrt{2}}$\tabularnewline
\hline
$\Xi_{cc}^{+}\to\Sigma_{c}^{0}\overline{K}^{0}\ell^{+}\nu_{\ell}$  & $a_{3}V_{cs}^*$ &  & \tabularnewline
\hline
\end{tabular}
\end{table}


The $c\to q\bar\ell\nu$ transition is induced by the effective Hamiltonian:
\begin{eqnarray}
{\cal H}_{eff}&=&\frac{G_F}{\sqrt2} \left[V_{cq}^* \bar q  \gamma^\mu(1-\gamma_5)c \bar \nu_{\ell}\gamma_\mu(1-\gamma_5) \ell\right] +h.c.,
\end{eqnarray}
where $q=d,s$ and the $V_{cd}$ and  $V_{cs}$    are   CKM matrix elements. The heavy-to-light quark operators
 will form an SU(3) triplet,  denoted as $H_{  3}$ with the components $(H_{  3})^1=0,~(H_{  3})^2=V_{cd}^*,~(H_{  3})^3=V_{cs}^*$.
At the hadron level, the effective Hamiltonian  for   decays of $\Xi_{cc}$ and $\Omega_{cc}$ into a singly charmed baryon and a light meson is constructed as:
\begin{eqnarray}
  {\cal H}_{{eff}}&=& a_1 (T_{cc})^i (H_{  3})^j  (\overline T_{\bf{c\bar 3}})_{[ik]}M^k_j~\bar\nu_\ell \ell+a_2 (T_{cc})^i (H_{  3})^j  (\overline T_{\bf{c\bar 3}})_{[kj]}M^k_i~\bar\nu_\ell \ell\nonumber \\
 &&+a_3 (T_{cc})^i (H_{  3})^j  (\overline T_{\bf{c6}})_{[ik]}M^k_j~\bar\nu_\ell \ell+a_4 (T_{cc})^i (H_{  3})^j  (\overline T_{\bf{c6}})_{[kj]}M^k_i~\bar\nu_\ell \ell\,.\label{eq:ccq_semi}
\end{eqnarray}
Here the $a_i$ are   SU(3) irreducible     amplitudes.  

The decay amplitudes for different channels can be deduced from the  Hamiltonian in Eq.~\eqref{eq:ccq_semi}, and given in Tab.~\ref{tab:ccq_bar3cqqand8mson}. The channels with the CKM factor $V_{cs}$ can have branching fractions about a few percents, while the $c\to d$ induced channels have the branching fractions at the order of $10^{-3}$.  From these amplitudes, we can find the relations for decay widths in the SU(3) symmetry limit. For   decays into a singly charmed baryon (anti-triplet), we have
\begin{eqnarray}
    \Gamma(\Xi_{cc}^{++}\to\Lambda_c^+\pi^0 \ell^+\nu_{\ell})= \frac{1}{2}\Gamma(\Xi_{cc}^{+}\to\Lambda_c^+\pi^- \ell^+\nu_{\ell})= \frac{1}{2}\Gamma(\Xi_{cc}^{+}\to\Xi_c^0K^0 \ell^+\nu_{\ell}),\nonumber\\
    \Gamma(\Xi_{cc}^{++}\to\Xi_c^+\pi^0 \ell^+\nu_{\ell})= \frac{1}{2}\Gamma(\Xi_{cc}^{+}\to\Xi_c^+\pi^- \ell^+\nu_{\ell})=\frac{1}{2}\Gamma(\Xi_{cc}^{++}\to\Xi_c^0\pi^+ \ell^+\nu_{\ell}),\nonumber\\
    \Gamma(\Omega_{cc}^{+}\to\Xi_c^+\pi^- \ell^+\nu_{\ell})= { }\Gamma(\Xi_{cc}^{++}\to\Xi_c^+K^0 \ell^+\nu_{\ell})= 2\Gamma(\Omega_{cc}^{+}\to\Xi_c^0\pi^0 \ell^+\nu_{\ell}),\nonumber\\
           \Gamma(\Xi_{cc}^{++}\to\Lambda_c^+\overline K^0 \ell^+\nu_{\ell})= { }\Gamma(\Xi_{cc}^{+}\to\Lambda_c^+K^- \ell^+\nu_{\ell}),\quad
     \Gamma(\Xi_{cc}^{++}\to\Xi_c^0K^+ \ell^+\nu_{\ell})= { }\Gamma(\Omega_{cc}^{+}\to\Lambda_c^+K^- \ell^+\nu_{\ell}),\nonumber\\
       \Gamma(\Xi_{cc}^{+}\to\Xi_c^0\pi^0 \ell^+\nu_{\ell})= \frac{1}{2}\Gamma(\Xi_{cc}^{+}\to\Xi_c^+\pi^- \ell^+\nu_{\ell}),
         \Gamma(\Omega_{cc}^{+}\to\Xi_c^0\overline K^0 \ell^+\nu_{\ell})= { }\Gamma(\Omega_{cc}^{+}\to\Xi_c^+K^- \ell^+\nu_{\ell}).\nonumber
\end{eqnarray}

Relations for    decays into a singly charmed baryon (sextet) are given as:
\begin{eqnarray}
    \Gamma(\Xi_{cc}^{++}\to\Sigma_{c}^{++}\pi^- \ell^+\nu_{\ell})= 2\Gamma(\Xi_{cc}^{++}\to\Xi_{c}^{\prime+}K^0 \ell^+\nu_{\ell})=2\Gamma(\Omega_{cc}^{+}\to\Xi_{c}^{\prime+}\pi^- \ell^+\nu_{\ell})= { }\Gamma(\Omega_{cc}^{+}\to\Omega_{c}^{0}K^0 \ell^+\nu_{\ell})\nonumber\\=4\Gamma(\Omega_{cc}^{+}\to\Xi_{c}^{\prime0}\pi^0 \ell^+\nu_{\ell}), \nonumber\\
     \Gamma(\Xi_{cc}^{++}\to\Sigma_{c}^{+}\overline K^0 \ell^+\nu_{\ell})= \frac{1}{2}\Gamma(\Xi_{cc}^{++}\to\Sigma_{c}^{++}K^- \ell^+\nu_{\ell})=\frac{1}{2}\Gamma(\Xi_{cc}^{+}\to\Sigma_{c}^{0}\overline K^0 \ell^+\nu_{\ell})= { }\Gamma(\Xi_{cc}^{+}\to\Sigma_{c}^{+}K^- \ell^+\nu_{\ell}),\nonumber\\
      \Gamma(\Xi_{cc}^{++}\to\Sigma_{c}^{0}\pi^+ \ell^+\nu_{\ell})= 2\Gamma(\Xi_{cc}^{++}\to\Xi_{c}^{\prime0}K^+ \ell^+\nu_{\ell})=2\Gamma(\Omega_{cc}^{+}\to\Sigma_{c}^{+}K^- \ell^+\nu_{\ell})= { }\Gamma(\Omega_{cc}^{+}\to\Sigma_{c}^{0}\overline K^0 \ell^+\nu_{\ell}),\nonumber\\
      \Gamma(\Xi_{cc}^{++}\to\Xi_{c}^{\prime+}\pi^0 \ell^+\nu_{\ell})= \frac{1}{4}\Gamma(\Xi_{cc}^{++}\to\Omega_{c}^{0}K^+ \ell^+\nu_{\ell})= \frac{1}{2}\Gamma(\Xi_{cc}^{+}\to\Xi_{c}^{\prime+}\pi^- \ell^+\nu_{\ell})= \frac{1}{4}\Gamma(\Xi_{cc}^{+}\to\Omega_{c}^{0}K^0 \ell^+\nu_{\ell})\nonumber\\=\frac{1}{2}\Gamma(\Xi_{cc}^{++}\to\Xi_{c}^{\prime0}\pi^+ \ell^+\nu_{\ell})=\Gamma(\Xi_{cc}^{+}\to\Xi_{c}^{\prime0}\pi^0 \ell^+\nu_{\ell}),\nonumber\\
       \Gamma(\Xi_{cc}^{+}\to\Sigma_{c}^{+}\pi^- \ell^+\nu_{\ell})= 6\Gamma(\Xi_{cc}^{++}\to\Sigma_{c}^{+}\eta \ell^+\nu_{\ell})= 3\Gamma(\Xi_{cc}^{+}\to\Sigma_{c}^{0}\eta \ell^+\nu_{\ell})={ }\Gamma(\Xi_{cc}^{+}\to\Xi_{c}^{\prime0}K^0 \ell^+\nu_{\ell})\nonumber\\=\Gamma(\Xi_{cc}^{+}\to\Sigma_{c}^{0}\pi^0 \ell^+\nu_{\ell}),\nonumber\\
              \Gamma(\Omega_{cc}^{+}\to\Xi_{c}^{\prime+}K^- \ell^+\nu_{\ell})=\Gamma(\Omega_{cc}^{+}\to\Xi_{c}^{\prime0}\overline K^0 \ell^+\nu_{\ell})= \frac{3}{4}\Gamma(\Omega_{cc}^{+}\to\Omega_{c}^{0}\eta \ell^+\nu_{\ell}).\nonumber
 \end{eqnarray}


\subsection{Semileptonic $\Xi_{bb}$ and $\Omega_{bb}$ decays}

The $b$ quark decay is governed  by the electro-weak   Hamiltonian
\begin{eqnarray}
 {\cal H}_{eff} &=& \frac{G_F}{\sqrt2} \left[V_{q'b} \bar q' \gamma^\mu(1-\gamma_5)b \bar  \ell\gamma_\mu(1-\gamma_5) \nu_{\ell}\right] +h.c.,
\end{eqnarray}
with $q'=u,c$.  The $b\to c$ transition is an SU(3) singlet, while the $b\to u$ transition forms an SU(3) triplet $H_{3}'$ with $(H_3')^1=1$ and $(H_3')^{2,3}=0$.  The hadron level  Hamiltonian for  semileptonic $\Xi_{bb}$ and $\Omega_{bb}$ decays  is constructed as
\begin{eqnarray}
{\cal H}_{{eff}}&=&a_5 (T_{bb})^i (\overline T_{bc})_jM^j_i~\bar \ell \nu_{\ell} \,+ a_6 (T_{bb})^i (H_{  3}')^j  (\overline T_{\bf{b\bar 3}})_{[ik]}M^k_j~\bar \ell\nu_{\ell} + a_7 (T_{bb})^i (H_{  3}')^j  (\overline T_{\bf{b\bar 3}})_{[jk]}M^k_i~\bar \ell\nu_{\ell} \nonumber \\
&&+ a_8 (T_{bb})^i  (H_{  3}')^j  (\overline T_{\bf{b6}})_{\{ik\}}M^k_j~\bar \ell \nu_{\ell}+ a_9 (T_{bb})^i  (H_{  3}')^j  (\overline T_{\bf{b6}})_{\{jk\}}M^k_i~\bar \ell \nu_{\ell}. \label{eq:bbq_semi}
\end{eqnarray}
The decay amplitudes can be deduced from this Hamiltonian, and the results are given in Tab.~\ref{tab:bbq_semi}.

For decays into a $bcq$, we have the relations for decay widths
\begin{eqnarray}
    \Gamma(\Xi_{bb}^{0}\to\Xi_{bc}^{+}\pi^0 \ell^-\bar\nu_{\ell})= 3\Gamma(\Xi_{bb}^{0}\to\Xi_{bc}^{+}\eta \ell^-\bar\nu_{\ell})
    = \frac{1}{2}\Gamma(\Xi_{bb}^{0}\to\Omega_{bc}^{0}K^+ \ell^-\bar\nu_{\ell})= \frac{1}{2}\Gamma(\Xi_{bb}^{-}\to\Xi_{bc}^{+}\pi^- \ell^-\bar\nu_{\ell})
    \nonumber\\= 3\Gamma(\Xi_{bb}^{-}\to\Xi_{bc}^{0}\eta \ell^-\bar\nu_{\ell})= \frac{1}{2}\Gamma(\Xi_{bb}^{-}\to\Omega_{bc}^{0}K^0 \ell^-\bar\nu_{\ell})
    = \frac{1}{2}\Gamma(\Omega_{bb}^{-}\to\Xi_{bc}^{+}K^- \ell^-\bar\nu_{\ell})= \frac{1}{2}\Gamma(\Omega_{bb}^{-}\to\Xi_{bc}^{0}\overline K^0 \ell^-\bar\nu_{\ell})
    \nonumber\\= \frac{3}{4}\Gamma(\Omega_{bb}^{-}\to\Omega_{bc}^{0}\eta \ell^-\bar\nu_{\ell})=\frac{1}{2}\Gamma(\Xi_{bb}^{0}\to\Xi_{bc}^{0}\pi^+ \ell^-\bar\nu_{\ell})=\Gamma(\Xi_{bb}^{-}\to\Xi_{bc}^{0}\pi^0 \ell^-\bar\nu_{\ell}).\nonumber
  \end{eqnarray}

For   $\Xi_{bb}$ and $\Omega_{bb}$ decays into a singly bottom baryon, we have the relations for decay widths:
\begin{eqnarray}
    \Gamma(\Xi_{bb}^{0}\to\Lambda_b^0\pi^+ \ell^-\bar\nu_{\ell})= { }\Gamma(\Xi_{bb}^{0}\to\Xi_b^0K^+ \ell^-\bar\nu_{\ell})= 2\Gamma(\Xi_{bb}^{-}\to\Lambda_b^0\pi^0 \ell^-\bar\nu_{\ell}),\nonumber\\
     \Gamma(\Omega_{bb}^{-}\to\Xi_b^-\pi^+ \ell^-\bar\nu_{\ell})= 2\Gamma(\Omega_{bb}^{-}\to\Xi_b^0\pi^0 \ell^-\bar\nu_{\ell})= { }\Gamma(\Xi_{bb}^{-}\to\Xi_b^-K^+ \ell^-\bar\nu_{\ell}),\nonumber\\
     \Gamma(\Xi_{bb}^{-}\to\Xi_b^0K^0 \ell^-\bar\nu_{\ell})= { }\Gamma(\Omega_{bb}^{-}\to\Lambda_b^0\overline K^0 \ell^-\bar\nu_{\ell}),\nonumber \\
    \Gamma(\Xi_{bb}^{0}\to\Sigma_{b}^{+}\pi^0 \ell^-\bar\nu_{\ell})= 3\Gamma(\Xi_{bb}^{0}\to\Sigma_{b}^{+}\eta \ell^-\bar\nu_{\ell})= { }\Gamma(\Xi_{bb}^{0}\to\Xi_{b}^{\prime0}K^+ \ell^-\bar\nu_{\ell})= 6\Gamma(\Xi_{bb}^{-}\to\Sigma_{b}^{0}\eta \ell^-\bar\nu_{\ell})\nonumber\\=\Gamma(\Xi_{bb}^{0}\to\Sigma_{b}^{0}\pi^+ \ell^-\bar\nu_{\ell}),\nonumber\\
          \Gamma(\Xi_{bb}^{-}\to\Sigma_{b}^{+}\pi^- \ell^-\bar\nu_{\ell})= 2\Gamma(\Xi_{bb}^{-}\to\Xi_{b}^{\prime0}K^0 \ell^-\bar\nu_{\ell})={ }\Gamma(\Omega_{bb}^{-}\to\Sigma_{b}^{+}K^- \ell^-\bar\nu_{\ell})= 2\Gamma(\Omega_{bb}^{-}\to\Sigma_{b}^{0}\overline K^0 \ell^-\bar\nu_{\ell}),\nonumber\\
           \Gamma(\Xi_{bb}^{-}\to\Sigma_{b}^{-}\pi^+ \ell^-\bar\nu_{\ell})= 2\Gamma(\Xi_{bb}^{-}\to\Xi_{b}^{\prime-}K^+ \ell^-\bar\nu_{\ell})= 4\Gamma(\Omega_{bb}^{-}\to\Xi_{b}^{\prime0}\pi^0 \ell^-\bar\nu_{\ell})={ }\Gamma(\Omega_{bb}^{-}\to\Omega_{b}^{-}K^+ \ell^-\bar\nu_{\ell})\nonumber\\=2\Gamma(\Omega_{bb}^{-}\to\Xi_{b}^{\prime-}\pi^+ \ell^-\bar\nu_{\ell}).\nonumber
  \end{eqnarray}

 \begin{table}
\caption{Amplitudes for doubly bottom baryons $\Xi_{bb}$ and $\Omega_{bb}$ decays into a triplet bcq and a light meson}\label{tab:bbq_semi}\begin{tabular}{|cc|cc|cc|}
\hline
channel  & amplitude  & channel  & amplitude & channel  & amplitude \tabularnewline
\hline
$\Xi_{bb}^{0}\to\Xi_{bc}^{+}\pi^{0}\ell^{-}\bar{\nu}_{\ell}$  & $\frac{a_{5}V_{cb}}{\sqrt{2}}$ & $\Xi_{bb}^{-}\to\Xi_{bc}^{0}\eta \ell^{-}\bar{\nu}_{\ell}$  & $\frac{a_{5}V_{cb}}{\sqrt{6}}$ & $\Xi_{bb}^{0}\to\Xi_{bc}^{+}\eta \ell^{-}\bar{\nu}_{\ell}$  & $\frac{a_{5}V_{cb}}{\sqrt{6}}$ \tabularnewline
\hline
$\Xi_{bb}^{-}\to\Omega_{bc}^{0}K^{0}\ell^{-}\bar{\nu}_{\ell}$  & $a_{5}V_{cb}$ & $\Xi_{bb}^{0}\to\Xi_{bc}^{0}\pi^{+}\ell^{-}\bar{\nu}_{\ell}$  & $a_{5}V_{cb}$ & $\Omega_{bb}^{-}\to\Xi_{bc}^{+}K^{-}\ell^{-}\bar{\nu}_{\ell}$  & $a_{5}V_{cb}$ \tabularnewline
\hline
$\Xi_{bb}^{0}\to\Omega_{bc}^{0}K^{+}\ell^{-}\bar{\nu}_{\ell}$  & $a_{5}V_{cb}$ & $\Omega_{bb}^{-}\to\Xi_{bc}^{0}\overline{K}^{0}\ell^{-}\bar{\nu}_{\ell}$  & $a_{5}V_{cb}$& $\Xi_{bb}^{-}\to\Xi_{bc}^{+}\pi^{-}\ell^{-}\bar{\nu}_{\ell}$  & $a_{5}V_{cb}$ \tabularnewline
\hline
$\Omega_{bb}^{-}\to\Omega_{bc}^{0}\eta \ell^{-}\bar{\nu}_{\ell}$  & $-\sqrt{\frac{2}{3}}a_{5}V_{cb}$& $\Xi_{bb}^{-}\to\Xi_{bc}^{0}\pi^{0}\ell^{-}\bar{\nu}_{\ell}$  & $-\frac{a_{5}V_{cb}}{\sqrt{2}}$ && \tabularnewline
\hline
\hline
$\Xi_{bb}^{0}\to\Lambda_{b}^{0}\pi^{+}\ell^{-}\bar{\nu}_{\ell}$  & $\left(a_{6}+a_{7}\right)V_{ub}$ & $\Xi_{bb}^{-}\to\Xi_{b}^{-}K^{+}\ell^{-}\bar{\nu}_{\ell}$  & $a_{6}V_{ub}$& $\Xi_{bb}^{0}\to\Xi_{b}^{0}K^{+}\ell^{-}\bar{\nu}_{\ell}$  & $\left(a_{6}+a_{7}\right)V_{ub}$ \tabularnewline
\hline
$\Omega_{bb}^{-}\to\Lambda_{b}^{0}\overline{K}^{0}\ell^{-}\bar{\nu}_{\ell}$  & $a_{7}V_{ub}$ & $\Xi_{bb}^{-}\to\Lambda_{b}^{0}\pi^{0}\ell^{-}\bar{\nu}_{\ell}$  & $-\frac{\left(a_{6}+a_{7}\right)V_{ub}}{\sqrt{2}}$ & $\Omega_{bb}^{-}\to\Xi_{b}^{0}\pi^{0}\ell^{-}\bar{\nu}_{\ell}$  & $-\frac{a_{6}V_{ub}}{\sqrt{2}}$\tabularnewline
\hline
$\Xi_{bb}^{-}\to\Lambda_{b}^{0}\eta \ell^{-}\bar{\nu}_{\ell}$  & $-\frac{\left(a_{6}-a_{7}\right)V_{ub}}{\sqrt{6}}$ & $\Omega_{bb}^{-}\to\Xi_{b}^{0}\eta \ell^{-}\bar{\nu}_{\ell}$  & $-\frac{\left(a_{6}+2a_{7}\right)V_{ub}}{\sqrt{6}}$ &&\tabularnewline
\hline
$\Xi_{bb}^{-}\to\Xi_{b}^{0}K^{0}\ell^{-}\bar{\nu}_{\ell}$  & $a_{7}V_{ub}$ & $\Omega_{bb}^{-}\to\Xi_{b}^{-}\pi^{+}\ell^{-}\bar{\nu}_{\ell}$  & $-a_{6}V_{ub}$&&\tabularnewline
\hline
\hline
$\Xi_{bb}^{0}\to\Sigma_{b}^{+}\pi^{0}\ell^{-}\bar{\nu}_{\ell}$  & $\frac{\left(a_{8}+a_{9}\right)V_{ub}}{\sqrt{2}}$ & $\Xi_{bb}^{-}\to\Xi_{b}^{\prime0}K^{0}\ell^{-}\bar{\nu}_{\ell}$  & $\frac{a_{9}V_{ub}}{\sqrt{2}}$ & $\Xi_{bb}^{0}\to\Sigma_{b}^{+}\eta \ell^{-}\bar{\nu}_{\ell}$  & $\frac{\left(a_{8}+a_{9}\right)V_{ub}}{\sqrt{6}}$ \tabularnewline
\hline
 $\Xi_{bb}^{-}\to\Xi_{b}^{\prime-}K^{+}\ell^{-}\bar{\nu}_{\ell}$  & $\frac{a_{8}V_{ub}}{\sqrt{2}}$& $\Xi_{bb}^{0}\to\Sigma_{b}^{0}\pi^{+}\ell^{-}\bar{\nu}_{\ell}$  & $\frac{\left(a_{8}+a_{9}\right)V_{ub}}{\sqrt{2}}$ & $\Omega_{bb}^{-}\to\Sigma_{b}^{+}K^{-}\ell^{-}\bar{\nu}_{\ell}$  & $a_{9}V_{ub}$\tabularnewline
\hline
$\Xi_{bb}^{0}\to\Xi_{b}^{\prime0}K^{+}\ell^{-}\bar{\nu}_{\ell}$  & $\frac{\left(a_{8}+a_{9}\right)V_{ub}}{\sqrt{2}}$ & $\Omega_{bb}^{-}\to\Sigma_{b}^{0}\overline{K}^{0}\ell^{-}\bar{\nu}_{\ell}$  & $\frac{a_{9}V_{ub}}{\sqrt{2}}$ & $\Xi_{bb}^{-}\to\Sigma_{b}^{+}\pi^{-}\ell^{-}\bar{\nu}_{\ell}$  & $a_{9}V_{ub}$  \tabularnewline
\hline
 $\Omega_{bb}^{-}\to\Xi_{b}^{\prime0}\pi^{0}\ell^{-}\bar{\nu}_{\ell}$  & $\frac{a_{8}V_{ub}}{2}$ &
$\Xi_{bb}^{-}\to\Sigma_{b}^{0}\pi^{0}\ell^{-}\bar{\nu}_{\ell}$  & $\frac{1}{2}\left(a_{8}-a_{9}\right)V_{ub}$ & $\Omega_{bb}^{-}\to\Xi_{b}^{\prime0}\eta \ell^{-}\bar{\nu}_{\ell}$  & $\frac{\left(a_{8}-2a_{9}\right)V_{ub}}{2\sqrt{3}}$\tabularnewline
\hline
$\Xi_{bb}^{-}\to\Sigma_{b}^{0}\eta \ell^{-}\bar{\nu}_{\ell}$  & $\frac{\left(a_{8}+a_{9}\right)V_{ub}}{2\sqrt{3}}$ & $\Omega_{bb}^{-}\to\Xi_{b}^{\prime-}\pi^{+}\ell^{-}\bar{\nu}_{\ell}$  & $\frac{a_{8}V_{ub}}{\sqrt{2}}$&&\tabularnewline
\hline
$\Xi_{bb}^{-}\to\Sigma_{b}^{-}\pi^{+}\ell^{-}\bar{\nu}_{\ell}$  & $a_{8}V_{ub}$ & $\Omega_{bb}^{-}\to\Omega_{b}^{-}K^{+}\ell^{-}\bar{\nu}_{\ell}$  & $a_{8}V_{ub}$&&\tabularnewline
\hline
\end{tabular}
\end{table}


\subsection{Semileptonic $\Xi_{bc}$ and $\Omega_{bc}$ decays }

For the charm quark decays in $\Xi_{bc}$ and $\Omega_{bc}$, one can obtain the decay amplitudes from those for $\Xi_{cc}$ and $\Omega_{cc}$ decays with the replacement of $T_{cc}\to T_{bc}$, $T_c\to T_b$ and $D\to B$.
For the bottom quark decay, one can obtain them  from those for $\Xi_{bb}$ and $\Omega_{bb}$ decays with $T_{bb}\to T_{bc}$, $T_b\to T_c$ and $B\to D$. Thus we do not repeat the tedious results here.


\section{Non-Leptonic $\Xi_{cc}$ and $\Omega_{cc}$ decays}
\label{sec:ccq_nonleptonic}

Usually the charm quark decays into light quarks are classified  into three groups: Cabibbo allowed, singly Cabibbo suppressed, and doubly Cabibbo suppressed:
\begin{eqnarray}
 c\to s \bar d u,  \;\;\; c\to u \bar dd/\bar ss, \;\;\; c\to  d \bar s u.
\end{eqnarray}
Under the flavor SU(3) symmetry,
the tree  operators like $\bar sc \bar ud$    transform
 as ${\bf  3}\otimes {\bf\bar 3}\otimes {\bf
3}={\bf  3}\oplus {\bf  3}\oplus {\bf\bar 6}\oplus {\bf {15}}$. So the hadron-level Hamiltonian  can
be decomposed in terms of a vector $(H_3)$, a traceless
tensor antisymmetric in upper indices, $H_{\bf\overline6}$, and a
traceless tensor symmetric in   upper indices,
$H_{\bf {15}}$. As we will show in the following, the   representation $H_3$ will vanishes from the unitarity of CKM matrix.

For the $c\to s  u \bar d$ transition, we have the nonzero matrix element:
\begin{eqnarray}
(H_{\overline 6})^{31}_2=-(H_{\overline 6})^{13}_2=1,\;\;\;
 (H_{15})^{31}_2= (H_{15})^{13}_2=1,\label{eq:H3615_c_allowed}
\end{eqnarray}
while for the doubly Cabibbo suppressed transition  $c\to d  u \bar s$, we have
\begin{eqnarray}
(H_{\overline 6})^{21}_3=-(H_{\overline 6})^{12}_3=\sin^2\theta_C,\;\;
 (H_{15})^{21}_3= (H_{15})^{12}_3=\sin^2\theta_C. \label{eq:H3615_c_doubly_suprressed}
\end{eqnarray}

For  the transition $c\to u \bar dd$, we have
\begin{eqnarray}
 (H_{3})^1=1,\;\;\;(H_{\overline 6})^{21}_2=-(H_{\overline 6})^{12}_2=(H_{\overline 6})^{13}_3=-(H_{\overline 6})^{31}_3=\frac{1}{2},\nonumber\\
 \frac{1}{3}(H_{15})^{21}_2= \frac{1}{3}(H_{15})^{12}_2=-\frac{1}{2}(H_{15})^{11}_1=
 -(H_{15})^{13}_3=-(H_{15})^{31}_3=\frac{1}{4},\label{eq:H3615_cc_d}
\end{eqnarray}
with all other remaining entries zero.  The overall CKM  factor is $V_{cd}^*V_{ud} \simeq -\sin(\theta_C)$. While
for  the transition $c\to u \bar ss$, we have
\begin{eqnarray}
 (H_{3})^1=1,\;\;\;(H_{\overline 6})^{31}_3=-(H_{\overline 6})^{13}_3=(H_{\overline 6})^{12}_2=-(H_{\overline 6})^{21}_2=\frac{1}{2},\nonumber\\
 \frac{1}{3}(H_{15})^{31}_3= \frac{1}{3}(H_{15})^{13}_3=-\frac{1}{2}(H_{15})^{11}_1=
 -(H_{15})^{12}_2=-(H_{15})^{21}_2=\frac{1}{4},\label{eq:H3615_cc_s}
\end{eqnarray}
with all other remaining entries zero.  The overall CKM factor is $V_{cs}^*V_{us} \simeq \sin(\theta_C)$. Since the CKM factors for  $c\to u \bar dd$ and $c\to u \bar ss$ are almost equal in magnitudes, we combine the two transitions. Thus the singly Cabibbo-suppressed channel has the following hadron-level Hamiltonian:
\begin{eqnarray}
(H_{\overline 6})^{31}_3 =-(H_{\overline 6})^{13}_3 =(H_{\overline 6})^{12}_2 =-(H_{\overline 6})^{21}_2 =\sin(\theta_C),\nonumber\\
 (H_{15})^{31}_3= (H_{15})^{13}_3=-(H_{15})^{12}_2=-(H_{15})^{21}_2= \sin(\theta_C).\label{eq:H3615_cc_singly_suppressed}
\end{eqnarray}

\subsection{Decays into a charmed baryon and two light mesons}



\begin{table}
\tiny
\caption{Doubly charmed baryon decays into an anti-triplet $cqq$ and two light mesons.  }\label{tab:ccq_cqq3_qqbarand8meson_1}
\centering
\begin{tabular}{|cc|cc|cccccccc}\hline
channel & amplitude &channel & amplitude \\\hline
\hline
{\tiny{}$\Xi_{cc}^{++}\to\Lambda_{c}^{+}\pi^{+}\overline{K}^{0}$ } & {$b_{2}+b_{4}+2b_{8}-b_{10}+b_{11}$} & {\tiny{}$\Xi_{cc}^{++}\to\Xi_{c}^{+}\pi^{+}\pi^{0}$ } & {\tiny{}$\frac{b_{2}+b_{4}+b_{10}-b_{11}}{\sqrt{2}}$}\tabularnewline
\hline
{\tiny{}$\Xi_{cc}^{++}\to\Xi_{c}^{+}\pi^{+}\eta$ } & {\tiny{}$\frac{-2b_{1}+b_{2}-4b_{3}-b_{4}-2b_{8}+2b_{9}+b_{10}-3b_{11}}{\sqrt{6}}$} & {\tiny{}$\Xi_{cc}^{++}\to\Xi_{c}^{+}K^{+}\overline{K}^{0}$ } & {\tiny{}$-b_{1}+b_{2}-2b_{3}+b_{8}+b_{9}-b_{10}$}\tabularnewline
\hline
{\tiny{}$\Xi_{cc}^{++}\to\Xi_{c}^{0}\pi^{+}\pi^{+}$ } & {\tiny{}$2\left(b_{2}+b_{4}+b_{10}-b_{11}\right)$} & {\tiny{}$\Xi_{cc}^{+}\to\Lambda_{c}^{+}\pi^{+}K^{-}$ } & {\tiny{}$-b_{1}-b_{6}+2b_{7}-b_{8}-b_{9}+b_{12}$}\tabularnewline
\hline
{\tiny{}$\Xi_{cc}^{+}\to\Lambda_{c}^{+}\pi^{0}\overline{K}^{0}$ } & {\tiny{}$\frac{b_{1}-b_{2}-b_{4}+b_{6}-2b_{7}-b_{8}+b_{9}+b_{10}-b_{11}-b_{12}}{\sqrt{2}}$} & {\tiny{}$\Xi_{cc}^{+}\to\Lambda_{c}^{+}\overline{K}^{0}\eta$ } & {\tiny{}$\frac{b_{1}+b_{2}+b_{4}+b_{6}-2b_{7}-b_{8}+b_{9}-b_{10}+b_{11}-b_{12}}{\sqrt{6}}$}\tabularnewline
\hline
{\tiny{}$\Xi_{cc}^{+}\to\Xi_{c}^{+}\pi^{+}\pi^{-}$ } & {\tiny{}$b_{2}-2b_{3}-4b_{5}-b_{6}+b_{10}-b_{12}$} & {\tiny{}$\Xi_{cc}^{+}\to\Xi_{c}^{+}\pi^{0}\pi^{0}$ } & {\tiny{}$-2b_{3}-b_{4}-4b_{5}-b_{6}+b_{11}-b_{12}$}\tabularnewline
\hline
{\tiny{}$\Xi_{cc}^{+}\to\Xi_{c}^{+}\pi^{0}\eta$ } & {\tiny{}$\frac{2b_{3}+b_{4}-b_{6}+2b_{7}+b_{11}-b_{12}}{\sqrt{3}}$} & {\tiny{}$\Xi_{cc}^{+}\to\Xi_{c}^{+}K^{+}K^{-}$ } & {\tiny{}$-2\left(2b_{5}+b_{6}-b_{7}\right)$}\tabularnewline
\hline
{\tiny{}$\Xi_{cc}^{+}\to\Xi_{c}^{+}K^{0}\overline{K}^{0}$ } & {\tiny{}$b_{2}-2b_{3}-4b_{5}-b_{6}-b_{10}+b_{12}$} & {\tiny{}$\Xi_{cc}^{+}\to\Xi_{c}^{+}\eta\eta$ } & {\tiny{}$\frac{1}{3}\left(-2b_{3}-b_{4}-12b_{5}-5b_{6}+4b_{7}-3b_{11}+3b_{12}\right)$}\tabularnewline
\hline
{\tiny{}$\Xi_{cc}^{+}\to\Xi_{c}^{0}\pi^{+}\pi^{0}$ } & {\tiny{}$-\frac{b_{2}+b_{4}+b_{10}-b_{11}}{\sqrt{2}}$} & {\tiny{}$\Xi_{cc}^{+}\to\Xi_{c}^{0}\pi^{+}\eta$ } & {\tiny{}$\frac{-2b_{1}+b_{2}+b_{4}-2b_{6}+4b_{7}-2b_{8}+2b_{9}+b_{10}-b_{11}-2b_{12}}{\sqrt{6}}$}\tabularnewline
\hline
{\tiny{}$\Xi_{cc}^{+}\to\Xi_{c}^{0}K^{+}\overline{K}^{0}$ } & {\tiny{}$-b_{1}-b_{6}+2b_{7}+b_{8}+b_{9}-b_{12}$} & {\tiny{}$\Omega_{cc}^{+}\to\Lambda_{c}^{+}\overline{K}^{0}\overline{K}^{0}$ } & {\tiny{}$2\left(b_{2}+b_{4}-b_{10}+b_{11}\right)$}\tabularnewline
\hline
{\tiny{}$\Omega_{cc}^{+}\to\Xi_{c}^{+}\pi^{+}K^{-}$ } & {\tiny{}$-b_{1}+b_{2}-2b_{3}-b_{8}-b_{9}+b_{10}$} & {\tiny{}$\Omega_{cc}^{+}\to\Xi_{c}^{+}\pi^{0}\overline{K}^{0}$ } & {\tiny{}$\frac{b_{1}+2b_{3}+b_{4}-b_{8}+b_{9}-b_{11}}{\sqrt{2}}$}\tabularnewline
\hline
{\tiny{}$\Omega_{cc}^{+}\to\Xi_{c}^{+}\overline{K}^{0}\eta$ } & {\tiny{}$\frac{b_{1}-2b_{2}+2b_{3}-b_{4}-b_{8}+b_{9}+2b_{10}-3b_{11}}{\sqrt{6}}$} & {\tiny{}$\Omega_{cc}^{+}\to\Xi_{c}^{0}\pi^{+}\overline{K}^{0}$ } & {\tiny{}$b_{2}+b_{4}-2b_{8}+b_{10}-b_{11}$}\tabularnewline
\hline
\hline
{\tiny{}$\Xi_{cc}^{++}\to\Lambda_{c}^{+}\pi^{+}K^{0}$ } & {\tiny{}$-\sin^{2}(\theta_{c})\left(b_{1}-b_{2}+2b_{3}-b_{8}-b_{9}+b_{10}\right)$} & {\tiny{}$\Xi_{cc}^{++}\to\Lambda_{c}^{+}\pi^{0}K^{+}$ } & {\tiny{}$-\frac{\left(b_{1}-b_{2}+2b_{3}+b_{8}-b_{9}-b_{10}+2b_{11}\right)\sin^{2}(\theta_{c})}{\sqrt{2}}$}\tabularnewline
\hline
{\tiny{}$\Xi_{cc}^{++}\to\Lambda_{c}^{+}K^{+}\eta$ } & {\tiny{}$\frac{\left(b_{1}+b_{2}+2b_{3}+2b_{4}+b_{8}-b_{9}+b_{10}\right)\sin^{2}(\theta_{c})}{\sqrt{6}}$} & {\tiny{}$\Xi_{cc}^{++}\to\Xi_{c}^{+}K^{+}K^{0}$ } & {\tiny{}$\left(b_{2}+b_{4}+2b_{8}-b_{10}+b_{11}\right)\sin^{2}(\theta_{c})$}\tabularnewline
\hline
{\tiny{}$\Xi_{cc}^{++}\to\Xi_{c}^{0}K^{+}K^{+}$ } & {\tiny{}$-2\left(b_{2}+b_{4}+b_{10}-b_{11}\right)\sin^{2}(\theta_{c})$} & {\tiny{}$\Xi_{cc}^{+}\to\Lambda_{c}^{+}\pi^{0}K^{0}$ } & {\tiny{}$\frac{\left(b_{1}-b_{2}+2b_{3}-b_{8}+b_{9}+b_{10}-2b_{11}\right)\sin^{2}(\theta_{c})}{\sqrt{2}}$}\tabularnewline
\hline
{\tiny{}$\Xi_{cc}^{+}\to\Lambda_{c}^{+}\pi^{-}K^{+}$ } & {\tiny{}$-\sin^{2}(\theta_{c})\left(b_{1}-b_{2}+2b_{3}+b_{8}+b_{9}-b_{10}\right)$} & {\tiny{}$\Xi_{cc}^{+}\to\Lambda_{c}^{+}K^{0}\eta$ } & {\tiny{}$\frac{\left(b_{1}+b_{2}+2b_{3}+2b_{4}-b_{8}+b_{9}-b_{10}\right)\sin^{2}(\theta_{c})}{\sqrt{6}}$}\tabularnewline
\hline
{\tiny{}$\Xi_{cc}^{+}\to\Xi_{c}^{+}K^{0}K^{0}$ } & {\tiny{}$2\left(b_{2}+b_{4}-b_{10}+b_{11}\right)\sin^{2}(\theta_{c})$} & {\tiny{}$\Xi_{cc}^{+}\to\Xi_{c}^{0}K^{+}K^{0}$ } & {\tiny{}$\left(b_{2}+b_{4}-2b_{8}+b_{10}-b_{11}\right)\left(-\sin^{2}(\theta_{c})\right)$}\tabularnewline
\hline
{\tiny{}$\Omega_{cc}^{+}\to\Lambda_{c}^{+}\pi^{+}\pi^{-}$ } & {\tiny{}$-2\left(2b_{5}+b_{6}-b_{7}\right)\sin^{2}(\theta_{c})$} & {\tiny{}$\Omega_{cc}^{+}\to\Lambda_{c}^{+}\pi^{0}\pi^{0}$ } & {\tiny{}$-2\left(2b_{5}+b_{6}-b_{7}\right)\sin^{2}(\theta_{c})$}\tabularnewline
\hline
{\tiny{}$\Omega_{cc}^{+}\to\Lambda_{c}^{+}\pi^{0}\eta$ } & {\tiny{}$\frac{2\left(b_{11}-b_{12}\right)\sin^{2}(\theta_{c})}{\sqrt{3}}$} & {\tiny{}$\Omega_{cc}^{+}\to\Lambda_{c}^{+}K^{+}K^{-}$ } & {\tiny{}$\left(b_{2}-2b_{3}-4b_{5}-b_{6}+b_{10}-b_{12}\right)\sin^{2}(\theta_{c})$}\tabularnewline
\hline
{\tiny{}$\Omega_{cc}^{+}\to\Lambda_{c}^{+}K^{0}\overline{K}^{0}$ } & {\tiny{}$\left(b_{2}-2b_{3}-4b_{5}-b_{6}-b_{10}+b_{12}\right)\sin^{2}(\theta_{c})$} & {\tiny{}$\Omega_{cc}^{+}\to\Lambda_{c}^{+}\eta\eta$ } & {\tiny{}$-\frac{2}{3}\left(4b_{3}+2b_{4}+6b_{5}+b_{6}+b_{7}\right)\sin^{2}(\theta_{c})$}\tabularnewline
\hline
{\tiny{}$\Omega_{cc}^{+}\to\Xi_{c}^{+}\pi^{0}K^{0}$ } & {\tiny{}$\frac{\left(b_{1}+b_{6}-2b_{7}-b_{8}+b_{9}-b_{12}\right)\sin^{2}(\theta_{c})}{\sqrt{2}}$} & {\tiny{}$\Omega_{cc}^{+}\to\Xi_{c}^{+}\pi^{-}K^{+}$ } & {\tiny{}$\left(b_{1}+b_{6}-2b_{7}+b_{8}+b_{9}-b_{12}\right)\left(-\sin^{2}(\theta_{c})\right)$}\tabularnewline
\hline
{\tiny{}$\Omega_{cc}^{+}\to\Xi_{c}^{+}K^{0}\eta$ } & {\tiny{}$\frac{\left(b_{1}-2b_{2}-2b_{4}+b_{6}-2b_{7}-b_{8}+b_{9}+2b_{10}-2b_{11}-b_{12}\right)\sin^{2}(\theta_{c})}{\sqrt{6}}$} & {\tiny{}$\Omega_{cc}^{+}\to\Xi_{c}^{0}\pi^{+}K^{0}$ } & {\tiny{}$\left(b_{1}+b_{6}-2b_{7}-b_{8}-b_{9}+b_{12}\right)\sin^{2}(\theta_{c})$}\tabularnewline
\hline
{\tiny{}$\Omega_{cc}^{+}\to\Xi_{c}^{0}\pi^{0}K^{+}$ } & {\tiny{}$\frac{\left(b_{1}+b_{6}-2b_{7}+b_{8}-b_{9}+b_{12}\right)\sin^{2}(\theta_{c})}{\sqrt{2}}$} & {\tiny{}$\Omega_{cc}^{+}\to\Xi_{c}^{0}K^{+}\eta$ } & {\tiny{}$-\frac{\left(b_{1}-2b_{2}-2b_{4}+b_{6}-2b_{7}+b_{8}-b_{9}-2b_{10}+2b_{11}+b_{12}\right)\sin^{2}(\theta_{c})}{\sqrt{6}}$}\tabularnewline
\hline
\hline
\end{tabular}
\end{table}

\begin{table}
\caption{Doubly charmed baryon decays into an anti-triplet $cqq$ and two light mesons.  }\label{tab:ccq_cqq3_qqbarand8meson_2}
\tiny
\begin{tabular}{|cc|}
\hline
channel  & amplitude \tabularnewline
\hline
\hline
$\Xi_{cc}^{++}\to\Lambda_{c}^{+}\pi^{+}\pi^{0}$ & $\sqrt{2}\left(b_{8}-b_{10}+b_{11}\right)\sin(\theta_{c})$\tabularnewline
\hline
$\Xi_{cc}^{++}\to\Lambda_{c}^{+}\pi^{+}\eta$ & $\sqrt{\frac{2}{3}}\left(b_{1}-2b_{2}+2b_{3}-b_{4}-2b_{8}-b_{9}+b_{10}\right)\sin(\theta_{c})$\tabularnewline
\hline
$\Xi_{cc}^{++}\to\Lambda_{c}^{+}K^{+}\overline{K}^{0}$ & $\left(b_{1}+2b_{3}+b_{4}+b_{8}-b_{9}+b_{11}\right)\sin(\theta_{c})$\tabularnewline
\hline
$\Xi_{cc}^{++}\to\Xi_{c}^{+}\pi^{+}K^{0}$ & $\left(b_{1}+2b_{3}+b_{4}+b_{8}-b_{9}+b_{11}\right)(-\sin(\theta_{c}))$\tabularnewline
\hline
$\Xi_{cc}^{++}\to\Xi_{c}^{+}\pi^{0}K^{+}$ & $-\frac{\left(b_{1}-2b_{2}+2b_{3}-b_{4}-b_{8}-b_{9}+b_{11}\right)\sin(\theta_{c})}{\sqrt{2}}$\tabularnewline
\hline
$\Xi_{cc}^{++}\to\Xi_{c}^{+}K^{+}\eta$ & $\frac{\left(b_{1}-2b_{2}+2b_{3}-b_{4}-5b_{8}-b_{9}+4b_{10}-3b_{11}\right)\sin(\theta_{c})}{\sqrt{6}}$\tabularnewline
\hline
$\Xi_{cc}^{+}\to\Lambda_{c}^{+}\pi^{+}\pi^{-}$ & $(b_{1}-b_{2}+2b_{3}+4b_{5}+2b_{6}-2b_{7}+b_{8}+b_{9}-b_{10})\sin(\theta_{c})$\tabularnewline
\hline
$\Xi_{cc}^{+}\to\Lambda_{c}^{+}\pi^{0}\pi^{0}$ & $(b_{1}-b_{2}+2b_{3}+4b_{5}+2b_{6}-2b_{7}-b_{8}+b_{9}+b_{10}-2b_{11})\sin(\theta_{c})$\tabularnewline
\hline
$\Xi_{cc}^{+}\to\Lambda_{c}^{+}\pi^{0}\eta$ & $-\frac{\left(b_{1}-2b_{2}+2b_{3}-b_{4}-b_{8}+b_{9}+2b_{10}-b_{11}-2b_{12}\right)\sin(\theta_{c})}{\sqrt{3}}$\tabularnewline
\hline
$\Xi_{cc}^{+}\to\Lambda_{c}^{+}K^{+}K^{-}$ & $\left(b_{1}-4b_{5}-b_{6}+b_{8}+b_{9}-b_{12}\right)(-\sin(\theta_{c}))$\tabularnewline
\hline
$\Xi_{cc}^{+}\to\Lambda_{c}^{+}\eta\eta$ & $\frac{1}{3}(-3b_{1}-3b_{2}+2b_{3}-2b_{4}+12b_{5}+2b_{6}+2b_{7}+3b_{8}-3b_{9}+3b_{10})\sin(\theta_{c})$\tabularnewline
\hline
$\Xi_{cc}^{+}\to\Xi_{c}^{+}\pi^{0}K^{0}$ & $\frac{\left(b_{2}+2b_{3}+2b_{4}-b_{6}+2b_{7}-b_{10}+b_{12}\right)\sin(\theta_{c})}{\sqrt{2}}$\tabularnewline
\hline
$\Xi_{cc}^{+}\to\Xi_{c}^{+}K^{0}\eta$ & $-\frac{\left(3b_{2}-2b_{3}+2b_{4}+b_{6}-2b_{7}-3b_{10}+4b_{11}-b_{12}\right)\sin(\theta_{c})}{\sqrt{6}}$\tabularnewline
\hline
$\Xi_{cc}^{+}\to\Xi_{c}^{0}\pi^{+}K^{0}$ & $(b_{1}-b_{2}-b_{4}+b_{6}-2b_{7}+b_{8}-b_{9}-b_{10}+b_{11}+b_{12})(-\sin(\theta_{c}))$\tabularnewline
\hline
$\Xi_{cc}^{++}\to\Xi_{c}^{0}\pi^{+}K^{+}$ & $2\left(b_{2}+b_{4}+b_{10}-b_{11}\right)\sin(\theta_{c})$\tabularnewline
\hline
$\Omega_{cc}^{+}\to\Xi_{c}^{+}K^{0}\overline{K}^{0}$ & $\left(2b_{3}+b_{4}+4b_{5}+b_{6}+b_{11}-b_{12}\right)(-\sin(\theta_{c}))$\tabularnewline
\hline
$\Omega_{cc}^{+}\to\Lambda_{c}^{+}\pi^{0}\overline{K}^{0}$ & $\frac{\left(b_{2}-2b_{3}+b_{6}-2b_{7}-b_{10}+2b_{11}-b_{12}\right)\sin(\theta_{c})}{\sqrt{2}}$\tabularnewline
\hline
$\Omega_{cc}^{+}\to\Lambda_{c}^{+}\overline{K}^{0}\eta$ & $-\frac{\left(3b_{2}+2b_{3}+4b_{4}-b_{6}+2b_{7}-3b_{10}+2b_{11}+b_{12}\right)\sin(\theta_{c})}{\sqrt{6}}$\tabularnewline
\hline
$\Omega_{cc}^{+}\to\Xi_{c}^{+}\pi^{+}\pi^{-}$ & $\left(b_{1}-4b_{5}-b_{6}+b_{8}+b_{9}-b_{12}\right)\sin(\theta_{c})$\tabularnewline
\hline
$\Omega_{cc}^{+}\to\Xi_{c}^{+}\pi^{0}\pi^{0}$ & $\left(b_{1}-4b_{5}-b_{6}-b_{8}+b_{9}-b_{12}\right)\sin(\theta_{c})$\tabularnewline
\hline
$\Omega_{cc}^{+}\to\Xi_{c}^{+}\pi^{0}\eta$ & $-\frac{\left(b_{1}+b_{2}+b_{4}+b_{6}-2b_{7}-b_{8}+b_{9}-b_{10}-b_{11}+b_{12}\right)\sin(\theta_{c})}{\sqrt{3}}$\tabularnewline
\hline
$\Omega_{cc}^{+}\to\Lambda_{c}^{+}\pi^{+}K^{-}$ & $\left(b_{2}-2b_{3}+b_{6}-2b_{7}+b_{10}-b_{12}\right)(-\sin(\theta_{c}))$\tabularnewline
\hline
$\Omega_{cc}^{+}\to\Xi_{c}^{0}K^{+}\overline{K}^{0}$ & $(b_{1}-b_{2}-b_{4}+b_{6}-2b_{7}+b_{8}-b_{9}-b_{10}+b_{11}+b_{12})(-\sin(\theta_{c}))$\tabularnewline
\hline
$\Omega_{cc}^{+}\to\Xi_{c}^{0}\pi^{+}\eta$ & $-\sqrt{\frac{2}{3}}(b_{1}+b_{2}+b_{4}+b_{6}-2b_{7}-2b_{8}-b_{9}+b_{10}-b_{11}+b_{12})\sin(\theta_{c})$\tabularnewline
\hline
$\Xi_{cc}^{+}\to\Xi_{c}^{0}\pi^{0}K^{+}$ & $-\frac{\left(b_{1}+b_{2}+b_{4}+b_{6}-2b_{7}-b_{8}-b_{9}+b_{10}-b_{11}+b_{12}\right)\sin(\theta_{c})}{\sqrt{2}}$\tabularnewline
\hline
$\Xi_{cc}^{+}\to\Xi_{c}^{0}K^{+}\eta$ & $\frac{\left(b_{1}+b_{2}+b_{4}+b_{6}-2b_{7}-5b_{8}-b_{9}+b_{10}-b_{11}+b_{12}\right)\sin(\theta_{c})}{\sqrt{6}}$\tabularnewline
\hline
$\Omega_{cc}^{+}\to\Xi_{c}^{+}\eta\eta$ & $\frac{1}{3}(-3b_{1}+6b_{2}-8b_{3}+2b_{4}-12b_{5}-5b_{6}+4b_{7}+3b_{8}-3b_{9}-6b_{10}+6b_{11}+3b_{12})\sin(\theta_{c})$\tabularnewline
\hline
$\Xi_{cc}^{+}\to\Lambda_{c}^{+}K^{0}\overline{K}^{0}$ & $\left(2b_{3}+b_{4}+4b_{5}+b_{6}+b_{11}-b_{12}\right)\sin(\theta_{c})$\tabularnewline
\hline
$\Xi_{cc}^{+}\to\Xi_{c}^{+}\pi^{-}K^{+}$ & $\left(b_{2}-2b_{3}+b_{6}-2b_{7}+b_{10}-b_{12}\right)\sin(\theta_{c})$\tabularnewline
\hline
$\Omega_{cc}^{+}\to\Xi_{c}^{+}K^{+}K^{-}$ & $(b_{1}-b_{2}+2b_{3}+4b_{5}+2b_{6}-2b_{7}+b_{8}+b_{9}-b_{10})(-\sin(\theta_{c}))$\tabularnewline
\hline
$\Omega_{cc}^{+}\to\Xi_{c}^{0}\pi^{+}\pi^{0}$ & $-\sqrt{2}b_{8}\sin(\theta_{c})$\tabularnewline
\hline \hline
\end{tabular}
\end{table}


\begin{table}
\begin{minipage}{0.85\linewidth}
\newcommand{\tabincell}[2]{\begin{tabular}{@{}#1@{}}#2\end{tabular}}
\caption{Doubly charmed baryon decays into a sextet $cqq$ and two light mesons. }\label{tab:ccq_cqq6_qqbarand8meson_1}\begin{tabular}{|cc|}
\hline
channel  & amplitude \tabularnewline
\hline
\hline
$\Xi_{cc}^{++}\to\Sigma_{c}^{++}\pi^{+}K^{-}$  & $b_{1}+b_{2}+b_{5}+b_{9}+b_{11}$ \tabularnewline
\hline
$\Xi_{cc}^{++}\to\Sigma_{c}^{++}\pi^{0}\overline{K}^{0}$  & $\frac{b_{1}-b_{2}+b_{3}-b_{9}-b_{10}}{\sqrt{2}}$ \tabularnewline
\hline
$\Xi_{cc}^{++}\to\Sigma_{c}^{++}\overline{K}^{0}\eta$  & $\frac{b_{1}-b_{2}+b_{3}-b_{9}-b_{10}}{\sqrt{6}}$ \tabularnewline
\hline
$\Xi_{cc}^{++}\to\Sigma_{c}^{+}\pi^{+}\overline{K}^{0}$  & $\frac{2b_{1}+b_{3}+b_{5}-b_{10}+b_{11}}{\sqrt{2}}$ \tabularnewline
\hline
$\Xi_{cc}^{++}\to\Xi_{c}^{\prime+}\pi^{+}\pi^{0}$  & $\frac{1}{2}\left(b_{3}+b_{5}+b_{10}-b_{11}\right)$ \tabularnewline
\hline
$\Xi_{cc}^{++}\to\Xi_{c}^{\prime+}\pi^{+}\eta$  & $\frac{-2b_{1}+2b_{2}+b_{3}+4b_{4}-b_{5}-2b_{9}+b_{10}-3b_{11}}{2\sqrt{3}}$ \tabularnewline
\hline
$\Xi_{cc}^{++}\to\Xi_{c}^{\prime+}K^{+}\overline{K}^{0}$  & $\frac{b_{1}+b_{2}+b_{3}+2b_{4}-b_{9}-b_{10}}{\sqrt{2}}$ \tabularnewline
\hline
$\Xi_{cc}^{++}\to\Xi_{c}^{\prime0}\pi^{+}\pi^{+}$  & $\sqrt{2}\left(b_{3}+b_{5}+b_{10}-b_{11}\right)$ \tabularnewline
\hline
$\Xi_{cc}^{++}\to\Omega_{c}^{0}\pi^{+}K^{+}$  & $b_{3}+b_{5}+b_{10}-b_{11}$ \tabularnewline
\hline
$\Xi_{cc}^{+}\to\Sigma_{c}^{++}\pi^{0}K^{-}$  & $-\frac{b_{5}-b_{7}-2b_{8}+b_{11}+b_{12}}{\sqrt{2}}$ \tabularnewline
\hline
$\Xi_{cc}^{+}\to\Sigma_{c}^{++}\pi^{-}\overline{K}^{0}$  & $b_{3}+b_{7}-b_{10}-b_{12}$ \tabularnewline
\hline
$\Xi_{cc}^{+}\to\Sigma_{c}^{++}K^{-}\eta$  & $\frac{b_{5}-b_{7}+2b_{8}+b_{11}+b_{12}}{\sqrt{6}}$ \tabularnewline
\hline
$\Xi_{cc}^{+}\to\Sigma_{c}^{+}\pi^{+}K^{-}$  & $\frac{b_{1}+b_{2}+b_{7}+2b_{8}+b_{9}-b_{12}}{\sqrt{2}}$ \tabularnewline
\hline
$\Xi_{cc}^{+}\to\Sigma_{c}^{+}\pi^{0}\overline{K}^{0}$  & $\frac{1}{2}\left(b_{1}-b_{2}-b_{3}-b_{5}-b_{7}+2b_{8}-b_{9}+b_{10}-b_{11}+b_{12}\right)$ \tabularnewline
\hline
$\Xi_{cc}^{+}\to\Sigma_{c}^{+}\overline{K}^{0}\eta$  & $\frac{b_{1}-b_{2}+b_{3}+b_{5}-b_{7}+2b_{8}-b_{9}-b_{10}+b_{11}+b_{12}}{2\sqrt{3}}$ \tabularnewline
\hline
$\Xi_{cc}^{+}\to\Sigma_{c}^{0}\pi^{+}\overline{K}^{0}$  & $2\left(b_{1}+b_{8}\right)$ \tabularnewline
\hline
$\Xi_{cc}^{+}\to\Xi_{c}^{\prime+}\pi^{+}\pi^{-}$  & $\frac{b_{3}+2b_{4}+4b_{6}+b_{7}+b_{10}+b_{12}}{\sqrt{2}}$ \tabularnewline
\hline
$\Xi_{cc}^{+}\to\Omega_{c}^{0}\pi^{+}K^{0}$  & $b_{3}+b_{7}+b_{10}+b_{12}$\tabularnewline
\hline
$\Xi_{cc}^{+}\to\Omega_{c}^{0}\pi^{0}K^{+}$  & $\frac{-b_{5}+b_{7}+b_{11}+b_{12}}{\sqrt{2}}$\tabularnewline
\hline
$\Omega_{cc}^{+}\to\Sigma_{c}^{++}\overline{K}^{0}K^{-}$  & $b_{3}+b_{5}-b_{10}+b_{11}$\tabularnewline
\hline
$\Omega_{cc}^{+}\to\Sigma_{c}^{+}\overline{K}^{0}\overline{K}^{0}$  & $\sqrt{2}\left(b_{3}+b_{5}-b_{10}+b_{11}\right)$\tabularnewline
\hline
$\Omega_{cc}^{+}\to\Xi_{c}^{\prime+}\pi^{+}K^{-}$  & $\frac{b_{1}+b_{2}+b_{3}+2b_{4}+b_{9}+b_{10}}{\sqrt{2}}$\tabularnewline
\hline
$\Omega_{cc}^{+}\to\Xi_{c}^{\prime+}\pi^{0}\overline{K}^{0}$  & $\frac{1}{2}\left(b_{1}-b_{2}-2b_{4}+b_{5}-b_{9}-b_{11}\right)$\tabularnewline
\hline
$\Omega_{cc}^{+}\to\Xi_{c}^{\prime+}\overline{K}^{0}\eta$  & $\frac{b_{1}-b_{2}-2b_{3}-2b_{4}-b_{5}-b_{9}+2b_{10}-3b_{11}}{2\sqrt{3}}$\tabularnewline
\hline
$\Omega_{cc}^{+}\to\Xi_{c}^{\prime0}\pi^{+}\overline{K}^{0}$  & $\frac{2b_{1}+b_{3}+b_{5}+b_{10}-b_{11}}{\sqrt{2}}$\tabularnewline
\hline
$\Omega_{cc}^{+}\to\Omega_{c}^{0}\pi^{+}\eta$  & $-\sqrt{\frac{2}{3}}\left(b_{1}-b_{2}+b_{3}+b_{9}+b_{10}\right)$\tabularnewline
\hline
$\Omega_{cc}^{+}\to\Omega_{c}^{0}K^{+}\overline{K}^{0}$  & $b_{1}+b_{2}+b_{5}-b_{9}-b_{11}$\tabularnewline
\hline
$\Xi_{cc}^{+}\to\Xi_{c}^{\prime+}\pi^{0}\pi^{0}$  & $\frac{2b_{4}-b_{5}+4b_{6}+b_{7}+b_{11}+b_{12}}{\sqrt{2}}$\tabularnewline
\hline
$\Xi_{cc}^{+}\to\Xi_{c}^{\prime+}\pi^{0}\eta$  & $\frac{-2b_{4}+b_{5}+b_{7}-2b_{8}+b_{11}+b_{12}}{\sqrt{6}}$\tabularnewline
\hline
$\Xi_{cc}^{+}\to\Xi_{c}^{\prime+}K^{+}K^{-}$  & $\sqrt{2}\left(2b_{6}+b_{7}+b_{8}\right)$\tabularnewline
\hline
$\Xi_{cc}^{+}\to\Xi_{c}^{\prime+}K^{0}\overline{K}^{0}$  & $\frac{b_{3}+2b_{4}+4b_{6}+b_{7}-b_{10}-b_{12}}{\sqrt{2}}$\tabularnewline
\hline
$\Xi_{cc}^{+}\to\Xi_{c}^{\prime0}\pi^{+}\pi^{0}$  & $\frac{1}{2}\left(-b_{3}-b_{5}-b_{10}+b_{11}\right)$\tabularnewline
\hline
$\Xi_{cc}^{+}\to\Xi_{c}^{\prime0}\pi^{+}\eta$  & $\frac{-2b_{1}+2b_{2}+b_{3}+b_{5}+2b_{7}-4b_{8}-2b_{9}+b_{10}-b_{11}+2b_{12}}{2\sqrt{3}}$\tabularnewline
\hline
$\Xi_{cc}^{+}\to\Xi_{c}^{\prime0}K^{+}\overline{K}^{0}$  & $\frac{b_{1}+b_{2}+b_{7}+2b_{8}-b_{9}+b_{12}}{\sqrt{2}}$\tabularnewline
\hline
\hline
\end{tabular}
\end{minipage}
\end{table}

\begin{table}
\scriptsize
\begin{minipage}{0.85\linewidth}
\newcommand{\tabincell}[2]{\begin{tabular}{@{}#1@{}}#2\end{tabular}}
\caption{Doubly charmed baryon decays into a sextet $cqq$ and two light mesons. }\label{tab:ccq_cqq6_qqbarand8meson_1sin}\begin{tabular}{|cc|}
\hline
channel  & amplitude \tabularnewline
\hline
\hline
$\Xi_{cc}^{++}\to\Sigma_{c}^{++}\pi^{+}\pi^{-}$  & $\left(b_{1}+b_{2}+b_{5}+b_{9}+b_{11}\right)(-\sin(\theta_{c}))$ \tabularnewline
\hline
$\Xi_{cc}^{++}\to\Sigma_{c}^{++}\pi^{0}\pi^{0}$  & $\left(b_{1}-b_{2}+b_{3}-b_{9}-b_{10}\right)\sin(\theta_{c})$ \tabularnewline
\hline
$\Xi_{cc}^{++}\to\Sigma_{c}^{++}\pi^{0}\eta$  & $-\frac{\left(b_{1}-b_{2}+b_{3}-b_{9}-b_{10}\right)\sin(\theta_{c})}{\sqrt{3}}$ \tabularnewline
\hline
$\Xi_{cc}^{++}\to\Sigma_{c}^{++}K^{+}K^{-}$  & $\left(b_{1}+b_{2}+b_{5}+b_{9}+b_{11}\right)\sin(\theta_{c})$ \tabularnewline
\hline
$\Xi_{cc}^{++}\to\Sigma_{c}^{++}\eta\eta$  & $\left(b_{1}-b_{2}+b_{3}-b_{9}-b_{10}\right)(-\sin(\theta_{c}))$ \tabularnewline
\hline
$\Xi_{cc}^{++}\to\Sigma_{c}^{+}\pi^{+}\pi^{0}$  & $\left(b_{1}-b_{10}+b_{11}\right)\sin(\theta_{c})$ \tabularnewline
\hline
$\Xi_{cc}^{++}\to\Sigma_{c}^{+}\pi^{+}\eta$  & $-\frac{\left(2b_{1}+b_{2}+2b_{3}+2b_{4}+b_{5}-b_{9}-b_{10}\right)\sin(\theta_{c})}{\sqrt{3}}$ \tabularnewline
\hline
$\Xi_{cc}^{++}\to\Sigma_{c}^{+}K^{+}\overline{K}^{0}$  & $\frac{\left(b_{1}-b_{2}-2b_{4}+b_{5}+b_{9}+b_{11}\right)\sin(\theta_{c})}{\sqrt{2}}$ \tabularnewline
\hline
$\Xi_{cc}^{++}\to\Sigma_{c}^{0}\pi^{+}\pi^{+}$  & $-2\left(b_{3}+b_{5}+b_{10}-b_{11}\right)\sin(\theta_{c})$ \tabularnewline
\hline
$\Xi_{cc}^{++}\to\Xi_{c}^{\prime+}\pi^{+}K^{0}$  & $-\frac{\left(b_{1}-b_{2}-2b_{4}+b_{5}+b_{9}+b_{11}\right)\sin(\theta_{c})}{\sqrt{2}}$ \tabularnewline
\hline
$\Xi_{cc}^{++}\to\Xi_{c}^{\prime+}\pi^{0}K^{+}$  & $\frac{1}{2}\left(b_{1}+b_{2}+2b_{3}+2b_{4}+b_{5}-b_{9}-b_{11}\right)\sin(\theta_{c})$ \tabularnewline
\hline
$\Xi_{cc}^{++}\to\Omega_{c}^{0}K^{+}K^{+}$  & $2\left(b_{3}+b_{5}+b_{10}-b_{11}\right)\sin(\theta_{c})$ \tabularnewline
\hline
$\Xi_{cc}^{+}\to\Sigma_{c}^{++}\pi^{0}\pi^{-}$  & $\frac{\left(b_{3}+b_{5}-2b_{8}-b_{10}+b_{11}\right)\sin(\theta_{c})}{\sqrt{2}}$ \tabularnewline
\hline
$\Xi_{cc}^{+}\to\Sigma_{c}^{++}\pi^{-}\eta$  & $-\frac{\left(3b_{3}+b_{5}+2b_{7}+2b_{8}-3b_{10}+b_{11}-2b_{12}\right)\sin(\theta_{c})}{\sqrt{6}}$ \tabularnewline
\hline
$\Xi_{cc}^{+}\to\Sigma_{c}^{++}K^{0}K^{-}$  & $\left(b_{5}-b_{7}+b_{11}+b_{12}\right)\sin(\theta_{c})$ \tabularnewline
\hline
$\Xi_{cc}^{+}\to\Sigma_{c}^{+}\pi^{+}\pi^{-}$  & $-\frac{\left(b_{1}+b_{2}+b_{3}+2b_{4}+4b_{6}+2b_{7}+2b_{8}+b_{9}+b_{10}\right)\sin(\theta_{c})}{\sqrt{2}}$ \tabularnewline
\hline
$\Xi_{cc}^{+}\to\Sigma_{c}^{+}\pi^{0}\pi^{0}$  & $\frac{\left(b_{1}-b_{2}-b_{3}-2b_{4}-4b_{6}-2b_{7}+2b_{8}-b_{9}+b_{10}-2b_{11}\right)\sin(\theta_{c})}{\sqrt{2}}$ \tabularnewline
\hline
$\Xi_{cc}^{+}\to\Sigma_{c}^{+}\pi^{0}\eta$  & $-\frac{\left(b_{1}-b_{2}-2b_{3}-2b_{4}-b_{5}-b_{9}+2b_{10}-b_{11}+2b_{12}\right)\sin(\theta_{c})}{\sqrt{6}}$ \tabularnewline
\hline
$\Xi_{cc}^{+}\to\Sigma_{c}^{+}K^{+}K^{-}$  & $\frac{\left(b_{1}+b_{2}-4b_{6}-b_{7}+b_{9}-b_{12}\right)\sin(\theta_{c})}{\sqrt{2}}$ \tabularnewline
\hline
$\Xi_{cc}^{+}\to\Sigma_{c}^{+}K^{0}\overline{K}^{0}$  & $-\frac{\left(2b_{4}-b_{5}+4b_{6}+b_{7}-b_{11}-b_{12}\right)\sin(\theta_{c})}{\sqrt{2}}$ \tabularnewline
\hline
$\Xi_{cc}^{+}\to\Sigma_{c}^{+}\eta\eta$  & $-\frac{\left(3b_{1}-3b_{2}+3b_{3}+2b_{4}+2b_{5}+12b_{6}+2b_{7}+2b_{8}-3b_{9}-3b_{10}\right)\sin(\theta_{c})}{3\sqrt{2}}$ \tabularnewline
\hline
$\Xi_{cc}^{+}\to\Sigma_{c}^{0}\pi^{+}\pi^{0}$  & $\frac{\left(2b_{1}+b_{3}+b_{5}+2b_{8}+b_{10}-b_{11}\right)\sin(\theta_{c})}{\sqrt{2}}$ \tabularnewline
\hline
$\Xi_{cc}^{+}\to\Sigma_{c}^{0}\pi^{+}\eta$  & $-\frac{\left(4b_{1}+2b_{2}+b_{3}+b_{5}+2b_{7}+2b_{8}-2b_{9}+b_{10}-b_{11}+2b_{12}\right)\sin(\theta_{c})}{\sqrt{6}}$ \tabularnewline
\hline
$\Xi_{cc}^{+}\to\Sigma_{c}^{0}K^{+}\overline{K}^{0}$  & $\left(b_{1}-b_{2}-b_{7}+b_{9}-b_{12}\right)\sin(\theta_{c})$ \tabularnewline
\hline
$\Omega_{cc}^{+}\to\Sigma_{c}^{+}\overline{K}^{0}\eta$  & $-\frac{\left(3b_{3}-2b_{4}+4b_{5}+b_{7}-2b_{8}-3b_{10}+2b_{11}-b_{12}\right)\sin(\theta_{c})}{2\sqrt{3}}$\tabularnewline
\hline
$\Omega_{cc}^{+}\to\Sigma_{c}^{0}\pi^{+}\overline{K}^{0}$  & $\left(b_{3}+b_{5}-2b_{8}+b_{10}-b_{11}\right)(-\sin(\theta_{c}))$\tabularnewline
\hline
$\Omega_{cc}^{+}\to\Xi_{c}^{\prime+}\pi^{+}\pi^{-}$  & $-\frac{\left(b_{1}+b_{2}-4b_{6}-b_{7}+b_{9}-b_{12}\right)\sin(\theta_{c})}{\sqrt{2}}$\tabularnewline
\hline
$\Omega_{cc}^{+}\to\Xi_{c}^{\prime+}\pi^{0}\pi^{0}$  & $\frac{\left(b_{1}-b_{2}+4b_{6}+b_{7}-b_{9}+b_{12}\right)\sin(\theta_{c})}{\sqrt{2}}$\tabularnewline
\hline
$\Omega_{cc}^{+}\to\Xi_{c}^{\prime+}\pi^{0}\eta$  & $-\frac{\left(b_{1}-b_{2}+b_{3}+b_{5}-b_{7}+2b_{8}-b_{9}-b_{10}-b_{11}-b_{12}\right)\sin(\theta_{c})}{\sqrt{6}}$\tabularnewline
\hline
$\Omega_{cc}^{+}\to\Xi_{c}^{\prime+}K^{+}K^{-}$  & $\frac{\left(b_{1}+b_{2}+b_{3}+2b_{4}+4b_{6}+2b_{7}+2b_{8}+b_{9}+b_{10}\right)\sin(\theta_{c})}{\sqrt{2}}$\tabularnewline
\hline
$\Omega_{cc}^{+}\to\Xi_{c}^{\prime+}K^{0}\overline{K}^{0}$  & $\frac{\left(2b_{4}-b_{5}+4b_{6}+b_{7}-b_{11}-b_{12}\right)\sin(\theta_{c})}{\sqrt{2}}$\tabularnewline
\hline
$\Omega_{cc}^{+}\to\Xi_{c}^{\prime+}\eta\eta$  & $-\frac{\left(3b_{1}-3b_{2}-6b_{3}-8b_{4}-2b_{5}-12b_{6}-5b_{7}+4b_{8}-3b_{9}+6b_{10}-6b_{11}+3b_{12}\right)\sin(\theta_{c})}{3\sqrt{2}}$\tabularnewline
\hline
$\Omega_{cc}^{+}\to\Xi_{c}^{\prime0}\pi^{+}\pi^{0}$  & $b_{1}\sin(\theta_{c})$\tabularnewline
\hline
$\Omega_{cc}^{+}\to\Xi_{c}^{\prime0}\pi^{+}\eta$  & $-\frac{\left(2b_{1}+b_{2}-b_{3}+b_{5}-b_{7}+2b_{8}-b_{9}-b_{10}-b_{11}-b_{12}\right)\sin(\theta_{c})}{\sqrt{3}}$\tabularnewline
\hline
$\Omega_{cc}^{+}\to\Xi_{c}^{\prime0}K^{+}\overline{K}^{0}$  & $\frac{\left(b_{1}-b_{2}+b_{3}-b_{5}+b_{7}+2b_{8}+b_{9}-b_{10}+b_{10}+b_{11}+b_{12}\right)\sin(\theta_{c})}{\sqrt{2}}$\tabularnewline
\hline
$\Omega_{cc}^{+}\to\Omega_{c}^{0}\pi^{+}K^{0}$  & $\left(b_{1}-b_{2}-b_{7}+b_{9}-b_{12}\right)(-\sin(\theta_{c}))$\tabularnewline
\hline
$\Omega_{cc}^{+}\to\Omega_{c}^{0}\pi^{0}K^{+}$  & $\frac{\left(b_{1}+b_{2}+b_{7}-b_{9}+b_{12}\right)\sin(\theta_{c})}{\sqrt{2}}$\tabularnewline
\hline
$\Xi_{cc}^{+}\to\Xi_{c}^{\prime+}\pi^{0}K^{0}$  & $\frac{1}{2}\left(b_{3}-2b_{4}+2b_{5}+b_{7}-2b_{8}-b_{10}-b_{12}\right)\sin(\theta_{c})$\tabularnewline
\hline
$\Xi_{cc}^{+}\to\Xi_{c}^{\prime+}\pi^{-}K^{+}$  & $\frac{\left(b_{3}+2b_{4}-b_{7}-2b_{8}+b_{10}+b_{12}\right)\sin(\theta_{c})}{\sqrt{2}}$\tabularnewline
\hline
$\Xi_{cc}^{+}\to\Xi_{c}^{\prime+}K^{0}\eta$  & $-\frac{\left(3b_{3}+2b_{4}+2b_{5}-b_{7}+2b_{8}-3b_{10}+4b_{11}+b_{12}\right)\sin(\theta_{c})}{2\sqrt{3}}$\tabularnewline
\hline
$\Xi_{cc}^{+}\to\Xi_{c}^{\prime0}\pi^{+}K^{0}$  & $-\frac{\left(b_{1}-b_{2}+b_{3}-b_{5}+b_{7}+2b_{8}+b_{9}+b_{10}+b_{11}+b_{12}\right)\sin(\theta_{c})}{\sqrt{2}}$\tabularnewline
\hline
$\Xi_{cc}^{+}\to\Xi_{c}^{\prime0}\pi^{0}K^{+}$  & $\frac{1}{2}\left(b_{1}+b_{2}-b_{3}+b_{5}-b_{7}+2b_{8}-b_{9}-b_{10}-b_{11}-b_{12}\right)\sin(\theta_{c})$\tabularnewline
\hline
$\Xi_{cc}^{+}\to\Xi_{c}^{\prime0}K^{+}\eta$  & $-\frac{\left(5b_{1}+b_{2}-b_{3}+b_{5}-b_{7}+2b_{8}-b_{9}-b_{10}-b_{11}-b_{12}\right)\sin(\theta_{c})}{2\sqrt{3}}$\tabularnewline
\hline
$\Omega_{cc}^{+}\to\Sigma_{c}^{++}\pi^{0}K^{-}$  & $\frac{\left(b_{3}+b_{7}+2b_{8}-b_{10}-b_{12}\right)\sin(\theta_{c})}{\sqrt{2}}$\tabularnewline
\hline
$\Omega_{cc}^{+}\to\Sigma_{c}^{++}\pi^{-}\overline{K}^{0}$  & $\left(b_{5}-b_{7}+b_{11}+b_{12}\right)(-\sin(\theta_{c}))$\tabularnewline
\hline
$\Omega_{cc}^{+}\to\Sigma_{c}^{++}K^{-}\eta$  & $-\frac{\left(3b_{3}+2b_{5}+b_{7}-2b_{8}-3b_{10}+2b_{11}-b_{12}\right)\sin(\theta_{c})}{\sqrt{6}}$\tabularnewline
\hline
$\Omega_{cc}^{+}\to\Sigma_{c}^{+}\pi^{+}K^{-}$  & $-\frac{\left(b_{3}+2b_{4}-b_{7}-2b_{8}+b_{10}+b_{12}\right)\sin(\theta_{c})}{\sqrt{2}}$\tabularnewline
\hline
$\Omega_{cc}^{+}\to\Sigma_{c}^{+}\pi^{0}\overline{K}^{0}$  & $\frac{1}{2}\left(b_{3}+2b_{4}-b_{7}+2b_{8}-b_{10}+2b_{11}+b_{12}\right)\sin(\theta_{c})$\tabularnewline
\hline
\hline
 \end{tabular}
\end{minipage}
\end{table}

\begin{table}
\scriptsize
\begin{minipage}{0.85\linewidth}
\newcommand{\tabincell}[2]{\begin{tabular}{@{}#1@{}}#2\end{tabular}}
\caption{Doubly charmed baryon decays into a sextet $cqq$ and two light mesons. }\label{tab:ccq_cqq6_qqbarand8meson_1sin2}\begin{tabular}{|cc|}
\hline
channel  & amplitude \tabularnewline
\hline
\hline
$\Xi_{cc}^{++}\to\Sigma_{c}^{++}\pi^{0}K^{0}$  & $\frac{\left(b_{1}-b_{2}+b_{3}-b_{9}-b_{10}\right)\sin^{2}(\theta_{c})}{\sqrt{2}}$ \tabularnewline
\hline
$\Xi_{cc}^{++}\to\Sigma_{c}^{++}\pi^{-}K^{+}$  & $\left(b_{1}+b_{2}+b_{5}+b_{9}+b_{11}\right)\sin^{2}(\theta_{c})$ \tabularnewline
\hline
$\Xi_{cc}^{++}\to\Sigma_{c}^{++}K^{0}\eta$  & $\frac{\left(b_{1}-b_{2}+b_{3}-b_{9}-b_{10}\right)\sin^{2}(\theta_{c})}{\sqrt{6}}$ \tabularnewline
\hline
$\Xi_{cc}^{++}\to\Sigma_{c}^{+}\pi^{+}K^{0}$  & $\frac{\left(b_{1}+b_{2}+b_{3}+2b_{4}-b_{9}-b_{10}\right)\sin^{2}(\theta_{c})}{\sqrt{2}}$ \tabularnewline
\hline
$\Xi_{cc}^{++}\to\Sigma_{c}^{+}\pi^{0}K^{+}$  & $-\frac{1}{2}\left(b_{1}-b_{2}-b_{3}-2b_{4}+b_{9}-b_{10}+2b_{11}\right)\sin^{2}(\theta_{c})$ \tabularnewline
\hline
$\Xi_{cc}^{++}\to\Sigma_{c}^{+}K^{+}\eta$  & $\frac{\left(b_{1}-b_{2}+b_{3}-2b_{4}+2b_{5}+b_{9}+b_{10}\right)\sin^{2}(\theta_{c})}{2\sqrt{3}}$ \tabularnewline
\hline
$\Xi_{cc}^{++}\to\Sigma_{c}^{0}\pi^{+}K^{+}$  & $\left(b_{3}+b_{5}+b_{10}-b_{11}\right)\sin^{2}(\theta_{c})$ \tabularnewline
\hline
$\Xi_{cc}^{++}\to\Xi_{c}^{\prime+}K^{+}K^{0}$  & $\frac{\left(2b_{1}+b_{3}+b_{5}-b_{10}+b_{11}\right)\sin^{2}(\theta_{c})}{\sqrt{2}}$ \tabularnewline
\hline
$\Xi_{cc}^{++}\to\Xi_{c}^{\prime0}K^{+}K^{+}$  & $\sqrt{2}\left(b_{3}+b_{5}+b_{10}-b_{11}\right)\sin^{2}(\theta_{c})$ \tabularnewline
\hline
$\Xi_{cc}^{+}\to\Sigma_{c}^{++}\pi^{-}K^{0}$  & $\left(b_{3}+b_{5}-b_{10}+b_{11}\right)\sin^{2}(\theta_{c})$ \tabularnewline
\hline
$\Xi_{cc}^{+}\to\Sigma_{c}^{+}\pi^{0}K^{0}$  & $\frac{1}{2}\left(b_{1}-b_{2}-b_{3}-2b_{4}-b_{9}+b_{10}-2b_{11}\right)\sin^{2}(\theta_{c})$ \tabularnewline
\hline
$\Xi_{cc}^{+}\to\Sigma_{c}^{+}\pi^{-}K^{+}$  & $\frac{\left(b_{1}+b_{2}+b_{3}+2b_{4}+b_{9}+b_{10}\right)\sin^{2}(\theta_{c})}{\sqrt{2}}$ \tabularnewline
\hline
$\Xi_{cc}^{+}\to\Sigma_{c}^{+}K^{0}\eta$  & $\frac{\left(b_{1}-b_{2}+b_{3}-2b_{4}+2b_{5}-b_{9}-b_{10}\right)\sin^{2}(\theta_{c})}{2\sqrt{3}}$ \tabularnewline
\hline
$\Xi_{cc}^{+}\to\Sigma_{c}^{0}\pi^{+}K^{0}$  & $\left(b_{1}+b_{2}+b_{5}-b_{9}-b_{11}\right)\sin^{2}(\theta_{c})$ \tabularnewline
\hline
$\Xi_{cc}^{+}\to\Sigma_{c}^{0}\pi^{0}K^{+}$  & $-\frac{\left(b_{1}-b_{2}+b_{3}+b_{9}+b_{10}\right)\sin^{2}(\theta_{c})}{\sqrt{2}}$ \tabularnewline
\hline
$\Xi_{cc}^{+}\to\Sigma_{c}^{0}K^{+}\eta$  & $\frac{\left(b_{1}-b_{2}+b_{3}+b_{9}+b_{10}\right)\sin^{2}(\theta_{c})}{\sqrt{6}}$ \tabularnewline
\hline
$\Xi_{cc}^{+}\to\Xi_{c}^{\prime+}K^{0}K^{0}$  & $\sqrt{2}\left(b_{3}+b_{5}-b_{10}+b_{11}\right)\sin^{2}(\theta_{c})$ \tabularnewline
\hline
$\Xi_{cc}^{+}\to\Xi_{c}^{\prime0}K^{+}K^{0}$  & $\frac{\left(2b_{1}+b_{3}+b_{5}+b_{10}-b_{11}\right)\sin^{2}(\theta_{c})}{\sqrt{2}}$ \tabularnewline
\hline
$\Omega_{cc}^{+}\to\Sigma_{c}^{+}\eta\eta$  & $\frac{1}{3}\sqrt{2}\left(4b_{4}-2b_{5}+6b_{6}+b_{7}+b_{8}\right)\sin^{2}(\theta_{c})$\tabularnewline
\hline
$\Omega_{cc}^{+}\to\Sigma_{c}^{0}\pi^{+}\pi^{0}$  & $-\sqrt{2}b_{8}\sin^{2}(\theta_{c})$\tabularnewline
\hline
$\Omega_{cc}^{+}\to\Sigma_{c}^{0}\pi^{+}\eta$  & $-\sqrt{\frac{2}{3}}\left(b_{5}-b_{7}-b_{8}-b_{11}-b_{12}\right)\sin^{2}(\theta_{c})$\tabularnewline
\hline
$\Omega_{cc}^{+}\to\Sigma_{c}^{0}K^{+}\overline{K}^{0}$  & $\left(b_{3}+b_{7}+b_{10}+b_{12}\right)\sin^{2}(\theta_{c})$\tabularnewline
\hline
$\Omega_{cc}^{+}\to\Xi_{c}^{\prime+}\pi^{0}K^{0}$  & $\frac{1}{2}\left(b_{1}-b_{2}-b_{7}+2b_{8}-b_{9}+b_{12}\right)\sin^{2}(\theta_{c})$\tabularnewline
\hline
$\Omega_{cc}^{+}\to\Xi_{c}^{\prime+}\pi^{-}K^{+}$  & $\frac{\left(b_{1}+b_{2}+b_{7}+2b_{8}+b_{9}-b_{12}\right)\sin^{2}(\theta_{c})}{\sqrt{2}}$\tabularnewline
\hline
$\Omega_{cc}^{+}\to\Xi_{c}^{\prime+}K^{0}\eta$  & $\frac{\left(b_{1}-b_{2}-2b_{3}-2b_{5}-b_{7}+2b_{8}-b_{9}+2b_{10}-2b_{11}+b_{12}\right)\sin^{2}(\theta_{c})}{2\sqrt{3}}$\tabularnewline
\hline
$\Omega_{cc}^{+}\to\Xi_{c}^{\prime0}\pi^{+}K^{0}$  & $\frac{\left(b_{1}+b_{2}+b_{7}+2b_{8}-b_{9}+b_{12}\right)\sin^{2}(\theta_{c})}{\sqrt{2}}$\tabularnewline
\hline
$\Omega_{cc}^{+}\to\Xi_{c}^{\prime0}\pi^{0}K^{+}$  & $-\frac{1}{2}\left(b_{1}-b_{2}-b_{7}+2b_{8}+b_{9}-b_{12}\right)\sin^{2}(\theta_{c})$\tabularnewline
\hline
$\Omega_{cc}^{+}\to\Xi_{c}^{\prime0}K^{+}\eta$  & $\frac{\left(b_{1}-b_{2}-2b_{3}-2b_{5}-b_{7}+2b_{8}+b_{9}-2b_{10}+2b_{11}-b_{12}\right)\sin^{2}(\theta_{c})}{2\sqrt{3}}$\tabularnewline
\hline
$\Omega_{cc}^{+}\to\Omega_{c}^{0}K^{+}K^{0}$  & $2\left(b_{1}+b_{8}\right)\sin^{2}(\theta_{c})$\tabularnewline
\hline
$\Omega_{cc}^{+}\to\Sigma_{c}^{++}\pi^{0}\pi^{-}$  & $\sqrt{2}b_{8}\sin^{2}(\theta_{c})$\tabularnewline
\hline
$\Omega_{cc}^{+}\to\Sigma_{c}^{++}\pi^{-}\eta$  & $-\sqrt{\frac{2}{3}}\left(b_{5}-b_{7}-b_{8}+b_{11}+b_{12}\right)\sin^{2}(\theta_{c})$\tabularnewline
\hline
$\Omega_{cc}^{+}\to\Sigma_{c}^{++}K^{0}K^{-}$  & $\left(b_{3}+b_{7}-b_{10}-b_{12}\right)\sin^{2}(\theta_{c})$\tabularnewline
\hline
$\Omega_{cc}^{+}\to\Sigma_{c}^{+}\pi^{+}\pi^{-}$  & $\sqrt{2}\left(2b_{6}+b_{7}+b_{8}\right)\sin^{2}(\theta_{c})$\tabularnewline
\hline
$\Omega_{cc}^{+}\to\Sigma_{c}^{+}\pi^{0}\pi^{0}$  & $\sqrt{2}\left(2b_{6}+b_{7}-b_{8}\right)\sin^{2}(\theta_{c})$\tabularnewline
\hline
$\Omega_{cc}^{+}\to\Sigma_{c}^{+}\pi^{0}\eta$  & $\sqrt{\frac{2}{3}}\left(b_{11}+b_{12}\right)\sin^{2}(\theta_{c})$\tabularnewline
\hline
$\Omega_{cc}^{+}\to\Sigma_{c}^{+}K^{+}K^{-}$  & $\frac{\left(b_{3}+2b_{4}+4b_{6}+b_{7}+b_{10}+b_{12}\right)\sin^{2}(\theta_{c})}{\sqrt{2}}$\tabularnewline
\hline
\hline
\end{tabular}
\end{minipage}
\end{table}

With the above expressions, one may derive the effective Hamiltonian for decays involving the anti-triplet heavy baryons  as
\begin{eqnarray}
 {\cal H}_{eff}&=&  b_1(T_{cc})^i  (\overline T_{c\bar 3})_{[ij]}M^{j}_{k}M^{m}_{l}  (H_{\overline6})^{kl}_{m}+ b_2(T_{cc})^i  (\overline T_{c\bar 3})_{[jk]}M^{j}_{i} M^{m}_{l}  (H_{\overline6})^{kl}_{m}\nonumber\\
 &&+  b_3(T_{cc})^i  (\overline T_{c\bar 3})_{[lm]}M^{j}_{i} M^{k}_{j}  (H_{\overline6})^{lm}_{k} +  b_4(T_{cc})^i  (\overline T_{c\bar 3})_{[lm]}M^{j}_{i} M^{m}_{k}  (H_{\overline6})^{kl}_{j}\nonumber\\
 &&+ b_5(T_{cc})^i  (\overline T_{c\bar 3})_{[jk]}M^{l}_{m} M^{m}_{l}  (H_{\overline6})^{jk}_{i} + b_6(T_{cc})^i  (\overline T_{c\bar 3})_{[jl]}M^{l}_{m} M^{m}_{k}  (H_{\overline6})^{jk}_{i}\nonumber\\
 &&+ b_7(T_{cc})^i  (\overline T_{c\bar 3})_{[lm]}M^{l}_{j} M^{m}_{k}  (H_{\overline6})^{jk}_{i} + b_8(T_{cc})^i  (\overline T_{c\bar 3})_{[ij]}M^{j}_{k} M^{l}_{m}  (H_{15})^{km}_{l}\nonumber\\
 &&+ b_{9}(T_{cc})^i  (\overline T_{c\bar 3})_{[ij]}M^{l}_{m} M^{m}_{k}  (H_{15})^{jk}_{l} + b_{10}(T_{cc})^i  (\overline T_{c\bar 3})_{[jk]}M^{j}_{i} M^{m}_{l}  (H_{15})^{kl}_{m}\nonumber\\
 &&+ b_{11}(T_{cc})^i  (\overline T_{c\bar 3})_{[lm]}M^{j}_{i} M^{m}_{k}  (H_{15})^{kl}_{j} + b_{12}(T_{cc})^i  (\overline T_{c\bar 3})_{[jl]}M^{l}_{m} M^{m}_{k}  (H_{15})^{jk}_{i}\nonumber\\
 &&+\overline b_1(T_{cc})^i  (\overline T_{c\bar 3})_{[ij]}M^{l}_{m}M^{m}_{k}  (H_{\overline6})^{jk}_{l}.
\end{eqnarray}
For the sextet baryon, we have the Hamiltonian
\begin{eqnarray}
 {\cal H}_{eff}&=&  b_1(T_{cc})^i  (\overline T_{c6})_{[ij]}M^{j}_{k}M^{m}_{l}  (H_{15})^{kl}_{m}+ b_2(T_{cc})^i  (\overline T_{c6})_{[ij]}M^{l}_{m} M^{m}_{k}  (H_{15})^{jk}_{l}\nonumber\\
 &&+ b_3(T_{cc})^i  (\overline T_{c6})_{[jk]}M^{j}_{i} M^{m}_{l}  (H_{15})^{kl}_{m} +  b_4(T_{cc})^i  (\overline T_{c6})_{[lm]}M^{j}_{i} M^{k}_{j}  (H_{15})^{lm}_{k} \nonumber\\
 &&+  b_5(T_{cc})^i  (\overline T_{c6})_{[lm]}M^{j}_{i} M^{m}_{k}  (H_{15})^{kl}_{j}+ b_6(T_{cc})^i  (\overline T_{c6})_{[jk]}M^{l}_{m} M^{m}_{l}  (H_{15})^{jk}_{i}\nonumber\\
 &&+ b_7(T_{cc})^i  (\overline T_{c6})_{[jl]}M^{l}_{m} M^{m}_{k}  (H_{15})^{jk}_{i}+ b_8(T_{cc})^i  (\overline T_{c6})_{[lm]}M^{l}_{j} M^{m}_{k}  (H_{15})^{jk}_{i}\nonumber\\
 &&+b_9(T_{cc})^i  (\overline T_{c6})_{[ij]}M^{j}_{k}M^{l}_{m}  (H_{\overline6})^{km}_{l}+ b_{10}(T_{cc})^i  (\overline T_{c6})_{[jk]}M^{j}_{i} M^{m}_{l}  (H_{\overline6})^{kl}_{m}\nonumber\\
 &&+ b_{11}(T_{cc})^i  (\overline T_{c6})_{[lm]}M^{j}_{i} M^{m}_{k}  (H_{\overline6})^{kl}_{j}+ b_{12}(T_{cc})^i  (\overline T_{c6})_{[jl]}M^{l}_{m} M^{m}_{k}  (H_{\overline6})^{jk}_{i}\nonumber\\
 &&+\overline b_9(T_{cc})^i  (\overline T_{c6})_{[ij]}M^{l}_{m}M^{m}_{k}  (H_{\overline6})^{jk}_{l}.
\end{eqnarray}
We have checked that the $\overline b_{1}$ and $\overline b_{9}$ terms  give the same contribution as the $b_1$ and $b_9$, and the corresponding amplitudes always contain the factor $b_1-\overline b_{1}$ for anti-triplet and $b_9-\overline b_{9}$ for sextet. So we can remove $\overline b_{1}$ and $\overline b_{9}$ term in the expanded amplitude.

It should be mentioned that the dynamical mechanisms of these terms are not all the same. For the production of final two light mesons, some terms contain one QCD coupling while the others contain two QCD couplings.
Expanding the above equations, we will obtain the decay amplitudes given in Tab.~\ref{tab:ccq_cqq3_qqbarand8meson_1}, Tab.~\ref{tab:ccq_cqq3_qqbarand8meson_2} for the anti-triplet baryon and Tab.~\ref{tab:ccq_cqq6_qqbarand8meson_1}, Tab.~\ref{tab:ccq_cqq6_qqbarand8meson_1sin}, Tab.~\ref{tab:ccq_cqq6_qqbarand8meson_1sin2} for the sextet. Based on the expanded amplitudes,  we derive the   relations for decay widths collected in Appendix~\ref{sec:relation_cc_cqq3_2m_bar3}.


\subsection{Decays into a light   baryon, a charmed meson and a light meson}
\begin{table}
\scriptsize
\newcommand{\tabincell}[2]{\begin{tabular}{@{}#1@{}}#2\end{tabular}}
\caption{Doubly charmed baryon decays into a light baryon in the octet, a charmed meson and a light meson.}\label{tab:ccq_qqq8_cqbarand8meson_1}\begin{tabular}{|c|c|c|c|c|c|c|c}\hline
channel & amplitude  \\\hline
\hline
$\Xi_{cc}^{++}\to \Lambda^0  D^+  \pi^+  $ & \tabincell{c}{$ -\frac{1}{\sqrt{6}}(2 c_1+c_2+2 c_4+c_5-c_6-c_8+c_9-2 c_{10}-2 c_{11}+c_{12}$\\$+2 c_{18}+c_{19}+2 c_{21}+c_{22}-c_{23}-c_{24}-3 c_{25}-3 c_{26})$}\\\hline
$\Xi_{cc}^{++}\to \Sigma^+  D^0  \pi^+  $ & $ -c_2+c_3-c_6+2 c_7-c_9+2 c_{10}-c_{19}-c_{20}-c_{23}-c_{25}$\\\hline
$\Xi_{cc}^{++}\to \Sigma^+  D^+  \pi^0  $ & $ -\frac{1}{\sqrt{2}}(c_3+c_5+2 c_7+c_8-2 c_{11}+c_{12}-c_{20}+c_{22}+c_{24}+c_{26})$\\\hline
$\Xi_{cc}^{++}\to \Sigma^+  D^+  \eta  $ & $ \frac{1}{\sqrt{6}}(c_3-2 c_4-c_5+2 c_7+c_8+2 c_{11}-c_{12}-c_{20}+2 c_{21}-c_{22}-3 c_{24}-c_{26})$\\\hline
$\Xi_{cc}^{++}\to \Sigma^0  D^+  \pi^+  $ & $ \frac{1}{\sqrt{2}}(c_2+c_5+c_6+c_8+c_9-2 c_{10}-2 c_{11}+c_{12}+c_{19}+c_{22}+c_{23}+c_{24}+c_{25}+c_{26})$\\\hline
$\Xi_{cc}^{++}\to {p}  D^+  \overline K^0  $ & $ -c_1-c_4+c_6+c_8+c_{18}+c_{21}-c_{23}-c_{24}$\\\hline
$\Xi_{cc}^{++}\to \Xi^0  D^+_s  \pi^+  $ & $ -c_1-c_2-c_9+2 c_{10}-c_{18}-c_{19}+c_{25}$\\\hline
$\Xi_{cc}^{+}\to \Lambda^0  D^0  \pi^+  $ & \tabincell{c}{$ \frac{1}{\sqrt{6}}(2 c_1+c_2+c_3-c_6+2 c_7-2 c_{13}-c_{14}+c_{15}+2 c_{16}+c_{17}$\\$2 c_{18}+c_{19}+3 c_{20}-c_{23}-2 c_{27}-c_{28}+c_{29}-3 c_{30})$}\\\hline
$\Xi_{cc}^{+}\to \Lambda^0  D^+  \pi^0  $ & \tabincell{c}{$ -\frac{1}{2 \sqrt{3}}(c_3-2 c_4-c_5+2 c_7+c_8-c_9+2 c_{10}+2 c_{11}-c_{12}-2 c_{13}-c_{14}+c_{15}+2 c_{16}$\\$c_{17}+3 c_{20}-2 c_{21}-c_{22}+c_{24}+3 c_{25}+3 c_{26}-2 c_{27}-c_{28}+c_{29}-3 c_{30})$}\\\hline
$\Xi_{cc}^{+}\to \Lambda^0  D^+  \eta  $ & \tabincell{c}{$ \frac{1}{6} (c_3+4 c_4+5 c_5+2 c_7+c_8-c_9+2 c_{10}+2 c_{11}-c_{12}+4 c_{13}+5 c_{14}+c_{15}$\\$+2 c_{16}+c_{17}+3 c_{20}-3 c_{22}-3 c_{24}+3 c_{25}+3 c_{26}-3 c_{28}-3 c_{29}-3 c_{30})$}\\\hline
$\Xi_{cc}^{+}\to \Lambda^0  D^+_s  \overline K^0  $ & \tabincell{c}{$ \frac{1}{\sqrt{6}}(-c_1-2 c_2+c_3-c_6+2 c_7+c_{13}-c_{14}-2 c_{15}2 c_{16}$\\$c_{17}+c_{18}+2 c_{19}+3 c_{20}+c_{23}-c_{27}+c_{28}+2 c_{29}-3 c_{30})$}\\\hline
$\Xi_{cc}^{+}\to \Sigma^+  D^0  \pi^0  $ & $ \frac{1}{\sqrt{2}}(c_9-2 c_{10}-c_{14}-c_{15}+2 c_{16}+c_{17}+c_{25}-c_{28}-c_{29}+c_{30})$\\\hline
$\Xi_{cc}^{+}\to \Sigma^+  D^0  \eta  $ & $ -\frac{1}{\sqrt{6}}(c_9-2 c_{10}+2 c_{13}+c_{14}-c_{15}-2 c_{16}-c_{17}+c_{25}-2 c_{27}+c_{28}+3 c_{29}-c_{30})$\\\hline
$\Xi_{cc}^{+}\to \Sigma^+  D^+  \pi^-  $ & $ 2 c_{11}-c_{12}+2 c_{16}+c_{17}-c_{26}+c_{30}$\\\hline
$\Xi_{cc}^{+}\to \Sigma^0  D^0  \pi^+  $ & $ \frac{1}{\sqrt{2}}(c_2-c_3+c_6-2 c_7+c_{14}+c_{15}-2 c_{16}-c_{17}+c_{19}+c_{20}+c_{23}+c_{28}+c_{29}-c_{30})$\\\hline
$\Xi_{cc}^{+}\to \Sigma^0  D^+  \pi^0  $ & \tabincell{c}{$ \frac{1}{2} (c_3+c_5+2 c_7+c_8-c_9+2 c_{10}+2 c_{11}-c_{12}+c_{14}+c_{15}+2 c_{16}$\\$+c_{17}-c_{20}+c_{22}+c_{24}-c_{25}-c_{26}+c_{28}+c_{29}+c_{30})$}\\\hline
$\Xi_{cc}^{+}\to \Sigma^0  D^+  \eta  $ & \tabincell{c}{$ \frac{1}{2 \sqrt{3}}(-c_3+2 c_4+c_5-2 c_7-c_8+c_9-2 c_{10}-2 c_{11}+c_{12}+2 c_{13}+c_{14}-c_{15}$\\$-2 c_{16}-c_{17}+c_{20}-2 c_{21}+c_{22}+3 c_{24}+c_{25}+c_{26}-2 c_{27}+c_{28}+3 c_{29}-c_{30})$}\\\hline
$\Xi_{cc}^{+}\to \Sigma^-  D^+  \pi^+  $ & $ c_2+c_5+c_6+c_8+c_{14}+c_{15}+c_{19}+c_{22}+c_{23}+c_{24}+c_{28}+c_{29}$\\\hline
$\Xi_{cc}^{+}\to {p}  D^0  \overline K^0  $ & $ c_1+c_2-c_{13}+c_{15}-c_{18}-c_{19}+c_{27}-c_{29}$\\\hline
$\Xi_{cc}^{+}\to {p}  D^+  K^-  $ & $ c_4+c_5+c_{13}+c_{14}-c_{21}-c_{22}-c_{27}-c_{28}$\\\hline
$\Xi_{cc}^{+}\to \Xi^-  D^+_s  \pi^+  $ & $ c_1+c_2-c_{13}+c_{15}+c_{18}+c_{19}-c_{27}+c_{29}$\\\hline
$\Xi_{cc}^{+}\to \Xi^0  D^+  K^0  $ & $ 2 c_{11}-c_{12}+2 c_{16}+c_{17}+c_{26}-c_{30}$\\\hline
$\Xi_{cc}^{+}\to \Xi^0  D^+_s  \pi^0  $ & $ \frac{1}{\sqrt{2}}(c_9-2 c_{10}+c_{13}-c_{15}-c_{25}+c_{27}-c_{29})$\\\hline
$\Xi_{cc}^{+}\to \Xi^0  D^+_s  \eta  $ & $ \frac{1}{\sqrt{6}}(-c_9+2 c_{10}+c_{13}+2 c_{14}+c_{15}-4 c_{16}-2 c_{17}+c_{25}+c_{27}-2 c_{28}-3 c_{29}+2 c_{30})$\\\hline
$\Omega_{cc}^{+}\to \Sigma^+  D^0  \overline K^0  $ & $ -c_1-c_2-c_9+2 c_{10}+c_{18}+c_{19}-c_{25}$\\\hline
$\Omega_{cc}^{+}\to \Sigma^+  D^+  K^-  $ & $ -c_4-c_5+2 c_{11}-c_{12}+c_{21}+c_{22}-c_{26}$\\\hline
$\Omega_{cc}^{+}\to \Sigma^0  D^+  \overline K^0  $ & $ \frac{1}{\sqrt{2}}(c_1+c_2+c_4+c_5+c_9-2 c_{10}-2 c_{11}+c_{12}-c_{18}-c_{19}-c_{21}-c_{22}+c_{25}+c_{26})$\\\hline
$\Omega_{cc}^{+}\to \Xi^-  D^+  \pi^+  $ & $ -c_1-c_4+c_6+c_8-c_{18}-c_{21}+c_{23}+c_{24}$\\\hline
$\Omega_{cc}^{+}\to \Xi^0  D^0  \pi^+  $ & $ c_1+c_3-c_6+2 c_7+c_{18}+c_{20}-c_{23}$\\\hline
$\Omega_{cc}^{+}\to \Xi^0  D^+  \pi^0  $ & $ -\frac{1}{\sqrt{2}}c_3-c_4+2 c_7+c_8+c_{20}-c_{21}+c_{24}$\\\hline
\hline
\end{tabular}
\end{table}

\begin{table}
\tiny
\newcommand{\tabincell}[2]{\begin{tabular}{@{}#1@{}}#2\end{tabular}}
\caption{Doubly charmed baryon decays into a light baryon in the octet, a charmed meson and a light meson.}\label{tab:ccq_qqq8_cqbarand8meson_2}\begin{tabular}{|c|c|c|c|c|c|c|c}\hline
channel & amplitude \\\hline
\hline$\Xi_{cc}^{++}\to \Lambda^0  D^+  K^+  $ & \tabincell{c}{$ -\frac{1}{\sqrt{6}}(2 c_1+c_2-c_4-2 c_5-c_6-c_8+c_9-2 c_{10}+4 c_{11} $\\$-2 c_{12}+2 c_{18}+c_{19}-c_{21}-2 c_{22}-c_{23}-c_{24}-3 c_{25}) \sin(\theta_{c})$}\\\hline
$\Xi_{cc}^{++}\to \Lambda^0  D^+_s  \pi^+  $ & \tabincell{c}{$ \frac{1}{\sqrt{6}}(c_1+2 c_2-2 c_4-c_5+c_6+c_8+2 c_9-4 c_{10}+2 c_{11}$\\$-c_{12}+c_{18}+2 c_{19}-2 c_{21}-c_{22}+c_{23}+c_{24}+3 c_{26}) \sin(\theta_{c})$}\\\hline
$\Xi_{cc}^{++}\to \Sigma^+  D^0  K^+  $ & $ \left(c_2-c_3+c_6-2 c_7+c_9-2 c_{10}+c_{19}+c_{20}+c_{23}+c_{25}\right) (-\sin(\theta_{c}))$\\\hline
$\Xi_{cc}^{++}\to \Sigma^+  D^+  K^0  $ & $ \left(c_3-c_4+2 c_7+c_8-c_{20}+c_{21}-c_{24}\right) \sin(\theta_{c})$\\\hline
$\Xi_{cc}^{++}\to \Sigma^+  D^+_s  \pi^0  $ & $ \frac{1}{\sqrt{2}}\left(c_1-c_5-c_6-c_8+2 c_{11}-c_{12}-c_{18}-c_{22}+c_{23}-c_{24}-c_{26}\right) \sin(\theta_{c})$\\\hline
$\Xi_{cc}^{++}\to \Sigma^0  D^+  K^+  $ & $ \frac{1}{\sqrt{2}}\left(c_2-c_4+c_6+c_8+c_9-2 c_{10}+c_{19}-c_{21}+c_{23}+c_{24}+c_{25}+2 c_{26}\right) \sin(\theta_{c})$\\\hline
$\Xi_{cc}^{++}\to \Sigma^0  D^+_s  \pi^+  $ & $ \frac{1}{\sqrt{2}}\left(-c_1+c_5+c_6+c_8-2 c_{11}+c_{12}-c_{18}+c_{22}+c_{23}+c_{24}+2 c_{25}+c_{26}\right) \sin(\theta_{c})$\\\hline
$\Xi_{cc}^{++}\to {p}  D^0  \pi^+  $ & $ \left(c_2-c_3+c_6-2 c_7+c_9-2 c_{10}+c_{19}+c_{20}+c_{23}+c_{25}\right) (-\sin(\theta_{c}))$\\\hline
$\Xi_{cc}^{++}\to {p}  D^+  \pi^0  $ & $ -\frac{1}{\sqrt{2}}\left(c_1+c_3+c_4+c_5-c_6+2 c_7-2 c_{11}+c_{12}-c_{18}-c_{20}-c_{21}+c_{22}+c_{23}+2 c_{24}+c_{26}\right) \sin(\theta_{c})$\\\hline
$\Xi_{cc}^{++}\to {p}  D^+  \eta  $ & $ \frac{1}{\sqrt{6}}\left(3 c_1+c_3+c_4-c_5-3 c_6+2 c_7-2 c_8+2 c_{11}-c_{12}-3 c_{18}-c_{20}-c_{21}-c_{22}+3 c_{23}-c_{26}\right) \sin(\theta_{c})$\\\hline
$\Xi_{cc}^{++}\to {p}  D^+_s  \overline K^0  $ & $ \left(c_3-c_4+2 c_7+c_8-c_{20}+c_{21}-c_{24}\right) \sin(\theta_{c})$\\\hline
$\Xi_{cc}^{+}\to \Lambda^0  D^0  K^+  $ & \tabincell{c}{$ \frac{1}{\sqrt{6}}(2 c_1+c_2+c_3-c_6+2 c_7+c_{13}+2 c_{14}+c_{15}-4 c_{16}$\\$-2 c_{17}+2 c_{18}+c_{19}+3 c_{20}-c_{23}+c_{27}+2 c_{28}+c_{29}) \sin(\theta_{c})$}\\\hline
$\Xi_{cc}^{+}\to \Lambda^0  D^+  K^0  $ & \tabincell{c}{$ \frac{1}{\sqrt{6}}(c_3+c_4+2 c_5+2 c_7+c_8-c_9+2 c_{10}-4 c_{11}+2 c_{12}+c_{13}+2 c_{14}+c_{15}$\\$-4 c_{16}-2 c_{17}+3 c_{20}-c_{21}-2 c_{22}-c_{24}+3 c_{25}-c_{27}-2 c_{28}-c_{29}) \sin(\theta_{c})$}\\\hline
$\Xi_{cc}^{+}\to \Lambda^0  D^+_s  \pi^0  $ & \tabincell{c}{$ -\frac{1}{2\sqrt{3}}(c_1+2 c_2-2 c_4-c_5+c_6+c_8+2 c_9-4 c_{10}+2 c_{11}-c_{12}-c_{18}$\\$-2 c_{19}-2 c_{21}-c_{22}-c_{23}+c_{24}+3 c_{26}+2 c_{27}-2 c_{28}-4 c_{29}) \sin(\theta_{c})$}\\\hline
$\Xi_{cc}^{+}\to \Sigma^+  D^0  K^0  $ & $ \left(c_9-2 c_{10}+c_{13}-c_{15}+c_{25}-c_{27}+c_{29}\right) (-\sin(\theta_{c}))$\\\hline
$\Xi_{cc}^{+}\to \Sigma^+  D^+_s  \pi^-  $ & $ \left(2 c_{11}-c_{12}+c_{13}+c_{14}-c_{26}-c_{27}-c_{28}\right) \sin(\theta_{c})$\\\hline
$\Xi_{cc}^{+}\to \Sigma^0  D^0  K^+  $ & $ \frac{1}{\sqrt{2}}\left(c_2-c_3+c_6-2 c_7-c_{13}+c_{15}+c_{19}+c_{20}+c_{23}-c_{27}+c_{29}-2 c_{30}\right) \sin(\theta_{c})$\\\hline
$\Xi_{cc}^{+}\to \Sigma^0  D^+  K^0  $ & $ -\frac{1}{\sqrt{2}}\left(c_3-c_4+2 c_7+c_8-c_9+2 c_{10}-c_{13}+c_{15}-c_{20}+c_{21}-c_{24}-c_{25}-2 c_{26}+c_{27}-c_{29}+2 c_{30}\right) \sin(\theta_{c})$\\\hline
$\Xi_{cc}^{+}\to \Sigma^0  D^+_s  \pi^0  $ & $ -\frac{1}{2} \left(c_1-c_5-c_6-c_8-2 c_{11}+c_{12}-2 c_{13}-2 c_{14}-c_{18}-c_{22}+c_{23}-c_{24}+2 c_{25}+c_{26}\right) \sin(\theta_{c})$\\\hline
$\Xi_{cc}^{+}\to \Sigma^0  D^+_s  \eta  $ & \tabincell{c}{$ \frac{1}{2 \sqrt{3}}(3 c_1+2 c_3+2 c_4+c_5-3 c_6+4 c_7-c_8-2 c_{11}+c_{12}-3 c_{18}-2 c_{20}$\\$-2 c_{21}+c_{22}+3 c_{23}+3 c_{24}+2 c_{25}+c_{26}+2 c_{27}+2 c_{28}+4 c_{30}) \sin(\theta_{c})$}\\\hline
$\Xi_{cc}^{+}\to \Sigma^-  D^+  K^+  $ & $ \left(c_2-c_4+c_6+c_8-c_{13}+c_{15}+c_{19}-c_{21}+c_{23}+c_{24}-c_{27}+c_{29}\right) \sin(\theta_{c})$\\\hline
$\Xi_{cc}^{+}\to \Sigma^-  D^+_s  \pi^+  $ & $ \left(-c_1+c_5+c_6+c_8+c_{13}+c_{14}-c_{18}+c_{22}+c_{23}+c_{24}+c_{27}+c_{28}\right) \sin(\theta_{c})$\\\hline
$\Xi_{cc}^{+}\to {p}  D^0  \pi^0  $ & $ \frac{1}{\sqrt{2}}\left(c_1+c_2+c_9-2 c_{10}-c_{13}-c_{14}+2 c_{16}+c_{17}-c_{18}-c_{19}+c_{25}+c_{27}-c_{28}-2 c_{29}+c_{30}\right) \sin(\theta_{c})$\\\hline
$\Xi_{cc}^{+}\to {p}  D^0  \eta  $ & $ -\frac{1}{\sqrt{6}}\left(3 c_1+3 c_2+c_9-2 c_{10}-c_{13}+c_{14}+2 c_{15}-2 c_{16}-c_{17}-3 c_{18}-3 c_{19}+c_{25}+c_{27}+c_{28}-c_{30}\right) \sin(\theta_{c})$\\\hline
$\Xi_{cc}^{+}\to {p}  D^+  \pi^-  $ & $ \left(c_4+c_5-2 c_{11}+c_{12}+c_{13}+c_{14}-2 c_{16}-c_{17}-c_{21}-c_{22}+c_{26}-c_{27}-c_{28}-c_{30}\right) (-\sin(\theta_{c}))$\\\hline
$\Xi_{cc}^{+}\to {p}  D^+_s  K^-  $ & $ \left(c_4+c_5+2 c_{16}+c_{17}-c_{21}-c_{22}+c_{30}\right) \sin(\theta_{c})$\\\hline
$\Omega_{cc}^{+}\to \Lambda^0  D^0  \pi^+  $ & \tabincell{c}{$ -\frac{1}{\sqrt{6}}(c_1-c_2+2 c_3-2 c_6+4 c_7+2 c_{13}+c_{14}-c_{15}-2 c_{16}$\\$-c_{17}+c_{18}-c_{19}-2 c_{23}+2 c_{27}+c_{28}-c_{29}+3 c_{30}) \sin(\theta_{c})$}\\\hline
$\Omega_{cc}^{+}\to \Lambda^0  D^+  \pi^0  $ & \tabincell{c}{$ \frac{1}{2 \sqrt{3}}(c_1-c_2+2 c_3-2 c_6+4 c_7+2 c_{13}+c_{14}-c_{15}-2 c_{16}-c_{17}-c_{18}$\\$+c_{19}-2 c_{21}+2 c_{22}+2 c_{23}+4 c_{24}+2 c_{27}+c_{28}-c_{29}+3 c_{30}) \sin(\theta_{c})$}\\\hline
$\Omega_{cc}^{+}\to \Lambda^0  D^+  \eta  $ & \tabincell{c}{$ \frac{1}{6} (-3 c_1+3 c_2-2 c_3-2 c_4+2 c_5+6 c_6-4 c_7+4 c_8+2 c_9-4 c_{10}+8 c_{11}-4 c_{12}+4 c_{13}$\\$+5 c_{14}+c_{15}+2 c_{16}+c_{17}+3 c_{18}-3 c_{19}-6 c_{23}-6 c_{25}-3 c_{28}-3 c_{29}-3 c_{30}) \sin(\theta_{c})$}\\\hline
$\Omega_{cc}^{+}\to \Lambda^0  D^+_s  \overline K^0  $ & \tabincell{c}{$ -\frac{1}{\sqrt{6}}(2 c_3-c_4+c_5+4 c_7+2 c_8-2 c_9+4 c_{10}-2 c_{11}+c_{12}-c_{13}+c_{14}+2 c_{15}$\\$-2 c_{16}-c_{17}+c_{21}-c_{22}-2 c_{24}-3 c_{26}+c_{27}-c_{28}-2 c_{29}+3 c_{30}) \sin(\theta_{c})$}\\\hline
$\Omega_{cc}^{+}\to \Sigma^+  D^0  \pi^0  $ & $ -\frac{1}{\sqrt{2}}\left(c_1+c_2+c_{14}+c_{15}-2 c_{16}-c_{17}-c_{18}-c_{19}+c_{28}+c_{29}-c_{30}\right) \sin(\theta_{c})$\\\hline
$\Omega_{cc}^{+}\to \Sigma^+  D^0  \eta  $ & $ \frac{1}{\sqrt{6}}\left(3 c_1+3 c_2+2 c_9-4 c_{10}-2 c_{13}-c_{14}+c_{15}+2 c_{16}+c_{17}-3 c_{18}-3 c_{19}+2 c_{25}+2 c_{27}-c_{28}-3 c_{29}+c_{30}\right) \sin(\theta_{c})$\\\hline
$\Omega_{cc}^{+}\to \Sigma^+  D^+  \pi^-  $ & $ \left(c_4+c_5+2 c_{16}+c_{17}-c_{21}-c_{22}+c_{30}\right) \sin(\theta_{c})$\\\hline
$\Omega_{cc}^{+}\to \Sigma^+  D^+_s  K^-  $ & $ \left(c_4+c_5-2 c_{11}+c_{12}+c_{13}+c_{14}-2 c_{16}-c_{17}-c_{21}-c_{22}+c_{26}-c_{27}-c_{28}-c_{30}\right) (-\sin(\theta_{c}))$\\\hline
$\Omega_{cc}^{+}\to \Sigma^0  D^0  \pi^+  $ & $ \frac{1}{\sqrt{2}}\left(c_1+c_2+c_{14}+c_{15}-2 c_{16}-c_{17}+c_{18}+c_{19}+2 c_{20}+c_{28}+c_{29}-c_{30}\right) \sin(\theta_{c})$\\\hline
$\Omega_{cc}^{+}\to \Sigma^0  D^+  \pi^0  $ & $ \frac{1}{2} \left(c_1+c_2+2 c_4+2 c_5+c_{14}+c_{15}+2 c_{16}+c_{17}-c_{18}-c_{19}-2 c_{20}+c_{28}+c_{29}+c_{30}\right) \sin(\theta_{c})$\\\hline
$\Omega_{cc}^{+}\to \Sigma^0  D^+  \eta  $ & \tabincell{c}{$ -\frac{1}{2 \sqrt{3}}(3 c_1+3 c_2+2 c_9-4 c_{10}-2 c_{13}-c_{14}+c_{15}+2 c_{16}+c_{17}-3 c_{18}-3 c_{19}$\\$-2 c_{20}-2 c_{21}-2 c_{22}+2 c_{25}+4 c_{26}+2 c_{27}-c_{28}-3 c_{29}+c_{30}) \sin(\theta_{c})$}\\\hline
$\Omega_{cc}^{+}\to \Sigma^0  D^+_s  \overline K^0  $ & $ \frac{1}{\sqrt{2}}\left(c_4+c_5-2 c_{11}+c_{12}+c_{13}+c_{14}-2 c_{16}-c_{17}+2 c_{20}-c_{21}-c_{22}+2 c_{25}+c_{26}-c_{27}-c_{28}-c_{30}\right) \sin(\theta_{c})$\\\hline
$\Omega_{cc}^{+}\to \Sigma^-  D^+  \pi^+  $ & $ \left(c_1+c_2+c_4+c_5+c_{14}+c_{15}+c_{18}+c_{19}+c_{21}+c_{22}+c_{28}+c_{29}\right) \sin(\theta_{c})$\\\hline
$\Omega_{cc}^{+}\to {p}  D^0  \overline K^0  $ & $ \left(c_9-2 c_{10}+c_{13}-c_{15}+c_{25}-c_{27}+c_{29}\right) (-\sin(\theta_{c}))$\\\hline
$\Omega_{cc}^{+}\to {p}  D^+  K^-  $ & $ \left(2 c_{11}-c_{12}+c_{13}+c_{14}-c_{26}-c_{27}-c_{28}\right) \sin(\theta_{c})$\\\hline
$\Omega_{cc}^{+}\to \Xi^-  D^+  K^+  $ & $ \left(-c_1+c_5+c_6+c_8+c_{13}+c_{14}-c_{18}+c_{22}+c_{23}+c_{24}+c_{27}+c_{28}\right) \sin(\theta_{c})$\\\hline
$\Omega_{cc}^{+}\to \Xi^-  D^+_s  \pi^+  $ & $ \left(c_2-c_4+c_6+c_8-c_{13}+c_{15}+c_{19}-c_{21}+c_{23}+c_{24}-c_{27}+c_{29}\right) \sin(\theta_{c})$\\\hline
$\Omega_{cc}^{+}\to \Xi^0  D^0  K^+  $ & $ \left(c_1+c_3-c_6+2 c_7-c_{13}-c_{14}+2 c_{16}+c_{17}+c_{18}+c_{20}-c_{23}-c_{27}-c_{28}-c_{30}\right) \sin(\theta_{c})$\\\hline
$\Omega_{cc}^{+}\to \Xi^0  D^+  K^0  $ & $ \left(c_3+c_5+2 c_7+c_8+2 c_{16}+c_{17}+c_{20}-c_{22}-c_{24}-c_{30}\right) \sin(\theta_{c})$\\\hline
$\Omega_{cc}^{+}\to \Xi^0  D^+_s  \pi^0  $ & $ -\frac{1}{\sqrt{2}}\left(c_2-c_4+c_6+c_8-c_{13}+c_{15}-c_{19}-c_{21}-c_{23}+c_{24}-c_{27}+c_{29}\right) \sin(\theta_{c})$\\\hline
\hline
\end{tabular}
\end{table}

\begin{table}
\scriptsize
\newcommand{\tabincell}[2]{\begin{tabular}{@{}#1@{}}#2\end{tabular}}
\caption{Doubly charmed baryon decays into a light baryon in the octet, a charmed meson and a light meson.}\label{tab:ccq_qqq8_cqbarand8meson_3}\begin{tabular}{|c|c|c|c|c|c|c|c}\hline
channel & amplitude \\\hline
\hline
$\Xi_{cc}^{++}\to \Lambda^0  D^+_s  K^+  $ & \tabincell{c}{$ -\frac{1}{\sqrt{6}}(c_1+2 c_2+c_4+2 c_5+c_6+c_8+2 c_9-4 c_{10}-4 c_{11}+2 c_{12}$\\$+c_{18}+2 c_{19}+c_{21}+2 c_{22}+c_{23}+c_{24}) \sin^2(\theta_{c})$}\\\hline
$\Xi_{cc}^{++}\to {p}  D^0  K^+  $ & $ \left(c_2-c_3+c_6-2 c_7+c_9-2 c_{10}+c_{19}+c_{20}+c_{23}+c_{25}\right) \sin^2(\theta_{c})$\\\hline
$\Xi_{cc}^{++}\to {p}  D^+  K^0  $ & $ \left(c_1+c_3-c_6+2 c_7-c_{18}-c_{20}+c_{23}\right) \left(-\sin^2(\theta_{c})\right)$\\\hline
$\Xi_{cc}^{++}\to {p}  D^+_s  \pi^0  $ & $ \frac{1}{\sqrt{2}}\left(c_4+c_5-2 c_{11}+c_{12}-c_{21}+c_{22}+2 c_{24}+c_{26}\right) \sin^2(\theta_{c})$\\\hline
$\Xi_{cc}^{++}\to {p}  D^+_s  \eta  $ & $ \frac{1}{\sqrt{6}}\left(2 c_3-c_4+c_5+4 c_7+2 c_8-2 c_{11}+c_{12}-2 c_{20}+c_{21}+c_{22}+c_{26}\right) \sin^2(\theta_{c})$\\\hline
$\Xi_{cc}^{+}\to \Lambda^0  D^+_s  K^0  $ & $ -\frac{1}{\sqrt{6}}\left(c_1+2 c_2+c_4+2 c_5+c_6+c_8+2 c_9-4 c_{10}-4 c_{11}+2 c_{12}-c_{18}-2 c_{19}-c_{21}-2 c_{22}-c_{23}-c_{24}\right) \sin^2(\theta_{c})$\\\hline
$\Xi_{cc}^{+}\to {p}  D^0  K^0  $ & $ \left(c_1+c_2+c_9-2 c_{10}-c_{18}-c_{19}+c_{25}\right) \sin^2(\theta_{c})$\\\hline
$\Xi_{cc}^{+}\to {p}  D^+_s  \pi^-  $ & $ \left(c_4+c_5-2 c_{11}+c_{12}-c_{21}-c_{22}+c_{26}\right) \sin^2(\theta_{c})$\\\hline
$\Omega_{cc}^{+}\to \Lambda^0  D^0  K^+  $ & $ \frac{1}{\sqrt{6}}\left(c_1-c_2+2 c_3-2 c_6+4 c_7-c_{13}-2 c_{14}-c_{15}+4 c_{16}+2 c_{17}+c_{18}-c_{19}-2 c_{23}-c_{27}-2 c_{28}-c_{29}\right) \sin^2(\theta_{c})$\\\hline
$\Omega_{cc}^{+}\to \Lambda^0  D^+  K^0  $ & $ \frac{1}{\sqrt{6}}\left(c_1-c_2+2 c_3-2 c_6+4 c_7-c_{13}-2 c_{14}-c_{15}+4 c_{16}+2 c_{17}-c_{18}+c_{19}+2 c_{23}+c_{27}+2 c_{28}+c_{29}\right) \sin^2(\theta_{c})$\\\hline
$\Omega_{cc}^{+}\to \Lambda^0  D^+_s  \pi^0  $ & $ \frac{1}{\sqrt{3}}\left(c_{21}-c_{22}-2 c_{24}+c_{27}-c_{28}-2 c_{29}\right) \sin^2(\theta_{c})$\\\hline
$\Omega_{cc}^{+}\to \Lambda^0  D^+_s  \eta  $ & $ -\frac{1}{3} \left(2 c_3-c_4+c_5+4 c_7+2 c_8-2 c_9+4 c_{10}+4 c_{11}-2 c_{12}-c_{13}+c_{14}+2 c_{15}+4 c_{16}+2 c_{17}\right) \sin^2(\theta_{c})$\\\hline
$\Omega_{cc}^{+}\to \Sigma^+  D^0  K^0  $ & $ \left(c_1+c_2-c_{13}+c_{15}-c_{18}-c_{19}+c_{27}-c_{29}\right) \left(-\sin^2(\theta_{c})\right)$\\\hline
$\Omega_{cc}^{+}\to \Sigma^+  D^+_s  \pi^-  $ & $ \left(c_4+c_5+c_{13}+c_{14}-c_{21}-c_{22}-c_{27}-c_{28}\right) \left(-\sin^2(\theta_{c})\right)$\\\hline
$\Omega_{cc}^{+}\to \Sigma^0  D^0  K^+  $ & $ -\frac{1}{\sqrt{2}}\left(c_1+c_2-c_{13}+c_{15}+c_{18}+c_{19}+2 c_{20}-c_{27}+c_{29}-2 c_{30}\right) \sin^2(\theta_{c})$\\\hline
$\Omega_{cc}^{+}\to \Sigma^0  D^+  K^0  $ & $ \frac{1}{\sqrt{2}}\left(c_1+c_2-c_{13}+c_{15}-c_{18}-c_{19}-2 c_{20}+c_{27}-c_{29}+2 c_{30}\right) \sin^2(\theta_{c})$\\\hline
$\Omega_{cc}^{+}\to \Sigma^0  D^+_s  \pi^0  $ & $ \left(c_4+c_5+c_{13}+c_{14}\right) \left(-\sin^2(\theta_{c})\right)$\\\hline
$\Omega_{cc}^{+}\to \Sigma^-  D^+  K^+  $ & $ \left(c_1+c_2-c_{13}+c_{15}+c_{18}+c_{19}-c_{27}+c_{29}\right) \left(-\sin^2(\theta_{c})\right)$\\\hline
$\Omega_{cc}^{+}\to \Sigma^-  D^+_s  \pi^+  $ & $ \left(c_4+c_5+c_{13}+c_{14}+c_{21}+c_{22}+c_{27}+c_{28}\right) \left(-\sin^2(\theta_{c})\right)$\\\hline
$\Omega_{cc}^{+}\to {p}  D^0  \pi^0  $ & $ \frac{1}{\sqrt{2}}\left(c_{13}+c_{14}-2 c_{16}-c_{17}-c_{27}+c_{28}+2 c_{29}-c_{30}\right) \sin^2(\theta_{c})$\\\hline
$\Omega_{cc}^{+}\to {p}  D^0  \eta  $ & $ -\frac{1}{\sqrt{6}}\left(2 c_9-4 c_{10}+c_{13}-c_{14}-2 c_{15}+2 c_{16}+c_{17}+2 c_{25}-c_{27}-c_{28}+c_{30}\right) \sin^2(\theta_{c})$\\\hline
$\Omega_{cc}^{+}\to {p}  D^+  \pi^-  $ & $ \left(c_{13}+c_{14}-2 c_{16}-c_{17}-c_{27}-c_{28}-c_{30}\right) \sin^2(\theta_{c})$\\\hline
$\Omega_{cc}^{+}\to {p}  D^+_s  K^-  $ & $ \left(2 c_{11}-c_{12}+2 c_{16}+c_{17}-c_{26}+c_{30}\right) \left(-\sin^2(\theta_{c})\right)$\\\hline
\hline
\end{tabular}
\end{table}

The   hadron-level Hamiltonian for the decays of $T_{cc}$ into a light octet baryon, a charmed meson and a light meson is given as
\begin{eqnarray}
 {\cal H}_{eff}&=& c_1(T_{cc})^i  \overline D ^j  \epsilon_{ijk}M^{n}_{m} (T_8)^{k}_{l}  (H_{6})^{lm}_n +c_2(T_{cc})^i  \overline D ^l  \epsilon_{ijk}M^{n}_{m} (T_8)^{k}_{l}  (H_{6})^{jm}_n\nonumber\\
 &&+c_3(T_{cc})^i  \overline D ^l  \epsilon_{ijk}M^{m}_{l} (T_8)^{k}_{n}  (H_{6})^{nj}_m+c_4(T_{cc})^i  \overline D ^l  \epsilon_{ijk}M^{j}_{n} (T_8)^{k}_{m}  (H_{6})^{mn}_l\nonumber\\
 &&+c_5(T_{cc})^i  \overline D ^l  \epsilon_{ijk}M^{m}_{n} (T_8)^{k}_{m}  (H_{6})^{jn}_l+c_6(T_{cc})^l  \overline D ^i  \epsilon_{ijk}M^{n}_{m} (T_8)^{k}_{l}  (H_{6})^{jm}_n\nonumber\\
 &&+c_7(T_{cc})^l  \overline D ^m  \epsilon_{ijk}M^{n}_{m} (T_8)^{k}_{l}  (H_{6})^{ij}_n+c_8(T_{cc})^l  \overline D ^m  \epsilon_{ijk}M^{i}_{n} (T_8)^{k}_{l}  (H_{6})^{jn}_m\nonumber\\
 &&+c_9(T_{cc})^l  \overline D ^i  \epsilon_{ijk}M^{n}_{l} (T_8)^{k}_{m}  (H_{6})^{jm}_n+c_{10}(T_{cc})^l  \overline D ^m  \epsilon_{ijk}M^{n}_{l} (T_8)^{k}_{m}  (H_{6})^{ij}_n\nonumber\\
 &&+c_{11}(T_{cc})^l  \overline D ^m  \epsilon_{ijk}M^{n}_{l} (T_8)^{k}_{n}  (H_{6})^{ij}_m+c_{12}(T_{cc})^l  \overline D ^m  \epsilon_{ijk}M^{i}_{l} (T_8)^{k}_{n}  (H_{6})^{jn}_m\nonumber\\
&&+c_{13}(T_{cc})^l  \overline D ^i  \epsilon_{ljk}M^{j}_{n} (T_8)^{k}_{m}  (H_{6})^{mn}_l+c_{14}(T_{cc})^l  \overline D ^i  \epsilon_{ijk}M^{m}_{n} (T_8)^{k}_{m}  (H_{6})^{jn}_l\nonumber\\
&&+c_{15}(T_{cc})^l  \overline D ^m  \epsilon_{ijk}M^{i}_{n} (T_8)^{k}_{m}  (H_{6})^{jn}_l +c_{16}(T_{cc})^l  \overline D ^m  \epsilon_{ijk}M^{n}_{m} (T_8)^{k}_{n}  (H_{6})^{ij}_l\nonumber\\
&& +c_{17}(T_{cc})^l  \overline D ^m  \epsilon_{ijk}M^{j}_{m} (T_8)^{k}_{n}  (H_{6})^{in}_l\nonumber\\
&&+c_{18}(T_{cc})^i  \overline D ^j  \epsilon_{ijk}M^{n}_{m} (T_8)^{k}_{l}  (H_{15})^{lm}_n+c_{19}(T_{cc})^i  \overline D ^l  \epsilon_{ijk}M^{n}_{m} (T_8)^{k}_{l}  (H_{15})^{jm}_n\nonumber\\
 &&+c_{20}(T_{cc})^i  \overline D ^l  \epsilon_{ijk}M^{m}_{l} (T_8)^{k}_{n}  (H_{15})^{nj}_m+c_{21}(T_{cc})^i  \overline D ^l  \epsilon_{ijk}M^{j}_{n} (T_8)^{k}_{m}  (H_{15})^{mn}_l\nonumber\\
 &&+c_{22}(T_{cc})^i  \overline D ^l  \epsilon_{ijk}M^{m}_{n} (T_8)^{k}_{m}  (H_{15})^{jn}_l+c_{23}(T_{cc})^l  \overline D ^i  \epsilon_{ijk}M^{n}_{m} (T_8)^{k}_{l}  (H_{15})^{jm}_n\nonumber\\
 &&+c_{24}(T_{cc})^l  \overline D ^m  \epsilon_{ijk}M^{i}_{n} (T_8)^{k}_{l}  (H_{15})^{jn}_m+c_{25}(T_{cc})^l  \overline D ^i  \epsilon_{ijk}M^{n}_{l} (T_8)^{k}_{m}  (H_{15})^{jm}_n\nonumber\\
&&+c_{26}(T_{cc})^l  \overline D ^m  \epsilon_{ijk}M^{i}_{l} (T_8)^{k}_{n}  (H_{15})^{jn}_m+c_{27}(T_{cc})^l  \overline D ^i  \epsilon_{ijk}M^{j}_{n} (T_8)^{k}_{m}  (H_{15})^{mn}_l\nonumber\\
&&+c_{28}(T_{cc})^l  \overline D ^i  \epsilon_{ijk}M^{n}_{m} (T_8)^{k}_{l}  (H_{15})^{jm}_i +c_{29}(T_{cc})^l  \overline D ^m  \epsilon_{ijk}M^{i}_{n} (T_8)^{k}_{m}  (H_{15})^{jn}_l\nonumber\\
&& +c_{30}(T_{cc})^l  \overline D ^m  \epsilon_{ijk}M^{i}_{n} (T_8)^{k}_{m}  (H_{15})^{jn}_l.
\end{eqnarray}

Expanding the above equations, we will obtain the decay amplitudes given in Tab.~\ref{tab:ccq_qqq8_cqbarand8meson_1}, Tab.~\ref{tab:ccq_qqq8_cqbarand8meson_2} and Tab.~\ref{tab:ccq_qqq8_cqbarand8meson_3}. This leads to the relations for decay widths: in Appendix~\ref{sec:relation_cc_baryon8_cm_lm}.




\begin{table}
{\scriptsize{}\caption{Doubly charmed baryon decays into a light baryon in the decuplet, a charmed meson and a light meson.}
\label{tab:ccq_qqq10_cqbar_1}}%
\begin{tabular}{|cc|cc|}
\hline
channel  & amplitude  & channel  & amplitude\tabularnewline
\hline
\hline
{\scriptsize{}$\Xi_{cc}^{++}\to\Delta^{++}D^{0}\overline{K}^{0}$} & {\scriptsize{}$c_{1}-c_{8}$} & {\scriptsize{}$\Omega_{cc}^{+}\to\Sigma^{\prime+}D^{+}\pi^{-}$} & {\scriptsize{}$-\frac{\left(c_{3}-2c_{7}\right)\sin(\theta_{c})}{\sqrt{3}}$}\tabularnewline
\hline
{\scriptsize{}$\Xi_{cc}^{++}\to\Delta^{++}D^{+}K^{-}$} & {\scriptsize{}$c_{3}$} & {\scriptsize{}$\Omega_{cc}^{+}\to\Delta^{++}D^{0}K^{-}$} & {\scriptsize{}$c_{6}\sin(\theta_{c})$}\tabularnewline
\hline
{\scriptsize{}$\Xi_{cc}^{++}\to\Delta^{+}D^{+}\overline{K}^{0}$} & {\scriptsize{}$\frac{c_{1}+c_{3}-c_{8}}{\sqrt{3}}$} & {\scriptsize{}$\Omega_{cc}^{+}\to\Delta^{+}D^{0}\overline{K}^{0}$} & {\scriptsize{}$\frac{\left(c_{6}-2c_{4}\right)\sin(\theta_{c})}{\sqrt{3}}$}\tabularnewline
\hline
{\scriptsize{}$\Xi_{cc}^{++}\to\Sigma^{\prime+}D^{0}\pi^{+}$} & {\scriptsize{}$\frac{c_{1}+2c_{2}+2c_{4}+c_{8}}{\sqrt{3}}$} & {\scriptsize{}$\Omega_{cc}^{+}\to\Delta^{+}D^{+}K^{-}$} & {\scriptsize{}$\frac{\left(c_{6}-2c_{5}\right)\sin(\theta_{c})}{\sqrt{3}}$}\tabularnewline
\hline
{\scriptsize{}$\Xi_{cc}^{++}\to\Sigma^{\prime+}D^{+}\pi^{0}$} & {\scriptsize{}$\frac{-2c_{2}+c_{3}+2c_{5}}{\sqrt{6}}$} & {\scriptsize{}$\Omega_{cc}^{+}\to\Delta^{0}D^{+}\overline{K}^{0}$} & {\scriptsize{}$\frac{\left(-2c_{4}-2c_{5}+c_{6}\right)\sin(\theta_{c})}{\sqrt{3}}$}\tabularnewline
\hline
{\scriptsize{}$\Xi_{cc}^{++}\to\Sigma^{\prime0}D^{+}\pi^{+}$} & {\scriptsize{}$\frac{c_{1}+c_{3}+2c_{4}+2c_{5}+c_{8}}{\sqrt{6}}$} & {\scriptsize{}$\Omega_{cc}^{+}\to\Sigma^{\prime+}D^{0}\pi^{0}$} & {\scriptsize{}$\frac{\left(c_{1}+c_{6}+2c_{7}-c_{8}\right)\sin(\theta_{c})}{\sqrt{6}}$}\tabularnewline
\hline
{\scriptsize{}$\Xi_{cc}^{++}\to\Xi^{\prime0}D_{s}^{+}\pi^{+}$} & {\scriptsize{}$\frac{c_{1}+2c_{4}+c_{8}}{\sqrt{3}}$} & {\scriptsize{}$\Omega_{cc}^{+}\to\Sigma^{\prime0}D^{0}\pi^{+}$} & {\scriptsize{}$-\frac{\left(c_{1}+2c_{2}-c_{6}-2c_{7}+c_{8}\right)\sin(\theta_{c})}{\sqrt{6}}$}\tabularnewline
\hline
{\scriptsize{}$\Xi_{cc}^{+}\to\Delta^{++}D^{0}K^{-}$} & {\scriptsize{}$c_{6}$} & {\scriptsize{}$\Omega_{cc}^{+}\to\Sigma^{\prime0}D^{+}\pi^{0}$} & {\scriptsize{}$\frac{\left(c_{1}+2c_{2}+c_{6}-2c_{7}-c_{8}\right)\sin(\theta_{c})}{2\sqrt{3}}$}\tabularnewline
\hline
{\scriptsize{}$\Xi_{cc}^{+}\to\Delta^{+}D^{0}\overline{K}^{0}$} & {\scriptsize{}$\frac{c_{1}+c_{6}-c_{8}}{\sqrt{3}}$} & {\scriptsize{}$\Omega_{cc}^{+}\to\Sigma^{\prime-}D^{+}\pi^{+}$} & {\scriptsize{}$-\frac{\left(c_{1}+c_{3}-c_{6}+c_{8}\right)\sin(\theta_{c})}{\sqrt{3}}$}\tabularnewline
\hline
{\scriptsize{}$\Xi_{cc}^{+}\to\Delta^{+}D^{+}K^{-}$} & {\scriptsize{}$\frac{c_{3}+c_{6}}{\sqrt{3}}$} & {\scriptsize{}$\Xi_{cc}^{+}\to\Sigma^{\prime0}D_{s}^{+}\pi^{0}$} & {\scriptsize{}$\frac{\left(c_{1}+c_{3}+2c_{4}-2c_{5}-c_{8}\right)\sin(\theta_{c})}{2\sqrt{3}}$}\tabularnewline
\hline
{\scriptsize{}$\Xi_{cc}^{+}\to\Delta^{0}D^{+}\overline{K}^{0}$} & {\scriptsize{}$\frac{c_{1}+c_{3}+c_{6}-c_{8}}{\sqrt{3}}$} & {\scriptsize{}$\Xi_{cc}^{+}\to\Delta^{+}D^{0}K^{0}$} & {\scriptsize{}$\frac{\left(c_{1}+2c_{4}-c_{8}\right)\sin^{2}(\theta_{c})}{\sqrt{3}}$ }\tabularnewline
\hline
{\scriptsize{}$\Xi_{cc}^{+}\to\Sigma^{\prime+}D^{0}\pi^{0}$} & {\scriptsize{}$\frac{-2c_{4}+c_{6}+2c_{7}}{\sqrt{6}}$} & {\scriptsize{}$\Xi_{cc}^{+}\to\Delta^{+}D_{s}^{+}\pi^{-}$} & {\scriptsize{}$\frac{\left(c_{3}+2c_{5}\right)\sin^{2}(\theta_{c})}{\sqrt{3}}$ }\tabularnewline
\hline
{\scriptsize{}$\Xi_{cc}^{+}\to\Sigma^{\prime+}D^{+}\pi^{-}$} & {\scriptsize{}$\frac{2\left(c_{5}+c_{7}\right)}{\sqrt{3}}$} & {\scriptsize{}$\Omega_{cc}^{+}\to\Delta^{-}D^{+}\pi^{+}$} & {\scriptsize{}$c_{6}\sin^{2}(\theta_{c})$}\tabularnewline
\hline
{\scriptsize{}$\Xi_{cc}^{+}\to\Sigma^{\prime0}D^{0}\pi^{+}$} & {\scriptsize{}$\frac{c_{1}+2c_{2}+c_{6}+2c_{7}+c_{8}}{\sqrt{6}}$} & {\scriptsize{}$\Omega_{cc}^{+}\to\Sigma^{\prime+}D_{s}^{+}\pi^{-}$} & {\scriptsize{}$\frac{\left(c_{3}+c_{6}\right)\sin^{2}(\theta_{c})}{\sqrt{3}}$}\tabularnewline
\hline
{\scriptsize{}$\Xi_{cc}^{+}\to\Sigma^{\prime0}D^{+}\pi^{0}$} & {\scriptsize{}$\frac{-2c_{2}+c_{3}-2c_{4}-2c_{5}+c_{6}-2c_{7}}{2\sqrt{3}}$} & {\scriptsize{}$\Omega_{cc}^{+}\to\Sigma^{\prime-}D_{s}^{+}\pi^{+}$} & {\scriptsize{}$\frac{\left(c_{3}+c_{6}\right)\sin^{2}(\theta_{c})}{\sqrt{3}}$}\tabularnewline
\hline
{\scriptsize{}$\Xi_{cc}^{+}\to\Sigma^{\prime-}D^{+}\pi^{+}$} & {\scriptsize{}$\frac{c_{1}+c_{3}+c_{6}+c_{8}}{\sqrt{3}}$} & {\scriptsize{}$\Xi_{cc}^{+}\to\Delta^{0}D^{0}K^{+}$} & {\scriptsize{}$\frac{\left(c_{1}+2c_{2}+c_{8}\right)\sin^{2}(\theta_{c})}{\sqrt{3}}$}\tabularnewline
\hline
{\scriptsize{}$\Omega_{cc}^{+}\to\Xi^{\prime0}D^{0}\pi^{+}$} & {\scriptsize{}$\frac{c_{1}+2c_{2}+c_{8}}{\sqrt{3}}$} & {\scriptsize{}$\Xi_{cc}^{+}\to\Delta^{0}D^{+}K^{0}$} & {\scriptsize{}$\frac{\left(c_{1}+2c_{2}+2c_{4}-c_{8}\right)\sin^{2}(\theta_{c})}{\sqrt{3}}$}\tabularnewline
\hline
{\scriptsize{}$\Omega_{cc}^{+}\to\Xi^{\prime0}D^{+}\pi^{0}$} & {\scriptsize{}$\frac{c_{3}-2c_{2}}{\sqrt{6}}$} & {\scriptsize{}$\Xi_{cc}^{+}\to\Delta^{0}D_{s}^{+}\pi^{0}$} & {\scriptsize{}$-\sqrt{\frac{2}{3}}c_{5}\sin^{2}(\theta_{c})$}\tabularnewline
\hline
{\scriptsize{}$\Omega_{cc}^{+}\to\Xi^{\prime-}D^{+}\pi^{+}$} & {\scriptsize{}$\frac{c_{1}+c_{3}+c_{8}}{\sqrt{3}}$} & {\scriptsize{}$\Xi_{cc}^{+}\to\Delta^{-}D^{+}K^{+}$} & {\scriptsize{}$\left(c_{1}+c_{8}\right)\sin^{2}(\theta_{c})$}\tabularnewline
\hline
{\scriptsize{}$\Xi_{cc}^{++}\to\Delta^{++}D^{0}\pi^{0}$} & {\scriptsize{}$\frac{\left(c_{1}-c_{8}\right)\sin(\theta_{c})}{\sqrt{2}}$} & {\scriptsize{}$\Xi_{cc}^{+}\to\Delta^{-}D_{s}^{+}\pi^{+}$} & {\scriptsize{}$c_{3}\sin^{2}(\theta_{c})$}\tabularnewline
\hline
{\scriptsize{}$\Xi_{cc}^{++}\to\Delta^{++}D^{+}\pi^{-}$} & {\scriptsize{}$c_{3}(-\sin(\theta_{c}))$} & {\scriptsize{}$\Omega_{cc}^{+}\to\Delta^{++}D^{0}\pi^{-}$} & {\scriptsize{}$c_{6}\sin^{2}(\theta_{c})$}\tabularnewline
\hline
{\scriptsize{}$\Xi_{cc}^{++}\to\Delta^{+}D^{0}\pi^{+}$} & {\scriptsize{}$-\frac{\left(c_{1}+2c_{2}+2c_{4}+c_{8}\right)\sin(\theta_{c})}{\sqrt{3}}$} & {\scriptsize{}$\Omega_{cc}^{+}\to\Delta^{+}D^{0}\pi^{0}$} & {\scriptsize{}$\sqrt{\frac{2}{3}}c_{7}\sin^{2}(\theta_{c})$}\tabularnewline
\hline
{\scriptsize{}$\Xi_{cc}^{++}\to\Delta^{+}D^{+}\pi^{0}$} & {\scriptsize{}$\frac{\left(c_{1}+2c_{2}-2c_{5}-c_{8}\right)\sin(\theta_{c})}{\sqrt{6}}$} & {\scriptsize{}$\Omega_{cc}^{+}\to\Delta^{+}D^{0}\eta$} & {\scriptsize{}$-\frac{1}{3}\sqrt{2}\left(2c_{4}-c_{6}-c_{7}\right)\sin^{2}(\theta_{c})$}\tabularnewline
\hline
{\scriptsize{}$\Xi_{cc}^{+}\to\Sigma^{\prime-}D_{s}^{+}\pi^{+}$} & {\scriptsize{}$-\frac{\left(c_{1}-c_{3}+c_{6}+c_{8}\right)\sin(\theta_{c})}{\sqrt{3}}$ } & {\scriptsize{}$\Omega_{cc}^{+}\to\Delta^{+}D^{+}\pi^{-}$} & {\scriptsize{}$\frac{\left(c_{6}+2c_{7}\right)\sin^{2}(\theta_{c})}{\sqrt{3}}$}\tabularnewline
\hline
{\scriptsize{}$\Xi_{cc}^{++}\to\Delta^{0}D^{+}\pi^{+}$} & {\scriptsize{}$-\frac{\left(c_{1}+c_{3}+2c_{4}+2c_{5}+c_{8}\right)\sin(\theta_{c})}{\sqrt{3}}$} & {\scriptsize{}$\Omega_{cc}^{+}\to\Delta^{+}D_{s}^{+}K^{-}$} & {\scriptsize{}$\frac{2\left(c_{5}+c_{7}\right)\sin^{2}(\theta_{c})}{\sqrt{3}}$}\tabularnewline
\hline
{\scriptsize{}$\Xi_{cc}^{++}\to\Sigma^{\prime+}D_{s}^{+}\pi^{0}$} & {\scriptsize{}$\frac{\left(c_{1}+c_{3}+2c_{5}-c_{8}\right)\sin(\theta_{c})}{\sqrt{6}}$} & {\scriptsize{}$\Omega_{cc}^{+}\to\Delta^{0}D^{0}\pi^{+}$} & {\scriptsize{}$\frac{\left(c_{6}+2c_{7}\right)\sin^{2}(\theta_{c})}{\sqrt{3}}$}\tabularnewline
\hline
{\scriptsize{}$\Xi_{cc}^{++}\to\Sigma^{\prime0}D_{s}^{+}\pi^{+}$} & {\scriptsize{}$-\frac{\left(c_{1}-c_{3}+2c_{4}-2c_{5}+c_{8}\right)\sin(\theta_{c})}{\sqrt{6}}$} & {\scriptsize{}$\Omega_{cc}^{+}\to\Delta^{0}D^{+}\pi^{0}$} & {\scriptsize{}$-\sqrt{\frac{2}{3}}c_{7}\sin^{2}(\theta_{c})$}\tabularnewline
\hline
{\scriptsize{}$\Xi_{cc}^{+}\to\Delta^{++}D^{0}\pi^{-}$} & {\scriptsize{}$c_{6}(-\sin(\theta_{c}))$} & {\scriptsize{}$\Omega_{cc}^{+}\to\Delta^{0}D^{+}\eta$} & {\scriptsize{}$-\frac{1}{3}\sqrt{2}\left(2c_{4}-c_{6}-c_{7}\right)\sin^{2}(\theta_{c})$}\tabularnewline
\hline
{\scriptsize{}$\Xi_{cc}^{+}\to\Delta^{+}D^{0}\pi^{0}$} & {\scriptsize{}$\frac{\left(c_{1}+2c_{4}-2c_{7}-c_{8}\right)\sin(\theta_{c})}{\sqrt{6}}$} & {\scriptsize{}$\Omega_{cc}^{+}\to\Delta^{0}D_{s}^{+}\overline{K}^{0}$} & {\scriptsize{}$\frac{2\left(c_{5}+c_{7}\right)\sin^{2}(\theta_{c})}{\sqrt{3}}$}\tabularnewline
\hline
{\scriptsize{}$\Xi_{cc}^{+}\to\Delta^{+}D^{0}\eta$} & {\scriptsize{}$-\frac{\left(3c_{1}+2c_{4}+2c_{6}+2c_{7}-3c_{8}\right)\sin(\theta_{c})}{3\sqrt{2}}$} & {\scriptsize{}$\Xi_{cc}^{++}\to\Delta^{++}D^{0}K^{0}$} & {\scriptsize{}$\left(c_{1}-c_{8}\right)\sin^{2}(\theta_{c})$}\tabularnewline
\hline
{\scriptsize{}$\Xi_{cc}^{+}\to\Delta^{+}D^{+}\pi^{-}$} & {\scriptsize{}$-\frac{\left(c_{3}+2c_{5}+c_{6}+2c_{7}\right)\sin(\theta_{c})}{\sqrt{3}}$} & {\scriptsize{}$\Xi_{cc}^{++}\to\Delta^{++}D_{s}^{+}\pi^{-}$} & {\scriptsize{}$c_{3}\sin^{2}(\theta_{c})$}\tabularnewline
\hline
{\scriptsize{}$\Xi_{cc}^{+}\to\Delta^{0}D^{0}\pi^{+}$} & {\scriptsize{}$-\frac{\left(c_{1}+2c_{2}+c_{6}+2c_{7}+c_{8}\right)\sin(\theta_{c})}{\sqrt{3}}$} & {\scriptsize{}$\Xi_{cc}^{++}\to\Delta^{+}D^{0}K^{+}$} & {\scriptsize{}$\frac{\left(c_{1}+2c_{2}+2c_{4}+c_{8}\right)\sin^{2}(\theta_{c})}{\sqrt{3}}$}\tabularnewline
\hline
{\scriptsize{}$\Xi_{cc}^{+}\to\Delta^{0}D^{+}\pi^{0}$} & {\scriptsize{}$\frac{\left(c_{1}+2c_{2}+2c_{4}+2c_{5}+2c_{7}-c_{8}\right)\sin(\theta_{c})}{\sqrt{6}}$} & {\scriptsize{}$\Xi_{cc}^{++}\to\Delta^{+}D^{+}K^{0}$} & {\scriptsize{}$\frac{\left(c_{1}+2c_{2}-c_{8}\right)\sin^{2}(\theta_{c})}{\sqrt{3}}$}\tabularnewline
\hline
{\scriptsize{}$\Xi_{cc}^{+}\to\Delta^{-}D^{+}\pi^{+}$} & {\scriptsize{}$\left(c_{1}+c_{3}+c_{6}+c_{8}\right)(-\sin(\theta_{c}))$} & {\scriptsize{}$\Xi_{cc}^{++}\to\Delta^{+}D_{s}^{+}\pi^{0}$} & {\scriptsize{}$\sqrt{\frac{2}{3}}c_{5}\sin^{2}(\theta_{c})$}\tabularnewline
\hline
{\scriptsize{}$\Xi_{cc}^{+}\to\Sigma^{\prime+}D_{s}^{+}\pi^{-}$} & {\scriptsize{}$\frac{\left(2c_{5}-c_{6}\right)\sin(\theta_{c})}{\sqrt{3}}$} & {\scriptsize{}$\Xi_{cc}^{++}\to\Delta^{0}D^{+}K^{+}$} & {\scriptsize{}$\frac{\left(c_{1}+2c_{4}+c_{8}\right)\sin^{2}(\theta_{c})}{\sqrt{3}}$}\tabularnewline
\hline
{\scriptsize{}$\Omega_{cc}^{+}\to\Xi^{\prime0}D_{s}^{+}\pi^{0}$} & {\scriptsize{}$\frac{\left(c_{1}+c_{3}+c_{6}-c_{8}\right)\sin(\theta_{c})}{\sqrt{6}}$} & {\scriptsize{}$\Xi_{cc}^{++}\to\Delta^{0}D_{s}^{+}\pi^{+}$} & {\scriptsize{}$\frac{\left(c_{3}+2c_{5}\right)\sin^{2}(\theta_{c})}{\sqrt{3}}$}\tabularnewline
\hline
{\scriptsize{}$\Omega_{cc}^{+}\to\Xi^{\prime-}D_{s}^{+}\pi^{+}$} & {\scriptsize{}$-\frac{\left(c_{1}-c_{3}-c_{6}+c_{8}\right)\sin(\theta_{c})}{\sqrt{3}}$} &  & \tabularnewline
\hline
\hline
\end{tabular}
\end{table}

For a light decuplet in the final state, the Hamiltonian is given as
\begin{eqnarray}
 {\cal H}_{eff}&=&  c_1(T_{cc})^l  \overline D ^m   (T_{10})_{iml} M^{j}_{n} (H_{15})^{in}_j +c_2(T_{cc})^l  \overline D ^m   (T_{10})_{ijl} M^{n}_{m} (H_{15})^{ij}_n \nonumber\\
 &&+c_3(T_{cc})^l  \overline D ^m   (T_{10})_{ijl} M^{i}_{n} (H_{15})^{nj}_m +c_4(T_{cc})^l  \overline D ^m   (T_{10})_{ijm} M^{n}_{l} (H_{15})^{ij}_n\nonumber\\
 &&+c_5(T_{cc})^l  \overline D ^m   (T_{10})_{ijn} M^{i}_{l} (H_{15})^{jn}_m+c_6(T_{cc})^l  \overline D ^m   (T_{10})_{ijm} M^{i}_{n} (H_{15})^{jn}_l\nonumber\\
 &&+c_7(T_{cc})^l  \overline D ^m   (T_{10})_{ijn} M^{i}_{m} (H_{15})^{jn}_l+c_8(T_{cc})^l  \overline D ^m   (T_{10})_{iml} M^{j}_{n} (H_{6})^{in}_j\nonumber\\
 &&+c_9(T_{cc})^l  \overline D ^m   (T_{10})_{ijl} M^{i}_{n} (H_{6})^{nj}_m+c_{10}(T_{cc})^l  \overline D ^m   (T_{10})_{ijm} M^{i}_{n} (H_{6})^{jn}_l.
\end{eqnarray}

The corresponding decay amplitudes are given in Tab.~\ref{tab:ccq_qqq10_cqbar_1}, and it leads to the relations for decay widths also collected  in Appendix~\ref{sec:relation_cc_baryon8_cm_lm}.

\section{Non-Leptonic $\Xi_{bb}$ and $\Omega_{bb}$ decays}
\label{sec:bbq_nonleptonic}

For the bottom quark decay, there are generically four kinds of quark-level transitions:
\begin{eqnarray}
b\to c\bar c d/s, \;
b\to c \bar u d/s, \;
b\to u \bar c d/s, \;
b\to q_1 \bar q_2 q_3,
\end{eqnarray}
with $q_{1,2,3}$ as  a  light quark.
Each of them can induce more than one types of decay modes at hadron level, which will be discussed in order.

\subsection{$b\to c\bar c d/s$: Decays into $J/\psi$, a bottom baryon and a light meson}

These decays  have the same topology  with  semileptonic $b\to s\ell^+\ell^-$ decays, and thus the SU(3) relations derived in this subsection are also applicable to semileptonic $b\to s\ell^+\ell^-$ decays.
The   transition operator $b\to c\bar cd/s$  can form an SU(3) triplet, which leads to the effective Hamiltonian:
\begin{eqnarray}
  {\cal H}_{{eff}}&= &  a_1 (T_{bb})^i (H_{  3})^j M^{k}_{j} (\overline T_{\bf{b\bar 3}})_{[ik]}~ J/\psi+ a_2 (T_{bb})^i (H_{  3})^j M^{k}_{i} (\overline T_{\bf{b\bar 3}})_{[jk]}~ J/\psi \nonumber\\
  &&+a_3 (T_{bb})^i (H_{  3})^j M^{k}_{j} (\overline T_{\bf{b6}})_{[ik]}~ J/\psi+a_4 (T_{bb})^i (H_{  3})^j M^{k}_{i} (\overline T_{\bf{b6}})_{[jk]}~ J/\psi,
\end{eqnarray}
with $(H_{  3})_{2}=V_{cd}^*$ and $(H_{  3})_{3}=V_{cs}^*$. Decay amplitudes are given in Tab.~\ref{tab:bbq_Jpsi_3bar}, from which  we derive  the relations for decay widths: Appendix~\ref{sec:relation_bb_Jpsi_baryon3_lm}.

 \begin{table}
\caption{Doubly bottom baryon decays into a $J/\psi$, light baryon  and a light meson.}\label{tab:bbq_Jpsi_3bar}
\begin{tabular}{|cc|cc|cc|}
\hline
channel  & amplitude  & channel  & amplitude  & channel  & amplitude \tabularnewline
\hline
\hline
$\Xi_{bb}^{0}\to\Lambda_{b}^{0}\pi^{0}J/\psi$  & $-\frac{\left(a_{1}+a_{2}\right)V_{cd}^{*}}{\sqrt{2}}$  & $\Xi_{bb}^{-}\to\Xi_{b}^{0}\pi^{-}J/\psi$  & $-a_{2}V_{cs}^{*}$ & $\Omega_{bb}^{-}\to\Lambda_{b}^{0}K^{-}J/\psi$  & $-a_{2}V_{cd}^{*}$\tabularnewline
\hline
$\Xi_{bb}^{0}\to\Lambda_{b}^{0}\overline{K}^{0}J/\psi$  & $a_{1}V_{cs}^{*}$  & $\Xi_{bb}^{-}\to\Xi_{b}^{-}\pi^{0}J/\psi$  & $\frac{a_{2}V_{cs}^{*}}{\sqrt{2}}$ & $\Omega_{bb}^{-}\to\Xi_{b}^{0}\pi^{-}J/\psi$  & $-a_{1}V_{cd}^{*}$\tabularnewline
\hline
$\Xi_{bb}^{0}\to\Lambda_{b}^{0}\eta J/\psi$  & $\frac{\left(a_{1}-a_{2}\right)V_{cd}^{*}}{\sqrt{6}}$  & $\Xi_{bb}^{-}\to\Xi_{b}^{-}K^{0}J/\psi$  & $\left(a_{1}+a_{2}\right)V_{cd}^{*}$ & $\Omega_{bb}^{-}\to\Xi_{b}^{0}K^{-}J/\psi$  & $-\left(a_{1}+a_{2}\right)V_{cs}^{*}$\tabularnewline
\hline
$\Xi_{bb}^{0}\to\Xi_{b}^{0}\pi^{0}J/\psi$  & $-\frac{a_{2}V_{cs}^{*}}{\sqrt{2}}$  & $\Xi_{bb}^{-}\to\Xi_{b}^{-}\eta J/\psi$  & $-\frac{\left(2a_{1}+a_{2}\right)V_{cs}^{*}}{\sqrt{6}}$ & $\Omega_{bb}^{-}\to\Xi_{b}^{-}\pi^{0}J/\psi$  & $\frac{a_{1}V_{cd}^{*}}{\sqrt{2}}$\tabularnewline
\hline
$\Xi_{bb}^{0}\to\Xi_{b}^{0}K^{0}J/\psi$  & $a_{1}V_{cd}^{*}$  & $\Xi_{bb}^{0}\to\Xi_{b}^{-}K^{+}J/\psi$  & $a_{2}V_{cd}^{*}$  & $\Omega_{bb}^{-}\to\Xi_{b}^{-}\overline{K}^{0}J/\psi$  & $-\left(a_{1}+a_{2}\right)V_{cs}^{*}$\tabularnewline
\hline
$\Xi_{bb}^{0}\to\Xi_{b}^{0}\eta J/\psi$  & $-\frac{\left(2a_{1}+a_{2}\right)V_{cs}^{*}}{\sqrt{6}}$  & $\Xi_{bb}^{-}\to\Lambda_{b}^{0}\pi^{-}J/\psi$  & $-\left(a_{1}+a_{2}\right)V_{cd}^{*}$  & $\Omega_{bb}^{-}\to\Xi_{b}^{-}\eta J/\psi$  & $-\frac{\left(a_{1}+2a_{2}\right)V_{cd}^{*}}{\sqrt{6}}$\tabularnewline
\hline
$\Xi_{bb}^{0}\to\Xi_{b}^{-}\pi^{+}J/\psi$  & $-a_{2}V_{cs}^{*}$  & $\Xi_{bb}^{-}\to\Lambda_{b}^{0}K^{-}J/\psi$  & $-a_{1}V_{cs}^{*}$  &  & \tabularnewline
\hline
\hline

$\Xi_{bb}^{0}\to\Sigma_{b}^{+}\pi^{-}J/\psi$  & $a_{3}V_{cd}^{*}$  & $\Xi_{bb}^{-}\to\Sigma_{b}^{-}\eta J/\psi$  & $\frac{\left(a_{3}+a_{4}\right)V_{cd}^{*}}{\sqrt{6}}$ & $\Omega_{bb}^{-}\to\Sigma_{b}^{0}K^{-}J/\psi$  & $\frac{a_{4}V_{cd}^{*}}{\sqrt{2}}$\tabularnewline
\hline
$\Xi_{bb}^{0}\to\Sigma_{b}^{+}K^{-}J/\psi$  & $a_{3}V_{cs}^{*}$  & $\Xi_{bb}^{-}\to\Xi_{b}^{\prime0}\pi^{-}J/\psi$  & $\frac{a_{4}V_{cs}^{*}}{\sqrt{2}}$ & $\Omega_{bb}^{-}\to\Sigma_{b}^{-}\overline{K}^{0}J/\psi$  & $a_{4}V_{cd}^{*}$\tabularnewline
\hline
$\Xi_{bb}^{0}\to\Sigma_{b}^{0}\pi^{0}J/\psi$  & $\frac{1}{2}\left(a_{4}-a_{3}\right)V_{cd}^{*}$  & $\Xi_{bb}^{-}\to\Xi_{b}^{\prime-}\pi^{0}J/\psi$  & $-\frac{1}{2}a_{4}V_{cs}^{*}$ & $\Omega_{bb}^{-}\to\Xi_{b}^{\prime0}\pi^{-}J/\psi$  & $\frac{a_{3}V_{cd}^{*}}{\sqrt{2}}$\tabularnewline
\hline
$\Xi_{bb}^{0}\to\Sigma_{b}^{0}\overline{K}^{0}J/\psi$  & $\frac{a_{3}V_{cs}^{*}}{\sqrt{2}}$  & $\Xi_{bb}^{-}\to\Xi_{b}^{\prime-}K^{0}J/\psi$  & $\frac{\left(a_{3}+a_{4}\right)V_{cd}^{*}}{\sqrt{2}}$ & $\Omega_{bb}^{-}\to\Xi_{b}^{\prime0}K^{-}J/\psi$  & $\frac{\left(a_{3}+a_{4}\right)V_{cs}^{*}}{\sqrt{2}}$\tabularnewline
\hline
$\Xi_{bb}^{0}\to\Sigma_{b}^{0}\eta J/\psi$  & $\frac{\left(a_{3}+a_{4}\right)V_{cd}^{*}}{2\sqrt{3}}$  & $\Xi_{bb}^{-}\to\Xi_{b}^{\prime-}\eta J/\psi$  & $\frac{\left(a_{4}-2a_{3}\right)V_{cs}^{*}}{2\sqrt{3}}$ & $\Omega_{bb}^{-}\to\Xi_{b}^{\prime-}\pi^{0}J/\psi$  & $-\frac{1}{2}a_{3}V_{cd}^{*}$\tabularnewline
\hline
$\Xi_{bb}^{0}\to\Sigma_{b}^{-}\pi^{+}J/\psi$  & $a_{4}V_{cd}^{*}$  & $\Xi_{bb}^{-}\to\Omega_{b}^{-}K^{0}J/\psi$  & $a_{4}V_{cs}^{*}$ & $\Omega_{bb}^{-}\to\Xi_{b}^{\prime-}\overline{K}^{0}J/\psi$  & $\frac{\left(a_{3}+a_{4}\right)V_{cs}^{*}}{\sqrt{2}}$\tabularnewline
\hline
$\Xi_{bb}^{0}\to\Xi_{b}^{\prime0}\pi^{0}J/\psi$  & $\frac{1}{2}a_{4}V_{cs}^{*}$  & $\Xi_{bb}^{0}\to\Omega_{b}^{-}K^{+}J/\psi$  & $a_{4}V_{cs}^{*}$  & $\Omega_{bb}^{-}\to\Xi_{b}^{\prime-}\eta J/\psi$  & $\frac{\left(a_{3}-2a_{4}\right)V_{cd}^{*}}{2\sqrt{3}}$\tabularnewline
\hline
$\Xi_{bb}^{0}\to\Xi_{b}^{\prime0}K^{0}J/\psi$  & $\frac{a_{3}V_{cd}^{*}}{\sqrt{2}}$  & $\Xi_{bb}^{-}\to\Sigma_{b}^{0}\pi^{-}J/\psi$  & $\frac{\left(a_{3}+a_{4}\right)V_{cd}^{*}}{\sqrt{2}}$  & $\Omega_{bb}^{-}\to\Omega_{b}^{-}K^{0}J/\psi$  & $a_{3}V_{cd}^{*}$\tabularnewline
\hline
$\Xi_{bb}^{0}\to\Xi_{b}^{\prime0}\eta J/\psi$  & $\frac{\left(a_{4}-2a_{3}\right)V_{cs}^{*}}{2\sqrt{3}}$  & $\Xi_{bb}^{-}\to\Sigma_{b}^{0}K^{-}J/\psi$  & $\frac{a_{3}V_{cs}^{*}}{\sqrt{2}}$  & $\Omega_{bb}^{-}\to\Omega_{b}^{-}\eta J/\psi$  & $-\sqrt{\frac{2}{3}}\left(a_{3}+a_{4}\right)V_{cs}^{*}$\tabularnewline
\hline
$\Xi_{bb}^{0}\to\Xi_{b}^{\prime-}\pi^{+}J/\psi$  & $\frac{a_{4}V_{cs}^{*}}{\sqrt{2}}$  & $\Xi_{bb}^{-}\to\Sigma_{b}^{-}\pi^{0}J/\psi$  & $-\frac{\left(a_{3}+a_{4}\right)V_{cd}^{*}}{\sqrt{2}}$  &  & \tabularnewline
\hline
$\Xi_{bb}^{0}\to\Xi_{b}^{\prime-}K^{+}J/\psi$  & $\frac{a_{4}V_{cd}^{*}}{\sqrt{2}}$  & $\Xi_{bb}^{-}\to\Sigma_{b}^{-}\overline{K}^{0}J/\psi$  & $a_{3}V_{cs}^{*}$  &  & \tabularnewline
\hline
\hline   \end{tabular}
\end{table}


\subsection{$b\to c\bar c d/s$:  Decays into a doubly heavy baryon $bcq$, an  anti-charmed  meson and a light meson  }

The   $b\to c\bar cd/s$   transition can induce another type of effective Hamiltonian:
\begin{eqnarray}
  {\cal H}_{{eff}}&=&   a_5  (T_{bb})^i  (H_{  3})^j  (\overline T_{bc})_i  D_k M^{k}_{j}+ a_6  (T_{bb})^i  (H_{  3})^j  (\overline T_{bc})_j  D_k M^{k}_{i} \nonumber\\
 &&+ a_7  (T_{bb})^i  (H_{  3})^j  (\overline T_{bc})_k  D_j M^{k}_{i}+ a_8  (T_{bb})^i  (H_{  3})^j  (\overline T_{bc})_k  D_i M^{k}_{j}.
\end{eqnarray}
This Hamiltonian denotes  the decays into doubly heavy baryon $bcq$  plus an  anti-charmed  meson.
Decay amplitudes are given in Tab.~\ref{tab:bbq_bcq_cbarq}. Thus we obtain the following relations for decay widths: Appendix~\ref{sec:relation_bb_bcq_cm_lm}.

\begin{table}
\caption{Doubly bottom baryon decays into a $bcq$, an anti-charmed meson and a light meson.}\label{tab:bbq_bcq_cbarq}
\tiny\begin{tabular}{cc|cc|ccc}\hline
channel  & amplitude  & channel  & amplitude & channel  & amplitude \tabularnewline
\hline
\hline
$\Xi_{bb}^{0}\to\Xi_{bc}^{+}\overline{D}^{0}\pi^{-}$  & $\left(a_{5}+a_{8}\right)V_{cd}^*$  & $\Omega_{bb}^{-}\to\Xi_{bc}^{0}\overline{D}^{0}K^{-}$  & $a_{6}V_{cd}^*$&
$\Xi_{bb}^{0}\to\Xi_{bc}^{+}\overline{D}^{0}K^{-}$  & $\left(a_{5}+a_{8}\right)V_{cs}^*$   \tabularnewline
\hline
$\Omega_{bb}^{-}\to\Xi_{bc}^{0}D^{-}\overline{K}^{0}$  & $\left(a_{6}+a_{7}\right)V_{cd}^*$&
$\Xi_{bb}^{0}\to\Xi_{bc}^{+}D^{-}\pi^{0}$  & $\frac{\left(a_{7}-a_{5}\right)V_{cd}^*}{\sqrt{2}}$  & $\Omega_{bb}^{-}\to\Xi_{bc}^{0}D_{s}^{-}\pi^{0}$  & $-\frac{a_{8}V_{cd}^*}{\sqrt{2}}$\tabularnewline
\hline
$\Xi_{bb}^{0}\to\Xi_{bc}^{+}D^{-}\overline{K}^{0}$  & $a_{5}V_{cs}^*$  & $\Omega_{bb}^{-}\to\Xi_{bc}^{0}D_{s}^{-}\overline{K}^{0}$  & $\left(a_{7}+a_{8}\right)V_{cs}^*$&
$\Xi_{bb}^{0}\to\Xi_{bc}^{+}D^{-}\eta$  & $\frac{\left(a_{5}+a_{7}\right)V_{cd}^*}{\sqrt{6}}$  \tabularnewline
\hline $\Omega_{bb}^{-}\to\Xi_{bc}^{0}D_{s}^{-}\eta$  & $\frac{\left(a_{8}-2a_{6}\right)V_{cd}^*}{\sqrt{6}}$&
$\Xi_{bb}^{0}\to\Xi_{bc}^{+}D_{s}^{-}\pi^{0}$  & $\frac{a_{7}V_{cs}^*}{\sqrt{2}}$  & $\Omega_{bb}^{-}\to\Omega_{bc}^{0}\overline{D}^{0}\pi^{-}$  & $a_{5}V_{cd}^*$\tabularnewline
\hline
$\Xi_{bb}^{0}\to\Xi_{bc}^{+}D_{s}^{-}K^{0}$  & $a_{5}V_{cd}^*$  & $\Omega_{bb}^{-}\to\Omega_{bc}^{0}\overline{D}^{0}K^{-}$  & $\left(a_{5}+a_{6}\right)V_{cs}^*$&
$\Xi_{bb}^{0}\to\Xi_{bc}^{+}D_{s}^{-}\eta$  & $\frac{\left(a_{7}-2a_{5}\right)V_{cs}^*}{\sqrt{6}}$  \tabularnewline
\hline $\Omega_{bb}^{-}\to\Omega_{bc}^{0}D^{-}\pi^{0}$  & $-\frac{a_{5}V_{cd}^*}{\sqrt{2}}$&
$\Xi_{bb}^{0}\to\Xi_{bc}^{0}\overline{D}^{0}\pi^{0}$  & $\frac{\left(a_{6}-a_{8}\right)V_{cd}^*}{\sqrt{2}}$  & $\Omega_{bb}^{-}\to\Omega_{bc}^{0}D^{-}\overline{K}^{0}$  & $\left(a_{5}+a_{6}\right)V_{cs}^*$\tabularnewline
\hline
$\Xi_{bb}^{0}\to\Xi_{bc}^{0}\overline{D}^{0}\overline{K}^{0}$  & $a_{8}V_{cs}^*$  & $\Omega_{bb}^{-}\to\Omega_{bc}^{0}D^{-}\eta$  & $\frac{\left(a_{5}-2a_{7}\right)V_{cd}^*}{\sqrt{6}}$&
$\Xi_{bb}^{0}\to\Xi_{bc}^{0}\overline{D}^{0}\eta$  & $\frac{\left(a_{6}+a_{8}\right)V_{cd}^*}{\sqrt{6}}$  \tabularnewline
\hline $\Omega_{bb}^{-}\to\Omega_{bc}^{0}D_{s}^{-}K^{0}$  & $\left(a_{5}+a_{8}\right)V_{cd}^*$&
$\Xi_{bb}^{0}\to\Xi_{bc}^{0}D^{-}\pi^{+}$  & $\left(a_{6}+a_{7}\right)V_{cd}^*$  & $\Omega_{bb}^{-}\to\Omega_{bc}^{0}D_{s}^{-}\eta$  & $-\sqrt{\frac{2}{3}}\left(a_{5}+a_{6}+a_{7}+a_{8}\right)V_{cs}^*$\tabularnewline
\hline
$\Xi_{bb}^{0}\to\Xi_{bc}^{0}D_{s}^{-}\pi^{+}$  & $a_{7}V_{cs}^*$  & $\Xi_{bb}^{-}\to\Xi_{bc}^{0}D_{s}^{-}K^{0}$  & $\left(a_{5}+a_{6}\right)V_{cd}^*$& $\Xi_{bb}^{0}\to\Xi_{bc}^{0}D_{s}^{-}K^{+}$  & $a_{6}V_{cd}^*$  \tabularnewline
\hline
 $\Xi_{bb}^{-}\to\Xi_{bc}^{0}D_{s}^{-}\eta$  & $\frac{\left(a_{7}-2a_{5}\right)V_{cs}^*}{\sqrt{6}}$&
$\Xi_{bb}^{0}\to\Omega_{bc}^{0}\overline{D}^{0}\pi^{0}$  & $\frac{a_{6}V_{cs}^*}{\sqrt{2}}$  & $\Xi_{bb}^{-}\to\Omega_{bc}^{0}\overline{D}^{0}\pi^{-}$  & $a_{6}V_{cs}^*$\tabularnewline
\hline
$\Xi_{bb}^{0}\to\Omega_{bc}^{0}\overline{D}^{0}K^{0}$  & $a_{8}V_{cd}^*$  & $\Xi_{bb}^{-}\to\Omega_{bc}^{0}D^{-}\pi^{0}$  & $-\frac{a_{6}V_{cs}^*}{\sqrt{2}}$&
$\Xi_{bb}^{0}\to\Omega_{bc}^{0}\overline{D}^{0}\eta$  & $\frac{\left(a_{6}-2a_{8}\right)V_{cs}^*}{\sqrt{6}}$  \tabularnewline
\hline $\Xi_{bb}^{-}\to\Omega_{bc}^{0}D^{-}K^{0}$  & $\left(a_{7}+a_{8}\right)V_{cd}^*$&
$\Xi_{bb}^{0}\to\Omega_{bc}^{0}D^{-}\pi^{+}$  & $a_{6}V_{cs}^*$  & $\Xi_{bb}^{-}\to\Omega_{bc}^{0}D^{-}\eta$  & $\frac{\left(a_{6}-2a_{8}\right)V_{cs}^*}{\sqrt{6}}$\tabularnewline
\hline
$\Xi_{bb}^{0}\to\Omega_{bc}^{0}D^{-}K^{+}$  & $a_{7}V_{cd}^*$  & $\Xi_{bb}^{-}\to\Omega_{bc}^{0}D_{s}^{-}K^{0}$  & $\left(a_{6}+a_{7}\right)V_{cs}^*$&
$\Xi_{bb}^{0}\to\Omega_{bc}^{0}D_{s}^{-}K^{+}$  & $\left(a_{6}+a_{7}\right)V_{cs}^*$ \tabularnewline
\hline $\Omega_{bb}^{-}\to\Xi_{bc}^{+}D^{-}K^{-}$  & $a_{7}V_{cd}^*$&
$\Xi_{bb}^{-}\to\Xi_{bc}^{+}D^{-}\pi^{-}$  & $\left(a_{7}+a_{8}\right)V_{cd}^*$  & $\Omega_{bb}^{-}\to\Xi_{bc}^{+}D_{s}^{-}\pi^{-}$  & $a_{8}V_{cd}^*$\tabularnewline
\hline
$\Xi_{bb}^{-}\to\Xi_{bc}^{+}D^{-}K^{-}$  & $a_{8}V_{cs}^*$  & $\Omega_{bb}^{-}\to\Xi_{bc}^{+}D_{s}^{-}K^{-}$  & $\left(a_{7}+a_{8}\right)V_{cs}^*$& $\Xi_{bb}^{-}\to\Xi_{bc}^{+}D_{s}^{-}\pi^{-}$  & $a_{7}V_{cs}^*$  \tabularnewline
\hline
 $\Xi_{bb}^{-}\to\Xi_{bc}^{0}D^{-}\overline{K}^{0}$  & $\left(a_{5}+a_{8}\right)V_{cs}^*$&
$\Xi_{bb}^{-}\to\Xi_{bc}^{0}\overline{D}^{0}\pi^{-}$  & $\left(a_{5}+a_{6}\right)V_{cd}^*$  & $\Xi_{bb}^{-}\to\Xi_{bc}^{0}D^{-}\eta$  & $\frac{\left(a_{5}+a_{6}+a_{7}+a_{8}\right)V_{cd}^*}{\sqrt{6}}$\tabularnewline
\hline
$\Xi_{bb}^{-}\to\Xi_{bc}^{0}\overline{D}^{0}K^{-}$  & $a_{5}V_{cs}^*$  & $\Xi_{bb}^{-}\to\Xi_{bc}^{0}D_{s}^{-}\pi^{0}$  & $-\frac{a_{7}V_{cs}^*}{\sqrt{2}}$  &
$\Xi_{bb}^{-}\to\Xi_{bc}^{0}D^{-}\pi^{0}$  & $-\frac{\left(a_{5}+a_{6}+a_{7}+a_{8}\right)V_{cd}^*}{\sqrt{2}}$   \tabularnewline
\hline
\hline
 \end{tabular}
\end{table}

\subsection{$b\to c \bar u d/s$ transition: Decays into a doubly heavy baryon $bcq$ plus two light mesons }

The operator to produce a charm quark  from the $b$-quark decay, $\bar c b \bar q u$, is given by
\begin{eqnarray}
{\cal H}_{eff} &=& \frac{G_{F}}{\sqrt{2}}
     V_{cb} V_{uq}^{*} \big[
     C_{1}  O^{\bar cu}_{1}
  +  C_{2}  O^{\bar cu}_{2}\Big] +{\rm h.c.} .
\end{eqnarray}
The light quarks in this effective Hamiltonian form an octet with the nonzero entry
$
(H_{{\bf8}})^2_1 =V_{ud}^*
$
for   the $b\to c\bar ud$ transition, and   $(H_{{\bf8}})^3_1 =V_{us}^*$ for  the $b\to c\bar
us$ transition.
The hadron-level effective Hamiltonian is then given as
\begin{eqnarray}
  {\cal H}_{{eff}}&=& a_9 (T_{bb})^i (\overline T_{bc})_i M^j_k M^l_j (H_{{\bf8}})^k_l+ a_{10} (T_{bb})^i (\overline T_{bc})_j M^j_i M^k_l (H_{{\bf8}})^l_k\nonumber \\
&&+a_{11} (T_{bb})^i (\overline T_{bc})_j M^l_i M^j_k (H_{{\bf8}})^k_l+a_{12} (T_{bb})^i (\overline T_{bc})_j M^l_i M^k_l (H_{{\bf8}})^j_k \nonumber \\
&&+a_{13} (T_{bb})^i (\overline T_{bc})_j M^j_k M^k_l (H_{{\bf8}})^l_i+a_{14} (T_{bb})^i (\overline T_{bc})_j M^k_l M^l_k (H_{{\bf8}})^j_i.
\end{eqnarray}
Decay amplitudes are expanded  in Tab.~\ref{tab:bbq_bcq_qqbar}, which  leads to the relations: Appendix~\ref{sec:relation_bb_bcq_2lm}.

 \begin{table}
\caption{Doubly bottom baryon decays into a $bcq$ and two light mesons.}\label{tab:bbq_bcq_qqbar}\begin{tabular}{|c  c|c  c|c|c|c|c|c|c}\hline
channel  & amplitude  & channel  & amplitude \tabularnewline
\hline
\hline
$\Xi_{bb}^{0}\to\Xi_{bc}^{+}\pi^{0}\pi^{-}$  & $\frac{\left(a_{10}+a_{11}\right)V_{ud}^*}{\sqrt{2}}$ & $\Xi_{bb}^{-}\to\Xi_{bc}^{0}K^{-}\eta$  & $\frac{\left(a_{10}-a_{9}\right)V_{us}^*}{\sqrt{6}}$\tabularnewline
\hline
$\Xi_{bb}^{0}\to\Xi_{bc}^{+}\pi^{0}K^{-}$  & $\frac{\left(a_{9}+a_{10}+a_{11}+a_{13}\right)V_{us}^*}{\sqrt{2}}$ & $\Xi_{bb}^{-}\to\Omega_{bc}^{0}\pi^{-}K^{0}$  & $\left(a_{10}+a_{11}\right)V_{ud}^*$\tabularnewline
\hline
$\Xi_{bb}^{0}\to\Xi_{bc}^{+}\pi^{-}\overline{K}^{0}$  & $\left(a_{9}+a_{13}\right)V_{us}^*$ & $\Xi_{bb}^{-}\to\Omega_{bc}^{0}\pi^{-}\eta$  & $\sqrt{\frac{2}{3}}\left(a_{12}-a_{11}\right)V_{us}^*$\tabularnewline
\hline
$\Xi_{bb}^{0}\to\Xi_{bc}^{+}\pi^{-}\eta$  & $\frac{\left(2a_{9}+a_{10}+a_{11}+2a_{13}\right)V_{ud}^*}{\sqrt{6}}$ & $\Xi_{bb}^{-}\to\Omega_{bc}^{0}K^{0}K^{-}$  & $\left(a_{10}+a_{12}\right)V_{us}^*$\tabularnewline
\hline
$\Xi_{bb}^{0}\to\Xi_{bc}^{+}K^{0}K^{-}$  & $\left(a_{9}+a_{13}\right)V_{ud}^*$ & $\Omega_{bb}^{-}\to\Xi_{bc}^{+}\pi^{-}K^{-}$  & $\left(a_{10}+a_{11}\right)V_{ud}^*$\tabularnewline
\hline
$\Xi_{bb}^{0}\to\Xi_{bc}^{+}K^{-}\eta$  & $\frac{\left(-a_{9}+a_{10}+a_{11}-a_{13}\right)V_{us}^*}{\sqrt{6}}$ & $\Omega_{bb}^{-}\to\Xi_{bc}^{+}K^{-}K^{-}$  & $2\left(a_{10}+a_{11}\right)V_{us}^*$\tabularnewline
\hline
$\Xi_{bb}^{0}\to\Xi_{bc}^{0}\pi^{+}\pi^{-}$  & $\left(a_{10}+a_{12}+a_{13}+2a_{14}\right)V_{ud}^*$ & $\Omega_{bb}^{-}\to\Xi_{bc}^{0}\pi^{0}K^{-}$  & $\frac{\left(a_{12}-a_{11}\right)V_{ud}^*}{\sqrt{2}}$\tabularnewline
\hline
$\Xi_{bb}^{0}\to\Xi_{bc}^{0}\pi^{+}K^{-}$  & $\left(a_{10}+a_{13}\right)V_{us}^*$ & $\Omega_{bb}^{-}\to\Xi_{bc}^{0}\pi^{-}\overline{K}^{0}$  & $\left(a_{10}+a_{12}\right)V_{ud}^*$\tabularnewline
\hline
$\Xi_{bb}^{0}\to\Xi_{bc}^{0}\pi^{0}\pi^{0}$  & $\left(-a_{11}+a_{12}+a_{13}+2a_{14}\right)V_{ud}^*$ & $\Omega_{bb}^{-}\to\Xi_{bc}^{0}\overline{K}^{0}K^{-}$  & $\left(a_{10}+a_{11}\right)V_{us}^*$\tabularnewline
\hline
$\Xi_{bb}^{0}\to\Xi_{bc}^{0}\pi^{0}\overline{K}^{0}$  & $\frac{\left(a_{11}-a_{13}\right)V_{us}^*}{\sqrt{2}}$ & $\Omega_{bb}^{-}\to\Xi_{bc}^{0}K^{-}\eta$  & $\frac{\left(a_{11}-a_{12}\right)V_{ud}^*}{\sqrt{6}}$\tabularnewline
\hline
$\Xi_{bb}^{0}\to\Xi_{bc}^{0}\pi^{0}\eta$  & $\frac{\left(a_{12}-a_{13}\right)V_{ud}^*}{\sqrt{3}}$ & $\Omega_{bb}^{-}\to\Omega_{bc}^{0}\pi^{0}K^{-}$  & $\frac{\left(a_{9}+a_{12}\right)V_{us}^*}{\sqrt{2}}$\tabularnewline
\hline
$\Xi_{bb}^{0}\to\Xi_{bc}^{0}K^{+}K^{-}$  & $\left(a_{12}+2a_{14}\right)V_{ud}^*$ & $\Omega_{bb}^{-}\to\Omega_{bc}^{0}\pi^{-}\overline{K}^{0}$  & $\left(a_{9}+a_{12}\right)V_{us}^*$\tabularnewline
\hline
$\Xi_{bb}^{0}\to\Xi_{bc}^{0}K^{0}\overline{K}^{0}$  & $\left(a_{13}+2a_{14}\right)V_{ud}^*$ & $\Omega_{bb}^{-}\to\Omega_{bc}^{0}\pi^{-}\eta$  & $\sqrt{\frac{2}{3}}\left(a_{9}-a_{10}\right)V_{ud}^*$\tabularnewline
\hline
$\Xi_{bb}^{0}\to\Xi_{bc}^{0}\overline{K}^{0}\eta$  & $\frac{\left(a_{11}-a_{13}\right)V_{us}^*}{\sqrt{6}}$ & $\Omega_{bb}^{-}\to\Omega_{bc}^{0}K^{0}K^{-}$  & $\left(a_{9}+a_{11}\right)V_{ud}^*$\tabularnewline
\hline
$\Xi_{bb}^{0}\to\Xi_{bc}^{0}\eta\eta$  & $\frac{1}{3}\left(a_{11}+a_{12}+a_{13}+6a_{14}\right)V_{ud}^*$ & $\Omega_{bb}^{-}\to\Omega_{bc}^{0}K^{-}\eta$  & $-\frac{\left(a_{9}+2a_{10}+2a_{11}+a_{12}\right)V_{us}^*}{\sqrt{6}}$\tabularnewline
\hline
$\Xi_{bb}^{0}\to\Omega_{bc}^{0}\pi^{+}\pi^{-}$  & $\left(a_{12}+2a_{14}\right)V_{us}^*$ & $\Xi_{bb}^{0}\to\Omega_{bc}^{0}\eta\eta$  & $\frac{1}{3}\left(-2a_{11}+a_{12}+4a_{13}+6a_{14}\right)V_{us}^*$\tabularnewline
\hline
$\Xi_{bb}^{0}\to\Omega_{bc}^{0}\pi^{0}\pi^{0}$  & $\left(a_{12}+2a_{14}\right)V_{us}^*$ & $\Xi_{bb}^{-}\to\Xi_{bc}^{+}\pi^{-}\pi^{-}$  & $2\left(a_{10}+a_{11}\right)V_{ud}^*$\tabularnewline
\hline
$\Xi_{bb}^{0}\to\Omega_{bc}^{0}\pi^{0}K^{0}$  & $\frac{\left(a_{11}-a_{13}\right)V_{ud}^*}{\sqrt{2}}$ & $\Xi_{bb}^{-}\to\Xi_{bc}^{+}\pi^{-}K^{-}$  & $\left(a_{10}+a_{11}\right)V_{us}^*$\tabularnewline
\hline
$\Xi_{bb}^{0}\to\Omega_{bc}^{0}\pi^{0}\eta$  & $\frac{\left(a_{12}-a_{11}\right)V_{us}^*}{\sqrt{3}}$ & $\Xi_{bb}^{-}\to\Xi_{bc}^{0}\pi^{0}\pi^{-}$  & $-\frac{\left(a_{10}+a_{11}\right)V_{ud}^*}{\sqrt{2}}$\tabularnewline
\hline
$\Xi_{bb}^{0}\to\Omega_{bc}^{0}\pi^{-}K^{+}$  & $\left(a_{10}+a_{13}\right)V_{ud}^*$ & $\Xi_{bb}^{-}\to\Xi_{bc}^{0}\pi^{0}K^{-}$  & $\frac{\left(a_{9}-a_{10}\right)V_{us}^*}{\sqrt{2}}$\tabularnewline
\hline
$\Xi_{bb}^{0}\to\Omega_{bc}^{0}K^{+}K^{-}$  & $\left(a_{10}+a_{12}+a_{13}+2a_{14}\right)V_{us}^*$ & $\Xi_{bb}^{-}\to\Xi_{bc}^{0}\pi^{-}\overline{K}^{0}$  & $\left(a_{9}+a_{11}\right)V_{us}^*$\tabularnewline
\hline
$\Xi_{bb}^{0}\to\Omega_{bc}^{0}K^{0}\overline{K}^{0}$  & $\left(a_{13}+2a_{14}\right)V_{us}^*$ & $\Xi_{bb}^{-}\to\Xi_{bc}^{0}\pi^{-}\eta$  & $\frac{\left(2a_{9}+a_{10}+a_{11}+2a_{12}\right)V_{ud}^*}{\sqrt{6}}$\tabularnewline
\hline
$\Xi_{bb}^{0}\to\Omega_{bc}^{0}K^{0}\eta$  & $\frac{\left(a_{11}-a_{13}\right)V_{ud}^*}{\sqrt{6}}$ & $\Xi_{bb}^{-}\to\Xi_{bc}^{0}K^{0}K^{-}$  & $\left(a_{9}+a_{12}\right)V_{ud}^*$\tabularnewline
\hline
\hline
\end{tabular}
\end{table}

\subsection{$b\to c \bar u d/s$ transition: Decays into a bottom  baryon $bqq$, a charmed meson and a light meson  }

The effective Hamiltonian from the operator $\bar c b \bar q u$  gives
\begin{eqnarray}
  {\cal H}_{{eff}}&=& a_{15} (T_{bb})^i (\overline T_{b\bar 3})_{[ij]} \overline D^j M^k_l (H_{{\bf8}})^l_k+ a_{16} (T_{bb})^i (\overline T_{b\bar 3})_{[ij]} \overline D^l M^j_k (H_{{\bf8}})^k_l  \nonumber\\
  &&+  a_{17} (T_{bb})^i (\overline T_{b\bar 3})_{[ij]} \overline D^l M^k_l (H_{{\bf8}})^j_k+ a_{18} (T_{bb})^i (\overline T_{b\bar 3})_{[jk]} \overline D^j M^l_i (H_{{\bf8}})^k_l \nonumber\\
&&+a_{19} (T_{bb})^i (\overline T_{b\bar 3})_{[jk]} \overline D^l M^j_i (H_{{\bf8}})^k_l+a_{20} (T_{bb})^i (\overline T_{b\bar 3})_{[jk]} \overline D^l M^k_l (H_{{\bf8}})^j_i\nonumber\\
&&+a_{21} (T_{bb})^i (\overline T_{b\bar 3})_{[kl]} \overline D^l M^k_j (H_{{\bf8}})^j_i+a_{22} (T_{bb})^i (\overline T_{b6})_{[ij]} \overline D^j M^k_l (H_{{\bf8}})^l_k \nonumber\\
&&+ a_{23} (T_{bb})^i (\overline T_{b6})_{[ij]} \overline D^l M^j_k (H_{{\bf8}})^k_l +  a_{24} (T_{bb})^i (\overline T_{b6})_{[ij]} \overline D^l M^k_l (H_{{\bf8}})^j_k \nonumber\\
&&+ a_{25} (T_{bb})^i (\overline T_{b6})_{[jk]} \overline D^j M^l_i (H_{{\bf8}})^k_l+a_{26} (T_{bb})^i (\overline T_{b6})_{[jk]} \overline D^l M^j_i (H_{{\bf8}})^k_l \nonumber\\
&&+a_{27} (T_{bb})^i (\overline T_{b6})_{[jk]} \overline D^l M^k_l (H_{{\bf8}})^j_i+a_{28} (T_{bb})^i (\overline T_{b6})_{[kl]} \overline D^l M^k_j (H_{{\bf8}})^j_i.
\end{eqnarray}

Results are given in Tab.~\ref{tab:bbq_bqq_cqbar_3bar} for anti-triplet and Tab.~\ref{tab:bbq_bqq_cqbar_6} for sextet, thus we have the relations for decay amplitudes: Appendix~\ref{sec:relation_bb_bqq6_cm_lm} for sextet. Actually, for the anti-triplet case there's no definite relations between the decay withs.


 \begin{table}
\caption{Doubly bottom baryon decays into a $bqq$(anti-triplet), a charmed meson and a light meson.}\label{tab:bbq_bqq_cqbar_3bar}\begin{tabular}{|c  c|c  c|c|c|c|c|c|c}\hline
channel  & amplitude  & channel  & amplitude \tabularnewline
\hline
\hline
$\Xi_{bb}^{0}\to\Lambda_{b}^{0}D^{0}\pi^{0}$  & $\frac{\left(-a_{16}+a_{17}+a_{18}+a_{19}-a_{20}+a_{21}\right)V_{ud}^*}{\sqrt{2}}$ & $\Omega_{bb}^{-}\to\Lambda_{b}^{0}D^{0}K^{-}$  & $\left(a_{18}+a_{19}\right)V_{ud}^*$\tabularnewline
\hline
$\Xi_{bb}^{0}\to\Lambda_{b}^{0}D^{0}\overline{K}^{0}$  & $\left(a_{16}-a_{21}\right)V_{us}^*$ & $\Omega_{bb}^{-}\to\Xi_{b}^{0}D^{0}\pi^{-}$  & $-\left(a_{15}+a_{16}\right)V_{ud}^*$\tabularnewline
\hline
$\Xi_{bb}^{0}\to\Lambda_{b}^{0}D^{0}\eta$  & $\frac{\left(a_{16}+a_{17}+a_{18}+a_{19}-a_{20}-a_{21}\right)V_{ud}^*}{\sqrt{6}}$ & $\Omega_{bb}^{-}\to\Xi_{b}^{0}D^{0}K^{-}$  & $\left(-a_{15}-a_{16}+a_{18}+a_{19}\right)V_{us}^*$\tabularnewline
\hline
$\Xi_{bb}^{0}\to\Lambda_{b}^{0}D^{+}\pi^{-}$  & $\left(a_{15}+a_{17}-a_{20}+a_{21}\right)V_{ud}^*$ & $\Omega_{bb}^{-}\to\Xi_{b}^{-}D^{0}\pi^{0}$  & $\frac{\left(a_{16}-a_{17}\right)V_{ud}^*}{\sqrt{2}}$\tabularnewline
\hline
$\Xi_{bb}^{0}\to\Lambda_{b}^{0}D^{+}K^{-}$  & $\left(a_{15}+a_{21}\right)V_{us}^*$ & $\Omega_{bb}^{-}\to\Xi_{b}^{-}D^{0}\overline{K}^{0}$  & $\left(a_{19}-a_{16}\right)V_{us}^*$\tabularnewline
\hline
$\Xi_{bb}^{0}\to\Lambda_{b}^{0}D_{s}^{+}K^{-}$  & $\left(a_{17}-a_{20}\right)V_{ud}^*$ & $\Omega_{bb}^{-}\to\Xi_{b}^{-}D^{0}\eta$  & $-\frac{\left(a_{16}+a_{17}-2a_{19}\right)V_{ud}^*}{\sqrt{6}}$\tabularnewline
\hline
$\Xi_{bb}^{0}\to\Xi_{b}^{0}D^{0}\pi^{0}$  & $\frac{\left(a_{17}+a_{18}+a_{19}-a_{20}\right)V_{us}^*}{\sqrt{2}}$ & $\Omega_{bb}^{-}\to\Xi_{b}^{-}D^{+}\pi^{-}$  & $-\left(a_{15}+a_{17}\right)V_{ud}^*$\tabularnewline
\hline
$\Xi_{bb}^{0}\to\Xi_{b}^{0}D^{0}K^{0}$  & $\left(a_{16}-a_{21}\right)V_{ud}^*$ & $\Omega_{bb}^{-}\to\Xi_{b}^{-}D^{+}K^{-}$  & $\left(a_{18}-a_{15}\right)V_{us}^*$\tabularnewline
\hline
$\Xi_{bb}^{0}\to\Xi_{b}^{0}D^{0}\eta$  & $\frac{\left(-2a_{16}+a_{17}+a_{18}+a_{19}-a_{20}+2a_{21}\right)V_{us}^*}{\sqrt{6}}$ & $\Omega_{bb}^{-}\to\Xi_{b}^{-}D_{s}^{+}K^{-}$  & $-\left(a_{17}+a_{18}\right)V_{ud}^*$\tabularnewline
\hline
$\Xi_{bb}^{0}\to\Xi_{b}^{0}D^{+}\pi^{-}$  & $\left(a_{17}-a_{20}\right)V_{us}^*$ & $\Xi_{bb}^{0}\to\Xi_{b}^{-}D_{s}^{+}\overline{K}^{0}$  & $\left(a_{21}-a_{20}\right)V_{us}^*$\tabularnewline
\hline
$\Xi_{bb}^{0}\to\Xi_{b}^{0}D_{s}^{+}\pi^{-}$  & $\left(a_{15}+a_{21}\right)V_{ud}^*$ & $\Xi_{bb}^{0}\to\Xi_{b}^{-}D_{s}^{+}\eta$  & $\frac{\left(-a_{18}-2a_{20}+a_{21}\right)V_{ud}^*}{\sqrt{6}}$\tabularnewline
\hline
$\Xi_{bb}^{0}\to\Xi_{b}^{0}D_{s}^{+}K^{-}$  & $\left(a_{15}+a_{17}-a_{20}+a_{21}\right)V_{us}^*$ & $\Xi_{bb}^{-}\to\Lambda_{b}^{0}D^{0}\pi^{-}$  & $\left(-a_{15}-a_{16}+a_{18}+a_{19}\right)V_{ud}^*$\tabularnewline
\hline
$\Xi_{bb}^{0}\to\Xi_{b}^{-}D^{0}\pi^{+}$  & $\left(a_{19}-a_{20}\right)V_{us}^*$ & $\Xi_{bb}^{-}\to\Lambda_{b}^{0}D^{0}K^{-}$  & $-\left(a_{15}+a_{16}\right)V_{us}^*$\tabularnewline
\hline
$\Xi_{bb}^{0}\to\Xi_{b}^{-}D^{0}K^{+}$  & $\left(a_{20}-a_{19}\right)V_{ud}^*$ & $\Xi_{bb}^{-}\to\Xi_{b}^{0}D^{0}\pi^{-}$  & $\left(a_{18}+a_{19}\right)V_{us}^*$\tabularnewline
\hline
$\Xi_{bb}^{0}\to\Xi_{b}^{-}D^{+}\pi^{0}$  & $\frac{\left(a_{18}+a_{20}\right)V_{us}^*}{\sqrt{2}}$ & $\Xi_{bb}^{-}\to\Xi_{b}^{-}D^{0}\pi^{0}$  & $\frac{\left(a_{17}-a_{19}\right)V_{us}^*}{\sqrt{2}}$\tabularnewline
\hline
$\Xi_{bb}^{0}\to\Xi_{b}^{-}D^{+}K^{0}$  & $\left(a_{20}-a_{21}\right)V_{ud}^*$ & $\Xi_{bb}^{-}\to\Xi_{b}^{-}D^{0}K^{0}$  & $\left(a_{16}-a_{19}\right)V_{ud}^*$\tabularnewline
\hline
$\Xi_{bb}^{0}\to\Xi_{b}^{-}D^{+}\eta$  & $\frac{\left(a_{18}-a_{20}+2a_{21}\right)V_{us}^*}{\sqrt{6}}$ & $\Xi_{bb}^{-}\to\Xi_{b}^{-}D^{0}\eta$  & $\frac{\left(-2a_{16}+a_{17}+a_{19}\right)V_{us}^*}{\sqrt{6}}$\tabularnewline
\hline
$\Xi_{bb}^{0}\to\Xi_{b}^{-}D_{s}^{+}\pi^{0}$  & $-\frac{\left(a_{18}+a_{21}\right)V_{ud}^*}{\sqrt{2}}$ & $\Xi_{bb}^{-}\to\Xi_{b}^{-}D^{+}\pi^{-}$  & $\left(a_{17}+a_{18}\right)V_{us}^*$\tabularnewline
\hline
$\Xi_{bb}^{-}\to\Xi_{b}^{-}D_{s}^{+}K^{-}$  & $\left(a_{15}+a_{17}\right)V_{us}^*$ & $\Xi_{bb}^{-}\to\Xi_{b}^{-}D_{s}^{+}\pi^{-}$  & $\left(a_{15}-a_{18}\right)V_{ud}^*$\tabularnewline
\hline
\hline \end{tabular}
\end{table}

\begin{table}
\caption{Doubly bottom baryon decays into a $bqq$(sextet), a charmed meson and a light meson.}\label{tab:bbq_bqq_cqbar_6}
\begin{tabular}{|cc|cc|}
\hline
channel  & amplitude  & channel  & amplitude \tabularnewline
\hline
\hline
$\Xi_{bb}^{0}\to\Sigma_{b}^{+}D^{0}\pi^{-}$  & $\left(a_{22}+a_{23}+a_{28}\right)V_{ud}^*$ & $\Omega_{bb}^{-}\to\Xi_{b}^{\prime-}D^{0}\eta$  & $\frac{\left(a_{23}+a_{24}-2a_{26}\right)V_{ud}^*}{2\sqrt{3}}$\tabularnewline
\hline
$\Xi_{bb}^{0}\to\Sigma_{b}^{+}D^{0}K^{-}$  & $\left(a_{22}+a_{23}+a_{28}\right)V_{us}^*$ & $\Omega_{bb}^{-}\to\Xi_{b}^{\prime-}D^{+}\pi^{-}$  & $\frac{\left(a_{22}+a_{24}\right)V_{ud}^*}{\sqrt{2}}$\tabularnewline
\hline
$\Xi_{bb}^{0}\to\Sigma_{b}^{0}D^{0}\pi^{0}$  & $\frac{1}{2}\left(-a_{23}+a_{24}+a_{25}+a_{26}+a_{27}-a_{28}\right)V_{ud}^*$ & $\Omega_{bb}^{-}\to\Xi_{b}^{\prime-}D^{+}K^{-}$  & $\frac{\left(a_{22}+a_{25}\right)V_{us}^*}{\sqrt{2}}$\tabularnewline
\hline
$\Xi_{bb}^{0}\to\Sigma_{b}^{0}D^{0}\overline{K}^{0}$  & $\frac{\left(a_{23}+a_{28}\right)V_{us}^*}{\sqrt{2}}$ & $\Omega_{bb}^{-}\to\Xi_{b}^{\prime-}D_{s}^{+}K^{-}$  & $\frac{\left(a_{24}+a_{25}\right)V_{ud}^*}{\sqrt{2}}$\tabularnewline
\hline
$\Xi_{bb}^{0}\to\Sigma_{b}^{0}D^{0}\eta$  & $\frac{\left(a_{23}+a_{24}+a_{25}+a_{26}+a_{27}+a_{28}\right)V_{ud}^*}{2\sqrt{3}}$ & $\Omega_{bb}^{-}\to\Omega_{b}^{-}D^{0}\pi^{0}$  & $\frac{a_{24}V_{us}^*}{\sqrt{2}}$\tabularnewline
\hline
$\Xi_{bb}^{0}\to\Sigma_{b}^{0}D^{+}\pi^{-}$  & $\frac{\left(a_{22}+a_{24}+a_{27}+a_{28}\right)V_{ud}^*}{\sqrt{2}}$ & $\Omega_{bb}^{-}\to\Omega_{b}^{-}D^{0}K^{0}$  & $a_{23}V_{ud}^*$\tabularnewline
\hline
$\Xi_{bb}^{0}\to\Sigma_{b}^{0}D^{+}K^{-}$  & $\frac{\left(a_{22}+a_{28}\right)V_{us}^*}{\sqrt{2}}$ & $\Omega_{bb}^{-}\to\Omega_{b}^{-}D^{0}\eta$  & $\frac{\left(-2a_{23}+a_{24}-2a_{26}\right)V_{us}^*}{\sqrt{6}}$\tabularnewline
\hline
$\Xi_{bb}^{0}\to\Sigma_{b}^{0}D_{s}^{+}K^{-}$  & $\frac{\left(a_{24}+a_{27}\right)V_{ud}^*}{\sqrt{2}}$ & $\Omega_{bb}^{-}\to\Omega_{b}^{-}D^{+}\pi^{-}$  & $a_{24}V_{us}^*$\tabularnewline
\hline
$\Xi_{bb}^{0}\to\Sigma_{b}^{-}D^{0}\pi^{+}$  & $\left(a_{26}+a_{27}\right)V_{ud}^*$ & $\Omega_{bb}^{-}\to\Omega_{b}^{-}D_{s}^{+}\pi^{-}$  & $a_{22}V_{ud}^*$\tabularnewline
\hline
$\Xi_{bb}^{0}\to\Sigma_{b}^{-}D^{+}\pi^{0}$  & $\frac{\left(a_{25}-a_{27}-a_{28}\right)V_{ud}^*}{\sqrt{2}}$ & $\Omega_{bb}^{-}\to\Omega_{b}^{-}D_{s}^{+}K^{-}$  & $\left(a_{22}+a_{24}+a_{25}\right)V_{us}^*$\tabularnewline
\hline
$\Xi_{bb}^{0}\to\Sigma_{b}^{-}D^{+}\overline{K}^{0}$  & $a_{28}V_{us}^*$ & $\Xi_{bb}^{-}\to\Xi_{b}^{\prime-}D_{s}^{+}K^{-}$  & $\frac{\left(a_{22}+a_{24}\right)V_{us}^*}{\sqrt{2}}$\tabularnewline
\hline
$\Xi_{bb}^{0}\to\Sigma_{b}^{-}D^{+}\eta$  & $\frac{\left(a_{25}+a_{27}+a_{28}\right)V_{ud}^*}{\sqrt{6}}$ & $\Xi_{bb}^{-}\to\Omega_{b}^{-}D^{0}K^{0}$  & $a_{26}V_{us}^*$\tabularnewline
\hline
$\Xi_{bb}^{0}\to\Sigma_{b}^{-}D_{s}^{+}\overline{K}^{0}$  & $a_{27}V_{ud}^*$ & $\Xi_{bb}^{-}\to\Omega_{b}^{-}D_{s}^{+}\pi^{-}$  & $a_{25}V_{us}^*$\tabularnewline
\hline
$\Xi_{bb}^{0}\to\Xi_{b}^{\prime0}D^{0}\pi^{0}$  & $\frac{1}{2}\left(a_{24}+a_{25}+a_{26}+a_{27}\right)V_{us}^*$ & $\Omega_{bb}^{-}\to\Sigma_{b}^{0}D^{0}K^{-}$  & $\frac{\left(a_{25}+a_{26}\right)V_{ud}^*}{\sqrt{2}}$\tabularnewline
\hline
$\Xi_{bb}^{0}\to\Xi_{b}^{\prime0}D^{0}K^{0}$  & $\frac{\left(a_{23}+a_{28}\right)V_{ud}^*}{\sqrt{2}}$ & $\Omega_{bb}^{-}\to\Sigma_{b}^{-}D^{0}\overline{K}^{0}$  & $a_{26}V_{ud}^*$\tabularnewline
\hline
$\Xi_{bb}^{0}\to\Xi_{b}^{\prime0}D^{0}\eta$  & $\frac{\left(-2a_{23}+a_{24}+a_{25}+a_{26}+a_{27}-2a_{28}\right)V_{us}^*}{2\sqrt{3}}$ & $\Omega_{bb}^{-}\to\Sigma_{b}^{-}D^{+}K^{-}$  & $a_{25}V_{ud}^*$\tabularnewline
\hline
$\Xi_{bb}^{0}\to\Xi_{b}^{\prime0}D^{+}\pi^{-}$  & $\frac{\left(a_{24}+a_{27}\right)V_{us}^*}{\sqrt{2}}$ & $\Omega_{bb}^{-}\to\Xi_{b}^{\prime0}D^{0}\pi^{-}$  & $\frac{\left(a_{22}+a_{23}\right)V_{ud}^*}{\sqrt{2}}$\tabularnewline
\hline
$\Xi_{bb}^{0}\to\Xi_{b}^{\prime0}D_{s}^{+}\pi^{-}$  & $\frac{\left(a_{22}+a_{28}\right)V_{ud}^*}{\sqrt{2}}$ & $\Omega_{bb}^{-}\to\Xi_{b}^{\prime0}D^{0}K^{-}$  & $\frac{\left(a_{22}+a_{23}+a_{25}+a_{26}\right)V_{us}^*}{\sqrt{2}}$\tabularnewline
\hline
$\Xi_{bb}^{0}\to\Xi_{b}^{\prime0}D_{s}^{+}K^{-}$  & $\frac{\left(a_{22}+a_{24}+a_{27}+a_{28}\right)V_{us}^*}{\sqrt{2}}$ & $\Omega_{bb}^{-}\to\Xi_{b}^{\prime-}D^{0}\pi^{0}$  & $\frac{1}{2}\left(a_{24}-a_{23}\right)V_{ud}^*$\tabularnewline
\hline
$\Xi_{bb}^{0}\to\Xi_{b}^{\prime-}D^{0}\pi^{+}$  & $\frac{\left(a_{26}+a_{27}\right)V_{us}^*}{\sqrt{2}}$ & $\Omega_{bb}^{-}\to\Xi_{b}^{\prime-}D^{0}\overline{K}^{0}$  & $\frac{\left(a_{23}+a_{26}\right)V_{us}^*}{\sqrt{2}}$\tabularnewline
\hline
$\Xi_{bb}^{0}\to\Xi_{b}^{\prime-}D^{0}K^{+}$  & $\frac{\left(a_{26}+a_{27}\right)V_{ud}^*}{\sqrt{2}}$ & $\Xi_{bb}^{-}\to\Sigma_{b}^{-}D^{0}\eta$  & $\frac{\left(a_{23}+a_{24}+a_{26}\right)V_{ud}^*}{\sqrt{6}}$\tabularnewline
\hline
$\Xi_{bb}^{0}\to\Xi_{b}^{\prime-}D^{+}\pi^{0}$  & $\frac{1}{2}\left(a_{25}-a_{27}\right)V_{us}^*$ & $\Xi_{bb}^{-}\to\Sigma_{b}^{-}D^{+}\pi^{-}$  & $\left(a_{22}+a_{24}+a_{25}\right)V_{ud}^*$\tabularnewline
\hline
$\Xi_{bb}^{0}\to\Xi_{b}^{\prime-}D^{+}K^{0}$  & $\frac{\left(a_{27}+a_{28}\right)V_{ud}^*}{\sqrt{2}}$ & $\Xi_{bb}^{-}\to\Sigma_{b}^{-}D^{+}K^{-}$  & $a_{22}V_{us}^*$\tabularnewline
\hline
$\Xi_{bb}^{0}\to\Xi_{b}^{\prime-}D^{+}\eta$  & $\frac{\left(a_{25}+a_{27}-2a_{28}\right)V_{us}^*}{2\sqrt{3}}$ & $\Xi_{bb}^{-}\to\Sigma_{b}^{-}D_{s}^{+}K^{-}$  & $a_{24}V_{ud}^*$\tabularnewline
\hline
$\Xi_{bb}^{0}\to\Xi_{b}^{\prime-}D_{s}^{+}\pi^{0}$  & $\frac{1}{2}\left(a_{25}-a_{28}\right)V_{ud}^*$ & $\Xi_{bb}^{-}\to\Xi_{b}^{\prime0}D^{0}\pi^{-}$  & $\frac{\left(a_{25}+a_{26}\right)V_{us}^*}{\sqrt{2}}$\tabularnewline
\hline
$\Xi_{bb}^{0}\to\Xi_{b}^{\prime-}D_{s}^{+}\overline{K}^{0}$  & $\frac{\left(a_{27}+a_{28}\right)V_{us}^*}{\sqrt{2}}$ & $\Xi_{bb}^{-}\to\Xi_{b}^{\prime-}D^{0}\pi^{0}$  & $\frac{1}{2}\left(a_{24}-a_{26}\right)V_{us}^*$\tabularnewline
\hline
$\Xi_{bb}^{0}\to\Xi_{b}^{\prime-}D_{s}^{+}\eta$  & $\frac{\left(a_{25}-2a_{27}+a_{28}\right)V_{ud}^*}{2\sqrt{3}}$ & $\Xi_{bb}^{-}\to\Xi_{b}^{\prime-}D^{0}K^{0}$  & $\frac{\left(a_{23}+a_{26}\right)V_{ud}^*}{\sqrt{2}}$\tabularnewline
\hline
$\Xi_{bb}^{0}\to\Omega_{b}^{-}D^{0}K^{+}$  & $\left(a_{26}+a_{27}\right)V_{us}^*$ & $\Xi_{bb}^{-}\to\Xi_{b}^{\prime-}D^{0}\eta$  & $\frac{\left(-2a_{23}+a_{24}+a_{26}\right)V_{us}^*}{2\sqrt{3}}$\tabularnewline
\hline
$\Xi_{bb}^{0}\to\Omega_{b}^{-}D^{+}K^{0}$  & $a_{27}V_{us}^*$ & $\Xi_{bb}^{-}\to\Xi_{b}^{\prime-}D^{+}\pi^{-}$  & $\frac{\left(a_{24}+a_{25}\right)V_{us}^*}{\sqrt{2}}$\tabularnewline
\hline
$\Xi_{bb}^{0}\to\Omega_{b}^{-}D_{s}^{+}\pi^{0}$  & $\frac{a_{25}V_{us}^*}{\sqrt{2}}$ & $\Xi_{bb}^{-}\to\Xi_{b}^{\prime-}D_{s}^{+}\pi^{-}$  & $\frac{\left(a_{22}+a_{25}\right)V_{ud}^*}{\sqrt{2}}$\tabularnewline
\hline
$\Xi_{bb}^{0}\to\Omega_{b}^{-}D_{s}^{+}K^{0}$  & $a_{28}V_{ud}^*$ & $\Xi_{bb}^{-}\to\Sigma_{b}^{0}D^{0}K^{-}$  & $\frac{\left(a_{22}+a_{23}\right)V_{us}^*}{\sqrt{2}}$\tabularnewline
\hline
$\Xi_{bb}^{0}\to\Omega_{b}^{-}D_{s}^{+}\eta$  & $\frac{\left(a_{25}-2\left(a_{27}+a_{28}\right)\right)V_{us}^*}{\sqrt{6}}$ & $\Xi_{bb}^{-}\to\Sigma_{b}^{-}D^{0}\pi^{0}$  & $-\frac{\left(a_{23}-a_{24}+a_{26}\right)V_{ud}^*}{\sqrt{2}}$\tabularnewline
\hline
$\Xi_{bb}^{-}\to\Sigma_{b}^{0}D^{0}\pi^{-}$  & $\frac{\left(a_{22}+a_{23}+a_{25}+a_{26}\right)V_{ud}^*}{\sqrt{2}}$ & $\Xi_{bb}^{-}\to\Sigma_{b}^{-}D^{0}\overline{K}^{0}$  & $a_{23}V_{us}^*$\tabularnewline
\hline
\hline
\end{tabular}
\end{table}

\subsection{$b\to u \bar c d/s$: Decays into a bottom baryon $bqq$ plus anti-charmed meson and a light meson  }

For the anti-charm production, the operator having the quark contents $(\bar ub)(\bar qc)$  is given by
\begin{eqnarray}
{\cal H}_{eff} &=& \frac{G_{F}}{\sqrt{2}}
     V_{ub} V_{cq}^{*} \big[
     C_{1}  O^{\bar uc}_{1}
  +  C_{2}  O^{\bar uc}_{2}\Big]+ {\rm h.c.}.
\end{eqnarray}
The two light anti-quarks form the ${\bf  \bar 3}$ and ${\bf  6}$ representations.
The anti-symmetric tensor $H_{\bar 3}''$ and the symmetric tensor
$H_{ 6}$ have nonzero components
$  (H_{\bar 3}'')^{13} =- (H_{\bar 3}'')^{31} =V_{cs}^*$, $ (H_{\bar 6})^{13}=(H_{\bar 6})^{31} =V_{cs}^*,
$
for the $b\to u\bar cs$ transition. For the transition $b\to
u\bar cd$ one requests the interchange of $2\leftrightarrow 3$ in the
subscripts, and $V_{cs}$ replaced by $V_{cd}$.

The effective Hamiltonian is constructed as
\begin{eqnarray}
  {\cal H}_{{eff}}&=& b_{1} (T_{bb})^i (\overline T_{b\bar 3})_{[ij]}   D_l M^j_k (H_{\bar 3}'')^{kl}+b_{2} (T_{bb})^i (\overline T_{b\bar 3})_{[ij]}   D_l M^l_k (H_{\bar  3}'')^{jk} \nonumber\\
 &&+b_{3} (T_{bb})^i (\overline T_{b\bar 3})_{[jk]}   D_i M^j_l (H_{\bar 3}'')^{kl} + b_{4} (T_{bb})^i (\overline T_{b\bar 3})_{[jk]}   D_l M^j_i (H_{\bar 3}'')^{kl} \nonumber\\
 &&+b_{5} (T_{bb})^i (\overline T_{b\bar 3})_{[kl]}   D_j M^j_i (H_{ \bar 3}'')^{kl}+b_{6} (T_{bb})^i (\overline T_{b\bar 3})_{[ij]}   D_l M^j_k (H_{6}'')^{kl} \nonumber\\
 &&+b_{7} (T_{bb})^i (\overline T_{b\bar 3})_{[ij]}   D_l M^l_k (H_{  6}'')^{jk} + b_{8} (T_{bb})^i (\overline T_{b\bar 3})_{[jk]}   D_i M^j_l (H_{  6}'')^{kl} \nonumber\\
 &&+b_{9} (T_{bb})^i (\overline T_{b\bar 3})_{[jk]}   D_l M^j_i (H_{  6}'')^{kl}+b_{10} (T_{bb})^i (\overline T_{b6})_{[ij]}   D_l M^j_k (H_{\bar 3}'')^{kl} \nonumber\\
 &&+b_{11} (T_{bb})^i (\overline T_{b6})_{[ij]}   D_l M^l_k (H_{\bar  3}'')^{jk}+b_{12} (T_{bb})^i (\overline T_{b6})_{[jk]}   D_i M^j_l (H_{\bar 3}'')^{kl}\nonumber\\
 && + b_{13} (T_{bb})^i (\overline T_{b6})_{[jk]}   D_l M^j_i (H_{\bar 3}'')^{kl}+b_{14} (T_{bb})^i (\overline T_{b6})_{[kl]}   D_j M^j_i (H_{6}'')^{kl}\nonumber\\
 &&+b_{15} (T_{bb})^i (\overline T_{b6})_{[ij]}   D_l M^j_k (H_{6}'')^{kl}+b_{16} (T_{bb})^i (\overline T_{b6})_{[ij]}   D_l M^l_k (H_{  6}'')^{jk} \nonumber\\
 &&+ b_{17} (T_{bb})^i (\overline T_{b6})_{[jk]}   D_i M^j_l (H_{  6}'')^{kl}+b_{18} (T_{bb})^i (\overline T_{b6})_{[jk]}   D_l M^j_i (H_{  6}'')^{kl} .
\end{eqnarray}

Decay amplitudes for different channels are given in Tab.~\ref{tab:bbq_bqq_qcbar_bar3} and Tab.~\ref{tab:bbq_bqq_qcbar_6}.  We derive   relations for decay amplitudes given in Appendix~\ref{sec:relation_bb_bqq3_acm_lm}.

\begin{table}
\scriptsize
\caption{Doubly bottom baryon decays into a $bqq$(anti-triplet), an anti-charmed meson and a light meson.}\label{tab:bbq_bqq_qcbar_bar3}\begin{tabular}{|c  c|c  c|c|c|c|c|c|c}\hline
channel  & amplitude  & channel  & amplitude \tabularnewline
\hline
\hline
$\Xi_{bb}^{0}\to\Lambda_{b}^{0}\overline{D}^{0}\pi^{0}$  & $\frac{\left(b_{1}-b_{2}-b_{4}+2b_{5}-b_{6}+b_{7}+2b_{8}+b_{9}\right)V_{cd}^*}{\sqrt{2}}$ & $\Omega_{bb}^{-}\to\Lambda_{b}^{0}D_{s}^{-}\pi^{0}$  & $\sqrt{2}b_{8}V_{cd}^*$\tabularnewline
\hline
$\Xi_{bb}^{0}\to\Lambda_{b}^{0}\overline{D}^{0}\overline{K}^{0}$  & $-\left(b_{1}+b_{3}-b_{6}+b_{8}\right)V_{cs}^*$ & $\Omega_{bb}^{-}\to\Lambda_{b}^{0}D_{s}^{-}\overline{K}^{0}$  & $-\left(b_{3}+b_{4}+b_{8}+b_{9}\right)V_{cs}^*$\tabularnewline
\hline
$\Xi_{bb}^{0}\to\Lambda_{b}^{0}\overline{D}^{0}\eta$  & $\frac{\left(-b_{1}-b_{2}-2b_{3}-b_{4}+2b_{5}+b_{6}+b_{7}+b_{9}\right)V_{cd}^*}{\sqrt{6}}$ & $\Omega_{bb}^{-}\to\Lambda_{b}^{0}D_{s}^{-}\eta$  & $-\sqrt{\frac{2}{3}}\left(b_{3}+2b_{5}\right)V_{cd}^*$\tabularnewline
\hline
$\Xi_{bb}^{0}\to\Lambda_{b}^{0}D^{-}\pi^{+}$  & $\left(b_{1}-b_{2}-b_{4}+2b_{5}+b_{6}+b_{7}-b_{9}\right)V_{cd}^*$ & $\Omega_{bb}^{-}\to\Xi_{b}^{0}\overline{D}^{0}\pi^{-}$  & $\left(b_{1}-b_{2}-b_{6}-b_{7}\right)V_{cd}^*$\tabularnewline
\hline
$\Xi_{bb}^{0}\to\Lambda_{b}^{0}D_{s}^{-}\pi^{+}$  & $\left(b_{1}-b_{4}+b_{6}-b_{9}\right)V_{cs}^*$ & $\Omega_{bb}^{-}\to\Xi_{b}^{0}\overline{D}^{0}K^{-}$  & $\left(b_{1}-b_{2}-b_{4}+2b_{5}-b_{6}-b_{7}+b_{9}\right)V_{cs}^*$\tabularnewline
\hline
$\Xi_{bb}^{0}\to\Lambda_{b}^{0}D_{s}^{-}K^{+}$  & $\left(-b_{2}+2b_{5}+b_{7}\right)V_{cd}^*$ & $\Omega_{bb}^{-}\to\Xi_{b}^{0}D^{-}\pi^{0}$  & $\frac{\left(-b_{1}+b_{2}-b_{6}+b_{7}\right)V_{cd}^*}{\sqrt{2}}$\tabularnewline
\hline
$\Xi_{bb}^{0}\to\Xi_{b}^{0}\overline{D}^{0}\pi^{0}$  & $\frac{\left(-b_{2}-b_{3}-b_{4}+2b_{5}+b_{7}+b_{8}+b_{9}\right)V_{cs}^*}{\sqrt{2}}$ & $\Omega_{bb}^{-}\to\Xi_{b}^{0}D^{-}\overline{K}^{0}$  & $-\left(b_{2}-2b_{5}+b_{7}\right)V_{cs}^*$\tabularnewline
\hline
$\Xi_{bb}^{0}\to\Xi_{b}^{0}\overline{D}^{0}K^{0}$  & $-\left(b_{1}+b_{3}-b_{6}+b_{8}\right)V_{cd}^*$ & $\Omega_{bb}^{-}\to\Xi_{b}^{0}D^{-}\eta$  & $-\frac{\left(b_{1}+b_{2}-2b_{4}+b_{6}+b_{7}-2b_{9}\right)V_{cd}^*}{\sqrt{6}}$\tabularnewline
\hline
$\Xi_{bb}^{0}\to\Xi_{b}^{0}\overline{D}^{0}\eta$  & $\frac{\left(2b_{1}-b_{2}+b_{3}-b_{4}+2b_{5}-2b_{6}+b_{7}+3b_{8}+b_{9}\right)V_{cs}^*}{\sqrt{6}}$ & $\Omega_{bb}^{-}\to\Xi_{b}^{0}D_{s}^{-}\pi^{0}$  & $-\frac{\left(b_{1}+b_{3}+b_{6}-b_{8}\right)V_{cs}^*}{\sqrt{2}}$\tabularnewline
\hline
$\Xi_{bb}^{0}\to\Xi_{b}^{0}D^{-}\pi^{+}$  & $\left(-b_{2}+2b_{5}+b_{7}\right)V_{cs}^*$ & $\Omega_{bb}^{-}\to\Xi_{b}^{0}D_{s}^{-}K^{0}$  & $-\left(b_{2}+b_{3}+b_{7}+b_{8}\right)V_{cd}^*$\tabularnewline
\hline
$\Xi_{bb}^{0}\to\Xi_{b}^{0}D^{-}K^{+}$  & $\left(b_{1}-b_{4}+b_{6}-b_{9}\right)V_{cd}^*$ & $\Omega_{bb}^{-}\to\Xi_{b}^{0}D_{s}^{-}\eta$  & $\frac{\left(-b_{1}+2b_{2}+b_{3}+2b_{4}-4b_{5}-b_{6}+2b_{7}+3b_{8}+2b_{9}\right)V_{cs}^*}{\sqrt{6}}$\tabularnewline
\hline
$\Xi_{bb}^{0}\to\Xi_{b}^{0}D_{s}^{-}K^{+}$  & $\left(b_{1}-b_{2}-b_{4}+2b_{5}+b_{6}+b_{7}-b_{9}\right)V_{cs}^*$ & $\Omega_{bb}^{-}\to\Xi_{b}^{-}\overline{D}^{0}\pi^{0}$  & $\frac{\left(-b_{1}+b_{2}+b_{6}-b_{7}\right)V_{cd}^*}{\sqrt{2}}$\tabularnewline
\hline
$\Xi_{bb}^{0}\to\Xi_{b}^{-}\overline{D}^{0}\pi^{+}$  & $\left(-b_{3}-b_{4}+b_{8}+b_{9}\right)V_{cs}^*$ & $\Omega_{bb}^{-}\to\Xi_{b}^{-}\overline{D}^{0}\overline{K}^{0}$  & $\left(b_{1}-b_{4}-b_{6}+b_{9}\right)V_{cs}^*$\tabularnewline
\hline
$\Xi_{bb}^{0}\to\Xi_{b}^{-}\overline{D}^{0}K^{+}$  & $\left(b_{3}+b_{4}-b_{8}-b_{9}\right)V_{cd}^*$ & $\Omega_{bb}^{-}\to\Xi_{b}^{-}\overline{D}^{0}\eta$  & $\frac{\left(b_{1}+b_{2}-2b_{4}-b_{6}-b_{7}+2b_{9}\right)V_{cd}^*}{\sqrt{6}}$\tabularnewline
\hline
$\Xi_{bb}^{-}\to\Lambda_{b}^{0}\overline{D}^{0}\pi^{-}$  & $\left(b_{1}-b_{2}-b_{4}+2b_{5}-b_{6}-b_{7}+b_{9}\right)V_{cd}^*$ & $\Omega_{bb}^{-}\to\Xi_{b}^{-}D^{-}\pi^{+}$  & $-\left(b_{1}-b_{2}+b_{6}+b_{7}\right)V_{cd}^*$\tabularnewline
\hline
$\Xi_{bb}^{-}\to\Lambda_{b}^{0}\overline{D}^{0}K^{-}$  & $\left(b_{1}-b_{2}-b_{6}-b_{7}\right)V_{cs}^*$ & $\Omega_{bb}^{-}\to\Xi_{b}^{-}D_{s}^{-}\pi^{+}$  & $-\left(b_{1}+b_{3}+b_{6}-b_{8}\right)V_{cs}^*$\tabularnewline
\hline
$\Xi_{bb}^{-}\to\Lambda_{b}^{0}D^{-}\pi^{0}$  & $\frac{\left(-b_{1}+b_{2}+b_{4}-2b_{5}-b_{6}+b_{7}+2b_{8}+b_{9}\right)V_{cd}^*}{\sqrt{2}}$ & $\Omega_{bb}^{-}\to\Xi_{b}^{-}D_{s}^{-}K^{+}$  & $\left(b_{2}+b_{3}-b_{7}-b_{8}\right)V_{cd}^*$\tabularnewline
\hline
$\Xi_{bb}^{-}\to\Lambda_{b}^{0}D^{-}\overline{K}^{0}$  & $-\left(b_{2}+b_{3}+b_{7}+b_{8}\right)V_{cs}^*$ & $\Xi_{bb}^{-}\to\Xi_{b}^{-}\overline{D}^{0}\pi^{0}$  & $\frac{\left(-b_{2}+b_{4}+b_{7}-b_{9}\right)V_{cs}^*}{\sqrt{2}}$\tabularnewline
\hline
$\Xi_{bb}^{-}\to\Lambda_{b}^{0}D^{-}\eta$  & $-\frac{\left(b_{1}+b_{2}+2b_{3}+b_{4}-2b_{5}+b_{6}+b_{7}+b_{9}\right)V_{cd}^*}{\sqrt{6}}$ & $\Xi_{bb}^{-}\to\Xi_{b}^{-}\overline{D}^{0}K^{0}$  & $\left(-b_{1}+b_{4}+b_{6}-b_{9}\right)V_{cd}^*$\tabularnewline
\hline
$\Xi_{bb}^{-}\to\Lambda_{b}^{0}D_{s}^{-}\pi^{0}$  & $\frac{\left(-b_{1}+b_{4}-b_{6}+b_{9}\right)V_{cs}^*}{\sqrt{2}}$ & $\Xi_{bb}^{-}\to\Xi_{b}^{-}\overline{D}^{0}\eta$  & $\frac{\left(2b_{1}-b_{2}-b_{4}-2b_{6}+b_{7}+b_{9}\right)V_{cs}^*}{\sqrt{6}}$\tabularnewline
\hline
$\Xi_{bb}^{-}\to\Lambda_{b}^{0}D_{s}^{-}K^{0}$  & $-\left(b_{2}-2b_{5}+b_{7}\right)V_{cd}^*$ & $\Xi_{bb}^{-}\to\Xi_{b}^{-}D^{-}\pi^{+}$  & $\left(-b_{2}-b_{3}+b_{7}+b_{8}\right)V_{cs}^*$\tabularnewline
\hline
$\Xi_{bb}^{-}\to\Lambda_{b}^{0}D_{s}^{-}\eta$  & $-\frac{\left(b_{1}-2b_{2}+b_{4}+b_{6}-2b_{7}+b_{9}\right)V_{cs}^*}{\sqrt{6}}$ & $\Xi_{bb}^{-}\to\Xi_{b}^{-}D^{-}K^{+}$  & $\left(b_{1}+b_{3}+b_{6}-b_{8}\right)V_{cd}^*$\tabularnewline
\hline
$\Xi_{bb}^{-}\to\Xi_{b}^{0}\overline{D}^{0}\pi^{-}$  & $\left(-b_{4}+2b_{5}+b_{9}\right)V_{cs}^*$ & $\Xi_{bb}^{-}\to\Xi_{b}^{-}D_{s}^{-}K^{+}$  & $\left(b_{1}-b_{2}+b_{6}+b_{7}\right)V_{cs}^*$\tabularnewline
\hline
$\Xi_{bb}^{-}\to\Xi_{b}^{0}D^{-}\pi^{0}$  & $\frac{\left(-b_{3}-2b_{5}+b_{8}\right)V_{cs}^*}{\sqrt{2}}$ & $\Omega_{bb}^{-}\to\Lambda_{b}^{0}\overline{D}^{0}K^{-}$  & $\left(-b_{4}+2b_{5}+b_{9}\right)V_{cd}^*$\tabularnewline
\hline
$\Xi_{bb}^{-}\to\Xi_{b}^{0}D^{-}K^{0}$  & $-\left(b_{3}+b_{4}+b_{8}+b_{9}\right)V_{cd}^*$ & $\Omega_{bb}^{-}\to\Lambda_{b}^{0}D^{-}\overline{K}^{0}$  & $-\left(b_{4}-2b_{5}+b_{9}\right)V_{cd}^*$\tabularnewline
\hline
$\Xi_{bb}^{-}\to\Xi_{b}^{0}D^{-}\eta$  & $\frac{\left(b_{3}+2b_{5}+3b_{8}\right)V_{cs}^*}{\sqrt{6}}$ & $\Xi_{bb}^{-}\to\Xi_{b}^{0}D_{s}^{-}K^{0}$  & $-\left(b_{4}-2b_{5}+b_{9}\right)V_{cs}^*$\tabularnewline
\hline
\hline
\end{tabular}
\end{table}

\begin{table}
\tiny
\caption{Doubly bottom baryon decays into a $bqq$(sextet), an anti-charmed meson and a light meson.}
\label{tab:bbq_bqq_qcbar_6}\begin{tabular}{|c  c|c  c|c|c|c|c|c|c}\hline
channel  & amplitude  & channel  & amplitude \tabularnewline
\hline
\hline
$\Xi_{bb}^{0}\to\Sigma_{b}^{+}\overline{D}^{0}\pi^{-}$  & $\left(-b_{10}+b_{11}+b_{12}+b_{14}+b_{15}+b_{16}\right)V_{cd}^*$ & $\Omega_{bb}^{-}\to\Xi_{b}^{\prime-}\overline{D}^{0}\overline{K}^{0}$  & $\frac{\left(-b_{10}-b_{13}+b_{14}+b_{17}\right)V_{cs}^*}{\sqrt{2}}$\tabularnewline
\hline
$\Xi_{bb}^{0}\to\Sigma_{b}^{+}\overline{D}^{0}K^{-}$  & $\left(-b_{10}+b_{11}+b_{12}+b_{14}+b_{15}+b_{16}\right)V_{cs}^*$ & $\Omega_{bb}^{-}\to\Xi_{b}^{\prime-}\overline{D}^{0}\eta$  & $\frac{\left(-b_{10}-b_{11}+2b_{13}+b_{14}+b_{15}-2b_{17}\right)V_{cd}^*}{2\sqrt{3}}$\tabularnewline
\hline
$\Xi_{bb}^{0}\to\Sigma_{b}^{+}D^{-}\pi^{0}$  & $\frac{\left(b_{10}-b_{11}+b_{13}+b_{14}-b_{15}+b_{17}\right)V_{cd}^*}{\sqrt{2}}$ & $\Omega_{bb}^{-}\to\Xi_{b}^{\prime-}D^{-}\pi^{+}$  & $\frac{\left(b_{10}-b_{11}+b_{14}+b_{15}\right)V_{cd}^*}{\sqrt{2}}$\tabularnewline
\hline
$\Xi_{bb}^{0}\to\Sigma_{b}^{+}D^{-}\overline{K}^{0}$  & $\left(b_{11}+b_{15}\right)V_{cs}^*$ & $\Omega_{bb}^{-}\to\Xi_{b}^{\prime-}D_{s}^{-}\pi^{+}$  & $\frac{\left(b_{10}-b_{12}+b_{14}+b_{16}\right)V_{cs}^*}{\sqrt{2}}$\tabularnewline
\hline
$\Xi_{bb}^{0}\to\Sigma_{b}^{+}D^{-}\eta$  & $\frac{\left(b_{10}+b_{11}+b_{13}+b_{14}+b_{15}+b_{17}\right)V_{cd}^*}{\sqrt{6}}$ & $\Omega_{bb}^{-}\to\Xi_{b}^{\prime-}D_{s}^{-}K^{+}$  & $\frac{\left(-b_{11}-b_{12}+b_{15}+b_{16}\right)V_{cd}^*}{\sqrt{2}}$\tabularnewline
\hline
$\Xi_{bb}^{0}\to\Sigma_{b}^{+}D_{s}^{-}\pi^{0}$  & $\frac{\left(b_{10}+b_{13}+b_{14}+b_{17}\right)V_{cs}^*}{\sqrt{2}}$ & $\Omega_{bb}^{-}\to\Omega_{b}^{-}\overline{D}^{0}\pi^{0}$  & $\frac{\left(b_{15}-b_{11}\right)V_{cs}^*}{\sqrt{2}}$\tabularnewline
\hline
$\Xi_{bb}^{0}\to\Sigma_{b}^{+}D_{s}^{-}K^{0}$  & $\left(b_{11}+b_{15}\right)V_{cd}^*$ & $\Omega_{bb}^{-}\to\Omega_{b}^{-}\overline{D}^{0}K^{0}$  & $\left(b_{14}-b_{10}\right)V_{cd}^*$\tabularnewline
\hline
$\Xi_{bb}^{0}\to\Sigma_{b}^{+}D_{s}^{-}\eta$  & $\frac{\left(b_{10}-2b_{11}+b_{13}+b_{14}-2b_{15}+b_{17}\right)V_{cs}^*}{\sqrt{6}}$ & $\Omega_{bb}^{-}\to\Omega_{b}^{-}\overline{D}^{0}\eta$  & $\frac{\left(2b_{10}-b_{11}+2b_{13}-2b_{14}+b_{15}-2b_{17}\right)V_{cs}^*}{\sqrt{6}}$\tabularnewline
\hline
$\Xi_{bb}^{0}\to\Sigma_{b}^{0}\overline{D}^{0}\pi^{0}$  & $\frac{1}{2}\left(b_{10}-b_{11}-2b_{12}-b_{13}-b_{14}+b_{15}+b_{17}+2b_{18}\right)V_{cd}^*$ & $\Omega_{bb}^{-}\to\Omega_{b}^{-}D^{-}\pi^{+}$  & $\left(b_{15}-b_{11}\right)V_{cs}^*$\tabularnewline
\hline
$\Xi_{bb}^{0}\to\Sigma_{b}^{0}\overline{D}^{0}\overline{K}^{0}$  & $\frac{\left(-b_{10}+b_{12}+b_{14}+b_{16}\right)V_{cs}^*}{\sqrt{2}}$ & $\Omega_{bb}^{-}\to\Omega_{b}^{-}D^{-}K^{+}$  & $\left(b_{10}+b_{14}\right)V_{cd}^*$\tabularnewline
\hline
$\Xi_{bb}^{0}\to\Sigma_{b}^{0}\overline{D}^{0}\eta$  & $\frac{\left(-b_{10}-b_{11}-b_{13}+b_{14}+b_{15}+2b_{16}+b_{17}+2b_{18}\right)V_{cd}^*}{2\sqrt{3}}$ & $\Omega_{bb}^{-}\to\Omega_{b}^{-}D_{s}^{-}K^{+}$  & $\left(b_{10}-b_{11}-b_{12}+b_{14}+b_{15}+b_{16}\right)V_{cs}^*$\tabularnewline
\hline
$\Xi_{bb}^{0}\to\Sigma_{b}^{0}D^{-}\pi^{+}$  & $\frac{\left(b_{10}-b_{11}+b_{13}+b_{14}+b_{15}+b_{17}+2b_{18}\right)V_{cd}^*}{\sqrt{2}}$ & $\Omega_{bb}^{-}\to\Sigma_{b}^{0}D_{s}^{-}\overline{K}^{0}$  & $\frac{\left(b_{12}+b_{13}+b_{16}+b_{17}\right)V_{cs}^*}{\sqrt{2}}$\tabularnewline
\hline
$\Xi_{bb}^{0}\to\Sigma_{b}^{0}D_{s}^{-}\pi^{+}$  & $\frac{\left(b_{10}+b_{13}+b_{14}+b_{17}\right)V_{cs}^*}{\sqrt{2}}$ & $\Omega_{bb}^{-}\to\Sigma_{b}^{0}D_{s}^{-}\eta$  & $\frac{\left(b_{16}-2b_{18}\right)V_{cd}^*}{\sqrt{3}}$\tabularnewline
\hline
$\Xi_{bb}^{0}\to\Sigma_{b}^{0}D_{s}^{-}K^{+}$  & $\frac{\left(-b_{11}+b_{15}+2b_{18}\right)V_{cd}^*}{\sqrt{2}}$ & $\Omega_{bb}^{-}\to\Sigma_{b}^{-}\overline{D}^{0}\overline{K}^{0}$  & $\left(b_{17}-b_{13}\right)V_{cd}^*$\tabularnewline
\hline
$\Xi_{bb}^{0}\to\Sigma_{b}^{-}\overline{D}^{0}\pi^{+}$  & $\left(-b_{12}-b_{13}+b_{16}+b_{17}\right)V_{cd}^*$ & $\Omega_{bb}^{-}\to\Sigma_{b}^{-}D_{s}^{-}\pi^{+}$  & $\left(b_{16}-b_{12}\right)V_{cd}^*$\tabularnewline
\hline
$\Xi_{bb}^{0}\to\Xi_{b}^{\prime0}\overline{D}^{0}\pi^{0}$  & $\frac{1}{2}\left(-b_{11}-b_{12}-b_{13}+b_{15}+b_{16}+b_{17}+2b_{18}\right)V_{cs}^*$ & $\Omega_{bb}^{-}\to\Xi_{b}^{\prime0}\overline{D}^{0}\pi^{-}$  & $\frac{\left(-b_{10}+b_{11}+b_{14}+b_{15}\right)V_{cd}^*}{\sqrt{2}}$\tabularnewline
\hline
$\Xi_{bb}^{0}\to\Xi_{b}^{\prime0}\overline{D}^{0}K^{0}$  & $\frac{\left(-b_{10}+b_{12}+b_{14}+b_{16}\right)V_{cd}^*}{\sqrt{2}}$ & $\Omega_{bb}^{-}\to\Xi_{b}^{\prime0}\overline{D}^{0}K^{-}$  & $\frac{\left(-b_{10}+b_{11}-b_{13}+b_{14}+b_{15}+b_{17}+2b_{18}\right)V_{cs}^*}{\sqrt{2}}$\tabularnewline
\hline
$\Xi_{bb}^{0}\to\Xi_{b}^{\prime0}\overline{D}^{0}\eta$  & $\frac{\left(2b_{10}-b_{11}-3b_{12}-b_{13}-2b_{14}+b_{15}-b_{16}+b_{17}+2b_{18}\right)V_{cs}^*}{2\sqrt{3}}$ & $\Omega_{bb}^{-}\to\Xi_{b}^{\prime0}D^{-}\pi^{0}$  & $\frac{1}{2}\left(b_{10}-b_{11}+b_{14}-b_{15}\right)V_{cd}^*$\tabularnewline
\hline
$\Xi_{bb}^{0}\to\Xi_{b}^{\prime0}D^{-}\pi^{+}$  & $\frac{\left(-b_{11}+b_{15}+2b_{18}\right)V_{cs}^*}{\sqrt{2}}$ & $\Omega_{bb}^{-}\to\Xi_{b}^{\prime0}D^{-}\overline{K}^{0}$  & $\frac{\left(b_{11}+b_{15}+2b_{18}\right)V_{cs}^*}{\sqrt{2}}$\tabularnewline
\hline
$\Xi_{bb}^{0}\to\Xi_{b}^{\prime0}D^{-}K^{+}$  & $\frac{\left(b_{10}+b_{13}+b_{14}+b_{17}\right)V_{cd}^*}{\sqrt{2}}$ & $\Omega_{bb}^{-}\to\Xi_{b}^{\prime0}D^{-}\eta$  & $\frac{\left(b_{10}+b_{11}-2b_{13}+b_{14}+b_{15}-2b_{17}\right)V_{cd}^*}{2\sqrt{3}}$\tabularnewline
\hline
$\Xi_{bb}^{0}\to\Xi_{b}^{\prime0}D_{s}^{-}K^{+}$  & $\frac{\left(b_{10}-b_{11}+b_{13}+b_{14}+b_{15}+b_{17}+2b_{18}\right)V_{cs}^*}{\sqrt{2}}$ & $\Omega_{bb}^{-}\to\Xi_{b}^{\prime0}D_{s}^{-}\pi^{0}$  & $\frac{1}{2}\left(b_{10}-b_{12}+b_{14}+b_{16}\right)V_{cs}^*$\tabularnewline
\hline
$\Xi_{bb}^{0}\to\Xi_{b}^{\prime-}\overline{D}^{0}\pi^{+}$  & $\frac{\left(-b_{12}-b_{13}+b_{16}+b_{17}\right)V_{cs}^*}{\sqrt{2}}$ & $\Omega_{bb}^{-}\to\Xi_{b}^{\prime0}D_{s}^{-}K^{0}$  & $\frac{\left(b_{11}+b_{12}+b_{15}+b_{16}\right)V_{cd}^*}{\sqrt{2}}$\tabularnewline
\hline
$\Xi_{bb}^{0}\to\Xi_{b}^{\prime-}\overline{D}^{0}K^{+}$  & $\frac{\left(-b_{12}-b_{13}+b_{16}+b_{17}\right)V_{cd}^*}{\sqrt{2}}$ & $\Omega_{bb}^{-}\to\Xi_{b}^{\prime0}D_{s}^{-}\eta$  & $\frac{\left(b_{10}-2b_{11}-3b_{12}-2b_{13}+b_{14}-2b_{15}-b_{16}-2b_{17}-4b_{18}\right)V_{cs}^*}{2\sqrt{3}}$\tabularnewline
\hline
$\Xi_{bb}^{0}\to\Omega_{b}^{-}\overline{D}^{0}K^{+}$  & $\left(-b_{12}-b_{13}+b_{16}+b_{17}\right)V_{cs}^*$ & $\Omega_{bb}^{-}\to\Xi_{b}^{\prime-}\overline{D}^{0}\pi^{0}$  & $\frac{1}{2}\left(b_{10}-b_{11}-b_{14}+b_{15}\right)V_{cd}^*$\tabularnewline
\hline
$\Xi_{bb}^{-}\to\Sigma_{b}^{+}D^{-}\pi^{-}$  & $\left(b_{12}+b_{13}+b_{16}+b_{17}\right)V_{cd}^*$ & $\Xi_{bb}^{-}\to\Xi_{b}^{\prime-}\overline{D}^{0}K^{0}$  & $\frac{\left(-b_{10}-b_{13}+b_{14}+b_{17}\right)V_{cd}^*}{\sqrt{2}}$\tabularnewline
\hline
$\Xi_{bb}^{-}\to\Sigma_{b}^{+}D^{-}K^{-}$  & $\left(b_{12}+b_{16}\right)V_{cs}^*$ & $\Xi_{bb}^{-}\to\Xi_{b}^{\prime-}\overline{D}^{0}\eta$  & $\frac{\left(2b_{10}-b_{11}-b_{13}-2b_{14}+b_{15}+b_{17}\right)V_{cs}^*}{2\sqrt{3}}$\tabularnewline
\hline
$\Xi_{bb}^{-}\to\Sigma_{b}^{+}D_{s}^{-}\pi^{-}$  & $\left(b_{13}+b_{17}\right)V_{cs}^*$ & $\Xi_{bb}^{-}\to\Xi_{b}^{\prime-}D^{-}\pi^{+}$  & $\frac{\left(-b_{11}-b_{12}+b_{15}+b_{16}\right)V_{cs}^*}{\sqrt{2}}$\tabularnewline
\hline
$\Xi_{bb}^{-}\to\Sigma_{b}^{0}\overline{D}^{0}\pi^{-}$  & $\frac{\left(-b_{10}+b_{11}-b_{13}+b_{14}+b_{15}+b_{17}+2b_{18}\right)V_{cd}^*}{\sqrt{2}}$ & $\Xi_{bb}^{-}\to\Xi_{b}^{\prime-}D^{-}K^{+}$  & $\frac{\left(b_{10}-b_{12}+b_{14}+b_{16}\right)V_{cd}^*}{\sqrt{2}}$\tabularnewline
\hline
$\Xi_{bb}^{-}\to\Sigma_{b}^{0}\overline{D}^{0}K^{-}$  & $\frac{\left(-b_{10}+b_{11}+b_{14}+b_{15}\right)V_{cs}^*}{\sqrt{2}}$ & $\Xi_{bb}^{-}\to\Xi_{b}^{\prime-}D_{s}^{-}K^{+}$  & $\frac{\left(b_{10}-b_{11}+b_{14}+b_{15}\right)V_{cs}^*}{\sqrt{2}}$\tabularnewline
\hline
$\Xi_{bb}^{-}\to\Sigma_{b}^{0}D^{-}\pi^{0}$  & $\frac{1}{2}\left(b_{10}-b_{11}-2b_{12}-b_{13}+b_{14}-b_{15}-b_{17}-2b_{18}\right)V_{cd}^*$ & $\Xi_{bb}^{-}\to\Omega_{b}^{-}\overline{D}^{0}K^{0}$  & $\left(b_{17}-b_{13}\right)V_{cs}^*$\tabularnewline
\hline
$\Xi_{bb}^{-}\to\Sigma_{b}^{0}D^{-}\overline{K}^{0}$  & $\frac{\left(b_{11}+b_{12}+b_{15}+b_{16}\right)V_{cs}^*}{\sqrt{2}}$ & $\Xi_{bb}^{-}\to\Omega_{b}^{-}D^{-}K^{+}$  & $\left(b_{16}-b_{12}\right)V_{cs}^*$\tabularnewline
\hline
$\Xi_{bb}^{-}\to\Sigma_{b}^{0}D^{-}\eta$  & $\frac{\left(b_{10}+b_{11}+b_{13}+b_{14}+b_{15}+2b_{16}+b_{17}+2b_{18}\right)V_{cd}^*}{2\sqrt{3}}$ & $\Omega_{bb}^{-}\to\Sigma_{b}^{+}D^{-}K^{-}$  & $\left(b_{13}+b_{17}\right)V_{cd}^*$\tabularnewline
\hline
$\Xi_{bb}^{-}\to\Sigma_{b}^{0}D_{s}^{-}\pi^{0}$  & $\frac{1}{2}\left(b_{10}-b_{13}+b_{14}-b_{17}\right)V_{cs}^*$ & $\Omega_{bb}^{-}\to\Sigma_{b}^{+}D_{s}^{-}\pi^{-}$  & $\left(b_{12}+b_{16}\right)V_{cd}^*$\tabularnewline
\hline
$\Xi_{bb}^{-}\to\Sigma_{b}^{0}D_{s}^{-}K^{0}$  & $\frac{\left(b_{11}+b_{15}+2b_{18}\right)V_{cd}^*}{\sqrt{2}}$ & $\Omega_{bb}^{-}\to\Sigma_{b}^{+}D_{s}^{-}K^{-}$  & $\left(b_{12}+b_{13}+b_{16}+b_{17}\right)V_{cs}^*$\tabularnewline
\hline
$\Xi_{bb}^{-}\to\Sigma_{b}^{0}D_{s}^{-}\eta$  & $\frac{\left(b_{10}-2b_{11}+b_{13}+b_{14}-2b_{15}+b_{17}\right)V_{cs}^*}{2\sqrt{3}}$ & $\Omega_{bb}^{-}\to\Sigma_{b}^{0}\overline{D}^{0}K^{-}$  & $\frac{\left(-b_{13}+b_{17}+2b_{18}\right)V_{cd}^*}{\sqrt{2}}$\tabularnewline
\hline
$\Xi_{bb}^{-}\to\Sigma_{b}^{-}\overline{D}^{0}\pi^{0}$  & $\frac{\left(b_{10}-b_{11}+b_{13}-b_{14}+b_{15}-b_{17}\right)V_{cd}^*}{\sqrt{2}}$ & $\Omega_{bb}^{-}\to\Sigma_{b}^{0}D^{-}\overline{K}^{0}$  & $\frac{\left(b_{13}+b_{17}+2b_{18}\right)V_{cd}^*}{\sqrt{2}}$\tabularnewline
\hline
$\Xi_{bb}^{-}\to\Sigma_{b}^{-}\overline{D}^{0}\overline{K}^{0}$  & $\left(b_{14}-b_{10}\right)V_{cs}^*$ & $\Omega_{bb}^{-}\to\Sigma_{b}^{0}D_{s}^{-}\pi^{0}$  & $-b_{12}V_{cd}^*$\tabularnewline
\hline
$\Xi_{bb}^{-}\to\Sigma_{b}^{-}\overline{D}^{0}\eta$  & $\frac{\left(-b_{10}-b_{11}-b_{13}+b_{14}+b_{15}+b_{17}\right)V_{cd}^*}{\sqrt{6}}$ & $\Xi_{bb}^{-}\to\Xi_{b}^{\prime0}D^{-}\pi^{0}$  & $-\frac{1}{2}\left(b_{12}-b_{16}+2b_{18}\right)V_{cs}^*$\tabularnewline
\hline
$\Xi_{bb}^{-}\to\Sigma_{b}^{-}D^{-}\pi^{+}$  & $\left(b_{10}-b_{11}-b_{12}+b_{14}+b_{15}+b_{16}\right)V_{cd}^*$ & $\Xi_{bb}^{-}\to\Xi_{b}^{\prime0}D^{-}K^{0}$  & $\frac{\left(b_{12}+b_{13}+b_{16}+b_{17}\right)V_{cd}^*}{\sqrt{2}}$\tabularnewline
\hline
$\Xi_{bb}^{-}\to\Sigma_{b}^{-}D_{s}^{-}\pi^{+}$  & $\left(b_{10}+b_{14}\right)V_{cs}^*$ & $\Xi_{bb}^{-}\to\Xi_{b}^{\prime0}D^{-}\eta$  & $-\frac{\left(3b_{12}+b_{16}-2b_{18}\right)V_{cs}^*}{2\sqrt{3}}$\tabularnewline
\hline
$\Xi_{bb}^{-}\to\Sigma_{b}^{-}D_{s}^{-}K^{+}$  & $\left(b_{15}-b_{11}\right)V_{cd}^*$ & $\Xi_{bb}^{-}\to\Xi_{b}^{\prime0}D_{s}^{-}K^{0}$  & $\frac{\left(b_{13}+b_{17}+2b_{18}\right)V_{cs}^*}{\sqrt{2}}$\tabularnewline
\hline
$\Xi_{bb}^{-}\to\Xi_{b}^{\prime0}\overline{D}^{0}\pi^{-}$  & $\frac{\left(-b_{13}+b_{17}+2b_{18}\right)V_{cs}^*}{\sqrt{2}}$ & $\Xi_{bb}^{-}\to\Xi_{b}^{\prime-}\overline{D}^{0}\pi^{0}$  & $\frac{1}{2}\left(-b_{11}+b_{13}+b_{15}-b_{17}\right)V_{cs}^*$\tabularnewline
\hline
\hline
\end{tabular}
\end{table}


The $\Xi_{bb}$ can decay into both $  D^0$ and $ D^0$.  The $D^0$ and $\overline D^0$ can form the CP eigenstates $D_{+}$ and $D_-$. Thus using the $\Xi_{bb}$ decays into the $ D_\pm$, one may construct the interference between the $b\to c\bar us$ and $b\to u\bar cs$. The CKM angle $\gamma$ can then be extracted from measuring decay widths of these channels, as in the case of $B\to DK$~\cite{Gronau:1991dp,Gronau:1990ra,Dunietz:1991yd,Atwood:1996ci,hep-ph/0008090,Giri:2003ty},  $B\to DK^*_{0,2}$~\cite{Wang:2011zw,Kim:2013ria} and others. This is also similar for the $\Omega_{bb}\to  D_\pm$ decays and the following $\Xi_{bc}\to D_\pm$ and $\Omega_{bc}\to   D_{\pm}$ channels.


\subsection{Charmless $b\to q_1 \bar q_2 q_3$ Decays: Decays into a bottom baryon and two light mesons}

The  charmless $b\to q$ ($q=d,s$) transition is controlled by the weak Hamiltonian ${\cal H}_{eff}$:
 \begin{eqnarray}
 {\cal H}_{eff} &=& \frac{G_{F}}{\sqrt{2}}
     \bigg\{ V_{ub} V_{uq}^{*} \big[
     C_{1}  O^{\bar uu}_{1}
  +  C_{2}  O^{\bar uu}_{2}\Big]- V_{tb} V_{tq}^{*} \big[{\sum\limits_{i=3}^{10}} C_{i}  O_{i} \Big]\bigg\}+ \mbox{h.c.} ,
 \label{eq:hamiltonian}
\end{eqnarray}
where  $O_{i}$ is a four-quark operator or a moment type operator. At the hadron level,  penguin operators  behave as the ${\bf  3}$ representation while  tree operators   can
be decomposed in terms of a vector $H_{\bf 3}$, a traceless
tensor antisymmetric in upper indices, $H_{\bf\overline6}$, and a
traceless tensor symmetric in   upper indices,
$H_{\bf{15}}$.
For the $\Delta S=0 (b\to d)$decays, the non-zero components of the effective Hamiltonian are~\cite{Savage:1989ub,He:2000ys,Hsiao:2015iiu}:
\begin{eqnarray}
 (H_3)^2=1,\;\;\;(H_{\overline6})^{12}_1=-(H_{\overline6})^{21}_1=(H_{\overline6})^{23}_3=-(H_{\overline6})^{32}_3=1,\nonumber\\
 2(H_{15})^{12}_1= 2(H_{15})^{21}_1=-3(H_{15})^{22}_2=
 -6(H_{15})^{23}_3=-6(H_{15})^{32}_3=6,\label{eq:H3615_bb}
\end{eqnarray}
and  all other remaining entries are zero. For the $\Delta S=1(b\to s)$
decays the nonzero entries in the $H_{\bf{3}}$, $H_{\bf\overline6}$,
$H_{\bf{15}}$ are obtained from Eq.~\eqref{eq:H3615_bb}
with the exchange  $2\leftrightarrow 3$.

The effective hadron-level Hamiltonian for decays into the bottom anti-triplet is constructed  as
\begin{eqnarray}
 {\cal H}_{eff}&=&c_1(T_{bb})^i  (\overline T_{b\bar 3})_{[ij]} M^{j}_{k} M^{k}_{l} (H_{3})^l +c_2(T_{bb})^i  (\overline T_{b\bar 3})_{[ij]} M^{l}_{k} M^{k}_{l} (H_{3})^j  \nonumber\\
 && + c_3(T_{bb})^i  (\overline T_{b\bar 3})_{[jk]} M^{j}_{i} M^{k}_{l} (H_{3})^l+ c_4(T_{bb})^i  (\overline T_{b\bar 3})_{[kl]} M^{j}_{i} M^{k}_{j} (H_{3})^l \nonumber\\
 &&+ c_5(T_{bb})^i  (\overline T_{b\bar 3})_{[ij]} M^{j}_{k} M^{m}_{l} (H_{\bar 6})^{kl}_{m} + c_6(T_{bb})^i  (\overline T_{b\bar 3})_{[jk]} M^{j}_{i} M^{m}_{l} (H_{\bar6})^{kl}_{m}\nonumber\\
 &&+ c_7(T_{bb})^i  (\overline T_{b\bar 3})_{[lm]} M^{j}_{i} M^{k}_{j} (H_{\bar6})^{lm}_{k}+ c_8(T_{bb})^i  (\overline T_{b\bar 3})_{[km]} M^{j}_{i} M^{m}_{l} (H_{\bar6})^{kl}_{j}\nonumber\\
 &&+ c_9(T_{bb})^i  (\overline T_{b\bar 3})_{[jk]} M^{l}_{m} M^{m}_{l} (H_{\bar6})^{jk}_{i} + c_{10}(T_{bb})^i  (\overline T_{b\bar 3})_{[jm]} M^{l}_{k} M^{m}_{l} (H_{\bar6})^{jk}_{i}\nonumber\\
 &&+ c_{11}(T_{bb})^i  (\overline T_{b\bar 3})_{[lm]} M^{l}_{j} M^{m}_{k} (H_{\bar6})^{jk}_{i}+ c_{12}(T_{bb})^i  (\overline T_{b\bar 3})_{[ij]} M^{j}_{k} M^{m}_{l} (H_{15})^{kl}_{m}\nonumber\\
 &&+ c_{13}(T_{bb})^i  (\overline T_{b\bar 3})_{[ij]} M^{l}_{m} M^{m}_{k} (H_{15})^{jk}_{l}+ c_{14}(T_{bb})^i  (\overline T_{b\bar 3})_{[jk]} M^{j}_{i} M^{m}_{l} (H_{15})^{kl}_{m}\nonumber\\
 &&+ c_{15}(T_{bb})^i  (\overline T_{b\bar 3})_{[km]} M^{j}_{i} M^{m}_{l} (H_{15})^{kl}_{j}+ c_{16}(T_{bb})^i  (\overline T_{b\bar 3})_{[jm]} M^{l}_{k} M^{m}_{l} (H_{15})^{jk}_{i}\nonumber\\
 &&+ \overline c_5(T_{bb})^i  (\overline T_{b\bar 3})_{[ij]} M^{l}_{m} M^{m}_{k} (H_{\bar 6})^{jk}_{l},
\end{eqnarray}
while for the sextet baryon, we have
\begin{eqnarray}
 {\cal H}_{eff}&=&c_1(T_{bb})^i  (\overline T_{b6})_{[ij]} M^{j}_{k} M^{k}_{l} (H_{3})^l +c_2(T_{bb})^i  (\overline T_{b6})_{[ij]} M^{l}_{k} M^{k}_{l} (H_{3})^j  \nonumber\\
 && + c_3(T_{bb})^i  (\overline T_{b6})_{[jk]} M^{j}_{i} M^{k}_{l} (H_{3})^l+ c_4(T_{bb})^i  (\overline T_{b6})_{[kl]} M^{j}_{i} M^{k}_{j} (H_{3})^l \nonumber\\
 &&+ c_5(T_{bb})^i  (\overline T_{b6})_{[ij]} M^{j}_{k} M^{m}_{l} (H_{\bar 6})^{kl}_{m} + c_6(T_{bb})^i  (\overline T_{b6})_{[jk]} M^{j}_{i} M^{m}_{l} (H_{\bar6})^{kl}_{m}\nonumber\\
 &&+ c_7(T_{bb})^i  (\overline T_{b6})_{[km]} M^{j}_{i} M^{m}_{l} (H_{\bar6})^{kl}_{j}+ c_8(T_{bb})^i  (\overline T_{b6})_{[jm]} M^{l}_{k} M^{m}_{l} (H_{\bar6})^{jk}_{i}\nonumber\\
 &&+ c_9(T_{bb})^i  (\overline T_{b6})_{[ij]} M^{j}_{k} M^{m}_{l} (H_{15})^{kl}_{m}+ c_{10}(T_{bb})^i  (\overline T_{b6})_{[ij]} M^{l}_{m} M^{m}_{k} (H_{15})^{jk}_{l}\nonumber\\
 &&+ c_{11}(T_{bb})^i  (\overline T_{b6})_{[jk]} M^{j}_{i} M^{m}_{l} (H_{15})^{kl}_{m}+ c_{12}(T_{bb})^i  (\overline T_{b6})_{[lm]} M^{j}_{i} M^{k}_{j} (H_{15})^{lm}_{k}\nonumber\\
 &&+ c_{13}(T_{bb})^i  (\overline T_{b6})_{[km]} M^{j}_{i} M^{m}_{l} (H_{15})^{kl}_{j}+ c_{14}(T_{bb})^i  (\overline T_{b6})_{[jk]} M^{l}_{m} M^{m}_{l} (H_{15})^{jk}_{i}\nonumber\\
 &&+ c_{15}(T_{bb})^i  (\overline T_{b6})_{[jm]} M^{l}_{k} M^{m}_{l} (H_{15})^{jk}_{i}+ c_{16}(T_{bb})^i  (\overline T_{b6})_{[lm]} M^{l}_{j} M^{m}_{k} (H_{15})^{jk}_{i}\nonumber\\
 &&+ \overline c_5(T_{bb})^i  (\overline T_{b\bar 3})_{[ij]} M^{l}_{m} M^{m}_{k} (H_{\bar 6})^{jk}_{l}.
\end{eqnarray}
There also exists an extra term with irreducible amplitude $\overline c_{5}$. We will remove  this term since it always appears in the amplitude in the form of $c_5-\overline c_{5}$.
Decay amplitudes for different channels are given in Tab.~\ref{tab:bbq_bqq_qqbar_vd_3bar_bd}, Tab.~\ref{tab:bbq_bqq_qqbar_vd_6_bd_1} and Tab.~\ref{tab:bbq_bqq_qqbar_vd_6_bd_2} for $b\to d$ transition;
Tab.~\ref{tab:bbq_bqq_qqbar_vs_3bar_bs}, Tab.~\ref{tab:bbq_bqq_qqbar_vs_6_bs_1} and Tab.~\ref{tab:bbq_bqq_qqbar_vs_6_bs_2} for $b\to s$ transition.
Thus, it leads to the relations for decay widths given in Appendix~\ref{sec:relation_bb_bqq3_2lm_b2d}.

\begin{table}
\scriptsize
\newcommand{\tabincell}[2]{\begin{tabular}{@{}#1@{}}#2\end{tabular}}
\caption{Doubly bottom baryon decays into a $bqq$(anti-triplet) induced by the charmless $b\to d$ transition and two light mesons. }\label{tab:bbq_bqq_qqbar_vd_3bar_bd}\begin{tabular}{|c c|c c|c|c|c|c|c|c}\hline
channel & amplitude \\\hline
\hline
$\Xi_{bb}^{0}\to \Lambda_b^0  \pi^+   \pi^-  $ & \tabincell{c}{$ c_1+2 c_2-c_3+c_4+c_5-c_6+2 c_7+4 c_9$\\$+2 c_{10}-2 c_{11}+3 c_{12}+c_{13}-3 c_{14}+2 c_{15}$}\\\hline
$\Xi_{bb}^{0}\to \Lambda_b^0  \pi^0   \pi^0  $ & \tabincell{c}{$ c_1+2 c_2-c_3+c_4+c_5-c_6+2 c_7+4 c_9$\\$+2 c_{10}-2 c_{11}-5 c_{12}+c_{13}+5 c_{14}-6 c_{15}$}\\\hline
$\Xi_{bb}^{0}\to \Lambda_b^0  \pi^0   \eta  $ & \tabincell{c}{$ \frac{1}{\sqrt{3}}(-c_1+c_4+c_5-2 c_6+2 c_7+c_8+c_{12}$\\$+5 c_{13}+4 c_{14}-3 c_{15}-6 c_{16})$}\\\hline
$\Xi_{bb}^{0}\to \Lambda_b^0  K^+   K^-  $ & $ 2 c_2+c_4+2 c_7-c_8+4 c_9+c_{10}+2 c_{13}+c_{15}-3 c_{16}$\\\hline
$\Xi_{bb}^{0}\to \Lambda_b^0  K^0   \overline K^0  $ & $ c_1+2 c_2-c_5+4 c_9+c_{10}-c_{12}-3 c_{13}+3 c_{16}$\\\hline
$\Xi_{bb}^{0}\to \Lambda_b^0  \eta   \eta  $ & \tabincell{c}{$ \frac{1}{3} (c_1+6 c_2+c_3+c_4-3 c_5-3 c_6+2 c_7+2 c_8$\\$+12 c_9+2 c_{10}+2 c_{11}+3 c_{12}-3 c_{13}+3 c_{14})$}\\\hline
$\Xi_{bb}^{0}\to \Xi_b^0  \pi^0   K^0  $ & \tabincell{c}{$ \frac{1}{\sqrt{2}}(-c_1+c_3-c_5-c_6+c_8-c_{10}+2 c_{11}$\\$+5 c_{12}+c_{13}-c_{14}+3 c_{15}-3 c_{16}$)}\\\hline
$\Xi_{bb}^{0}\to \Xi_b^0  \pi^-   K^+  $ & \tabincell{c}{$ c_1-c_3+c_5-c_6+c_8+c_{10}-2 c_{11}$\\$+3 c_{12}-c_{13}-3 c_{14}+c_{15}+3 c_{16}$}\\\hline
$\Xi_{bb}^{0}\to \Xi_b^0  K^0   \eta  $ & \tabincell{c}{$ \frac{1}{\sqrt{6}}(-c_1+c_3-c_5-c_6+c_8-c_{10}$\\$+2 c_{11}+5 c_{12}+c_{13}-c_{14}+3 c_{15}-3 c_{16})$}\\\hline
$\Xi_{bb}^{0}\to \Xi_b^-  \pi^+   K^0  $ & $ c_3-c_4-c_6+2 c_7-c_{10}+2 c_{11}-c_{14}-2 c_{15}+3 c_{16}$\\\hline
$\Xi_{bb}^{0}\to \Xi_b^-  \pi^0   K^+  $ & $ \frac{c_3-c_4+c_6+2 c_7-2 c_8-c_{10}+2 c_{11}-5 c_{14}+2 c_{15}+3 c_{16}}{\sqrt{2}}$\\\hline
$\Xi_{bb}^{0}\to \Xi_b^-  K^+   \eta  $ & $ \frac{-c_3+c_4+3 c_6-2 c_7-2 c_8+c_{10}-2 c_{11}-3 c_{14}+6 c_{15}-3 c_{16}}{\sqrt{6}}$\\\hline
$\Xi_{bb}^{-}\to \Lambda_b^0  \pi^0   \pi^-  $ & $ -4 \sqrt{2} \left(c_{12}-c_{14}+c_{15}\right)$\\\hline
$\Xi_{bb}^{-}\to \Lambda_b^0  \pi^-   \eta  $ & \tabincell{c}{$ \sqrt{\frac{2}{3}} (-c_1+c_4+c_5-2 c_6+2 c_7$\\$+c_8-3 c_{12}-3 c_{13}+c_{15}+2 c_{16})$}\\\hline
$\Xi_{bb}^{-}\to \Lambda_b^0  K^0   K^-  $ & $ -c_1+c_4+c_5+2 c_7-c_8+c_{12}-3 c_{13}+c_{15}+2 c_{16}$\\\hline
$\Xi_{bb}^{-}\to \Xi_b^0  \pi^-   K^0  $ & $ -2 c_6+2 c_8-4 c_{14}+4 c_{15}$\\\hline
$\Xi_{bb}^{-}\to \Xi_b^-  \pi^0   K^0  $ & $ \frac{-c_1+c_4-c_5+2 c_6-2 c_7-c_8+5 c_{12}+c_{13}-4 c_{14}+c_{15}+2 c_{16}}{\sqrt{2}}$\\\hline
$\Xi_{bb}^{-}\to \Xi_b^-  \pi^-   K^+  $ & $ c_1-c_4+c_5+2 c_7-c_8+3 c_{12}-c_{13}+3 c_{15}-2 c_{16}$\\\hline
$\Xi_{bb}^{-}\to \Xi_b^-  K^0   \eta  $ & $ \frac{-c_1+c_4-c_5+2 c_6-2 c_7-c_8+5 c_{12}+c_{13}-4 c_{14}+c_{15}+2 c_{16}}{\sqrt{6}}$\\\hline
$\Omega_{bb}^{-}\to \Lambda_b^0  \pi^0   K^-  $ & $ \frac{-c_3+c_4-c_6+2 c_7-c_{10}+2 c_{11}+5 c_{14}-6 c_{15}+c_{16}}{\sqrt{2}}$\\\hline
$\Omega_{bb}^{-}\to \Lambda_b^0  \pi^-   \overline K^0  $ & $ -c_3+c_4-c_6+2 c_7-c_{10}+2 c_{11}-3 c_{14}+2 c_{15}+c_{16}$\\\hline
$\Omega_{bb}^{-}\to \Lambda_b^0  K^-   \eta  $ & $ \frac{c_3-c_4-3 c_6-2 c_7+4 c_8+c_{10}-2 c_{11}+3 c_{14}-2 c_{15}-c_{16}}{\sqrt{6}}$\\\hline
$\Omega_{bb}^{-}\to \Xi_b^0  \pi^0   \pi^-  $ & $ -4 \sqrt{2} c_{12}$\\\hline
$\Omega_{bb}^{-}\to \Xi_b^0  \pi^-   \eta  $ & \tabincell{c}{$ \sqrt{\frac{2}{3}} (-c_1+c_3+c_5+c_6-c_8+c_{10}-2 c_{11}$\\$-3 c_{12}-3 c_{13}+3 c_{14}-c_{15}+c_{16})$}\\\hline
$\Omega_{bb}^{-}\to \Xi_b^0  K^0   K^-  $ & \tabincell{c}{$ -c_1+c_3+c_5-c_6+c_8+c_{10}-2 c_{11}$\\$+c_{12}-3 c_{13}-c_{14}+3 c_{15}+c_{16}$}\\\hline
$\Omega_{bb}^{-}\to \Xi_b^-  \pi^+   \pi^-  $ & $ -c_1-2 c_2-c_5+4 c_9+c_{10}-3 c_{12}-c_{13}+c_{16}$\\\hline
$\Omega_{bb}^{-}\to \Xi_b^-  \pi^0   \pi^0  $ & $ -c_1-2 c_2-c_5+4 c_9+c_{10}+5 c_{12}-c_{13}+c_{16}$\\\hline
$\Omega_{bb}^{-}\to \Xi_b^-  \pi^0   \eta  $ & \tabincell{c}{$ \frac{1}{\sqrt{3}}(c_1-c_3-c_5-c_6+c_8-c_{10}$\\$+2 c_{11}-c_{12}-5 c_{13}+5 c_{14}+c_{15}-c_{16})$}\\\hline
$\Omega_{bb}^{-}\to \Xi_b^-  K^+   K^-  $ & $ -2 c_2-c_4+2 c_7-c_8+4 c_9+c_{10}-2 c_{13}+3 c_{15}-c_{16}$\\\hline
$\Omega_{bb}^{-}\to \Xi_b^-  K^0   \overline K^0  $ & \tabincell{c}{$ -c_1-2 c_2+c_3-c_4+c_5-c_6+2 c_7+4 c_9+2 c_{10}$\\$-2 c_{11}+c_{12}+3 c_{13}-c_{14}-2 c_{15}$}\\\hline
$\Omega_{bb}^{-}\to \Xi_b^-  \eta   \eta  $ & \tabincell{c}{$ \frac{1}{3} (-c_1-6 c_2+2 c_3-4 c_4+3 c_5-6 c_6+8 c_7+2 c_8$\\$+12 c_9+5 c_{10}-4 c_{11}-3 c_{12}+3 c_{13}+6 c_{14}-6 c_{15}-3 c_{16})$}\\\hline
\hline
\end{tabular}
\end{table}

\begin{table}
\footnotesize
\newcommand{\tabincell}[2]{\begin{tabular}{@{}#1@{}}#2\end{tabular}}
\caption{Doubly bottom baryon decays into a $bqq$(anti-triplet) induced by the charmless $b\to s$ transition and two light mesons.}\label{tab:bbq_bqq_qqbar_vs_3bar_bs}\begin{tabular}{|c c|c c|c|c|c|c|c|c}\hline
channel & amplitude \\\hline
\hline
$\Xi_{bb}^{0}\to \Lambda_b^0  \pi^+   K^-  $ & \tabincell{c}{$ c_1-c_3+c_5-c_6+c_8+c_{10}-2 c_{11}$\\$+3 c_{12}-c_{13}-3 c_{14}+c_{15}+3 c_{16}$}\\\hline
$\Xi_{bb}^{0}\to \Lambda_b^0  \pi^0   \overline K^0  $ & $ \frac{-c_1+c_3-c_5-c_6+c_8-c_{10}+2 c_{11}+5 c_{12}+c_{13}-c_{14}+3 c_{15}-3 c_{16}}{\sqrt{2}}$\\\hline
$\Xi_{bb}^{0}\to \Lambda_b^0  \overline K^0   \eta  $ & $ \frac{-c_1+c_3-c_5-c_6+c_8-c_{10}+2 c_{11}+5 c_{12}+c_{13}-c_{14}+3 c_{15}-3 c_{16}}{\sqrt{6}}$\\\hline
$\Xi_{bb}^{0}\to \Xi_b^0  \pi^+   \pi^-  $ & $ 2 c_2+c_4+2 c_7-c_8+4 c_9+c_{10}+2 c_{13}+c_{15}-3 c_{16}$\\\hline
$\Xi_{bb}^{0}\to \Xi_b^0  \pi^0   \pi^0  $ & \tabincell{c}{$ 2 c_2+c_4-2 c_6+2 c_7+c_8+4 c_9+c_{10}$\\$+2 c_{13}+4 c_{14}-3 c_{15}-3 c_{16}$}\\\hline
$\Xi_{bb}^{0}\to \Xi_b^0  \pi^0   \eta  $ & \tabincell{c}{$ \frac{1}{\sqrt{3}}(-c_3+c_4+2 c_5-c_6+2 c_7+c_{10}$\\$-2 c_{11}-4 c_{12}+4 c_{13}+5 c_{14}-6 c_{15}-3 c_{16})$}\\\hline
$\Xi_{bb}^{0}\to \Xi_b^0  K^+   K^-  $ & \tabincell{c}{$ c_1+2 c_2-c_3+c_4+c_5-c_6+2 c_7+4 c_9$\\$+2 c_{10}-2 c_{11}+3 c_{12}+c_{13}-3 c_{14}+2 c_{15}$}\\\hline
$\Xi_{bb}^{0}\to \Xi_b^0  K^0   \overline K^0  $ & $ c_1+2 c_2-c_5+4 c_9+c_{10}-c_{12}-3 c_{13}+3 c_{16}$\\\hline
$\Xi_{bb}^{0}\to \Xi_b^0  \eta   \eta  $ & \tabincell{c}{$ \frac{1}{3} (4 c_1+6 c_2-2 c_3+c_4+2 c_7-c_8+12 c_9$\\$+5 c_{10}-4 c_{11}-12 c_{12}-6 c_{13}+6 c_{14}-9 c_{15}+9 c_{16})$}\\\hline
$\Xi_{bb}^{0}\to \Xi_b^-  \pi^+   \pi^0  $ & $ \sqrt{2} \left(-c_6+c_8+2 c_{14}-2 c_{15}\right)$\\\hline
$\Xi_{bb}^{0}\to \Xi_b^-  \pi^+   \eta  $ & $ \sqrt{\frac{2}{3}} \left(-c_3+c_4-2 c_7+c_8+c_{10}-2 c_{11}+3 c_{14}-3 c_{16}\right)$\\\hline
$\Xi_{bb}^{0}\to \Xi_b^-  K^+   \overline K^0  $ & $ -c_3+c_4+c_6-2 c_7+c_{10}-2 c_{11}+c_{14}+2 c_{15}-3 c_{16}$\\\hline
$\Xi_{bb}^{-}\to \Lambda_b^0  \pi^0   K^-  $ & $ \frac{-c_1+c_3+c_5+c_6-c_8+c_{10}-2 c_{11}-7 c_{12}-3 c_{13}+3 c_{14}-c_{15}+c_{16}}{\sqrt{2}}$\\\hline
$\Xi_{bb}^{-}\to \Lambda_b^0  \pi^-   \overline K^0  $ & \tabincell{c}{$ -c_1+c_3+c_5-c_6+c_8+c_{10}-2 c_{11}$\\$+c_{12}-3 c_{13}-c_{14}+3 c_{15}+c_{16}$}\\\hline
$\Xi_{bb}^{-}\to \Lambda_b^0  K^-   \eta  $ & $ \frac{c_1-c_3-c_5-c_6+c_8-c_{10}+2 c_{11}-9 c_{12}+3 c_{13}-3 c_{14}+c_{15}-c_{16}}{\sqrt{6}}$\\\hline
$\Xi_{bb}^{-}\to \Xi_b^0  \pi^0   \pi^-  $ & $ \sqrt{2} \left(-c_6+c_8+2 c_{14}-2 c_{15}\right)$\\\hline
$\Xi_{bb}^{-}\to \Xi_b^0  \pi^-   \eta  $ & $ \sqrt{\frac{2}{3}} \left(-c_3+c_4+2 c_7-c_8-c_{10}+2 c_{11}+3 c_{14}-4 c_{15}+c_{16}\right)$\\\hline
$\Xi_{bb}^{-}\to \Xi_b^0  K^0   K^-  $ & $ -c_3+c_4-c_6+2 c_7-c_{10}+2 c_{11}-3 c_{14}+2 c_{15}+c_{16}$\\\hline
$\Xi_{bb}^{-}\to \Xi_b^-  \pi^+   \pi^-  $ & $ 2 c_2+c_4-2 c_7+c_8-4 c_9-c_{10}+2 c_{13}-3 c_{15}+c_{16}$\\\hline
$\Xi_{bb}^{-}\to \Xi_b^-  \pi^0   \pi^0  $ &\tabincell{c}{$ 2 c_2+c_4+2 c_6-2 c_7-c_8$\\$-4 c_9-c_{10}+2 c_{13}-4 c_{14}+c_{15}+c_{16}$}\\\hline
$\Xi_{bb}^{-}\to \Xi_b^-  \pi^0   \eta  $ & \tabincell{c}{$ \frac{1}{\sqrt{3}}(c_3-c_4+2 c_5-c_6+2 c_7+c_{10}$\\$-2 c_{11}-4 c_{12}+4 c_{13}-c_{14}-2 c_{15}-c_{16})$}\\\hline
$\Xi_{bb}^{-}\to \Xi_b^-  K^+   K^-  $ & $ c_1+2 c_2+c_5-4 c_9-c_{10}+3 c_{12}+c_{13}-c_{16}$\\\hline
$\Xi_{bb}^{-}\to \Xi_b^-  K^0   \overline K^0  $ & \tabincell{c}{$ c_1+2 c_2-c_3+c_4-c_5+c_6-2 c_7-4 c_9$\\$-2 c_{10}+2 c_{11}-c_{12}-3 c_{13}+c_{14}+2 c_{15}$}\\\hline
$\Xi_{bb}^{-}\to \Xi_b^-  \eta   \eta  $ & \tabincell{c}{$ \frac{1}{3} (4 c_1+6 c_2-2 c_3+c_4-2 c_7+c_8-12 c_9$\\$-5 c_{10}+4 c_{11}-12 c_{12}-6 c_{13}+6 c_{14}+3 c_{15}-3 c_{16})$}\\\hline
$\Omega_{bb}^{-}\to \Lambda_b^0  \overline K^0   K^-  $ & $ -2 c_6+2 c_8-4 c_{14}+4 c_{15}$\\\hline
$\Omega_{bb}^{-}\to \Xi_b^0  \pi^0   K^-  $ & $ \frac{-c_1+c_4+c_5-2 c_6+2 c_7+c_8-7 c_{12}-3 c_{13}+4 c_{14}-3 c_{15}+2 c_{16}}{\sqrt{2}}$\\\hline
$\Omega_{bb}^{-}\to \Xi_b^0  \pi^-   \overline K^0  $ & $ -c_1+c_4+c_5+2 c_7-c_8+c_{12}-3 c_{13}+c_{15}+2 c_{16}$\\\hline
$\Omega_{bb}^{-}\to \Xi_b^0  K^-   \eta  $ & $ \frac{c_1-c_4-c_5+2 c_6-2 c_7-c_8-9 c_{12}+3 c_{13}+12 c_{14}-13 c_{15}-2 c_{16}}{\sqrt{6}}$\\\hline
$\Omega_{bb}^{-}\to \Xi_b^-  \pi^+   K^-  $ & $ -c_1+c_4-c_5-2 c_7+c_8-3 c_{12}+c_{13}-3 c_{15}+2 c_{16}$\\\hline
$\Omega_{bb}^{-}\to \Xi_b^-  \pi^0   \overline K^0  $ & $ \frac{c_1-c_4+c_5-2 c_6+2 c_7+c_8-5 c_{12}-c_{13}+4 c_{14}-c_{15}-2 c_{16}}{\sqrt{2}}$\\\hline
$\Omega_{bb}^{-}\to \Xi_b^-  \overline K^0   \eta  $ & $ \frac{c_1-c_4+c_5-2 c_6+2 c_7+c_8-5 c_{12}-c_{13}+4 c_{14}-c_{15}-2 c_{16}}{\sqrt{6}}$\\\hline
\hline
\end{tabular}
\end{table}

\begin{table}
\footnotesize
\newcommand{\tabincell}[2]{\begin{tabular}{@{}#1@{}}#2\end{tabular}}
\caption{Doubly bottom baryon decays into a $bqq$(sextet) induced by the charmless $b\to d$ transition and two light mesons.}\label{tab:bbq_bqq_qqbar_vd_6_bd_1}\begin{tabular}{|c c|c c|c|c|c|c|c|c}\hline
channel & amplitude \\\hline
\hline
$\Xi_{bb}^{0}\to \Sigma_{b}^{+}  \pi^0   \pi^-  $ & $ \frac{c_3+c_6+c_7+8 c_9+3 c_{11}+3 c_{13}+6 c_{16}}{\sqrt{2}}$\\\hline
$\Xi_{bb}^{0}\to \Sigma_{b}^{+}  \pi^-   \eta  $ & $ \frac{2 c_1+c_3-2 c_5+c_6+c_7+2 c_8+6 c_9+6 c_{10}+3 c_{11}+3 c_{13}+6 c_{15}+6 c_{16}}{\sqrt{6}}$\\\hline
$\Xi_{bb}^{0}\to \Sigma_{b}^{+}  K^0   K^-  $ & $ c_1-c_5+c_8-c_9+3 c_{10}+3 c_{15}$\\\hline
$\Xi_{bb}^{0}\to \Sigma_{b}^{0}  \pi^+   \pi^-  $ & $ \frac{c_1+2 c_2+c_3+c_4+c_5+c_6+3 c_9+c_{10}+3 c_{11}+6 c_{12}-2 c_{13}+12 c_{14}+6 c_{15}+6 c_{16}}{\sqrt{2}}$\\\hline
$\Xi_{bb}^{0}\to \Sigma_{b}^{0}  \pi^0   \pi^0  $ & $ \frac{c_1+2 c_2-c_3+c_4+c_5-c_6-2 c_7-5 c_9+c_{10}+5 c_{11}+6 c_{12}+12 c_{14}+6 c_{15}-6 c_{16}}{\sqrt{2}}$\\\hline
$\Xi_{bb}^{0}\to \Sigma_{b}^{0}  \pi^0   \eta  $ & $ \frac{-c_1+c_4+c_5-2 c_6-c_7-2 c_8+c_9+5 c_{10}+4 c_{11}+6 c_{12}+3 c_{13}}{\sqrt{6}}$\\\hline
$\Xi_{bb}^{0}\to \Sigma_{b}^{0}  K^+   K^-  $ & $ \frac{2 c_2+c_4+c_7-c_8+2 c_{10}+6 c_{12}-c_{13}+12 c_{14}+3 c_{15}}{\sqrt{2}}$\\\hline
$\Xi_{bb}^{0}\to \Sigma_{b}^{0}  K^0   \overline K^0  $ & $ \frac{c_1+2 c_2-c_5+c_8-c_9-3 c_{10}+12 c_{14}+3 c_{15}}{\sqrt{2}}$\\\hline
$\Xi_{bb}^{0}\to \Sigma_{b}^{0}  \eta   \eta  $ & $ \frac{c_1+6 c_2+c_3+c_4-3 c_5-3 c_6+3 c_9-3 c_{10}+3 c_{11}+6 c_{12}+6 c_{13}+36 c_{14}+6 c_{15}+6 c_{16}}{3 \sqrt{2}}$\\\hline
$\Xi_{bb}^{0}\to \Sigma_{b}^{-}  \pi^+   \pi^0  $ & $ -\frac{c_3+c_6+c_7-5 c_{11}-5 c_{13}+6 c_{16}}{\sqrt{2}}$\\\hline
$\Xi_{bb}^{0}\to \Sigma_{b}^{-}  \pi^+   \eta  $ & $ \frac{c_3+2 c_4-3 c_6-c_7-2 c_8+3 c_{11}-4 c_{12}+c_{13}+6 c_{15}+6 c_{16}}{\sqrt{6}}$\\\hline
$\Xi_{bb}^{0}\to \Sigma_{b}^{-}  K^+   \overline K^0  $ & $ c_4+c_7-c_8-2 c_{12}-c_{13}+3 c_{15}$\\\hline
$\Xi_{bb}^{0}\to \Xi_{b}^{\prime0}  \pi^0   K^0  $ & $ \frac{1}{2} \left(-c_1+c_3-c_5-c_6+c_7-c_8+5 c_9+c_{10}-c_{11}+3 c_{13}-3 c_{15}+6 c_{16}\right)$\\\hline
$\Xi_{bb}^{0}\to \Xi_{b}^{\prime0}  \pi^-   K^+  $ & $ \frac{c_1+c_3+c_5+c_6-c_7+c_8+3 c_9-c_{10}+3 c_{11}-c_{13}+3 c_{15}+6 c_{16}}{\sqrt{2}}$\\\hline
$\Xi_{bb}^{0}\to \Xi_{b}^{\prime0}  K^0   \eta  $ & $ -\frac{c_1-c_3+c_5+c_6-c_7+c_8-5 c_9-c_{10}+c_{11}-3 c_{13}+3 c_{15}-6 c_{16}}{2 \sqrt{3}}$\\\hline
$\Xi_{bb}^{0}\to \Xi_{b}^{\prime-}  \pi^+   K^0  $ & $ \frac{c_3+c_4-c_6-c_8-c_{11}-2 c_{12}-2 c_{13}+3 c_{15}+6 c_{16}}{\sqrt{2}}$\\\hline
$\Xi_{bb}^{0}\to \Xi_{b}^{\prime-}  \pi^0   K^+  $ & $ \frac{1}{2} \left(-c_3+c_4-c_6-c_8+5 c_{11}-2 c_{12}+4 c_{13}+3 c_{15}-6 c_{16}\right)$\\\hline
$\Xi_{bb}^{0}\to \Xi_{b}^{\prime-}  K^+   \eta  $ & $ \frac{c_3-c_4-3 c_6-4 c_7+c_8+3 c_{11}+2 c_{12}+4 c_{13}-3 c_{15}+6 c_{16}}{2 \sqrt{3}}$\\\hline
$\Xi_{bb}^{0}\to \Omega_{b}^{-}  K^+   K^0  $ & $ c_3-c_6-c_7-c_{11}-c_{13}+6 c_{16}$\\\hline
$\Xi_{bb}^{-}\to \Sigma_{b}^{+}  \pi^-   \pi^-  $ & $ 2 \left(c_3+c_6+c_7+3 c_{11}+3 c_{13}-2 c_{16}\right)$\\\hline
$\Xi_{bb}^{-}\to \Sigma_{b}^{0}  \pi^0   \pi^-  $ & $ -c_3-c_6-c_7+4 c_9+c_{11}+c_{13}+2 c_{16}$\\\hline
$\Xi_{bb}^{-}\to \Sigma_{b}^{0}  \pi^-   \eta  $ & $ \frac{c_1+c_3+c_4-c_5-c_6+3 c_9+3 c_{10}+3 c_{11}+6 c_{12}+2 c_{13}-2 c_{15}-2 c_{16}}{\sqrt{3}}$\\\hline
$\Xi_{bb}^{-}\to \Sigma_{b}^{0}  K^0   K^-  $ & $ \frac{c_1+c_4-c_5+c_7-c_9+3 c_{10}+6 c_{12}-c_{13}-2 c_{15}}{\sqrt{2}}$\\\hline
$\Xi_{bb}^{-}\to \Sigma_{b}^{-}  \pi^+   \pi^-  $ & $ c_1+2 c_2+c_4+c_5-c_7+3 c_9+c_{10}-2 c_{12}+3 c_{13}-4 c_{14}-2 c_{15}$\\\hline
$\Xi_{bb}^{-}\to \Sigma_{b}^{-}  \pi^0   \pi^0  $ & $ c_1+2 c_2+c_3+c_4+c_5+c_6-5 c_9+c_{10}-5 c_{11}-2 c_{12}-2 c_{13}-4 c_{14}-2 c_{15}-2 c_{16}$\\\hline
$\Xi_{bb}^{-}\to \Sigma_{b}^{-}  \pi^0   \eta  $ & $ \frac{-c_1-c_3-c_4+c_5+c_6+c_9+5 c_{10}+c_{11}+2 c_{12}+2 c_{13}+2 c_{15}+2 c_{16}}{\sqrt{3}}$\\\hline
$\Xi_{bb}^{-}\to \Sigma_{b}^{-}  K^+   K^-  $ & $ 2 \left(c_2+c_{10}-2 c_{14}\right)$\\\hline
$\Xi_{bb}^{-}\to \Sigma_{b}^{-}  K^0   \overline K^0  $ & $ c_1+2 c_2+c_4-c_5+c_7-c_9-3 c_{10}-2 c_{12}-c_{13}-4 c_{14}-2 c_{15}$\\\hline
$\Xi_{bb}^{-}\to \Sigma_{b}^{-}  \eta   \eta  $ & $ \frac{1}{3} \left(c_1+6 c_2+c_3+c_4-3 c_5-3 c_6+3 c_9-3 c_{10}+3 c_{11}-2 c_{12}-2 c_{13}-12 c_{14}-2 c_{15}-2 c_{16}\right)$\\\hline
$\Xi_{bb}^{-}\to \Xi_{b}^{\prime0}  \pi^-   K^0  $ & $ \sqrt{2} \left(c_3+c_{11}+c_{13}-2 c_{16}\right)$\\\hline
$\Xi_{bb}^{-}\to \Xi_{b}^{\prime-}  \pi^0   K^0  $ & $ \frac{1}{2} \left(-c_1-2 c_3-c_4-c_5+c_7+5 c_9+c_{10}+6 c_{11}+2 c_{12}+3 c_{13}+2 c_{15}+4 c_{16}\right)$\\\hline
$\Xi_{bb}^{-}\to \Xi_{b}^{\prime-}  \pi^-   K^+  $ & $ \frac{c_1+c_4+c_5-c_7+3 c_9-c_{10}-2 c_{12}+3 c_{13}-2 c_{15}}{\sqrt{2}}$\\\hline
$\Xi_{bb}^{-}\to \Xi_{b}^{\prime-}  K^0   \eta  $ & $ -\frac{c_1-2 c_3+c_4+c_5+4 c_6+3 c_7-5 c_9-c_{10}-2 c_{11}-2 c_{12}+c_{13}-2 c_{15}+4 c_{16}}{2 \sqrt{3}}$\\\hline
$\Xi_{bb}^{-}\to \Omega_{b}^{-}  K^0   K^0  $ & $ 2 \left(c_3-c_6-c_7-c_{11}-c_{13}-2 c_{16}\right)$\\\hline
$\Omega_{bb}^{-}\to \Sigma_{b}^{+}  \pi^-   K^-  $ & $ c_3+c_6+c_7+3 c_{11}+3 c_{13}-2 c_{16}$\\\hline
$\Omega_{bb}^{-}\to \Sigma_{b}^{0}  \pi^0   K^-  $ & $ \frac{1}{2} \left(-c_3+c_4-c_6-2 c_7+c_8+5 c_{11}+6 c_{12}-c_{15}+2 c_{16}\right)$\\\hline
$\Omega_{bb}^{-}\to \Sigma_{b}^{0}  \pi^-   \overline K^0  $ & $ \frac{c_3+c_4+c_6+c_8+3 c_{11}+6 c_{12}-2 c_{13}-c_{15}-2 c_{16}}{\sqrt{2}}$\\\hline
$\Omega_{bb}^{-}\to \Sigma_{b}^{0}  K^-   \eta  $ & $ \frac{c_3-c_4-3 c_6-2 c_7-c_8+3 c_{11}-6 c_{12}+8 c_{13}+c_{15}-2 c_{16}}{2 \sqrt{3}}$\\\hline
$\Omega_{bb}^{-}\to \Sigma_{b}^{-}  \pi^+   K^-  $ & $ c_4-c_7+c_8-2 c_{12}+3 c_{13}-c_{15}$\\\hline
$\Omega_{bb}^{-}\to \Sigma_{b}^{-}  \pi^0   \overline K^0  $ & $ -\frac{c_3+c_4+c_6+c_8-5 c_{11}-2 c_{12}-2 c_{13}-c_{15}-2 c_{16}}{\sqrt{2}}$\\\hline
$\Omega_{bb}^{-}\to \Sigma_{b}^{-}  \overline K^0   \eta  $ & $ \frac{c_3-c_4-3 c_6-2 c_7-c_8+3 c_{11}+2 c_{12}+c_{15}-2 c_{16}}{\sqrt{6}}$\\\hline
\hline
\end{tabular}
\end{table}

\begin{table}
\small
\newcommand{\tabincell}[2]{\begin{tabular}{@{}#1@{}}#2\end{tabular}}
\caption{Doubly bottom baryon decays into a $bqq$(sextet) induced by the charmless $b\to d$ transitionand two light mesons.}\label{tab:bbq_bqq_qqbar_vd_6_bd_2}\begin{tabular}{|c c|c c|c|c|c|c|c|c}\hline
channel & amplitude \\\hline
\hline
$\Omega_{bb}^{-}\to \Xi_{b}^{\prime0}  \pi^0   \pi^-  $ & $ 4 c_9$\\\hline
$\Omega_{bb}^{-}\to \Xi_{b}^{\prime0}  \pi^-   \eta  $ & $ \frac{c_1-c_3-c_5-c_6+c_7-c_8+3 c_9+3 c_{10}-3 c_{11}+c_{13}-c_{15}+2 c_{16}}{\sqrt{3}}$\\\hline
$\Omega_{bb}^{-}\to \Xi_{b}^{\prime0}  K^0   K^-  $ & $ \frac{c_1+c_3-c_5-c_6+c_7-c_8-c_9+3 c_{10}-c_{11}+3 c_{13}-c_{15}-2 c_{16}}{\sqrt{2}}$\\\hline
$\Omega_{bb}^{-}\to \Xi_{b}^{\prime-}  \pi^+   \pi^-  $ & $ \frac{c_1+2 c_2+c_5-c_8+3 c_9+c_{10}-4 c_{14}-c_{15}}{\sqrt{2}}$\\\hline
$\Omega_{bb}^{-}\to \Xi_{b}^{\prime-}  \pi^0   \pi^0  $ & $ \frac{c_1+2 c_2+c_5-c_8-5 c_9+c_{10}-4 c_{14}-c_{15}}{\sqrt{2}}$\\\hline
$\Omega_{bb}^{-}\to \Xi_{b}^{\prime-}  \pi^0   \eta  $ & $ \frac{-c_1+c_3+c_5+c_6-c_7+c_8+c_9+5 c_{10}-5 c_{11}-c_{13}+c_{15}-2 c_{16}}{\sqrt{6}}$\\\hline
$\Omega_{bb}^{-}\to \Xi_{b}^{\prime-}  K^+   K^-  $ & $ \frac{2 c_2+c_4-c_7+c_8+2 c_{10}-2 c_{12}+3 c_{13}-4 c_{14}-c_{15}}{\sqrt{2}}$\\\hline
$\Omega_{bb}^{-}\to \Xi_{b}^{\prime-}  K^0   \overline K^0  $ & $ \frac{c_1+2 c_2+c_3+c_4-c_5-c_6-c_9-3 c_{10}-c_{11}-2 c_{12}-2 c_{13}-4 c_{14}-2 c_{15}-2 c_{16}}{\sqrt{2}}$\\\hline
$\Omega_{bb}^{-}\to \Xi_{b}^{\prime-}  \eta   \eta  $ & $ \frac{c_1+6 c_2-2 c_3+4 c_4-3 c_5+6 c_6+6 c_7+3 c_8+3 c_9-3 c_{10}-6 c_{11}-8 c_{12}-2 c_{13}-12 c_{14}-5 c_{15}+4 c_{16}}{3 \sqrt{2}}$\\\hline
$\Omega_{bb}^{-}\to \Omega_{b}^{-}  \pi^0   K^0  $ & $ \frac{-c_1-c_5+c_8+5 c_9+c_{10}+c_{15}}{\sqrt{2}}$\\\hline
$\Omega_{bb}^{-}\to \Omega_{b}^{-}  \pi^-   K^+  $ & $ c_1+c_5-c_8+3 c_9-c_{10}-c_{15}$\\\hline
$\Omega_{bb}^{-}\to \Omega_{b}^{-}  K^0   \eta  $ & $ \frac{-c_1-2 c_3-c_5+2 c_6+2 c_7+c_8+5 c_9+c_{10}+2 c_{11}+2 c_{13}+c_{15}+4 c_{16}}{\sqrt{6}}$\\\hline
\hline
\end{tabular}
\end{table}

\begin{table}
\footnotesize
\newcommand{\tabincell}[2]{\begin{tabular}{@{}#1@{}}#2\end{tabular}}
\caption{Doubly bottom baryon decays into a $bqq$(sextet)  induced by the charmless $b\to s$ transition and  two light mesons.}\label{tab:bbq_bqq_qqbar_vs_6_bs_1}\begin{tabular}{|c c|c c|c|c|c|c|c|c}\hline
channel & amplitude \\\hline
\hline
$\Xi_{bb}^{0}\to \Sigma_{b}^{+}  \pi^0   K^-  $ & $ \frac{c_1+c_3-c_5+c_6+c_7+c_8+c_9+7 c_{10}+3 c_{11}+3 c_{12}+3 c_{14}+3 c_{16}+6 c_{17}}{\sqrt{2}}$\\\hline
$\Xi_{bb}^{0}\to \Sigma_{b}^{+}  \pi^-   \overline K^0  $ & $ c_1-c_5+c_6+c_9-c_{10}+3 c_{11}+3 c_{16}$\\\hline
$\Xi_{bb}^{0}\to \Sigma_{b}^{+}  K^-   \eta  $ & $ \frac{-c_1+c_3+c_5-c_6+c_7+c_8-c_9+9 c_{10}-3 c_{11}+3 c_{12}+3 c_{14}-3 c_{16}+6 c_{17}}{\sqrt{6}}$\\\hline
$\Xi_{bb}^{0}\to \Sigma_{b}^{0}  \pi^+   K^-  $ & \tabincell{c}{$ \frac{1}{\sqrt{2}}(c_1+c_3+c_5-c_6+c_7-c_8+c_9+3 c_{10}$\\$-c_{11}+3 c_{12}-c_{14}+3 c_{16}+6 c_{17})$}\\\hline
$\Xi_{bb}^{0}\to \Sigma_{b}^{0}  \pi^0   \overline K^0  $ & \tabincell{c}{$ \frac{1}{2} (-c_1+c_3-c_5+c_6-c_7+c_8-c_9+5 c_{10}$\\$+c_{11}-c_{12}+3 c_{14}-3 c_{16}+6 c_{17})$}\\\hline
$\Xi_{bb}^{0}\to \Sigma_{b}^{0}  \overline K^0   \eta  $ & $ -\frac{c_1-c_3+c_5-c_6+c_7-c_8+c_9-5 c_{10}-c_{11}+c_{12}-3 c_{14}+3 c_{16}-6 c_{17}}{2 \sqrt{3}}$\\\hline
$\Xi_{bb}^{0}\to \Sigma_{b}^{-}  \pi^+   \overline K^0  $ & $ c_3-c_7-c_8-c_{12}-c_{14}+6 c_{17}$\\\hline
$\Xi_{bb}^{0}\to \Xi_{b}^{\prime0}  \pi^+   \pi^-  $ & $ \frac{2 c_2+c_4+c_8-c_9+2 c_{11}-c_{14}+3 c_{16}}{\sqrt{2}}$\\\hline
$\Xi_{bb}^{0}\to \Xi_{b}^{\prime0}  \pi^0   \pi^0  $ & $ \frac{2 c_2+c_4-2 c_7-c_8-c_9+2 c_{11}+4 c_{12}+3 c_{14}+3 c_{16}}{\sqrt{2}}$\\\hline
$\Xi_{bb}^{0}\to \Xi_{b}^{\prime0}  \pi^0   \eta  $ & $ -\frac{c_3-c_4-2 c_5+2 c_6+c_7+2 c_8+c_9+4 c_{10}-4 c_{11}-5 c_{12}-3 c_{16}+6 c_{17}}{\sqrt{6}}$\\\hline
$\Xi_{bb}^{0}\to \Xi_{b}^{\prime0}  K^+   K^-  $ & $ \frac{c_1+2 c_2+c_3+c_4+c_5-c_6+c_7+3 c_{10}+c_{11}+3 c_{12}-2 c_{14}+6 c_{16}+6 c_{17}}{\sqrt{2}}$\\\hline
$\Xi_{bb}^{0}\to \Xi_{b}^{\prime0}  K^0   \overline K^0  $ & $ \frac{c_1+2 c_2-c_5+c_6+c_9-c_{10}-3 c_{11}+3 c_{16}}{\sqrt{2}}$\\\hline
$\Xi_{bb}^{0}\to \Xi_{b}^{\prime0}  \eta   \eta  $ & $ \frac{4 c_1+6 c_2-2 c_3+c_4-3 c_8+3 c_9-12 c_{10}-6 c_{11}+6 c_{12}-3 c_{14}+15 c_{16}-12 c_{17}}{3 \sqrt{2}}$\\\hline
$\Xi_{bb}^{0}\to \Xi_{b}^{\prime-}  \pi^+   \pi^0  $ & $ -c_7-c_8+2 \left(c_{12}+c_{14}\right)$\\\hline
$\Xi_{bb}^{0}\to \Xi_{b}^{\prime-}  \pi^+   \eta  $ & $ \frac{-c_3+c_4+c_8-c_9+3 c_{12}+2 c_{14}+3 c_{16}-6 c_{17}}{\sqrt{3}}$\\\hline
$\Xi_{bb}^{0}\to \Xi_{b}^{\prime-}  K^+   \overline K^0  $ & $ \frac{c_3+c_4-c_7-c_9-c_{12}-2 c_{14}+3 c_{16}+6 c_{17}}{\sqrt{2}}$\\\hline
$\Xi_{bb}^{0}\to \Omega_{b}^{-}  \pi^+   K^0  $ & $ c_4+c_8-c_9-c_{14}+3 c_{16}$\\\hline
$\Xi_{bb}^{0}\to \Omega_{b}^{-}  \pi^0   K^+  $ & $ \frac{c_4-2 c_7-c_8-c_9+4 c_{12}+3 c_{14}+3 c_{16}}{\sqrt{2}}$\\\hline
$\Xi_{bb}^{0}\to \Omega_{b}^{-}  K^+   \eta  $ & $ -\frac{2 c_3+c_4+c_8-c_9-6 c_{12}-7 c_{14}+3 c_{16}+12 c_{17}}{\sqrt{6}}$\\\hline
$\Xi_{bb}^{-}\to \Sigma_{b}^{+}  \pi^-   K^-  $ & $ c_3+c_7+c_8+3 c_{12}+3 c_{14}-2 c_{17}$\\\hline
$\Xi_{bb}^{-}\to \Sigma_{b}^{0}  \pi^0   K^-  $ & \tabincell{c}{$ \frac{1}{2} (c_1-c_3-c_5+c_6-c_7+c_8-c_9+7 c_{10}$\\$+3 c_{11}-3 c_{12}+c_{14}-c_{16}+2 c_{17})$}\\\hline
$\Xi_{bb}^{-}\to \Sigma_{b}^{0}  \pi^-   \overline K^0  $ & $ \frac{c_1+c_3-c_5+c_6-c_7+c_8-c_9-c_{10}+3 c_{11}-c_{12}+3 c_{14}-c_{16}-2 c_{17}}{\sqrt{2}}$\\\hline
$\Xi_{bb}^{-}\to \Sigma_{b}^{0}  K^-   \eta  $ & $ \frac{-c_1+c_3+c_5-c_6+c_7-c_8+c_9+9 c_{10}-3 c_{11}+3 c_{12}-c_{14}+c_{16}-2 c_{17}}{2 \sqrt{3}}$\\\hline
$\Xi_{bb}^{-}\to \Sigma_{b}^{-}  \pi^+   K^-  $ & $ c_1+c_5-c_6-c_9+3 c_{10}-c_{11}-c_{16}$\\\hline
$\Xi_{bb}^{-}\to \Sigma_{b}^{-}  \pi^0   \overline K^0  $ & $ \frac{-c_1-c_3-c_5+c_6+c_7+c_8+c_9+5 c_{10}+c_{11}+c_{12}+c_{14}+c_{16}+2 c_{17}}{\sqrt{2}}$\\\hline
$\Xi_{bb}^{-}\to \Sigma_{b}^{-}  \overline K^0   \eta  $ & $ -\frac{c_1-c_3+c_5-c_6+c_7+c_8-c_9-5 c_{10}-c_{11}+c_{12}+c_{14}-c_{16}+2 c_{17}}{\sqrt{6}}$\\\hline
$\Xi_{bb}^{-}\to \Xi_{b}^{\prime0}  \pi^0   \pi^-  $ & $ -c_7-c_8+2 \left(c_{12}+c_{14}\right)$\\\hline
$\Xi_{bb}^{-}\to \Xi_{b}^{\prime0}  \pi^-   \eta  $ & $ -\frac{c_3-c_4+c_8-c_9-3 c_{12}+2 c_{14}+c_{16}-2 c_{17}}{\sqrt{3}}$\\\hline
$\Xi_{bb}^{-}\to \Xi_{b}^{\prime0}  K^0   K^-  $ & $ \frac{c_3+c_4+c_7+c_9+3 c_{12}-2 c_{14}-c_{16}-2 c_{17}}{\sqrt{2}}$\\\hline
$\Xi_{bb}^{-}\to \Xi_{b}^{\prime-}  \pi^+   \pi^-  $ & $ \frac{2 c_2+c_4-c_8+c_9+2 c_{11}+3 c_{14}-c_{16}}{\sqrt{2}}$\\\hline
$\Xi_{bb}^{-}\to \Xi_{b}^{\prime-}  \pi^0   \pi^0  $ & $ \frac{2 c_2+c_4+2 c_7+c_8+c_9+2 c_{11}-4 c_{12}-c_{14}-c_{16}}{\sqrt{2}}$\\\hline
$\Xi_{bb}^{-}\to \Xi_{b}^{\prime-}  \pi^0   \eta  $ & $ \frac{c_3-c_4+2 c_5-2 c_6-c_7-2 c_8-c_9-4 c_{10}+4 c_{11}-c_{12}+c_{16}-2 c_{17}}{\sqrt{6}}$\\\hline
$\Xi_{bb}^{-}\to \Xi_{b}^{\prime-}  K^+   K^-  $ & $ \frac{c_1+2 c_2+c_5-c_6-c_9+3 c_{10}+c_{11}-c_{16}}{\sqrt{2}}$\\\hline
$\Xi_{bb}^{-}\to \Xi_{b}^{\prime-}  K^0   \overline K^0  $ & $ \frac{c_1+2 c_2+c_3+c_4-c_5+c_6-c_7-c_{10}-3 c_{11}-c_{12}-2 c_{14}-2 c_{16}-2 c_{17}}{\sqrt{2}}$\\\hline
$\Xi_{bb}^{-}\to \Xi_{b}^{\prime-}  \eta   \eta  $ & $ \frac{4 c_1+6 c_2-2 c_3+c_4+3 c_8-3 c_9-12 c_{10}-6 c_{11}+6 c_{12}+c_{14}-5 c_{16}+4 c_{17}}{3 \sqrt{2}}$\\\hline
$\Xi_{bb}^{-}\to \Omega_{b}^{-}  \pi^0   K^0  $ & $ -\frac{c_4+2 c_7+c_8+c_9-4 c_{12}-c_{14}-c_{16}}{\sqrt{2}}$\\\hline
$\Xi_{bb}^{-}\to \Omega_{b}^{-}  \pi^-   K^+  $ & $ c_4-c_8+c_9+3 c_{14}-c_{16}$\\\hline
$\Xi_{bb}^{-}\to \Omega_{b}^{-}  K^0   \eta  $ & $ \frac{-2 c_3-c_4+c_8-c_9+6 c_{12}+3 c_{14}+c_{16}+4 c_{17}}{\sqrt{6}}$\\\hline
\hline
\end{tabular}
\end{table}

\begin{table}
\small
\newcommand{\tabincell}[2]{\begin{tabular}{@{}#1@{}}#2\end{tabular}}
\caption{Doubly bottom baryon decays into a $bqq$(sextet)  induced by the charmless $b\to s$ transition and  two light mesons.}\label{tab:bbq_bqq_qqbar_vs_6_bs_2}\begin{tabular}{|c c|c c|c|c|c|c|c|c}\hline
channel & amplitude \\\hline
\hline
$\Omega_{bb}^{-}\to \Sigma_{b}^{+}  K^-   K^-  $ & $ 2 \left(c_3+c_7+c_8+3 c_{12}+3 c_{14}-2 c_{17}\right)$\\\hline
$\Omega_{bb}^{-}\to \Sigma_{b}^{0}  \overline K^0   K^-  $ & $ \sqrt{2} \left(c_3+c_{12}+c_{14}-2 c_{17}\right)$\\\hline
$\Omega_{bb}^{-}\to \Sigma_{b}^{-}  \overline K^0   \overline K^0  $ & $ 2 \left(c_3-c_7-c_8-c_{12}-c_{14}-2 c_{17}\right)$\\\hline
$\Omega_{bb}^{-}\to \Xi_{b}^{\prime0}  \pi^0   K^-  $ & \tabincell{c}{$ \frac{1}{2} (c_1+c_4-c_5+c_6-2 c_7-c_8+7 c_{10}$\\$+3 c_{11}+4 c_{12}+3 c_{14}-2 c_{16})$}\\\hline
$\Omega_{bb}^{-}\to \Xi_{b}^{\prime0}  \pi^-   \overline K^0  $ & $ \frac{c_1+c_4-c_5+c_6+c_8-c_{10}+3 c_{11}-c_{14}-2 c_{16}}{\sqrt{2}}$\\\hline
$\Omega_{bb}^{-}\to \Xi_{b}^{\prime0}  K^-   \eta  $ & $ -\frac{c_1+4 c_3+c_4-c_5+c_6+2 c_7+3 c_8-9 c_{10}+3 c_{11}-c_{14}-2 c_{16}-8 c_{17}}{2 \sqrt{3}}$\\\hline
$\Omega_{bb}^{-}\to \Xi_{b}^{\prime-}  \pi^+   K^-  $ & $ \frac{c_1+c_4+c_5-c_6-c_8+3 c_{10}-c_{11}+3 c_{14}-2 c_{16}}{\sqrt{2}}$\\\hline
$\Omega_{bb}^{-}\to \Xi_{b}^{\prime-}  \pi^0   \overline K^0  $ & \tabincell{c}{$ \frac{1}{2} (-c_1-c_4-c_5+c_6-2 c_7-c_8+5 c_{10}$\\$+c_{11}+4 c_{12}+c_{14}+2 c_{16})$}\\\hline
$\Omega_{bb}^{-}\to \Xi_{b}^{\prime-}  \overline K^0   \eta  $ & \tabincell{c}{$ \frac{1}{2 \sqrt{3}}(-c_1-4 c_3-c_4-c_5+c_6+2 c_7+3 c_8+5 c_{10}$\\$+c_{11}+8 c_{12}+5 c_{14}+2 c_{16}+8 c_{17})$}\\\hline
$\Omega_{bb}^{-}\to \Omega_{b}^{-}  \pi^+   \pi^-  $ & $ 2 \left(c_2+c_{11}\right)$\\\hline
$\Omega_{bb}^{-}\to \Omega_{b}^{-}  \pi^0   \pi^0  $ & $ 2 \left(c_2+c_{11}\right)$\\\hline
$\Omega_{bb}^{-}\to \Omega_{b}^{-}  \pi^0   \eta  $ & $ \frac{2 \left(c_5-c_6+c_7-2 c_{10}+2 c_{11}-2 c_{12}\right)}{\sqrt{3}}$\\\hline
$\Omega_{bb}^{-}\to \Omega_{b}^{-}  K^+   K^-  $ & $ c_1+2 c_2+c_4+c_5-c_6-c_8+3 c_{10}+c_{11}+3 c_{14}-2 c_{16}$\\\hline
$\Omega_{bb}^{-}\to \Omega_{b}^{-}  K^0   \overline K^0  $ & $ c_1+2 c_2+c_4-c_5+c_6+c_8-c_{10}-3 c_{11}-c_{14}-2 c_{16}$\\\hline
$\Omega_{bb}^{-}\to \Omega_{b}^{-}  \eta   \eta  $ & \tabincell{c}{$ \frac{2}{3} (2 c_1+3 c_2+2 c_3+2 c_4-6 c_{10}$\\$-3 c_{11}-6 c_{12}-4 c_{14}-4 c_{16}-4 c_{17})$}\\\hline
\hline
\end{tabular}
\end{table}



\subsection{Charmless $b\to q_1 \bar q_2 q_3$ Decays:  Decays into a bottom meson, a light baryon octet and a light meson}

The effective Hamiltonian is given as
\begin{eqnarray}
 {\cal H}_{eff}&=&d_1(T_{bb})^i  \overline B ^j  \epsilon_{ijk} (T_8)^{k}_{l} M^l_m (H_{3})^m + d_2(T_{bb})^i   \overline B ^l  \epsilon_{ijk} (T_8)^{k}_{m} M^j_l (H_{3})^m\nonumber\\
 &&+ d_3(T_{bb})^i   \overline B ^l  \epsilon_{ijk} (T_8)^{k}_{m} M^m_l (H_{3})^j+ d_4(T_{bb})^i   \overline B ^l  \epsilon_{ijk} (T_8)^{k}_{l} M^j_m (H_{3})^m\nonumber\\
 &&+ d_5(T_{bb})^l   \overline B ^i  \epsilon_{ijk} (T_8)^{k}_{l} M^j_m (H_{3})^m+ d_6(T_{bb})^l   \overline B ^m  \epsilon_{ijk} (T_8)^{k}_{l} M^i_m (H_{3})^j\nonumber\\
 &&+ d_7(T_{bb})^l   \overline B ^i  \epsilon_{ijk} (T_8)^{k}_{m} M^j_l (H_{3})^m+ d_8(T_{bb})^l   \overline B ^i  \epsilon_{ijk} (T_8)^{k}_{m} M^m_l (H_{3})^j\nonumber\\
 &&+ d_9(T_{bb})^l   \overline B ^m  \epsilon_{ijk} (T_8)^{k}_{m} M^i_l (H_{3})^j+ d_{10}(T_{bb})^i   \overline B ^j  \epsilon_{ijk} (T_8)^{k}_{l} M^n_m (H_{\bar6})^{lm}_{n}\nonumber\\
 &&+ d_{11}(T_{bb})^i  \overline B ^l  \epsilon_{ijk} (T_8)^{k}_{l} M^n_m (H_{\bar6})^{jm}_{n}+ d_{12}(T_{bb})^i   \overline B ^l  \epsilon_{ijk} (T_8)^{k}_{n}M^m_l (H_{\bar6})^{nj}_{m} \nonumber\\
 &&+ d_{13}(T_{bb})^i  \overline B ^l  \epsilon_{ijk} (T_8)^{k}_{m} M^j_n (H_{\bar6})^{mn}_{l}+ d_{14}(T_{bb})^i   \overline B ^l  \epsilon_{ijk} (T_8)^{k}_{m} M^m_n (H_{\bar6})^{jn}_{l}\nonumber\\
 &&+ d_{15}(T_{bb})^l  \overline B ^i  \epsilon_{ijk} (T_8)^{k}_{l} M^n_m (H_{\bar6})^{jm}_{n}+ d_{16}(T_{bb})^l   \overline B ^m  \epsilon_{ijk} (T_8)^{k}_{l} M^n_m (H_{\bar6})^{ij}_{n}\nonumber\\
 &&+ d_{17}(T_{bb})^l  \overline B ^m  \epsilon_{ijk} (T_8)^{k}_{l} M^i_n (H_{\bar6})^{jn}_{m}+ d_{18}(T_{bb})^l   \overline B ^i  \epsilon_{ijk} (T_8)^{k}_{m} M^n_l (H_{\bar6})^{jm}_{n}\nonumber\\
 &&+ d_{19}(T_{bb})^l  \overline B ^m  \epsilon_{ijk} (T_8)^{k}_{m} M^n_l (H_{\bar6})^{ij}_{n}+ d_{20}(T_{bb})^l   \overline B ^m  \epsilon_{ijk} (T_8)^{k}_{n} M^n_l (H_{\bar6})^{ij}_{m}\nonumber\\
 &&+ d_{21}(T_{bb})^l  \overline B ^m  \epsilon_{ijk} (T_8)^{k}_{n} M^i_l (H_{\bar6})^{jn}_{m}+ d_{22}(T_{bb})^l   \overline B ^i  \epsilon_{ijk} (T_8)^{k}_{m} M^j_n (H_{\bar6})^{mn}_{l}\nonumber\\
 &&+ d_{23}(T_{bb})^l  \overline B ^i  \epsilon_{ijk} (T_8)^{k}_{m} M^m_n (H_{\bar6})^{jn}_{l}+ d_{24}(T_{bb})^l   \overline B ^m  \epsilon_{ijk} (T_8)^{k}_{m} M^i_n (H_{\bar6})^{jn}_{l}\nonumber\\
 &&+ d_{25}(T_{bb})^l  \overline B ^m  \epsilon_{ijk} (T_8)^{k}_{n} M^n_m (H_{\bar6})^{ij}_{l}+ d_{26}(T_{bb})^l   \overline B ^m  \epsilon_{ijk} (T_8)^{k}_{n} M^j_m (H_{\bar6})^{in}_{l}\nonumber\\
 &&+ d_{27}(T_{bb})^i  \overline B ^j  \epsilon_{ijk} (T_8)^{k}_{l} M^n_m (H_{15})^{lm}_{n}+ d_{28}(T_{bb})^i   \overline B ^l  \epsilon_{ijk} (T_8)^{k}_{l} M^n_m (H_{15})^{jm}_{n}\nonumber\\
 &&+ d_{29}(T_{bb})^i  \overline B ^l  \epsilon_{ijk} (T_8)^{k}_{n} M^m_l (H_{15})^{nj}_{m}+ d_{30}(T_{bb})^i   \overline B ^l  \epsilon_{ijk} (T_8)^{k}_{m} M^j_n (H_{15})^{mn}_{l}\nonumber\\
 &&+ d_{31}(T_{bb})^i  \overline B ^l  \epsilon_{ijk} (T_8)^{k}_{m} M^m_n (H_{15})^{jn}_{l}+ d_{32}(T_{bb})^l   \overline B ^i  \epsilon_{ijk} (T_8)^{k}_{l} M^n_m (H_{15})^{jm}_{n}\nonumber\\
 &&+ d_{33}(T_{bb})^l  \overline B ^m  \epsilon_{ijk} (T_8)^{k}_{l} M^i_n (H_{15})^{jn}_{m}+ d_{34}(T_{bb})^l   \overline B ^i  \epsilon_{ijk} (T_8)^{k}_{m} M^n_l (H_{15})^{jm}_{n}\nonumber\\
 &&+ d_{35}(T_{bb})^l  \overline B ^m  \epsilon_{ijk} (T_8)^{k}_{n} M^i_l (H_{15})^{jn}_{m}+ d_{36}(T_{bb})^l   \overline B ^i  \epsilon_{ijk} (T_8)^{k}_{m} M^j_n (H_{15})^{mn}_{l}\nonumber\\
 &&+ d_{37}(T_{bb})^l  \overline B ^i  \epsilon_{ijk} (T_8)^{k}_{m} M^m_n (H_{15})^{jn}_{l}+ d_{38}(T_{bb})^l   \overline B ^m  \epsilon_{ijk} (T_8)^{k}_{m} M^i_n (H_{15})^{jn}_{l}\nonumber\\
 &&+ d_{39}(T_{bb})^l  \overline B ^m  \epsilon_{ijk} (T_8)^{k}_{n} M^j_m (H_{15})^{in}_{l}.
\end{eqnarray}
Decay amplitudes for different channels are given in Tab.~\ref{tab:bbq_qqq8_bqbarand8meson_vd_1} and Tab.~\ref{tab:bbq_qqq8_bqbarand8meson_vd_2} for $b\to d$ transition; Tab.~\ref{tab:bbq_qqq8_bqbarand8meson_vS_bs_1} and  Tab.~\ref{tab:bbq_qqq8_bqbarand8meson_vS_bs_2} for $b\to s$ transition respectively.

  \begin{table}
  \tiny
  \newcommand{\tabincell}[2]{\begin{tabular}{@{}#1@{}}#2\end{tabular}}
\caption{Doubly bottom baryon decays into a bottom meson, a light baryon(8)  induced by the charmless $b\to d$ transition and a light meson.}\label{tab:bbq_qqq8_bqbarand8meson_vd_1}\begin{tabular}{|c c|c c|c|c|c|c}\hline
channel & amplitude \\\hline
\hline
$\Xi_{bb}^{0}\to \Lambda^0  B^-  K^+  $ & \tabincell{c}{$ -\frac{1}{\sqrt{6}}(d_2+2 d_3+d_6+d_7+2 d_8+d_9-d_{12}-d_{13}-2 d_{14}$\\$-2 d_{16}-d_{17}+d_{18}-2 d_{19}+4 d_{20}-2 d_{21}-d_{22}-2 d_{23}-d_{24}+4 d_{25}$\\$+2 d_{26}-3 d_{29}+3 d_{30}+6 d_{31}+3 d_{33}-3 d_{34}+3 d_{36}+6 d_{37}+3 d_{38})$}\\\hline
$\Xi_{bb}^{0}\to \Lambda^0  \overline B^0  K^0  $ & \tabincell{c}{$ \frac{1}{\sqrt{6}}(-2 d_1-d_2-2 d_3-d_4+d_5-d_6+2 d_{10}+d_{11}+d_{12}-d_{15}+2 d_{16}$\\$+d_{22}+2 d_{23}+d_{24}-4 d_{25}-2 d_{26}+2 d_{27}+d_{28}+3 d_{29}+2 d_{30}+4 d_{31}$\\$-d_{32}+2 d_{33}+3 d_{36}+6 d_{37}+3 d_{38})$}\\\hline
$\Xi_{bb}^{0}\to \Lambda^0  \overline B^0_s  \pi^0  $ & \tabincell{c}{$ \frac{1}{2 \sqrt{3}}(d_1+2 d_4+d_5+d_7-d_8-2 d_9+d_{10}+2 d_{11}-2 d_{13}-d_{14}+d_{15}$\\$+d_{17}+2 d_{18}-4 d_{19}+2 d_{20}-d_{21}-5 d_{27}-10 d_{28}-2 d_{30}-d_{31}$\\$-5 d_{32}+d_{33}+3 d_{35}+6 d_{36}-6 d_{37}-12 d_{38})$}\\\hline
$\Xi_{bb}^{0}\to \Lambda^0  \overline B^0_s  \eta  $ & \tabincell{c}{$ \frac{1}{6} (-d_1+2 d_2+4 d_3-2 d_4-d_5+2 d_6+d_7-d_8-2 d_9+3 d_{10}+6 d_{11}$\\$-2 d_{12}+4 d_{13}+5 d_{14}+3 d_{15}-4 d_{16}+d_{17}+2 d_{18}-4 d_{19}+2 d_{20}-d_{21}$\\$-2 d_{22}+2 d_{23}+4 d_{24}+8 d_{25}+4 d_{26}-3 d_{27}-6 d_{28}-6 d_{29}-3 d_{31}$\\$-3 d_{32}-3 d_{33}+3 d_{35})$}\\\hline
$\Xi_{bb}^{0}\to \Sigma^+  B^-  K^0  $ & \tabincell{c}{$ -d_4-d_5+d_{11}-d_{13}+d_{15}+d_{17}-d_{22}$\\$+d_{24}+d_{28}-3 d_{30}+d_{32}+3 d_{33}-3 d_{36}+3 d_{38}$}\\\hline
$\Xi_{bb}^{0}\to \Sigma^+  \overline B^0_s  \pi^-  $ & \tabincell{c}{$ -d_1+d_5-d_{10}+d_{14}+d_{15}+d_{17}+d_{22}$\\$+d_{23}-3 d_{27}+d_{31}+3 d_{32}+d_{33}+3 d_{36}+3 d_{37}$}\\\hline
$\Xi_{bb}^{0}\to \Sigma^0  B^-  K^+  $ & \tabincell{c}{$ \frac{1}{\sqrt{2}}(d_2-d_6+d_7-d_9+d_{12}-d_{13}+2 d_{16}+d_{17}-d_{18}+2 d_{19}-d_{22}$\\$+d_{24}-d_{29}+3 d_{30}-3 d_{33}-d_{34}-6 d_{35}+3 d_{36}-3 d_{38}+6 d_{39})$}\\\hline
$\Xi_{bb}^{0}\to \Sigma^0  \overline B^0  K^0  $ & \tabincell{c}{$ \frac{1}{\sqrt{2}}(d_2+d_4+d_5-d_6-d_{11}+d_{12}-d_{15}+2 d_{16}+d_{22}$\\$-d_{24}-d_{28}-d_{29}-2 d_{30}-d_{32}+2 d_{33}+3 d_{36}-3 d_{38}+6 d_{39})$}\\\hline
$\Xi_{bb}^{0}\to \Sigma^0  \overline B^0_s  \pi^0  $ &\tabincell{c}{ $ \frac{1}{2} (-d_1+d_5-d_7-d_8-d_{10}+d_{14}+d_{15}+d_{17}+2 d_{20}-d_{21}$\\$+2 d_{22}+2 d_{23}+5 d_{27}+d_{31}-5 d_{32}+d_{33}-6 d_{34}-d_{35})$}\\\hline
$\Xi_{bb}^{0}\to \Sigma^0  \overline B^0_s  \eta  $ & \tabincell{c}{$ \frac{1}{2 \sqrt{3}}(d_1-2 d_2-d_5+2 d_6-d_7-d_8-3 d_{10}-2 d_{12}-2 d_{13}-d_{14}$\\$+3 d_{15}-4 d_{16}+d_{17}+2 d_{20}-d_{21}+3 d_{27}+2 d_{29}+2 d_{30}$\\$-d_{31}-3 d_{32}-3 d_{33}-6 d_{34}-d_{35}-6 d_{36}-6 d_{37}-12 d_{39})$}\\\hline
$\Xi_{bb}^{0}\to \Sigma^-  \overline B^0  K^+  $ & $ d_7-d_9-d_{18}+2 d_{19}-d_{22}+d_{24}-d_{34}+2 d_{35}+3 d_{36}-3 d_{38}$\\\hline
$\Xi_{bb}^{0}\to \Sigma^-  \overline B^0_s  \pi^+  $ & $ -d_7-d_8+2 d_{20}-d_{21}+d_{22}+d_{23}+2 d_{34}-d_{35}-3 d_{36}-3 d_{37}$\\\hline
$\Xi_{bb}^{0}\to {p}  B^-  \pi^0  $ & \tabincell{c}{$ \frac{1}{\sqrt{2}}(d_3-d_4-d_5+d_6+d_8+d_9-d_{11}+d_{12}-d_{13}-d_{14}-d_{15}$\\$+2 d_{16}-d_{18}+2 d_{19}+2 d_{20}-d_{21}-d_{22}-d_{23}+2 d_{25}+d_{26}+5 d_{28}$\\$+3 d_{29}-3 d_{30}+3 d_{31}+5 d_{32}+6 d_{33}+3 d_{34}+3 d_{35}-3 d_{36}+3 d_{37}+6 d_{38}-3 d_{39})$}\\\hline
$\Xi_{bb}^{0}\to {p}  B^-  \eta  $ & \tabincell{c}{$ \frac{1}{\sqrt{6}}(d_3+d_4+d_5+d_6+d_8+d_9-3 d_{11}+d_{12}+d_{13}-d_{14}-3 d_{15}$\\$+2 d_{16}-2 d_{17}-d_{18}+2 d_{19}+2 d_{20}-d_{21}+d_{22}-d_{23}-2 d_{24}+2 d_{25}$\\$+d_{26}+3 d_{28}+3 d_{29}+3 d_{30}+3 d_{31}+3 d_{32}+3 d_{34}+3 d_{35}+3 d_{36}+3 d_{37}-3 d_{39})$}\\\hline
$\Xi_{bb}^{0}\to {p}  \overline B^0  \pi^-  $ & \tabincell{c}{$ d_1+d_3-d_5+d_6+d_{10}+d_{12}-d_{15}+2 d_{16}-d_{22}-d_{23}$\\$+2 d_{25}+d_{26}+3 d_{27}+3 d_{29}-2 d_{31}-3 d_{32}-2 d_{33}-3 d_{36}-3 d_{37}-3 d_{39}$}\\\hline
$\Xi_{bb}^{0}\to {p}  \overline B^0_s  K^-  $ & \tabincell{c}{$ d_3+d_6+d_{12}+d_{14}+2 d_{16}+d_{17}$\\$+2 d_{25}+d_{26}+3 d_{29}-d_{31}-d_{33}-3 d_{39}$}\\\hline
$\Xi_{bb}^{0}\to \Xi^-  \overline B^0_s  K^+  $ & \tabincell{c}{$ -d_8-d_9-d_{18}+2 d_{19}+2 d_{20}$\\$-d_{21}+d_{23}+d_{24}+d_{34}+d_{35}-3 d_{37}-3 d_{38}$}\\\hline
$\Xi_{bb}^{0}\to \Xi^0  \overline B^0_s  K^0  $ & \tabincell{c}{$ -d_1-d_4+d_{10}+d_{11}+d_{13}+d_{14}+d_{23}$\\$+d_{24}+d_{27}+d_{28}+d_{30}+d_{31}+3 d_{37}+3 d_{38}$}\\\hline
$\Xi_{bb}^{-}\to \Lambda^0  B^-  K^0  $ & \tabincell{c}{$ \frac{1}{\sqrt{6}}(2 d_1+d_4-d_5-d_7-2 d_8-d_9-2 d_{10}-d_{11}+d_{13}+2 d_{14}+d_{15}$\\$+d_{17}-d_{18}+2 d_{19}-4 d_{20}+2 d_{21}-2 d_{27}-d_{28}+3 d_{30}+6 d_{31}$\\$+d_{32}+3 d_{33}+3 d_{34}+2 d_{36}+4 d_{37}+2 d_{38})$}\\\hline
$\Xi_{bb}^{-}\to \Lambda^0  \overline B^0_s  \pi^-  $ & \tabincell{c}{$ \frac{1}{\sqrt{6}}(d_1+2 d_4+d_5+d_7-d_8-2 d_9+d_{10}+2 d_{11}-2 d_{13}-d_{14}+d_{15}$\\$+d_{17}+2 d_{18}-4 d_{19}+2 d_{20}-d_{21}+3 d_{27}+6 d_{28}-2 d_{30}$\\$-d_{31}+3 d_{32}+d_{33}+3 d_{35}-2 d_{36}+2 d_{37}+4 d_{38})$}\\\hline
$\Xi_{bb}^{-}\to \Sigma^0  B^-  K^0  $ & \tabincell{c}{$ \frac{1}{\sqrt{2}}(d_4+d_5+d_7-d_9-d_{11}+d_{13}-d_{15}-d_{17}-d_{18}+2 d_{19}$\\$-d_{28}+3 d_{30}-d_{32}-3 d_{33}-d_{34}-6 d_{35}-2 d_{36}+2 d_{38})$}\\\hline
$\Xi_{bb}^{-}\to \Sigma^0  \overline B^0_s  \pi^-  $ & \tabincell{c}{$ \frac{1}{\sqrt{2}}(d_1-d_5-d_7-d_8+d_{10}-d_{14}-d_{15}-d_{17}+2 d_{20}-d_{21}$\\$+3 d_{27}-d_{31}-3 d_{32}-d_{33}-6 d_{34}-d_{35}+2 d_{36}+2 d_{37})$}\\\hline
$\Xi_{bb}^{-}\to \Sigma^-  B^-  K^+  $ & $ d_2-d_6+d_{12}-d_{13}+2 d_{16}+d_{17}-d_{29}+3 d_{30}-3 d_{33}-2 d_{39}$\\\hline
$\Xi_{bb}^{-}\to \Sigma^-  \overline B^0  K^0  $ & \tabincell{c}{$ d_2+d_4+d_5-d_6+d_7-d_9-d_{11}+d_{12}-d_{15}+2 d_{16}-d_{18}+2 d_{19}$\\$-d_{28}-d_{29}-2 d_{30}-d_{32}+2 d_{33}-d_{34}+2 d_{35}-2 d_{36}+2 d_{38}-2 d_{39}$}\\\hline
$\Xi_{bb}^{-}\to \Sigma^-  \overline B^0_s  \pi^0  $ & \tabincell{c}{$ \frac{1}{\sqrt{2}}(-d_1+d_5+d_7+d_8-d_{10}+d_{14}+d_{15}+d_{17}-2 d_{20}+d_{21}$\\$+5 d_{27}+d_{31}-5 d_{32}+d_{33}-2 d_{34}+d_{35}-2 d_{36}-2 d_{37})$}\\\hline
$\Xi_{bb}^{-}\to \Sigma^-  \overline B^0_s  \eta  $ &  \tabincell{c}{$ \frac{1}{\sqrt{6}}(d_1-2 d_2-d_5+2 d_6-d_7-d_8-3 d_{10}-2 d_{12}-2 d_{13}-d_{14}$\\$+3 d_{15}-4 d_{16}+d_{17}+2 d_{20}-d_{21}+3 d_{27}+2 d_{29}+2 d_{30}-d_{31}$\\$-3 d_{32}-3 d_{33}+2 d_{34}-d_{35}+2 d_{36}+2 d_{37}+4 d_{39})$}\\\hline
$\Xi_{bb}^{-}\to {p}  B^-  \pi^-  $ &  \tabincell{c}{$ -d_1-d_4+d_8+d_9-d_{10}-d_{11}-d_{13}-d_{14}-d_{18}+2 d_{19}$\\$+2 d_{20}-d_{21}-3 d_{27}-3 d_{28}-3 d_{30}-3 d_{31}+3 d_{34}+3 d_{35}-2 d_{37}-2 d_{38}$}\\\hline
$\Xi_{bb}^{-}\to \Xi^-  \overline B^0_s  K^0  $ &  \tabincell{c}{$ d_1+d_4-d_8-d_9-d_{10}-d_{11}-d_{13}-d_{14}-d_{18}+2 d_{19}$\\$+2 d_{20}-d_{21}-d_{27}-d_{28}-d_{30}-d_{31}+d_{34}+d_{35}+2 d_{37}+2 d_{38}$}\\\hline
$\Omega_{bb}^{-}\to \Lambda^0  B^-  \pi^0  $ & \tabincell{c}{$ -\frac{1}{2 \sqrt{3}}(d_1-d_2+d_3-d_4-2 d_5+2 d_6+d_{10}-d_{11}+2 d_{12}-2 d_{15}$\\$+4 d_{16}+2 d_{22}+d_{23}-d_{24}-2 d_{25}-d_{26}-5 d_{27}+5 d_{28}-6 d_{30}$\\$+6 d_{31}+10 d_{32}+12 d_{33}+2 d_{36}+d_{37}-d_{38}+3 d_{39})$}\\\hline
\hline
\end{tabular}
\end{table}
\begin{table}
\scriptsize
\newcommand{\tabincell}[2]{\begin{tabular}{@{}#1@{}}#2\end{tabular}}
\caption{Doubly bottom baryon decays into a bottom meson, a light baryon(8) induced by the charmless $b\to d$ transition and a light meson.}\label{tab:bbq_qqq8_bqbarand8meson_vd_2}
\begin{tabular}{|c c|c c|c|c|c|c}\hline
channel & amplitude \\\hline
\hline
$\Omega_{bb}^{-}\to \Lambda^0  B^-  \eta  $ & \tabincell{c}{$ \frac{1}{6} (d_1+d_2-d_3-d_4-2 d_5-2 d_6+2 d_7+4 d_8+2 d_9-3 d_{10}+3 d_{11}$\\$-2 d_{12}-2 d_{13}+2 d_{14}+6 d_{15}-4 d_{16}+4 d_{17}+2 d_{18}-4 d_{19}+8 d_{20}$\\$-4 d_{21}+4 d_{22}+5 d_{23}+d_{24}+2 d_{25}+d_{26}+3 d_{27}-3 d_{28}-6 d_{32}-6 d_{34}$\\$-3 d_{37}-3 d_{38}-3 d_{39})$}\\\hline
$\Omega_{bb}^{-}\to \Lambda^0  \overline B^0  \pi^-  $ & \tabincell{c}{$ \frac{1}{\sqrt{6}}(-d_1+d_2-d_3+d_4+2 d_5-2 d_6-d_{10}+d_{11}-2 d_{12}+2 d_{15}$\\$-4 d_{16}-2 d_{22}-d_{23}+d_{24}+2 d_{25}+d_{26}-3 d_{27}+3 d_{28}-2 d_{30}$\\$+2 d_{31}+6 d_{32}+4 d_{33}-2 d_{36}-d_{37}+d_{38}-3 d_{39})$}\\\hline
$\Omega_{bb}^{-}\to \Lambda^0  \overline B^0_s  K^-  $ & \tabincell{c}{$ \frac{1}{\sqrt{6}}(d_2-d_3-2 d_6+d_7-d_8-2 d_9-2 d_{12}+d_{13}-d_{14}-4 d_{16}$\\$-2 d_{17}+2 d_{18}-4 d_{19}+2 d_{20}-d_{21}+d_{22}-d_{23}-2 d_{24}+2 d_{25}$\\$+d_{26}-d_{30}+d_{31}+2 d_{33}+3 d_{35}-d_{36}+d_{37}+2 d_{38}-3 d_{39})$}\\\hline
$\Omega_{bb}^{-}\to \Sigma^+  B^-  \pi^-  $ & \tabincell{c}{$ d_1+d_4+d_{10}+d_{11}+d_{13}+d_{14}+d_{23}$\\$+d_{24}+3 d_{27}+3 d_{28}+3 d_{30}+3 d_{31}+d_{37}+d_{38}$}\\\hline
$\Omega_{bb}^{-}\to \Sigma^0  B^-  \pi^0  $ & \tabincell{c}{$ \frac{1}{2} (d_1-d_2-d_3+d_4+d_{10}+d_{11}+2 d_{13}+2 d_{14}+d_{23}$\\$+d_{24}+2 d_{25}+d_{26}-5 d_{27}-5 d_{28}-6 d_{29}+d_{37}+d_{38}+d_{39})$}\\\hline
$\Omega_{bb}^{-}\to \Sigma^0  B^-  \eta  $ & \tabincell{c}{$ -\frac{1}{2 \sqrt{3}}(d_1+d_2+d_3+d_4+2 d_7-2 d_9-3 d_{10}-3 d_{11}-2 d_{18}+4 d_{19}$\\$+2 d_{22}+d_{23}-d_{24}-2 d_{25}-d_{26}+3 d_{27}+3 d_{28}+6 d_{29}+6 d_{30}$\\$+6 d_{31}-2 d_{34}-12 d_{35}-2 d_{36}+d_{37}+3 d_{38}-d_{39})$}\\\hline
$\Omega_{bb}^{-}\to \Sigma^0  \overline B^0  \pi^-  $ & \tabincell{c}{$ -\frac{1}{\sqrt{2}}(d_1+d_2+d_3+d_4+d_{10}+d_{11}+d_{23}+d_{24}-2 d_{25}$\\$-d_{26}+3 d_{27}+3 d_{28}+6 d_{29}-2 d_{30}-2 d_{31}+d_{37}+d_{38}-d_{39})$}\\\hline
$\Omega_{bb}^{-}\to \Sigma^0  \overline B^0_s  K^-  $ & \tabincell{c}{$ -\frac{1}{\sqrt{2}}(d_2+d_3+d_7+d_8+d_{13}+d_{14}-2 d_{20}+d_{21}+d_{22}+d_{23}$\\$-2 d_{25}-d_{26}+6 d_{29}-d_{30}-d_{31}+6 d_{34}+d_{35}-d_{36}-d_{37}-d_{39})$}\\\hline
$\Omega_{bb}^{-}\to \Sigma^-  B^-  \pi^+  $ & \tabincell{c}{$ -d_2-d_3+d_{13}+d_{14}+2 d_{25}$\\$+d_{26}+2 d_{29}-3 d_{30}-3 d_{31}+d_{39}$}\\\hline
$\Omega_{bb}^{-}\to \Sigma^-  \overline B^0  \pi^0  $ & \tabincell{c}{$ \frac{1}{\sqrt{2}}(d_1+d_2+d_3+d_4+d_{10}+d_{11}+d_{23}+d_{24}-2 d_{25}-d_{26}$\\$-5 d_{27}-5 d_{28}-2 d_{29}-2 d_{30}-2d_{31}+d_{37}+d_{38}-d_{39})$}\\\hline
$\Omega_{bb}^{-}\to \Sigma^-  \overline B^0  \eta  $ & \tabincell{c}{$ -\frac{1}{\sqrt{6}}(d_1+d_2+d_3+d_4+2 d_7-2 d_9-3 d_{10}-3 d_{11}-2 d_{18}+4 d_{19}$\\$+2 d_{22}+d_{23}-d_{24}-2 d_{25}-d_{26}+3 d_{27}+3 d_{28}-2 d_{29}$\\$-2 d_{30}-2 d_{31}-2 d_{34}+4 d_{35}-2 d_{36}+d_{37}+3 d_{38}-d_{39})$}\\\hline
$\Omega_{bb}^{-}\to \Sigma^-  \overline B^0_s  \overline K^0  $ & \tabincell{c}{$ -d_2-d_3-d_7-d_8-d_{13}-d_{14}+2 d_{20}-d_{21}-d_{22}$\\$-d_{23}+2 d_{25}+d_{26}+2 d_{29}+d_{30}+d_{31}+2 d_{34}$\\$-d_{35}+d_{36}+d_{37}+d_{39}$}\\\hline
$\Omega_{bb}^{-}\to {p}  B^-  K^-  $ & \tabincell{c}{$ d_8+d_9-d_{18}+2 d_{19}+2 d_{20}-d_{21}$\\$+d_{23}+d_{24}+3 d_{34}+3 d_{35}-d_{37}-d_{38}$}\\\hline
$\Omega_{bb}^{-}\to \Xi^-  B^-  K^+  $ & \tabincell{c}{$ -d_3-d_6+d_{12}+d_{14}+2 d_{16}+d_{17}+2 d_{25}$\\$+d_{26}+d_{29}-3 d_{31}-3 d_{33}-d_{39}$}\\\hline
$\Omega_{bb}^{-}\to \Xi^-  \overline B^0  K^0  $ & \tabincell{c}{$ -d_1-d_3+d_5-d_6+d_{10}+d_{12}-d_{15}+2 d_{16}-d_{22}-d_{23}$\\$+2 d_{25}+d_{26}+d_{27}+d_{29}+2 d_{31}-d_{32}+2 d_{33}-d_{36}-d_{37}-d_{39}$}\\\hline
$\Omega_{bb}^{-}\to \Xi^-  \overline B^0_s  \pi^0  $ & \tabincell{c}{$ \frac{1}{\sqrt{2}}(d_4+d_5+d_{11}-d_{13}+d_{15}+d_{17}-d_{22}+d_{24}$\\$-5 d_{28}-d_{30}-5 d_{32}+d_{33}-d_{36}+d_{38})$}\\\hline
$\Omega_{bb}^{-}\to \Xi^-  \overline B^0_s  \eta  $ & \tabincell{c}{$ \frac{1}{\sqrt{6}}(2 d_3-d_4-d_5+2 d_6+2 d_8+2 d_9+3 d_{11}-2 d_{12}+d_{13}$\\$+2 d_{14}+3 d_{15}-4 d_{16}+d_{17}+2 d_{18}-4 d_{19}-4 d_{20}+2 d_{21}+d_{22}$\\$+2 d_{23}+d_{24}-4 d_{25}-2 d_{26}-3 d_{28}-2 d_{29}+d_{30}-2 d_{31}$\\$-3 d_{32}-3 d_{33}-2 d_{34}-2 d_{35}+d_{36}-2 d_{37}-3 d_{38}+2 d_{39})$}\\\hline
$\Omega_{bb}^{-}\to \Xi^0  B^-  K^0  $ & \tabincell{c}{$ d_1-d_5-d_{10}+d_{14}+d_{15}+d_{17}+d_{22}$\\$+d_{23}-d_{27}+3 d_{31}+d_{32}+3 d_{33}+d_{36}+d_{37}$}\\\hline
$\Omega_{bb}^{-}\to \Xi^0  \overline B^0_s  \pi^-  $ & \tabincell{c}{$ d_4+d_5+d_{11}-d_{13}+d_{15}+d_{17}-d_{22}$\\$+d_{24}+3 d_{28}-d_{30}+3 d_{32}+d_{33}-d_{36}+d_{38}$}\\\hline
\hline
\end{tabular}
\end{table}

\begin{table}
\tiny
  \newcommand{\tabincell}[2]{\begin{tabular}{@{}#1@{}}#2\end{tabular}}
\caption{Doubly bottom baryon decays into a bottom meson, a light baryon(8) induced by the charmless $b\to s$ transition and a light meson.}\label{tab:bbq_qqq8_bqbarand8meson_vS_bs_1}\begin{tabular}{|c|c|c|c|c|c|c|c}\hline
channel & amplitude \\\hline
\hline
$\Xi_{bb}^{0}\to \Lambda^0  B^-  \pi^+  $ & \tabincell{c}{$ -\frac{1}{\sqrt{6}}(2 d_2+d_3-d_6+2 d_7+d_8-d_9+d_{12}-2 d_{13}-d_{14}+2 d_{16}+d_{17}$\\$-d_{18}+2 d_{19}+2 d_{20}-d_{21}-2 d_{22}-d_{23}+d_{24}+2 d_{25}$\\$+d_{26}-3 d_{29}+6 d_{30}+3 d_{31}-3 d_{33}-3 d_{34}-9 d_{35}+6 d_{36}+3 d_{37}-3 d_{38}+9 d_{39})$}\\\hline
$\Xi_{bb}^{0}\to \Lambda^0  \overline B^0  \pi^0  $ & \tabincell{c}{
$ \frac{1}{2 \sqrt{3}}(2 d_2+d_3-d_6+2 d_7+d_8-d_9+4 d_{10}+2 d_{11}+d_{12}+2 d_{13}+d_{14}$\\$-2 d_{15}+2 d_{16}-d_{17}+d_{18}-2 d_{19}-2 d_{20}+d_{21}-2 d_{22}-d_{23}+d_{24}$\\$+2 d_{25}+d_{26}-8 d_{27}-4 d_{28}-3 d_{29}-2 d_{30}-d_{31}+4 d_{32}+d_{33}+9 d_{34}$\\$+3 d_{35}+6 d_{36}+3 d_{37}-3 d_{38}+9 d_{39})$}\\\hline
$\Xi_{bb}^{0}\to \Lambda^0  \overline B^0  \eta  $ & \tabincell{c}{
$ \frac{1}{6} (4 d_1-2 d_2-d_3+2 d_4-2 d_5+d_6+2 d_7+d_8-d_9-d_{12}-4 d_{13}$\\$-5 d_{14}-2 d_{16}-d_{17}+d_{18}-2 d_{19}-2 d_{20}+d_{21}-4 d_{22}-5 d_{23}$\\$--d_{24}2 d_{25}-d_{26}-12 d_{27}-6 d_{28}+3 d_{29}-3 d_{31}+6 d_{32}$\\$-3 d_{33}+9 d_{34}+3 d_{35}-9 d_{37}-9 d_{38}-9 d_{39})$}\\\hline
$\Xi_{bb}^{0}\to \Lambda^0  \overline B^0_s  \overline K^0  $ & \tabincell{c}{
$ \frac{1}{\sqrt{6}}(-d_1-2 d_2-d_3-2 d_4-d_5+d_6+d_{10}+2 d_{11}-d_{12}+d_{15}-2 d_{16}$\\$-d_{22}+d_{23}+2 d_{24}-2 d_{25}-d_{26}+d_{27}+2 d_{28}+3 d_{29}$\\$+4 d_{30}+2 d_{31}+d_{32}-2 d_{33}-3 d_{36}+3 d_{37}+6 d_{38}-9 d_{39})$}\\\hline
$\Xi_{bb}^{0}\to \Sigma^+  B^-  \pi^0  $ & \tabincell{c}{
$ -\frac{1}{\sqrt{2}}(d_3+d_6+d_8+d_9-2 d_{11}+d_{12}-d_{14}-2 d_{15}+2 d_{16}-d_{17}-d_{18}$\\$+2 d_{19}+2 d_{20}-d_{21}-d_{23}-d_{24}+2 d_{25}+d_{26}+4 d_{28}$\\$+3 d_{29}+3 d_{31}+4 d_{32}+3 d_{33}+3 d_{34}+3 d_{35}+3 d_{37}+3 d_{38}-3 d_{39})$}\\\hline
$\Xi_{bb}^{0}\to \Sigma^+  B^-  \eta  $ & \tabincell{c}{
$ -\frac{1}{\sqrt{6}}(d_3-2 d_4-2 d_5+d_6+d_8+d_9+d_{12}-2 d_{13}-d_{14}+2 d_{16}$\\$+d_{17}-d_{18}+2 d_{19}+2 d_{20}-d_{21}-2 d_{22}-d_{23}+d_{24}$\\$+2 d_{25}+d_{26}+6 d_{28}+3 d_{29}-6 d_{30}+3 d_{31}+6 d_{32}+9 d_{33}+3 d_{34}$\\$+3 d_{35}-6 d_{36}+3 d_{37}+9 d_{38}-3 d_{39})$}\\\hline
$\Xi_{bb}^{0}\to \Sigma^+  \overline B^0  \pi^-  $ & \tabincell{c}{
$ -d_3-d_6-d_{12}-d_{14}-2 d_{16}-d_{17}-2 d_{25}-d_{26}-3 d_{29}+d_{31}+d_{33}+3 d_{39}$}\\\hline
$\Xi_{bb}^{0}\to \Sigma^+  \overline B^0_s  K^-  $ & \tabincell{c}{
$ -d_1-d_3+d_5-d_6-d_{10}-d_{12}+d_{15}-2 d_{16}+d_{22}$\\$+d_{23}-2 d_{25}-d_{26}-3 d_{27}-3 d_{29}+2 d_{31}$\\$+3 d_{32}+2 d_{33}+3 d_{36}+3 d_{37}+3 d_{39}$}\\\hline
$\Xi_{bb}^{0}\to \Sigma^0  B^-  \pi^+  $ & \tabincell{c}{
$ \frac{1}{\sqrt{2}}(d_3+d_6+d_8+d_9-d_{12}-d_{14}-2 d_{16}-d_{17}+d_{18}$\\$-2 d_{19}+2 d_{20}-d_{21}-d_{23}-d_{24}+2 d_{25}+d_{26}-d_{29}+3 d_{31}$\\$+3 d_{33}-d_{34}+3 d_{35}+3 d_{37}+3 d_{38}-3 d_{39})$}\\\hline
$\Xi_{bb}^{0}\to \Sigma^0  \overline B^0  \pi^0  $ & \tabincell{c}{
$ \frac{1}{2} (-d_3-d_6+d_8+d_9-2 d_{11}+d_{12}-d_{14}-2 d_{15}+2 d_{16}-d_{17}-d_{18}$\\$+2 d_{19}-2 d_{20}+d_{21}-d_{23}-d_{24}-2 d_{25}-d_{26}+4 d_{28}+d_{29}$\\$+d_{31}+4 d_{32}+d_{33}+3 d_{34}-d_{35}+3 d_{37}+3 d_{38}+3 d_{39})$}\\\hline
$\Xi_{bb}^{0}\to \Sigma^0  \overline B^0  \eta  $ & \tabincell{c}{
$ \frac{1}{2 \sqrt{3}}(d_3-2 d_4-2 d_5+d_6+d_8+d_9-d_{12}+2 d_{13}+d_{14}-2 d_{16}-d_{17}$\\$-d_{18}+2 d_{19}-2 d_{20}+d_{21}-2 d_{22}-d_{23}+d_{24}+2 d_{25}+d_{26}+6 d_{28}-d_{29}$\\$+2 d_{30}-d_{31}+6 d_{32}-3 d_{33}+3 d_{34}-d_{35}-6 d_{36}+3 d_{37}+9 d_{38}-3 d_{39})$}\\\hline
$\Xi_{bb}^{0}\to \Sigma^0  \overline B^0_s  \overline K^0  $ & \tabincell{c}{
$ \frac{1}{\sqrt{2}}(d_1+d_3-d_5+d_6-d_{10}-d_{12}+d_{15}-2 d_{16}-d_{22}-d_{23}+2 d_{25}$\\$+d_{26}-d_{27}-d_{29}-2 d_{31}+d_{32}-2 d_{33}-3 d_{36}-3 d_{37}-3 d_{39})$}\\\hline
$\Xi_{bb}^{0}\to \Sigma^-  \overline B^0  \pi^+  $ & \tabincell{c}{
$ d_8+d_9+d_{18}-2 d_{19}-2 d_{20}+d_{21}-d_{23}-d_{24}-d_{34}-d_{35}+3 d_{37}+3 d_{38}$}\\\hline
$\Xi_{bb}^{0}\to {p}  B^-  \overline K^0  $ & \tabincell{c}{
$ d_4+d_5-d_{11}+d_{13}-d_{15}-d_{17}+d_{22}-d_{24}-d_{28}+3 d_{30}-d_{32}-3 d_{33}+3 d_{36}-3 d_{38}$}\\\hline
$\Xi_{bb}^{0}\to {p}  \overline B^0  K^-  $ & \tabincell{c}{
$ d_1-d_5+d_{10}-d_{14}-d_{15}-d_{17}-d_{22}-d_{23}+3 d_{27}-d_{31}-3 d_{32}-d_{33}-3 d_{36}-3 d_{37}$}\\\hline
$\Xi_{bb}^{0}\to \Xi^-  \overline B^0  K^+  $ & \tabincell{c}{
$ d_7+d_8-2 d_{20}+d_{21}-d_{22}-d_{23}-2 d_{34}+d_{35}+3 d_{36}+3 d_{37}$}\\\hline
$\Xi_{bb}^{0}\to \Xi^-  \overline B^0_s  \pi^+  $ & \tabincell{c}{
$ -d_7+d_9+d_{18}-2 d_{19}+d_{22}-d_{24}+d_{34}-2 d_{35}-3 d_{36}+3 d_{38}$}\\\hline
$\Xi_{bb}^{0}\to \Xi^0  B^-  K^+  $ & \tabincell{c}{
$ -d_2-d_3-d_7-d_8+d_{13}+d_{14}-2 d_{20}+d_{21}+d_{22}+d_{23}-2 d_{25}$\\$-d_{26}+2 d_{29}-3 d_{30}-3 d_{31}+2 d_{34}+3 d_{35}-3 d_{36}-3 d_{37}-3 d_{39}$}\\\hline
$\Xi_{bb}^{0}\to \Xi^0  \overline B^0  K^0  $ & $ -d_2-d_3-d_{13}-d_{14}-2 d_{25}-d_{26}+2 d_{29}+d_{30}+d_{31}-3 d_{39}$\\\hline
$\Xi_{bb}^{0}\to \Xi^0  \overline B^0_s  \pi^0  $ & \tabincell{c}{$ \frac{1}{\sqrt{2}}(d_7-d_9+2 d_{10}+2 d_{11}+d_{18}-2 d_{19}-d_{22}+d_{24}-4 d_{27}-4 d_{28}$\\$+3 d_{34}+2 d_{35}+3 d_{36}-3 d_{38})$}\\\hline
$\Xi_{bb}^{0}\to \Xi^0  \overline B^0_s  \eta  $ & \tabincell{c}{
$ \frac{1}{\sqrt{6}}(2 d_1+2 d_2+2 d_3+2 d_4+d_7-d_9+d_{18}-2 d_{19}-d_{22}-2 d_{23}-d_{24}$\\$+4 d_{25}+2 d_{26}-6 d_{27}-6 d_{28}-4 d_{29}-4 d_{30}-4 d_{31}+3 d_{34}+2 d_{35}$\\$+3 d_{36}-6 d_{37}-9 d_{38}+6 d_{39})$}\\\hline
$\Xi_{bb}^{-}\to \Lambda^0  B^-  \pi^0  $ & \tabincell{c}{
$ \frac{1}{2 \sqrt{3}}(2 d_2+d_3-d_6+2 d_7+d_8-d_9-4 d_{10}-2 d_{11}-d_{12}-2 d_{13}-d_{14}$\\$+2 d_{15}-2 d_{16}+d_{17}-d_{18}+2 d_{19}+2 d_{20}-d_{21}+2 d_{22}+d_{23}$\\$-d_{24}-2 d_{25}-d_{26}+8 d_{27}+4 d_{28}+9 d_{29}+6 d_{30}+3 d_{31}$\\$-4 d_{32}-3 d_{33}-3 d_{34}-9 d_{35}-2 d_{36}-d_{37}+d_{38}-3 d_{39})$}\\\hline
$\Xi_{bb}^{-}\to \Lambda^0  B^-  \eta  $ & \tabincell{c}{$ \frac{1}{6} (-4 d_1+2 d_2+d_3-2 d_4+2 d_5-d_6-2 d_7-d_8+d_9-d_{12}-4 d_{13}$\\$-5 d_{14}-2 d_{16}-d_{17}+d_{18}-2 d_{19}-2 d_{20}+d_{21}-4 d_{22}-5 d_{23}-d_{24}$\\$-2 d_{25}-d_{26}+12 d_{27}+6 d_{28}+9 d_{29}-9 d_{31}-6 d_{32}$\\$-9 d_{33}+3 d_{34}+9 d_{35}-3 d_{37}-3 d_{38}-3 d_{39})$}\\\hline
$\Xi_{bb}^{-}\to \Lambda^0  \overline B^0  \pi^-  $ & \tabincell{c}{
$ \frac{1}{\sqrt{6}}(2 d_2+d_3-d_6+2 d_7+d_8-d_9-d_{12}+2 d_{13}+d_{14}-2 d_{16}$\\$-d_{17}+d_{18}-2 d_{19}-2 d_{20}+d_{21}+2 d_{22}+d_{23}-d_{24}$\\$-2 d_{25}-d_{26}+9 d_{29}-2 d_{30}-d_{31}+d_{33}+9 d_{34}$\\$+3 d_{35}-2 d_{36}-d_{37}+d_{38}-3 d_{39})$}\\\hline
$\Xi_{bb}^{-}\to \Lambda^0  \overline B^0_s  K^-  $ & \tabincell{c}{$ \frac{1}{\sqrt{6}}(d_1+2 d_2+d_3+2 d_4+d_5-d_6+d_{10}+2 d_{11}-d_{12}+d_{15}-2 d_{16}$\\$-d_{22}+d_{23}+2 d_{24}-2 d_{25}-d_{26}+3 d_{27}+6 d_{28}+9 d_{29}-4 d_{30}-2 d_{31}$\\$+3 d_{32}+2 d_{33}-d_{36}+d_{37}+2 d_{38}-3 d_{39})$}\\\hline
$\Xi_{bb}^{-}\to \Sigma^+  B^-  \pi^-  $ & \tabincell{c}{$ -d_8-d_9+d_{18}-2 d_{19}-2 d_{20}+d_{21}-d_{23}-d_{24}-3 d_{34}-3 d_{35}+d_{37}+d_{38}$}\\\hline
$\Xi_{bb}^{-}\to \Sigma^0  B^-  \pi^0  $ & \tabincell{c}{$ \frac{1}{2} (d_3+d_6-d_8-d_9-2 d_{11}+d_{12}-d_{14}-2 d_{15}+2 d_{16}-d_{17}-d_{18}$\\$+2 d_{19}-2 d_{20}+d_{21}-d_{23}-d_{24}-2 d_{25}-d_{26}+4 d_{28}+3 d_{29}+3 d_{31}$\\$+4 d_{32}+3 d_{33}+d_{34}-3 d_{35}+d_{37}+d_{38}+d_{39})$}\\\hline
\hline
\end{tabular}
\end{table}
\begin{table}
\tiny
\newcommand{\tabincell}[2]{\begin{tabular}{@{}#1@{}}#2\end{tabular}}
\caption{Doubly bottom baryon decays into a bottom meson, a light baryon(8)  induced by the charmless $b\to s$ transition and a light meson.}\label{tab:bbq_qqq8_bqbarand8meson_vS_bs_2}
\begin{tabular}{|c|c|c|c|c|c|c|c}\hline
channel & amplitude \\\hline
\hline
$\Xi_{bb}^{-}\to \Sigma^0  B^-  \eta  $ & \tabincell{c}{$ \frac{1}{2 \sqrt{3}}(d_3-2 d_4-2 d_5+d_6+d_8+d_9+d_{12}-2 d_{13}-d_{14}+2 d_{16}+d_{17}$\\$+d_{18}-2 d_{19}+2 d_{20}-d_{21}+2 d_{22}+d_{23}-d_{24}-2 d_{25}-d_{26}+6 d_{28}+3 d_{29}$\\$-6 d_{30}+3 d_{31}+6 d_{32}+9 d_{33}-d_{34}+3 d_{35}+2 d_{36}-d_{37}-3 d_{38}+d_{39})$}\\\hline
$\Xi_{bb}^{-}\to \Sigma^0  \overline B^0  \pi^-  $ & \tabincell{c}{$ \frac{1}{\sqrt{2}}(d_3+d_6+d_8+d_9+d_{12}+d_{14}+2 d_{16}+d_{17}-d_{18}+2 d_{19}$\\$-2 d_{20}+d_{21}+d_{23}+d_{24}-2 d_{25}-d_{26}+3 d_{29}-d_{31}-d_{33}$\\$+3 d_{34}-d_{35}-d_{37}-d_{38}+d_{39})$}\\\hline
$\Xi_{bb}^{-}\to \Sigma^0  \overline B^0_s  K^-  $ & \tabincell{c}{$ \frac{1}{\sqrt{2}}(d_1+d_3-d_5+d_6+d_{10}+d_{12}-d_{15}+2 d_{16}+d_{22}+d_{23}-2 d_{25}$\\$-d_{26}+3 d_{27}+3 d_{29}-2 d_{31}-3 d_{32}-2 d_{33}$\\$+d_{36}+d_{37}+d_{39})$}\\\hline
$\Xi_{bb}^{-}\to \Sigma^-  B^-  \pi^+  $ & \tabincell{c}{$ d_3+d_6-d_{12}-d_{14}-2 d_{16}-d_{17}-2 d_{25}-d_{26}-d_{29}$\\$+3 d_{31}+3 d_{33}+d_{39}$}\\\hline
$\Xi_{bb}^{-}\to \Sigma^-  \overline B^0  \pi^0  $ & \tabincell{c}{$ -\frac{1}{\sqrt{2}}(d_3+d_6+d_8+d_9+2 d_{11}-d_{12}+d_{14}+2 d_{15}-2 d_{16}+d_{17}$\\$+d_{18}-2 d_{19}-2 d_{20}+d_{21}+d_{23}+d_{24}-2 d_{25}-d_{26}-4 d_{28}-d_{29}$\\$-d_{31}-4 d_{32}-d_{33}-d_{34}-d_{35}-d_{37}-d_{38}+d_{39})$}\\\hline
$\Xi_{bb}^{-}\to \Sigma^-  \overline B^0  \eta  $ & \tabincell{c}{$ \frac{1}{\sqrt{6}}(d_3-2 d_4-2 d_5+d_6+d_8+d_9-d_{12}+2 d_{13}+d_{14}-2 d_{16}-d_{17}$\\$+d_{18}-2 d_{19}-2 d_{20}+d_{21}+2 d_{22}+d_{23}-d_{24}-2 d_{25}-d_{26}+6 d_{28}-d_{29}$\\$+2 d_{30}-d_{31}+6 d_{32}-3 d_{33}-d_{34}-d_{35}+2 d_{36}-d_{37}-3 d_{38}+d_{39})$}\\\hline
$\Xi_{bb}^{-}\to \Sigma^-  \overline B^0_s  \overline K^0  $ & \tabincell{c}{$ d_1+d_3-d_5+d_6-d_{10}-d_{12}+d_{15}-2 d_{16}+d_{22}+d_{23}-2 d_{25}$\\$-d_{26}-d_{27}-d_{29}-2 d_{31}+d_{32}-2 d_{33}+d_{36}+d_{37}+d_{39}$}\\\hline
$\Xi_{bb}^{-}\to {p}  B^-  K^-  $ & \tabincell{c}{$ -d_1-d_4-d_{10}-d_{11}-d_{13}-d_{14}-d_{23}-d_{24}-3 d_{27}-3 d_{28}$\\$-3 d_{30}-3 d_{31}-d_{37}-d_{38}$}\\\hline
$\Xi_{bb}^{-}\to \Xi^-  B^-  K^+  $ & $ d_2+d_3-d_{13}-d_{14}-2 d_{25}-d_{26}-2 d_{29}+3 d_{30}+3 d_{31}-d_{39}$\\\hline
$\Xi_{bb}^{-}\to \Xi^-  \overline B^0  K^0  $ & \tabincell{c}{$ d_2+d_3+d_7+d_8+d_{13}+d_{14}-2 d_{20}+d_{21}+d_{22}+d_{23}-2 d_{25}-d_{26}$\\$-2 d_{29}-d_{30}-d_{31}-2 d_{34}+d_{35}-d_{36}-d_{37}-d_{39}$}\\\hline
$\Xi_{bb}^{-}\to \Xi^-  \overline B^0_s  \pi^0  $ & \tabincell{c}{$ \frac{1}{\sqrt{2}}(d_7-d_9-2 d_{10}-2 d_{11}-d_{18}+2 d_{19}+d_{22}-d_{24}+4 d_{27}$\\$+4 d_{28}-d_{34}+2 d_{35}-d_{36}+d_{38})$}\\\hline
$\Xi_{bb}^{-}\to \Xi^-  \overline B^0_s  \eta  $ & \tabincell{c}{$ \frac{1}{\sqrt{6}}(-2 d_1-2 d_2-2 d_3-2 d_4-d_7+d_9+d_{18}-2 d_{19}-d_{22}-2 d_{23}$\\$-d_{24}+4 d_{25}+2 d_{26}+6 d_{27}+6 d_{28}+4 d_{29}+4 d_{30}+4 d_{31}+d_{34}-2 d_{35}$\\$+d_{36}-2 d_{37}-3 d_{38}+2 d_{39})$}\\\hline
$\Xi_{bb}^{-}\to \Xi^0  B^-  K^0  $ & $ -d_7-d_8-2 d_{20}+d_{21}-d_{22}-d_{23}+2 d_{34}+3 d_{35}+d_{36}+d_{37}$\\\hline
$\Xi_{bb}^{-}\to \Xi^0  \overline B^0_s  \pi^-  $ & $ d_7-d_9+d_{18}-2 d_{19}+d_{22}-d_{24}+3 d_{34}+2 d_{35}-d_{36}+d_{38}$\\\hline
$\Omega_{bb}^{-}\to \Lambda^0  B^-  \overline K^0  $ & \tabincell{c}{$ \frac{1}{\sqrt{6}}(d_1-d_4-2 d_5-2 d_7-d_8+d_9-d_{10}+d_{11}-d_{13}+d_{14}+2 d_{15}$\\$+2 d_{17}+d_{18}-2 d_{19}-2 d_{20}+d_{21}-d_{27}+d_{28}-3 d_{30}+3 d_{31}+2 d_{32}$\\$+6 d_{33}+3 d_{34}+9 d_{35}+4 d_{36}+2 d_{37}-2 d_{38})$}\\\hline
$\Omega_{bb}^{-}\to \Lambda^0  \overline B^0  K^-  $ & \tabincell{c}{$ \frac{1}{\sqrt{6}}(-d_1+d_4+2 d_5+2 d_7+d_8-d_9-d_{10}+d_{11}-d_{13}+d_{14}+2 d_{15}$\\$+2 d_{17}+d_{18}-2 d_{19}-2 d_{20}+d_{21}-3 d_{27}+3 d_{28}-d_{30}+d_{31}$\\$+6 d_{32}+2 d_{33}+9 d_{34}+3 d_{35}-4 d_{36}-2 d_{37}+2 d_{38})$}\\\hline
$\Omega_{bb}^{-}\to \Sigma^+  B^-  K^-  $ & \tabincell{c}{$ d_1+d_4-d_8-d_9+d_{10}+d_{11}+d_{13}+d_{14}+d_{18}-2 d_{19}-2 d_{20}$\\$+d_{21}+3 d_{27}+3 d_{28}+3 d_{30}+3 d_{31}-3 d_{34}-3 d_{35}+2 d_{37}+2 d_{38}$}\\\hline
$\Omega_{bb}^{-}\to \Sigma^0  B^-  \overline K^0  $ & \tabincell{c}{$ \frac{1}{\sqrt{2}}(-d_1-d_4+d_8+d_9+d_{10}+d_{11}-d_{13}-d_{14}+d_{18}-2 d_{19}$\\$+2 d_{20}-d_{21}+d_{27}+d_{28}-3 d_{30}-3 d_{31}-d_{34}+3 d_{35}-2 d_{37}-2 d_{38})$}\\\hline
$\Omega_{bb}^{-}\to \Sigma^0  \overline B^0  K^-  $ & \tabincell{c}{$ \frac{1}{\sqrt{2}}(-d_1-d_4+d_8+d_9-d_{10}-d_{11}+d_{13}+d_{14}-d_{18}+2 d_{19}-2 d_{20}$\\$+d_{21}-3 d_{27}-3 d_{28}+d_{30}+d_{31}+3 d_{34}-d_{35}-2 d_{37}-2 d_{38})$}\\\hline
$\Omega_{bb}^{-}\to \Sigma^-  \overline B^0  \overline K^0  $ & \tabincell{c}{$ -d_1-d_4+d_8+d_9+d_{10}+d_{11}+d_{13}+d_{14}+d_{18}-2 d_{19}-2 d_{20}+d_{21}$\\$+d_{27}+d_{28}+d_{30}+d_{31}-d_{34}-d_{35}-2 d_{37}-2 d_{38}$}\\\hline
$\Omega_{bb}^{-}\to \Xi^-  B^-  \pi^+  $ & $ -d_2+d_6-d_{12}+d_{13}-2 d_{16}-d_{17}+d_{29}-3 d_{30}+3 d_{33}+2 d_{39}$\\\hline
$\Omega_{bb}^{-}\to \Xi^-  \overline B^0  \pi^0  $ & \tabincell{c}{$ \frac{1}{\sqrt{2}}(d_2-d_6+2 d_{10}+d_{12}+d_{13}-2 d_{15}+2 d_{16}-d_{17}-4 d_{27}$\\$-d_{29}-d_{30}+4 d_{32}+d_{33}-2 d_{39})$}\\\hline
$\Omega_{bb}^{-}\to \Xi^-  \overline B^0  \eta  $ & \tabincell{c}{$ \frac{1}{\sqrt{6}}(2 d_1-d_2-2 d_5+d_6-2 d_7-2 d_8-d_{12}-d_{13}-2 d_{14}-2 d_{16}-d_{17}$\\$+4 d_{20}-2 d_{21}-6 d_{27}+d_{29}+d_{30}-2 d_{31}+6 d_{32}-3 d_{33}+4 d_{34}-2 d_{35}$\\$+4 d_{36}+4 d_{37}+2 d_{39})$}\\\hline
$\Omega_{bb}^{-}\to \Xi^-  \overline B^0_s  \overline K^0  $ & \tabincell{c}{$ -d_2-d_4-d_5+d_6-d_7+d_9+d_{11}-d_{12}+d_{15}-2 d_{16}+d_{18}-2 d_{19}$\\$+d_{28}+d_{29}+2 d_{30}+d_{32}-2 d_{33}+d_{34}-2 d_{35}+2 d_{36}-2 d_{38}+2 d_{39}$}\\\hline
$\Omega_{bb}^{-}\to \Xi^0  B^-  \pi^0  $ & \tabincell{c}{$ \frac{1}{\sqrt{2}}(d_2-d_6-2 d_{10}-d_{12}-d_{13}+2 d_{15}-2 d_{16}+d_{17}+4 d_{27}$\\$+3 d_{29}+3 d_{30}-4 d_{32}-3 d_{33}-2 d_{39})$}\\\hline
$\Omega_{bb}^{-}\to \Xi^0  B^-  \eta  $ & \tabincell{c}{$ -\frac{1}{\sqrt{6}}(2 d_1-d_2-2 d_5+d_6-2 d_7-2 d_8+d_{12}+d_{13}+2 d_{14}+2 d_{16}+d_{17}$\\$-4 d_{20}+2 d_{21}-6 d_{27}-3 d_{29}-3 d_{30}+6 d_{31}+6 d_{32}+9 d_{33}+4 d_{34}$\\$+6 d_{35}+4 d_{36}+4 d_{37}+2 d_{39})$}\\\hline
$\Omega_{bb}^{-}\to \Xi^0  \overline B^0  \pi^-  $ & $ d_2-d_6-d_{12}+d_{13}-2 d_{16}-d_{17}+3 d_{29}-d_{30}+d_{33}-2 d_{39}$\\\hline
$\Omega_{bb}^{-}\to \Xi^0  \overline B^0_s  K^-  $ & \tabincell{c}{$ d_2+d_4+d_5-d_6+d_7-d_9+d_{11}-d_{12}+d_{15}-2 d_{16}+d_{18}-2 d_{19}$\\$+3 d_{28}+3 d_{29}-2 d_{30}+3 d_{32}+2 d_{33}+3 d_{34}+2 d_{35}-2 d_{36}+2 d_{38}-2 d_{39}$}\\\hline
\hline
\end{tabular}
\end{table}


\subsection{Charmless $b\to q_1 \bar q_2 q_3$ Decays:  Decays into a bottom meson, a light baryon decuplet and a light meson}

The effective Hamiltonian is given as
\begin{eqnarray}
 {\cal H}_{eff}&=& f_1(T_{bb})^i   \overline B ^j (T_{10})_{ijk} M^k_l (H_{3})^l +f_2(T_{bb})^i   \overline B ^j (T_{10})_{ikl} M^k_j (H_{3})^l \nonumber \\
 && +f_3(T_{bb})^i   \overline B ^j (T_{10})_{jkl} M^k_i (H_{3})^l +f_4(T_{bb})^i  \overline B ^j  (T_{10})_{ijk} M^m_l (H_{15})^{kl}_m\nonumber \\
&&+f_5(T_{bb})^i  \overline B ^j  (T_{10})_{ikl} M^m_j (H_{15})^{kl}_m +f_6(T_{bb})^i  \overline B ^j  (T_{10})_{ilm} M^m_k (H_{15})^{kl}_j\nonumber \\
&&+f_7(T_{bb})^i  \overline B ^j  (T_{10})_{jkl} M^m_i (H_{15})^{kl}_m +f_8(T_{bb})^i  \overline B ^j  (T_{10})_{klm} M^m_i (H_{15})^{kl}_j\nonumber \\
&&+f_9(T_{bb})^i  \overline B ^j  (T_{10})_{jkl} M^l_m (H_{15})^{km}_i +f_{10}(T_{bb})^i  \overline B ^j  (T_{10})_{klm} M^m_j (H_{15})^{kl}_i \nonumber \\
&&+f_{11}(T_{bb})^i  \overline B ^j  (T_{10})_{ijk} M^m_l (H_{\bar6})^{kl}_m+f_{12}(T_{bb})^i  \overline B ^j  (T_{10})_{ilm} M^m_k (H_{\bar6})^{kl}_j\nonumber \\
&&+f_{13}(T_{bb})^i  \overline B ^j  (T_{10})_{jkl} M^l_m (H_{\bar6})^{km}_i.
\end{eqnarray}
Decay amplitudes for different channels are given in Tab.~\ref{tab:bbq_qqq10_bqbarand8meson_vd_1} and Tab.~\ref{tab:bbq_qqq10_bqbarand8meson_vd_2} for $b\to d$ transition; Tab.~\ref{tab:bbq_qqq10_bqbarand8meson_vs_1} and Tab.~\ref{tab:bbq_qqq10_bqbarand8meson_vs_2} for $b\to s$ transition.
We summarize the corresponding relations for decay widths in Appendix~\ref{sec:relation_bb_baryon10_bm_lm_b2d}.

\begin{table}
\footnotesize
\caption{Doubly bottom baryon decays (induced by the $b\to d$ transition) into a bottom meson, a light baryon(10)  and a light meson.}\label{tab:bbq_qqq10_bqbarand8meson_vd_1}\begin{tabular}{|c|c|c|c|c|c|c|c}\hline
channel & amplitude \\\hline
\hline
$\Xi_{bb}^{0}\to \Delta^{++}  B^-  \pi^-  $ & $ f_1+3 f_4+3 f_6+3 f_9+f_{11}-f_{12}+f_{13}$\\\hline
$\Xi_{bb}^{0}\to \Delta^{+}  B^-  \pi^0  $ & $ \frac{-f_1+f_2+f_3+5 f_4+6 f_5+6 f_7+6 f_8+6 f_{10}-f_{11}+2 f_{12}-2 f_{13}}{\sqrt{6}}$\\\hline
$\Xi_{bb}^{0}\to \Delta^{+}  B^-  \eta  $ & $ \frac{f_1+f_2+f_3+3 f_4+6 f_5+6 f_6+6 f_7+6 f_8+6 f_9+6 f_{10}-3 f_{11}}{3 \sqrt{2}}$\\\hline
$\Xi_{bb}^{0}\to \Delta^{+}  \overline B^0  \pi^-  $ & $ \frac{f_1+f_2+3 f_4+6 f_5-2 f_6+3 f_9+6 f_{10}+f_{11}+f_{13}}{\sqrt{3}}$\\\hline
$\Xi_{bb}^{0}\to \Delta^{+}  \overline B^0_s  K^-  $ & $ \frac{f_2+6 f_5-f_6+6 f_{10}-f_{12}}{\sqrt{3}}$\\\hline
$\Xi_{bb}^{0}\to \Delta^{0}  B^-  \pi^+  $ & $ \frac{f_2+f_3-2 f_5+3 f_6-2 f_7+6 f_8+3 f_9+6 f_{10}+f_{12}-f_{13}}{\sqrt{3}}$\\\hline
$\Xi_{bb}^{0}\to \Delta^{0}  \overline B^0  \pi^0  $ & $ -\frac{f_1+f_2-f_3-5 f_4-2 f_5-2 f_6-6 f_7+2 f_8+6 f_{10}+f_{11}+2 f_{13}}{\sqrt{6}}$\\\hline
$\Xi_{bb}^{0}\to \Delta^{0}  \overline B^0  \eta  $ & $ \frac{f_1+f_2+f_3+3 f_4-2 f_5-2 f_6+6 f_7-2 f_8+6 f_9+6 f_{10}-3 f_{11}}{3 \sqrt{2}}$\\\hline
$\Xi_{bb}^{0}\to \Delta^{0}  \overline B^0_s  \overline K^0  $ & $ \frac{f_2-2 f_5-f_6+6 f_{10}-f_{12}}{\sqrt{3}}$\\\hline
$\Xi_{bb}^{0}\to \Delta^{-}  \overline B^0  \pi^+  $ & $ f_3-2 f_7-2 f_8+3 f_9-f_{13}$\\\hline
$\Xi_{bb}^{0}\to \Sigma^{\prime+}  B^-  K^0  $ & $ \frac{f_1-f_4+3 f_6+3 f_9-f_{11}-f_{12}+f_{13}}{\sqrt{3}}$\\\hline
$\Xi_{bb}^{0}\to \Sigma^{\prime+}  \overline B^0_s  \pi^-  $ & $ \frac{f_1+3 f_4-f_6+3 f_9+f_{11}+f_{12}+f_{13}}{\sqrt{3}}$\\\hline
$\Xi_{bb}^{0}\to \Sigma^{\prime0}  B^-  K^+  $ & $ \frac{f_2+f_3-2 f_5+3 f_6-2 f_7+6 f_8+3 f_9+6 f_{10}+f_{12}-f_{13}}{\sqrt{6}}$\\\hline
$\Xi_{bb}^{0}\to \Sigma^{\prime0}  \overline B^0  K^0  $ & $ \frac{f_1+f_2-f_4-2 f_5-2 f_6+3 f_9+6 f_{10}-f_{11}+f_{13}}{\sqrt{6}}$\\\hline
$\Xi_{bb}^{0}\to \Sigma^{\prime0}  \overline B^0_s  \pi^0  $ & $ -\frac{f_1-f_3-5 f_4-f_6-6 f_7+2 f_8+f_{11}+f_{12}+2 f_{13}}{2 \sqrt{3}}$\\\hline
$\Xi_{bb}^{0}\to \Sigma^{\prime0}  \overline B^0_s  \eta  $ & $ \frac{1}{6} \left(f_1-2 f_2+f_3+3 f_4+4 f_5+f_6+6 f_7-2 f_8+6 f_9-12 f_{10}-3 f_{11}+3 f_{12}\right)$\\\hline
$\Xi_{bb}^{0}\to \Sigma^{\prime-}  \overline B^0  K^+  $ & $ \frac{f_3-2 f_7-2 f_8+3 f_9-f_{13}}{\sqrt{3}}$\\\hline
$\Xi_{bb}^{0}\to \Sigma^{\prime-}  \overline B^0_s  \pi^+  $ & $ \frac{f_3-2 f_7-2 f_8+3 f_9-f_{13}}{\sqrt{3}}$\\\hline
$\Xi_{bb}^{0}\to \Xi^{\prime0}  \overline B^0_s  K^0  $ & $ \frac{f_1-f_4-f_6+3 f_9-f_{11}+f_{12}+f_{13}}{\sqrt{3}}$\\\hline
$\Xi_{bb}^{0}\to \Xi^{\prime-}  \overline B^0_s  K^+  $ & $ \frac{f_3-2 f_7-2 f_8+3 f_9-f_{13}}{\sqrt{3}}$\\\hline
$\Xi_{bb}^{-}\to \Delta^{+}  B^-  \pi^-  $ & $ \frac{f_1+f_3+3 f_4+3 f_6+6 f_7+6 f_8-2 f_9+f_{11}-f_{12}}{\sqrt{3}}$\\\hline
$\Xi_{bb}^{-}\to \Delta^{0}  B^-  \pi^0  $ & $ -\frac{f_1-f_2+f_3-5 f_4-6 f_5-2 f_7+6 f_8-2 f_9+2 f_{10}+f_{11}-2 f_{12}}{\sqrt{6}}$\\\hline
$\Xi_{bb}^{-}\to \Delta^{0}  B^-  \eta  $ & $ \frac{f_1+f_2+f_3+3 f_4+6 f_5+6 f_6-2 f_7+6 f_8-2 f_9-2 f_{10}-3 f_{11}}{3 \sqrt{2}}$\\\hline
$\Xi_{bb}^{-}\to \Delta^{0}  \overline B^0  \pi^-  $ & $ \frac{f_1+f_2+f_3+3 f_4+6 f_5-2 f_6+6 f_7-2 f_8-2 f_9-2 f_{10}+f_{11}}{\sqrt{3}}$\\\hline
$\Xi_{bb}^{-}\to \Delta^{0}  \overline B^0_s  K^-  $ & $ \frac{f_2+6 f_5-f_6-2 f_{10}-f_{12}}{\sqrt{3}}$\\\hline
$\Xi_{bb}^{-}\to \Delta^{-}  B^-  \pi^+  $ & $ f_2-2 f_5+3 f_6-2 f_{10}+f_{12}$\\\hline
$\Xi_{bb}^{-}\to \Delta^{-}  \overline B^0  \pi^0  $ & $ -\frac{f_1+f_2+f_3-5 f_4-2 f_5-2 f_6-2 f_7-2 f_8-2 f_9-2 f_{10}+f_{11}}{\sqrt{2}}$\\\hline
$\Xi_{bb}^{-}\to \Delta^{-}  \overline B^0  \eta  $ & $ \frac{f_1+f_2+f_3+3 f_4-2 f_5-2 f_6-2 f_7-2 f_8-2 f_9-2 f_{10}-3 f_{11}}{\sqrt{6}}$\\\hline
$\Xi_{bb}^{-}\to \Delta^{-}  \overline B^0_s  \overline K^0  $ & $ f_2-2 f_5-f_6-2 f_{10}-f_{12}$\\\hline
$\Xi_{bb}^{-}\to \Sigma^{\prime0}  B^-  K^0  $ & $ \frac{f_1+f_3-f_4+3 f_6-2 f_7+6 f_8-2 f_9-f_{11}-f_{12}}{\sqrt{6}}$\\\hline
$\Xi_{bb}^{-}\to \Sigma^{\prime0}  \overline B^0_s  \pi^-  $ & $ \frac{f_1+f_3+3 f_4-f_6+6 f_7-2 f_8-2 f_9+f_{11}+f_{12}}{\sqrt{6}}$\\\hline
$\Xi_{bb}^{-}\to \Sigma^{\prime-}  B^-  K^+  $ & $ \frac{f_2-2 f_5+3 f_6-2 f_{10}+f_{12}}{\sqrt{3}}$\\\hline
$\Xi_{bb}^{-}\to \Sigma^{\prime-}  \overline B^0  K^0  $ & $ \frac{f_1+f_2+f_3-f_4-2 f_5-2 f_6-2 f_7-2 f_8-2 f_9-2 f_{10}-f_{11}}{\sqrt{3}}$\\\hline
$\Xi_{bb}^{-}\to \Sigma^{\prime-}  \overline B^0_s  \pi^0  $ & $ -\frac{f_1+f_3-5 f_4-f_6-2 f_7-2 f_8-2 f_9+f_{11}+f_{12}}{\sqrt{6}}$\\\hline
$\Xi_{bb}^{-}\to \Sigma^{\prime-}  \overline B^0_s  \eta  $ & $ \frac{f_1-2 f_2+f_3+3 f_4+4 f_5+f_6-2 f_7-2 f_8-2 f_9+4 f_{10}-3 f_{11}+3 f_{12}}{3 \sqrt{2}}$\\\hline
$\Xi_{bb}^{-}\to \Xi^{\prime-}  \overline B^0_s  K^0  $ & $ \frac{f_1+f_3-f_4-f_6-2 f_7-2 f_8-2 f_9-f_{11}+f_{12}}{\sqrt{3}}$\\\hline
$\Omega_{bb}^{-}\to \Delta^{+}  B^-  K^-  $ & $ \frac{f_3+6 f_7+6 f_8-f_9+f_{13}}{\sqrt{3}}$\\\hline
$\Omega_{bb}^{-}\to \Delta^{0}  B^-  \overline K^0  $ & $ \frac{f_3-2 f_7+6 f_8-f_9+f_{13}}{\sqrt{3}}$\\\hline
$\Omega_{bb}^{-}\to \Delta^{0}  \overline B^0  K^-  $ & $ \frac{f_3+6 f_7-2 f_8-f_9+f_{13}}{\sqrt{3}}$\\\hline
$\Omega_{bb}^{-}\to \Delta^{-}  \overline B^0  \overline K^0  $ & $ f_3-2 f_7-2 f_8-f_9+f_{13}$\\\hline
$\Omega_{bb}^{-}\to \Sigma^{\prime+}  B^-  \pi^-  $ & $ \frac{f_1+3 f_4+3 f_6-f_9+f_{11}-f_{12}-f_{13}}{\sqrt{3}}$\\\hline
\hline
\end{tabular}
\end{table}
\begin{table}
\small
\caption{Doubly bottom baryon decays into a bottom meson, a light baryon(10) induced by the charmless $b\to d$ transition and a light meson.}\label{tab:bbq_qqq10_bqbarand8meson_vd_2}\begin{tabular}{|c|c|c|c|c|c|c|c}\hline
channel & amplitude \\\hline
\hline
$\Omega_{bb}^{-}\to \Sigma^{\prime0}  B^-  \pi^0  $ & $ \frac{-f_1+f_2+5 f_4+6 f_5+f_9-2 f_{10}-f_{11}+2 f_{12}+f_{13}}{2 \sqrt{3}}$\\\hline
$\Omega_{bb}^{-}\to \Sigma^{\prime0}  B^-  \eta  $ & $ \frac{1}{6} \left(f_1+f_2-2 f_3+3 f_4+6 f_5+6 f_6+4 f_7-12 f_8+f_9-2 f_{10}-3 f_{11}-3 f_{13}\right)$\\\hline
$\Omega_{bb}^{-}\to \Sigma^{\prime0}  \overline B^0  \pi^-  $ & $ \frac{f_1+f_2+3 f_4+6 f_5-2 f_6-f_9-2 f_{10}+f_{11}-f_{13}}{\sqrt{6}}$\\\hline
$\Omega_{bb}^{-}\to \Sigma^{\prime0}  \overline B^0_s  K^-  $ & $ \frac{f_2+f_3+6 f_5-f_6+6 f_7-2 f_8-f_9-2 f_{10}-f_{12}+f_{13}}{\sqrt{6}}$\\\hline
$\Omega_{bb}^{-}\to \Sigma^{\prime-}  B^-  \pi^+  $ & $ \frac{f_2-2 f_5+3 f_6-2 f_{10}+f_{12}}{\sqrt{3}}$\\\hline
$\Omega_{bb}^{-}\to \Sigma^{\prime-}  \overline B^0  \pi^0  $ & $ \frac{-f_1-f_2+5 f_4+2 f_5+2 f_6+f_9+2 f_{10}-f_{11}+f_{13}}{\sqrt{6}}$\\\hline
$\Omega_{bb}^{-}\to \Sigma^{\prime-}  \overline B^0  \eta  $ & $ \frac{f_1+f_2-2 f_3+3 f_4-2 f_5-2 f_6+4 f_7+4 f_8+f_9-2 f_{10}-3 f_{11}-3 f_{13}}{3 \sqrt{2}}$\\\hline
$\Omega_{bb}^{-}\to \Sigma^{\prime-}  \overline B^0_s  \overline K^0  $ & $ \frac{f_2+f_3-2 f_5-f_6-2 f_7-2 f_8-f_9-2 f_{10}-f_{12}+f_{13}}{\sqrt{3}}$\\\hline
$\Omega_{bb}^{-}\to \Xi^{\prime0}  B^-  K^0  $ & $ \frac{f_1-f_4+3 f_6-f_9-f_{11}-f_{12}-f_{13}}{\sqrt{3}}$\\\hline
$\Omega_{bb}^{-}\to \Xi^{\prime0}  \overline B^0_s  \pi^-  $ & $ \frac{f_1+3 f_4-f_6-f_9+f_{11}+f_{12}-f_{13}}{\sqrt{3}}$\\\hline
$\Omega_{bb}^{-}\to \Xi^{\prime-}  B^-  K^+  $ & $ \frac{f_2-2 f_5+3 f_6-2 f_{10}+f_{12}}{\sqrt{3}}$\\\hline
$\Omega_{bb}^{-}\to \Xi^{\prime-}  \overline B^0  K^0  $ & $ \frac{f_1+f_2-f_4-2 f_5-2 f_6-f_9-2 f_{10}-f_{11}-f_{13}}{\sqrt{3}}$\\\hline
$\Omega_{bb}^{-}\to \Xi^{\prime-}  \overline B^0_s  \pi^0  $ & $ \frac{-f_1+5 f_4+f_6+f_9-f_{11}-f_{12}+f_{13}}{\sqrt{6}}$\\\hline
$\Omega_{bb}^{-}\to \Xi^{\prime-}  \overline B^0_s  \eta  $ & $ \frac{f_1-2 f_2-2 f_3+3 f_4+4 f_5+f_6+4 f_7+4 f_8+f_9+4 f_{10}-3 f_{11}+3 f_{12}-3 f_{13}}{3 \sqrt{2}}$\\\hline
$\Omega_{bb}^{-}\to \Omega^-  \overline B^0_s  K^0  $ & $ f_1-f_4-f_6-f_9-f_{11}+f_{12}-f_{13}$\\\hline
\hline
\end{tabular}
\end{table}

\begin{table}
\footnotesize
\caption{Doubly bottom baryon decays into a bottom meson, a light baryon(10) induced by the charmless $b\to s$ transition and a light meson.}\label{tab:bbq_qqq10_bqbarand8meson_vs_1}\begin{tabular}{|c|c|c|c|c|c|c|c}\hline
channel & amplitude \\\hline
\hline
$\Xi_{bb}^{0}\to \Delta^{++}  B^-  K^-  $ & $ f_1+3 f_4+3 f_6+3 f_9+f_{11}-f_{12}+f_{13}$\\\hline
$\Xi_{bb}^{0}\to \Delta^{+}  B^-  \overline K^0  $ & $ \frac{f_1-f_4+3 f_6+3 f_9-f_{11}-f_{12}+f_{13}}{\sqrt{3}}$\\\hline
$\Xi_{bb}^{0}\to \Delta^{+}  \overline B^0  K^-  $ & $ \frac{f_1+3 f_4-f_6+3 f_9+f_{11}+f_{12}+f_{13}}{\sqrt{3}}$\\\hline
$\Xi_{bb}^{0}\to \Delta^{0}  \overline B^0  \overline K^0  $ & $ \frac{f_1-f_4-f_6+3 f_9-f_{11}+f_{12}+f_{13}}{\sqrt{3}}$\\\hline
$\Xi_{bb}^{0}\to \Sigma^{\prime+}  B^-  \pi^0  $ & $ \frac{f_2+f_3+4 f_4+6 f_5+3 f_6+6 f_7+6 f_8+3 f_9+6 f_{10}-2 f_{11}+f_{12}-f_{13}}{\sqrt{6}}$\\\hline
$\Xi_{bb}^{0}\to \Sigma^{\prime+}  B^-  \eta  $ & $ \frac{-2 f_1+f_2+f_3+6 f_4+6 f_5-3 f_6+6 f_7+6 f_8-3 f_9+6 f_{10}+3 f_{12}-3 f_{13}}{3 \sqrt{2}}$\\\hline
$\Xi_{bb}^{0}\to \Sigma^{\prime+}  \overline B^0  \pi^-  $ & $ \frac{f_2+6 f_5-f_6+6 f_{10}-f_{12}}{\sqrt{3}}$\\\hline
$\Xi_{bb}^{0}\to \Sigma^{\prime+}  \overline B^0_s  K^-  $ & $ \frac{f_1+f_2+3 f_4+6 f_5-2 f_6+3 f_9+6 f_{10}+f_{11}+f_{13}}{\sqrt{3}}$\\\hline
$\Xi_{bb}^{0}\to \Sigma^{\prime0}  B^-  \pi^+  $ & $ \frac{f_2+f_3-2 f_5+3 f_6-2 f_7+6 f_8+3 f_9+6 f_{10}+f_{12}-f_{13}}{\sqrt{6}}$\\\hline
$\Xi_{bb}^{0}\to \Sigma^{\prime0}  \overline B^0  \pi^0  $ & $ \frac{-f_2+f_3+4 f_4+2 f_5+f_6+6 f_7-2 f_8+3 f_9-6 f_{10}-2 f_{11}+f_{12}-f_{13}}{2 \sqrt{3}}$\\\hline
$\Xi_{bb}^{0}\to \Sigma^{\prime0}  \overline B^0  \eta  $ & $ \frac{1}{6} \left(-2 f_1+f_2+f_3+6 f_4-2 f_5+f_6+6 f_7-2 f_8-3 f_9+6 f_{10}-3 f_{12}-3 f_{13}\right)$\\\hline
$\Xi_{bb}^{0}\to \Sigma^{\prime0}  \overline B^0_s  \overline K^0  $ & $ \frac{f_1+f_2-f_4-2 f_5-2 f_6+3 f_9+6 f_{10}-f_{11}+f_{13}}{\sqrt{6}}$\\\hline
$\Xi_{bb}^{0}\to \Sigma^{\prime-}  \overline B^0  \pi^+  $ & $ \frac{f_3-2 f_7-2 f_8+3 f_9-f_{13}}{\sqrt{3}}$\\\hline
$\Xi_{bb}^{0}\to \Xi^{\prime0}  B^-  K^+  $ & $ \frac{f_2+f_3-2 f_5+3 f_6-2 f_7+6 f_8+3 f_9+6 f_{10}+f_{12}-f_{13}}{\sqrt{3}}$\\\hline
$\Xi_{bb}^{0}\to \Xi^{\prime0}  \overline B^0  K^0  $ & $ \frac{f_2-2 f_5-f_6+6 f_{10}-f_{12}}{\sqrt{3}}$\\\hline
$\Xi_{bb}^{0}\to \Xi^{\prime0}  \overline B^0_s  \pi^0  $ & $ \frac{f_3+4 f_4+6 f_7-2 f_8+3 f_9-2 f_{11}-f_{13}}{\sqrt{6}}$\\\hline
$\Xi_{bb}^{0}\to \Xi^{\prime0}  \overline B^0_s  \eta  $ & $ \frac{-2 f_1-2 f_2+f_3+6 f_4+4 f_5+4 f_6+6 f_7-2 f_8-3 f_9-12 f_{10}-3 f_{13}}{3 \sqrt{2}}$\\\hline
$\Xi_{bb}^{0}\to \Xi^{\prime-}  \overline B^0  K^+  $ & $ \frac{f_3-2 f_7-2 f_8+3 f_9-f_{13}}{\sqrt{3}}$\\\hline
$\Xi_{bb}^{0}\to \Xi^{\prime-}  \overline B^0_s  \pi^+  $ & $ \frac{f_3-2 f_7-2 f_8+3 f_9-f_{13}}{\sqrt{3}}$\\\hline
$\Xi_{bb}^{0}\to \Omega^-  \overline B^0_s  K^+  $ & $ f_3-2 f_7-2 f_8+3 f_9-f_{13}$\\\hline
$\Xi_{bb}^{-}\to \Delta^{+}  B^-  K^-  $ & $ \frac{f_1+3 f_4+3 f_6-f_9+f_{11}-f_{12}-f_{13}}{\sqrt{3}}$\\\hline
$\Xi_{bb}^{-}\to \Delta^{0}  B^-  \overline K^0  $ & $ \frac{f_1-f_4+3 f_6-f_9-f_{11}-f_{12}-f_{13}}{\sqrt{3}}$\\\hline
$\Xi_{bb}^{-}\to \Delta^{0}  \overline B^0  K^-  $ & $ \frac{f_1+3 f_4-f_6-f_9+f_{11}+f_{12}-f_{13}}{\sqrt{3}}$\\\hline
$\Xi_{bb}^{-}\to \Delta^{-}  \overline B^0  \overline K^0  $ & $ f_1-f_4-f_6-f_9-f_{11}+f_{12}-f_{13}$\\\hline
$\Xi_{bb}^{-}\to \Sigma^{\prime+}  B^-  \pi^-  $ & $ \frac{f_3+6 f_7+6 f_8-f_9+f_{13}}{\sqrt{3}}$\\\hline
$\Xi_{bb}^{-}\to \Sigma^{\prime0}  B^-  \pi^0  $ & $ \frac{f_2-f_3+4 f_4+6 f_5+3 f_6+2 f_7-6 f_8+f_9-2 f_{10}-2 f_{11}+f_{12}-f_{13}}{2 \sqrt{3}}$\\\hline
$\Xi_{bb}^{-}\to \Sigma^{\prime0}  B^-  \eta  $ & $ \frac{1}{6} \left(-2 f_1+f_2+f_3+6 f_4+6 f_5-3 f_6-2 f_7+6 f_8+f_9-2 f_{10}+3 f_{12}+3 f_{13}\right)$\\\hline
$\Xi_{bb}^{-}\to \Sigma^{\prime0}  \overline B^0  \pi^-  $ & $ \frac{f_2+f_3+6 f_5-f_6+6 f_7-2 f_8-f_9-2 f_{10}-f_{12}+f_{13}}{\sqrt{6}}$\\\hline
$\Xi_{bb}^{-}\to \Sigma^{\prime0}  \overline B^0_s  K^-  $ & $ \frac{f_1+f_2+3 f_4+6 f_5-2 f_6-f_9-2 f_{10}+f_{11}-f_{13}}{\sqrt{6}}$\\\hline
$\Xi_{bb}^{-}\to \Sigma^{\prime-}  B^-  \pi^+  $ & $ \frac{f_2-2 f_5+3 f_6-2 f_{10}+f_{12}}{\sqrt{3}}$\\\hline
$\Xi_{bb}^{-}\to \Sigma^{\prime-}  \overline B^0  \pi^0  $ & $ \frac{-f_2-f_3+4 f_4+2 f_5+f_6+2 f_7+2 f_8+f_9+2 f_{10}-2 f_{11}+f_{12}-f_{13}}{\sqrt{6}}$\\\hline
$\Xi_{bb}^{-}\to \Sigma^{\prime-}  \overline B^0  \eta  $ & $ \frac{-2 f_1+f_2+f_3+6 f_4-2 f_5+f_6-2 f_7-2 f_8+f_9-2 f_{10}-3 f_{12}+3 f_{13}}{3 \sqrt{2}}$\\\hline
$\Xi_{bb}^{-}\to \Sigma^{\prime-}  \overline B^0_s  \overline K^0  $ & $ \frac{f_1+f_2-f_4-2 f_5-2 f_6-f_9-2 f_{10}-f_{11}-f_{13}}{\sqrt{3}}$\\\hline
$\Xi_{bb}^{-}\to \Xi^{\prime0}  B^-  K^0  $ & $ \frac{f_3-2 f_7+6 f_8-f_9+f_{13}}{\sqrt{3}}$\\\hline
$\Xi_{bb}^{-}\to \Xi^{\prime0}  \overline B^0_s  \pi^-  $ & $ \frac{f_3+6 f_7-2 f_8-f_9+f_{13}}{\sqrt{3}}$\\\hline
$\Xi_{bb}^{-}\to \Xi^{\prime-}  B^-  K^+  $ & $ \frac{f_2-2 f_5+3 f_6-2 f_{10}+f_{12}}{\sqrt{3}}$\\\hline
$\Xi_{bb}^{-}\to \Xi^{\prime-}  \overline B^0  K^0  $ & $ \frac{f_2+f_3-2 f_5-f_6-2 f_7-2 f_8-f_9-2 f_{10}-f_{12}+f_{13}}{\sqrt{3}}$\\\hline
$\Xi_{bb}^{-}\to \Xi^{\prime-}  \overline B^0_s  \pi^0  $ & $ \frac{-f_3+4 f_4+2 f_7+2 f_8+f_9-2 f_{11}-f_{13}}{\sqrt{6}}$\\\hline
$\Xi_{bb}^{-}\to \Xi^{\prime-}  \overline B^0_s  \eta  $ & $ \frac{-2 f_1-2 f_2+f_3+6 f_4+4 f_5+4 f_6-2 f_7-2 f_8+f_9+4 f_{10}+3 f_{13}}{3 \sqrt{2}}$\\\hline
$\Xi_{bb}^{-}\to \Omega^-  \overline B^0_s  K^0  $ & $ f_3-2 f_7-2 f_8-f_9+f_{13}$\\\hline
$\Omega_{bb}^{-}\to \Sigma^{\prime+}  B^-  K^-  $ & $ \frac{f_1+f_3+3 f_4+3 f_6+6 f_7+6 f_8-2 f_9+f_{11}-f_{12}}{\sqrt{3}}$\\\hline
\end{tabular}
\end{table}

\begin{table}
\small
\caption{Doubly bottom baryon decays into a bottom meson, a light baryon(10) induced by the charmless $b\to s$ transition and a light meson.}\label{tab:bbq_qqq10_bqbarand8meson_vs_2}\begin{tabular}{|c|c|c|c|c|c|c|c}\hline
channel & amplitude \\\hline
\hline
$\Omega_{bb}^{-}\to \Sigma^{\prime0}  B^-  \overline K^0  $ & $ \frac{f_1+f_3-f_4+3 f_6-2 f_7+6 f_8-2 f_9-f_{11}-f_{12}}{\sqrt{6}}$\\\hline
$\Omega_{bb}^{-}\to \Sigma^{\prime0}  \overline B^0  K^-  $ & $ \frac{f_1+f_3+3 f_4-f_6+6 f_7-2 f_8-2 f_9+f_{11}+f_{12}}{\sqrt{6}}$\\\hline
$\Omega_{bb}^{-}\to \Sigma^{\prime-}  \overline B^0  \overline K^0  $ & $ \frac{f_1+f_3-f_4-f_6-2 f_7-2 f_8-2 f_9-f_{11}+f_{12}}{\sqrt{3}}$\\\hline
$\Omega_{bb}^{-}\to \Xi^{\prime0}  B^-  \pi^0  $ & $ \frac{f_2+4 f_4+6 f_5+3 f_6-2 f_{10}-2 f_{11}+f_{12}}{\sqrt{6}}$\\\hline
$\Omega_{bb}^{-}\to \Xi^{\prime0}  B^-  \eta  $ & $ \frac{-2 f_1+f_2-2 f_3+6 f_4+6 f_5-3 f_6+4 f_7-12 f_8+4 f_9-2 f_{10}+3 f_{12}}{3 \sqrt{2}}$\\\hline
$\Omega_{bb}^{-}\to \Xi^{\prime0}  \overline B^0  \pi^-  $ & $ \frac{f_2+6 f_5-f_6-2 f_{10}-f_{12}}{\sqrt{3}}$\\\hline
$\Omega_{bb}^{-}\to \Xi^{\prime0}  \overline B^0_s  K^-  $ & $ \frac{f_1+f_2+f_3+3 f_4+6 f_5-2 f_6+6 f_7-2 f_8-2 f_9-2 f_{10}+f_{11}}{\sqrt{3}}$\\\hline
$\Omega_{bb}^{-}\to \Xi^{\prime-}  B^-  \pi^+  $ & $ \frac{f_2-2 f_5+3 f_6-2 f_{10}+f_{12}}{\sqrt{3}}$\\\hline
$\Omega_{bb}^{-}\to \Xi^{\prime-}  \overline B^0  \pi^0  $ & $ \frac{-f_2+4 f_4+2 f_5+f_6+2 f_{10}-2 f_{11}+f_{12}}{\sqrt{6}}$\\\hline
$\Omega_{bb}^{-}\to \Xi^{\prime-}  \overline B^0  \eta  $ & $ \frac{-2 f_1+f_2-2 f_3+6 f_4-2 f_5+f_6+4 f_7+4 f_8+4 f_9-2 f_{10}-3 f_{12}}{3 \sqrt{2}}$\\\hline
$\Omega_{bb}^{-}\to \Xi^{\prime-}  \overline B^0_s  \overline K^0  $ & $ \frac{f_1+f_2+f_3-f_4-2 f_5-2 f_6-2 f_7-2 f_8-2 f_9-2 f_{10}-f_{11}}{\sqrt{3}}$\\\hline
$\Omega_{bb}^{-}\to \Omega^-  B^-  K^+  $ & $ f_2-2 f_5+3 f_6-2 f_{10}+f_{12}$\\\hline
$\Omega_{bb}^{-}\to \Omega^-  \overline B^0  K^0  $ & $ f_2-2 f_5-f_6-2 f_{10}-f_{12}$\\\hline
$\Omega_{bb}^{-}\to \Omega^-  \overline B^0_s  \pi^0  $ & $ \sqrt{2} \left(2 f_4-f_{11}\right)$\\\hline
$\Omega_{bb}^{-}\to \Omega^-  \overline B^0_s  \eta  $ & $ -\sqrt{\frac{2}{3}} \left(f_1+f_2+f_3-3 f_4-2 f_5-2 f_6-2 f_7-2 f_8-2 f_9-2 f_{10}\right)$\\\hline
\hline
\end{tabular}
\end{table}


\section{Non-Leptonic $\Xi_{bc}$ and $\Omega_{bc}$ decays}
\label{sec:bcq_nonleptonic}

Decays of $\Xi_{bc}$ and $\Omega_{bc}$ can proceed via the $b$ quark decay or the $c$ quark decay.
As we have shown in the semileptonic channels, for the charm quark decays, one can obtain the decay amplitudes from those for $\Xi_{cc}$ and $\Omega_{cc}$ decays with the replacement of $T_{cc}\to T_{bc}$, $T_c\to T_b$ and $D\to B$.
For the bottom quark decay, one can obtain them  from those for $\Xi_{bb}$ and $\Omega_{bb}$ decays with $T_{bb}\to T_{bc}$, $T_b\to T_c$ and $B\to D$. Thus we do not present the tedious results again.

\section{Conclusions}
\label{sec:conclusions}

Quite recently, the LHCb collaboration has observed the   $\Xi_{cc}^{++}$  in the   final state $\Lambda_c K^-\pi^+\pi^+$.  Such an important observation will undoubtedly promote the research on both hadron spectroscopy and weak decays of   doubly  heavy baryons.

In this paper, we have analyzed   weak decays of doubly heavy baryons
$\Xi_{cc}$,  $\Omega_{cc}$, $\Xi_{bc}^{(\prime)}$, $\Omega_{bc}^{(\prime)}$, $\Xi_{bb}$ and $\Omega_{bb}$ under the flavor SU(3) symmetry, where the final states involve one or two light mesons.  This is inspired by the experimental fact that the $\Xi_{cc}^{++}\to \Lambda_c K^-\pi^+\pi^+$ is not dominated by any two-body intermediate state.
 Decay amplitudes for various semileptonic and nonleptonic decays have been parametrized in terms of a few  SU(3) irreducible amplitudes.  We have found a number of relations or sum rules between   decay widths, which can be examined in future measurements at experimental facilities like LHC~\cite{Bediaga:2012py}, Belle II~\cite{Aushev:2010bq} and CEPC~\cite{CEPC_preCDR}.  On the one hand, at first sight the number of relations is desperately large.   On the other hand, once a few decay branching fractions were measured in future,  these relations can  provide richful important clues  for the exploration  of other decay modes.

It should be stressed that our  analysis in this work using the flavor SU(3) symmetry is only applicable to   non-resonant contributions. For a complete exploration of three-body decays, one should also take into account resonant   contributions from  two-body states and this has been given in Ref.~\cite{Wang:2017azm}. Relative phases between them can  be obtained in a Dalitz plot analysis or measurements of invariant mass distributions.  In addition, SU(3) symmetry breaking effects might also be relevant. Such effects in the phase space can be incorporated once  masses of all involved hadrons are known.  This will remedy the relations for decay widths we derived. Actually, we have removed the channels kinematically prohibited.  Further deviations,  if found by experimentalists in future, would have the indications on  decay dynamics in the doubly heavy baryon system. We hope this analysis together with experimental measurements in future will help to establish a QCD-rooted approach to handle the production and decays of doubly heavy baryons.

\section*{Acknowledgements}

The authors are   grateful to   Jibo He,  Xiao-Hui Hu,  Cai-Dian L\"u,  Fu-Sheng Yu,  Zhen-Xing Zhao  for useful discussions.   W.W. thanks Cai-Dian L\"u,  and Qiang Zhao for their hospitality when this work is finalized at IHEP, CAS.
This work is supported  in part   by National  Natural
Science Foundation of China under Grant
 No.11575110, 11655002, 11735010,  Natural  Science Foundation of Shanghai under Grant  No.~15DZ2272100 and No.~15ZR1423100, Shanghai Key Laboratory for Particle Physics and Cosmology, and  by  MOE  Key Laboratory for Particle Physics, Astrophysics and Cosmology.

\begin{appendix}

\section{Relations between Nonleptonic Decay Widths }
\subsection{Doubly charmed baryon decays into a charmed baryon and two light mesons}

\label{sec:relation_cc_cqq3_2m_bar3}

For   decays into an anti-triplet baryon, we have thee relations:
\begin{eqnarray}
    \Gamma(\Xi_{cc}^{++}\to\Xi_c^0\pi^+ \pi^+ )= 4\Gamma(\Xi_{cc}^{++}\to\Xi_c^+\pi^+ \pi^0 )= 4\Gamma(\Xi_{cc}^{+}\to\Xi_c^0\pi^+ \pi^0 ),\nonumber\\
          \Gamma(\Xi_{cc}^{++}\to\Xi_c^+\pi^+ K^0 )= { }\Gamma(\Xi_{cc}^{++}\to\Lambda_c^+K^+ \overline K^0 ),\quad
     \Gamma(\Xi_{cc}^{+}\to\Xi_c^+\pi^- K^+ )= { }\Gamma(\Omega_{cc}^{+}\to\Lambda_c^+\pi^+ K^- ),\nonumber\\
     \Gamma(\Xi_{cc}^{+}\to\Xi_c^0\pi^+ K^0 )= { }\Gamma(\Omega_{cc}^{+}\to\Xi_c^0K^+ \overline K^0 ),\qquad
      \Gamma(\Omega_{cc}^{+}\to\Lambda_c^+\pi^0 \pi^0 )= \frac{1}{2}\Gamma(\Omega_{cc}^{+}\to\Lambda_c^+\pi^+ \pi^- ),\nonumber\\
       \Gamma(\Omega_{cc}^{+}\to\Xi_c^+\pi^+ \pi^- )= { }\Gamma(\Xi_{cc}^{+}\to\Lambda_c^+K^- K^+ ),\quad
     \Gamma(\Xi_{cc}^{+}\to\Lambda_c^+K^0 \overline K^0 )= { }\Gamma(\Omega_{cc}^{+}\to\Xi_c^+K^0 \overline K^0 ),\nonumber\\
     \Gamma(\Xi_{cc}^{+}\to\Lambda_c^+\pi^+ \pi^- )= { }\Gamma(\Omega_{cc}^{+}\to\Xi_c^+K^+ K^- ).\nonumber
    \end{eqnarray}

We derive the relations for decays into a sextet baryon:
\begin{eqnarray}
        \Gamma(\Xi_{cc}^{++}\to\Sigma_{c}^{++}\pi^0 \pi^0 )= { }\Gamma(\Xi_{cc}^{++}\to\Sigma_{c}^{++}\eta \eta )=\frac{3}{2}\Gamma(\Xi_{cc}^{++}\to\Sigma_{c}^{++}\eta \pi^0 ),\nonumber\\
                   \Gamma(\Xi_{cc}^{++}\to\Xi_{c}^{\prime0}\pi^+ \pi^+ )= \Gamma(\Xi_{cc}^{++}\to\Omega_{c}^{0}\pi^+ K^+ )= 4\Gamma(\Xi_{cc}^{+}\to\Xi_{c}^{\prime0}\pi^+ \pi^0 )=4\Gamma(\Xi_{cc}^{++}\to\Xi_{c}^{\prime+}\pi^+ \pi^0 ),\nonumber\\
        \Gamma(\Xi_{cc}^{+}\to\Sigma_{c}^{+}\pi^+ \pi^- )= { }\Gamma(\Omega_{cc}^{+}\to\Xi_{c}^{\prime+}K^- K^+ )= 4\Gamma(\Xi_{cc}^{++}\to\Xi_{c}^{\prime+}\pi^0 \pi^+ ),\nonumber\\
         \Gamma(\Xi_{cc}^{++}\to\Xi_{c}^{\prime+}\pi^+ K^0 )= { }\Gamma(\Xi_{cc}^{++}\to\Sigma_{c}^{+}K^+ \overline K^0 ),\qquad
 \Gamma(\Xi_{cc}^{++}\to\Sigma_{c}^{++}\pi^+ \pi^- )= { }\Gamma(\Xi_{cc}^{++}\to\Sigma_{c}^{++}K^- K^+ ),\nonumber\\
       \Gamma(\Xi_{cc}^{++}\to\Sigma_{c}^{++}\pi^0 K^0 )= 3\Gamma(\Xi_{cc}^{++}\to\Sigma_{c}^{++}K^0 \eta ),\qquad
       \Gamma(\Xi_{cc}^{++}\to\Sigma_{c}^{++}\pi^0 \overline K^0 )= 3\Gamma(\Xi_{cc}^{++}\to\Sigma_{c}^{++}\overline K^0 \eta ),\nonumber\\
        \Gamma(\Xi_{cc}^{++}\to\Sigma_{c}^{0}\pi^+ \pi^+ )= { }\Gamma(\Xi_{cc}^{++}\to\Omega_{c}^{0}K^+ K^+ ),\qquad
        \Gamma(\Xi_{cc}^{++}\to\Xi_{c}^{\prime+}\pi^+ \pi^0 )= \frac{1}{4}\Gamma(\Xi_{cc}^{++}\to\Omega_{c}^{0}\pi^+ K^+ ),\nonumber\\
             \Gamma(\Xi_{cc}^{+}\to\Sigma_{c}^{0}\pi^0 K^+ )= 3\Gamma(\Xi_{cc}^{+}\to\Sigma_{c}^{0}K^+ \eta ),\qquad
             \Gamma(\Xi_{cc}^{+}\to\Xi_{c}^{\prime+}\pi^- K^+ )= { }\Gamma(\Omega_{cc}^{+}\to\Sigma_{c}^{+}\pi^+ K^- ),\nonumber\\
                 \Gamma(\Omega_{cc}^{+}\to\Sigma_{c}^{+}\overline K^0 \overline K^0 )= 1\Gamma(\Omega_{cc}^{+}\to\Sigma_{c}^{++}\overline K^0 K^- ),\qquad
                  \Gamma(\Omega_{cc}^{+}\to\Xi_{c}^{\prime+}\pi^+ \pi^- )= { }\Gamma(\Xi_{cc}^{+}\to\Sigma_{c}^{+}K^- K^+ ),\nonumber\\
                   \Gamma(\Omega_{cc}^{+}\to\Omega_{c}^{0}\pi^+ K^0 )= { }\Gamma(\Xi_{cc}^{+}\to\Sigma_{c}^{0}K^+ \overline K^0 ),\qquad
    \Gamma(\Xi_{cc}^{++}\to\Sigma_{c}^{0}\pi^+ K^+ )= 1\Gamma(\Xi_{cc}^{++}\to\Xi_{c}^{\prime0}K^+ K^+ ),\nonumber\\
           \Gamma(\Xi_{cc}^{+}\to\Sigma_{c}^{+}K^0 \overline K^0 )= { }\Gamma(\Omega_{cc}^{+}\to\Xi_{c}^{\prime+}K^0 \overline K^0 ),\qquad
       \Gamma(\Xi_{cc}^{+}\to\Sigma_{c}^{++}\pi^- K^0 )= 1\Gamma(\Xi_{cc}^{+}\to\Xi_{c}^{\prime+}K^0 K^0 ),\nonumber\\
             \Gamma(\Xi_{cc}^{+}\to\Xi_{c}^{\prime0}\pi^+ K^0 )= { }\Gamma(\Omega_{cc}^{+}\to\Xi_{c}^{\prime0}K^+ \overline K^0 ),\qquad
      \Gamma(\Omega_{cc}^{+}\to\Sigma_{c}^{0}\pi^+ \pi^0 )= { }\Gamma(\Omega_{cc}^{+}\to\Sigma_{c}^{++}\pi^0 \pi^- ),\nonumber\\
      \Gamma(\Omega_{cc}^{+}\to\Sigma_{c}^{++}\pi^- \overline K^0 )= { }\Gamma(\Xi_{cc}^{+}\to\Sigma_{c}^{++}K^0 K^- ).\nonumber
     \end{eqnarray}

\subsection{$\Xi_{cc}$ and $\Omega_{c}$ decays into a octet baryon , a charmed meson and a light meson}
\label{sec:relation_cc_baryon8_cm_lm}

The following relations are derived for
decays into an octet baryon:
\begin{eqnarray}
    \Gamma(\Xi_{cc}^{++}\to\Sigma^+D^+K^0 )= { }\Gamma(\Xi_{cc}^{++}\to{p}D^+_s\overline K^0 ),\qquad
    \Gamma(\Xi_{cc}^{++}\to{p}D^0\pi^+ )= { }\Gamma(\Xi_{cc}^{++}\to\Sigma^+D^0K^+ ),\nonumber\\
    \Gamma(\Xi_{cc}^{+}\to\Sigma^+D^0K^0 )= { }\Gamma(\Omega_{cc}^{+}\to{p}D^0\overline K^0 ),\qquad
    \Gamma(\Xi_{cc}^{+}\to\Sigma^+D^+_s\pi^- )= { }\Gamma(\Omega_{cc}^{+}\to{p}D^+K^- ),\nonumber\\
    \Gamma(\Xi_{cc}^{+}\to\Sigma^-D^+_s\pi^+ )= { }\Gamma(\Omega_{cc}^{+}\to\Xi^-D^+K^+ ),\qquad
     \Gamma(\Xi_{cc}^{+}\to{p}D^+\pi^- )= { }\Gamma(\Omega_{cc}^{+}\to\Sigma^+D^+_sK^- ),\nonumber\\
      \Gamma(\Omega_{cc}^{+}\to\Sigma^+D^+\pi^- )= { }\Gamma(\Xi_{cc}^{+}\to{p}D^+_sK^- ),\qquad
       \Gamma(\Omega_{cc}^{+}\to\Xi^-D^+_s\pi^+ )= { }\Gamma(\Xi_{cc}^{+}\to\Sigma^-D^+K^+ ).\nonumber
     \end{eqnarray}
For a decuplet baryon, we have
\begin{eqnarray}
     \Gamma(\Xi_{cc}^{++}\to\Delta^{0}D^+_s\pi^+ )= { }\Gamma(\Xi_{cc}^{+}\to\Delta^{+}D^+_s\pi^- ),\qquad
       \Gamma(\Xi_{cc}^{++}\to\Sigma^{\prime0}D^+_s\pi^+ )= { }\Gamma(\Xi_{cc}^{++}\to\Sigma^{\prime0}D^+K^+ ),\nonumber\\
            \Gamma(\Xi_{cc}^{+}\to\Delta^{++}D^0\pi^- )= { }\Gamma(\Omega_{cc}^{+}\to\Delta^{++}D^0K^- ),\qquad
       \Gamma(\Xi_{cc}^{+}\to\Delta^{-}D^+_s\pi^+ )= { }\Gamma(\Xi_{cc}^{++}\to\Delta^{++}D^+_s\pi^- ),\nonumber\\
        \Gamma(\Xi_{cc}^{+}\to\Sigma^{\prime+}D^+_s\pi^- )= { }\Gamma(\Omega_{cc}^{+}\to\Delta^{+}D^+K^- ),\qquad
              \Gamma(\Omega_{cc}^{+}\to\Delta^{0}D^0\pi^+ )= { }\Gamma(\Omega_{cc}^{+}\to\Delta^{+}D^+\pi^- ),\nonumber\\
               \Gamma(\Omega_{cc}^{+}\to\Delta^{0}D^+_s\overline K^0 )= { }\Gamma(\Omega_{cc}^{+}\to\Delta^{+}D^+_sK^- ),\qquad
                \Gamma(\Omega_{cc}^{+}\to\Delta^{-}D^+\pi^+ )= { }\Gamma(\Omega_{cc}^{+}\to\Delta^{++}D^0\pi^- ),\nonumber\\
                  \Gamma(\Omega_{cc}^{+}\to\Sigma^{\prime-}D^+_s\pi^+ )= { }\Gamma(\Omega_{cc}^{+}\to\Sigma^{\prime+}D^+_s\pi^- ),\qquad
                  \Gamma(\Omega_{cc}^{+}\to\Xi^{\prime-}D^+_s\pi^+ )= { }\Gamma(\Xi_{cc}^{+}\to\Sigma^{\prime-}D^+K^+ ),\nonumber\\
    \Gamma(\Omega_{cc}^{+}\to\Delta^{+}D^0\pi^0 )= \frac{1}{4}\Gamma(\Omega_{cc}^{+}\to\Delta^{0}D^+\pi^0 ),\qquad
     \Gamma(\Omega_{cc}^{+}\to\Delta^{+}D^0\eta )= { }\Gamma(\Omega_{cc}^{+}\to\Delta^{0}D^+\eta ),\nonumber\\
    \Gamma(\Xi_{cc}^{++}\to\Delta^{+}D^+_s\pi^0 )= { }\Gamma(\Xi_{cc}^{+}\to\Delta^{0}D^+_s\pi^0 ).\nonumber
    \end{eqnarray}

\subsection{$\Xi_{bb}$ and $\Omega_{bb}$ decays into a bottom baryon, $J/\psi$ and a light meson}
\label{sec:relation_bb_Jpsi_baryon3_lm}

Decays into an anti-triplet baryon have the relations:
\begin{eqnarray}
    \Gamma(\Xi_{bb}^{0}\to\Lambda_b^0\pi^0 J/\psi)= \frac{1}{2}\Gamma(\Xi_{bb}^{-}\to\Lambda_b^0\pi^- J/\psi)= \frac{1}{2}\Gamma(\Xi_{bb}^{-}\to\Xi_b^-K^0 J/\psi),\nonumber\\
     \Gamma(\Omega_{bb}^{-}\to\Xi_b^0\pi^- J/\psi)= { }\Gamma(\Xi_{bb}^{0}\to\Xi_b^0K^0 J/\psi)=2\Gamma(\Omega_{bb}^{-}\to\Xi_b^-\pi^0 J/\psi),\nonumber\\
       \Gamma(\Xi_{bb}^{0}\to\Xi_b^0\pi^0 J/\psi)=\frac{1}{2}\Gamma(\Xi_{bb}^{0}\to\Xi_b^-\pi^+ J/\psi)=\frac{1}{2}\Gamma(\Xi_{bb}^{-}\to\Xi_b^0\pi^- J/\psi)=\Gamma(\Xi_{bb}^{-}\to\Xi_b^-\pi^0 J/\psi),\nonumber\\
         \Gamma(\Xi_{bb}^{0}\to\Xi_b^-K^+ J/\psi)= { }\Gamma(\Omega_{bb}^{-}\to\Lambda_b^0K^- J/\psi),\qquad
        \Gamma(\Xi_{bb}^{-}\to\Lambda_b^0\pi^- J/\psi)= { }\Gamma(\Xi_{bb}^{-}\to\Xi_b^-K^0 J/\psi),\nonumber\\
            \Gamma(\Xi_{bb}^{0}\to\Lambda_b^0\overline K^0 J/\psi)= { }\Gamma(\Xi_{bb}^{-}\to\Lambda_b^0K^- J/\psi),\qquad
              \Gamma(\Omega_{bb}^{-}\to\Xi_b^-\overline K^0 J/\psi)= { }\Gamma(\Omega_{bb}^{-}\to\Xi_b^0K^- J/\psi).\nonumber
    \end{eqnarray}
Decays into a sextet baryon have the relations:
\begin{eqnarray}
    \Gamma(\Xi_{bb}^{0}\to\Sigma_{b}^{+}\pi^- J/\psi)= 2\Gamma(\Xi_{bb}^{0}\to\Xi_{b}^{\prime0}K^0 J/\psi)= { }\Gamma(\Omega_{bb}^{-}\to\Omega_{b}^{-}K^0 J/\psi)=4\Gamma(\Omega_{bb}^{-}\to\Xi_{b}^{\prime-}\pi^0 J/\psi)\nonumber\\=2\Gamma(\Omega_{bb}^{-}\to\Xi_{b}^{\prime0}\pi^- J/\psi)=\frac{1}{2}\Gamma(\Omega_{bb}^{-}\to\Xi_{b}^{\prime0}\pi^- J/\psi),\nonumber\\
    \Gamma(\Xi_{bb}^{0}\to\Xi_{b}^{\prime0}\pi^0 J/\psi)= \frac{1}{4}\Gamma(\Xi_{bb}^{0}\to\Omega_{b}^{-}K^+ J/\psi)= \frac{1}{2}\Gamma(\Xi_{bb}^{-}\to\Xi_{b}^{\prime0}\pi^- J/\psi)= \frac{1}{4}\Gamma(\Xi_{bb}^{-}\to\Omega_{b}^{-}K^0 J/\psi)\nonumber\\=\frac{1}{2}\Gamma(\Xi_{bb}^{0}\to\Xi_{b}^{\prime-}\pi^+ J/\psi)=\Gamma(\Xi_{bb}^{-}\to\Xi_{b}^{\prime-}\pi^0 J/\psi),\nonumber\\
    \Gamma(\Xi_{bb}^{-}\to\Sigma_{b}^{0}\pi^- J/\psi)= 6\Gamma(\Xi_{bb}^{0}\to\Sigma_{b}^{0}\eta J/\psi)= 3\Gamma(\Xi_{bb}^{-}\to\Sigma_{b}^{-}\eta J/\psi)={ }\Gamma(\Xi_{bb}^{-}\to\Xi_{b}^{\prime-}K^0 J/\psi)\nonumber\\=\Gamma(\Xi_{bb}^{-}\to\Sigma_{b}^{-}\pi^0 J/\psi),\nonumber\\
     \Gamma(\Xi_{bb}^{0}\to\Sigma_{b}^{0}\overline K^0 J/\psi)= \frac{1}{2}\Gamma(\Xi_{bb}^{0}\to\Sigma_{b}^{+}K^- J/\psi)= { }\Gamma(\Xi_{bb}^{-}\to\Sigma_{b}^{0}K^- J/\psi)=\frac{1}{2}\Gamma(\Xi_{bb}^{-}\to\Sigma_{b}^{-}\overline K^0 J/\psi),\nonumber\\
      \Gamma(\Xi_{bb}^{0}\to\Sigma_{b}^{-}\pi^+ J/\psi)= 2\Gamma(\Xi_{bb}^{0}\to\Xi_{b}^{\prime-}K^+ J/\psi)=2\Gamma(\Omega_{bb}^{-}\to\Sigma_{b}^{0}K^- J/\psi)= { }\Gamma(\Omega_{bb}^{-}\to\Sigma_{b}^{-}\overline K^0 J/\psi),\nonumber\\
        \Gamma(\Omega_{bb}^{-}\to\Xi_{b}^{\prime0}K^- J/\psi)= \frac{3}{4}\Gamma(\Omega_{bb}^{-}\to\Omega_{b}^{-}\eta J/\psi)=\Gamma(\Omega_{bb}^{-}\to\Xi_{b}^{\prime-}\overline K^0 J/\psi).\nonumber
     \end{eqnarray}

\subsection{$\Xi_{bb}$ and $\Omega_{bb}$ decays into a $bcq$, an anti-charmed meson and a light meson}
\label{sec:relation_bb_bcq_cm_lm}
\begin{eqnarray}
     \Gamma(\Xi_{bb}^{0}\to\Xi_{bc}^{+}D^-_s\pi^0 )= \frac{1}{2}\Gamma(\Xi_{bb}^{-}\to\Xi_{bc}^{+}D^-_s\pi^- )=\frac{1}{2}\Gamma(\Xi_{bb}^{0}\to\Xi_{bc}^{0}D^-_s\pi^+ )= { }\Gamma(\Xi_{bb}^{-}\to\Xi_{bc}^{0}D^-_s\pi^0 ),\nonumber\\
     \Gamma(\Xi_{bb}^{0}\to\Omega_{bc}^{0}\overline D^0\pi^0 )= \frac{1}{2}\Gamma(\Xi_{bb}^{-}\to\Omega_{bc}^{0}\overline D^0\pi^- )= \frac{1}{4}\Gamma(\Xi_{bb}^{-}\to\Omega_{bc}^{0}D^-\pi^0 )=\frac{1}{2}\Gamma(\Xi_{bb}^{0}\to\Omega_{bc}^{0}D^-\pi^+ ),\nonumber\\
     \Gamma(\Omega_{bb}^{-}\to\Xi_{bc}^{+}D^-_s\pi^- )= { }\Gamma(\Xi_{bb}^{0}\to\Omega_{bc}^{0}\overline D^0K^0 )=2\Gamma(\Omega_{bb}^{-}\to\Xi_{bc}^{0}D^-_s\pi^0 ),\nonumber\\
      \Gamma(\Omega_{bb}^{-}\to\Omega_{bc}^{0}\overline D^0\pi^- )= { }\Gamma(\Xi_{bb}^{0}\to\Xi_{bc}^{+}D^-_sK^0 )=2\Gamma(\Omega_{bb}^{-}\to\Omega_{bc}^{0}D^-\pi^0 ),\nonumber\\
    \Gamma(\Xi_{bb}^{0}\to\Xi_{bc}^{+}\overline D^0\pi^- )= { }\Gamma(\Omega_{bb}^{-}\to\Omega_{bc}^{0}D^-_sK^0 ),\qquad
    \Gamma(\Xi_{bb}^{0}\to\Xi_{bc}^{+}D^-\overline K^0 )= { }\Gamma(\Xi_{bb}^{-}\to\Xi_{bc}^{0}\overline D^0K^- ),\nonumber\\
      \Gamma(\Xi_{bb}^{0}\to\Xi_{bc}^{0}\overline D^0\overline K^0 )= { }\Gamma(\Xi_{bb}^{-}\to\Xi_{bc}^{+}D^-K^- ),\qquad
       \Gamma(\Xi_{bb}^{0}\to\Xi_{bc}^{0}D^-\pi^+ )= { }\Gamma(\Omega_{bb}^{-}\to\Xi_{bc}^{0}D^-\overline K^0 ),\nonumber\\
           \Gamma(\Xi_{bb}^{0}\to\Omega_{bc}^{0}D^-K^+ )= { }\Gamma(\Omega_{bb}^{-}\to\Xi_{bc}^{+}D^-K^- ),\qquad
              \Gamma(\Xi_{bb}^{0}\to\Omega_{bc}^{0}D^-_sK^+ )= { }\Gamma(\Xi_{bb}^{-}\to\Omega_{bc}^{0}D^-_sK^0 ),\nonumber\\
               \Gamma(\Xi_{bb}^{-}\to\Xi_{bc}^{+}D^-\pi^- )= { }\Gamma(\Xi_{bb}^{-}\to\Omega_{bc}^{0}D^-K^0 ),\qquad
               \Gamma(\Xi_{bb}^{-}\to\Xi_{bc}^{0}\overline D^0\pi^- )= { }\Gamma(\Xi_{bb}^{-}\to\Xi_{bc}^{0}D^-_sK^0 ),\nonumber\\
                \Gamma(\Xi_{bb}^{-}\to\Xi_{bc}^{0}D^-\pi^0 )= 3\Gamma(\Xi_{bb}^{-}\to\Xi_{bc}^{0}D^-\eta ),\qquad
                \Gamma(\Xi_{bb}^{-}\to\Xi_{bc}^{0}D^-\overline K^0 )= { }\Gamma(\Xi_{bb}^{0}\to\Xi_{bc}^{+}\overline D^0K^- ),\nonumber\\
                 \Gamma(\Omega_{bb}^{-}\to\Xi_{bc}^{0}D^-_s\overline K^0 )= { }\Gamma(\Omega_{bb}^{-}\to\Xi_{bc}^{+}D^-_sK^- ),\qquad
                 \Gamma(\Omega_{bb}^{-}\to\Omega_{bc}^{0}D^-\overline K^0 )= { }\Gamma(\Omega_{bb}^{-}\to\Omega_{bc}^{0}\overline D^0K^- ),\nonumber\\
     \Gamma(\Xi_{bb}^{0}\to\Omega_{bc}^{0}\overline D^0\eta )= { }\Gamma(\Xi_{bb}^{-}\to\Omega_{bc}^{0}D^-\eta ),\qquad
     \Gamma(\Xi_{bb}^{0}\to\Xi_{bc}^{0}D^-_sK^+ )= { }\Gamma(\Omega_{bb}^{-}\to\Xi_{bc}^{0}\overline D^0K^- ),\nonumber\\
     \Gamma(\Xi_{bb}^{0}\to\Xi_{bc}^{+}D^-_s\eta )= { }\Gamma(\Xi_{bb}^{-}\to\Xi_{bc}^{0}D^-_s\eta ).\nonumber
    \end{eqnarray}

\subsection{$\Xi_{bb}$ and $\Omega_{bb}$ decays into a $bcq$ and two light mesons}
\label{sec:relation_bb_bcq_2lm}
\begin{eqnarray}
    \Gamma(\Xi_{bb}^{0}\to\Xi_{bc}^{+}\pi^0 \pi^- )= \frac{1}{2}\Gamma(\Xi_{bb}^{-}\to\Omega_{bc}^{0}\pi^- K^0 )= \frac{1}{2}\Gamma(\Omega_{bb}^{-}\to\Xi_{bc}^{+}\pi^- K^- )= \frac{1}{4}\Gamma(\Xi_{bb}^{-}\to\Xi_{bc}^{+}\pi^- \pi^- ) \nonumber\\=\frac{1}{2}\Gamma(\Xi_{bb}^{-}\to\Omega_{bc}^{0}K^0 \pi^- )= \frac{1}{2}\Gamma(\Omega_{bb}^{-}\to\Xi_{bc}^{+}K^- \pi^- )= { }\Gamma(\Xi_{bb}^{-}\to\Xi_{bc}^{0}\pi^0 \pi^- ),\nonumber\\
     \Gamma(\Xi_{bb}^{0}\to\Xi_{bc}^{0}\pi^0 \overline K^0 )= 3\Gamma(\Xi_{bb}^{0}\to\Xi_{bc}^{0}\overline K^0 \eta )
     = 3\Gamma(\Xi_{bb}^{0}\to\Xi_{bc}^{0}\eta \overline K^0 ),\nonumber\\
     \Gamma(\Omega_{bb}^{-}\to\Xi_{bc}^{0}\pi^0 K^- )= 3\Gamma(\Omega_{bb}^{-}\to\Xi_{bc}^{0}K^- \eta )= 3\Gamma(\Omega_{bb}^{-}\to\Xi_{bc}^{0}\eta K^- ),\nonumber\\
     \Gamma(\Xi_{bb}^{-}\to\Xi_{bc}^{0}\pi^0 K^- )= 3\Gamma(\Xi_{bb}^{-}\to\Xi_{bc}^{0}K^- \eta )= 3\Gamma(\Xi_{bb}^{-}\to\Xi_{bc}^{0}\eta K^- ),\nonumber\\
      \Gamma(\Omega_{bb}^{-}\to\Xi_{bc}^{0}\overline K^0 K^- )= \frac{1}{2}\Gamma(\Omega_{bb}^{-}\to\Xi_{bc}^{+}K^- K^-)=\Gamma(\Xi_{bb}^{-}\to\Xi_{bc}^{+}\pi^- K^- ),\nonumber\\
       \Gamma(\Xi_{bb}^{0}\to\Omega_{bc}^{0}\pi^0 K^0 )= 3\Gamma(\Xi_{bb}^{0}\to\Omega_{bc}^{0}K^0 \eta )= 3\Gamma(\Xi_{bb}^{0}\to\Omega_{bc}^{0}\eta K^0 ),\nonumber\\
      \Gamma(\Xi_{bb}^{0}\to\Omega_{bc}^{0}\pi^0 \pi^0 )= \frac{1}{2}\Gamma(\Xi_{bb}^{0}\to\Omega_{bc}^{0}\pi^+ \pi^- ),
         \Gamma(\Omega_{bb}^{-}\to\Omega_{bc}^{0}\pi^- \overline K^0 )= 2\Gamma(\Omega_{bb}^{-}\to\Omega_{bc}^{0}\pi^0 K^- ),\nonumber\\
       \Gamma(\Xi_{bb}^{0}\to\Omega_{bc}^{0}\pi^0 \eta )= \frac{1}{2}\Gamma(\Xi_{bb}^{-}\to\Omega_{bc}^{0}\pi^- \eta ).\nonumber
     \end{eqnarray}

\subsection{$\Xi_{bb}$ and $\Omega_{bb}$ decays into a bottom sextet,  a charmed meson and a light meson}
\label{sec:relation_bb_bqq6_cm_lm}
\begin{eqnarray}
    \Gamma(\Xi_{bb}^{0}\to\Sigma_{b}^{-}D^0\pi^+ )= 2\Gamma(\Xi_{bb}^{0}\to\Xi_{b}^{\prime-}D^0K^+ ),\nonumber\\
     \Gamma(\Xi_{bb}^{0}\to\Xi_{b}^{\prime-}D^0\pi^+ )= \frac{1}{2}\Gamma(\Xi_{bb}^{0}\to\Omega_{b}^{-}D^0K^+ ),\nonumber\\
     \Gamma(\Xi_{bb}^{0}\to\Omega_{b}^{-}D^+_s\pi^0 )= \frac{1}{2}\Gamma(\Xi_{bb}^{-}\to\Omega_{b}^{-}D^+_s\pi^- ),\nonumber\\
      \Gamma(\Omega_{bb}^{-}\to\Omega_{b}^{-}D^0\pi^0 )= \frac{1}{2}\Gamma(\Omega_{bb}^{-}\to\Omega_{b}^{-}D^+\pi^- ).\nonumber
     \end{eqnarray}

\subsection{$\Xi_{bb}$ and $\Omega_{bb}$ decays into a bottom baryon, an anti-charmed meson and a light meson}
\label{sec:relation_bb_bqq3_acm_lm}

Channels involving an anti-triplet baryon have
\begin{eqnarray}
    \Gamma(\Xi_{bb}^{0}\to\Lambda_b^0D^-_s\pi^+ )= 2\Gamma(\Xi_{bb}^{-}\to\Lambda_b^0D^-_s\pi^0 ),\nonumber\\
    \Gamma(\Omega_{bb}^{-}\to\Xi_b^-D^-_s\pi^+ )= 2\Gamma(\Omega_{bb}^{-}\to\Xi_b^0D^-_s\pi^0 ), \nonumber
     \end{eqnarray}
while decay modes involving a sextet baryon have
\begin{eqnarray}
    \Gamma(\Xi_{bb}^{0}\to\Sigma_{b}^{0}D^-_s\pi^+ )= { }\Gamma(\Xi_{bb}^{0}\to\Sigma_{b}^{+}D^-_s\pi^0 ),\nonumber\\
    \Gamma(\Xi_{bb}^{0}\to\Sigma_{b}^{-}\overline D^0\pi^+ )= 2\Gamma(\Xi_{bb}^{0}\to\Xi_{b}^{\prime-}\overline D^0K^+ ),\nonumber\\ \Gamma(\Xi_{bb}^{0}\to\Xi_{b}^{\prime-}\overline D^0\pi^+ )= \frac{1}{2}\Gamma(\Xi_{bb}^{0}\to\Omega_{b}^{-}\overline D^0K^+ ),\nonumber\\ \Gamma(\Xi_{bb}^{-}\to\Sigma_{b}^{+}D^-\pi^- )= 2\Gamma(\Xi_{bb}^{-}\to\Xi_{b}^{\prime0}D^-K^0 ),\nonumber\\
    \Gamma(\Omega_{bb}^{-}\to\Sigma_{b}^{0}D^-_s\overline K^0 )= \frac{1}{2}\Gamma(\Omega_{bb}^{-}\to\Sigma_{b}^{+}D^-_sK^- ),\nonumber\\ \Gamma(\Omega_{bb}^{-}\to\Xi_{b}^{\prime-}D^-_s\pi^+ )= 2\Gamma(\Omega_{bb}^{-}\to\Xi_{b}^{\prime0}D^-_s\pi^0 ),\nonumber\\
    \Gamma(\Xi_{bb}^{0}\to\Sigma_{b}^{+}D^-_s\eta )= 2\Gamma(\Xi_{bb}^{-}\to\Sigma_{b}^{0}D^-_s\eta ).\nonumber
     \end{eqnarray}

\subsection{$\Xi_{bb}$ and $\Omega_{bb}$ decays into a bottom baryon and two light mesons}
\label{sec:relation_bb_bqq3_2lm_b2d}

For the anti-triplet baryon, we have
\begin{eqnarray}
    \Gamma(\Xi_{bb}^{0}\to\Xi_b^0\pi^0 K^0 )= 3\Gamma(\Xi_{bb}^{0}\to\Xi_b^0K^0 \eta ),
    \Gamma(\Xi_{bb}^{-}\to\Xi_b^-\pi^0 K^0 )= 3\Gamma(\Xi_{bb}^{-}\to\Xi_b^-K^0 \eta ).\nonumber
    \end{eqnarray}
\begin{eqnarray}
    \Gamma(\Xi_{bb}^{0}\to\Lambda_b^0\pi^0 \overline K^0 )= 3\Gamma(\Xi_{bb}^{0}\to\Lambda_b^0\overline K^0 \eta ),
    \Gamma(\Xi_{bb}^{0}\to\Xi_b^-\pi^+ \pi^0 )= { }\Gamma(\Xi_{bb}^{-}\to\Xi_b^0\pi^0 \pi^- ),\nonumber\\
     \Gamma(\Omega_{bb}^{-}\to\Xi_b^-\pi^0 \overline K^0 )= 3\Gamma(\Omega_{bb}^{-}\to\Xi_b^-\overline K^0 \eta ).\nonumber
     \end{eqnarray}


Decays into a sextet heavy baryon  have
\begin{eqnarray}
    \Gamma(\Xi_{bb}^{0}\to\Xi_{b}^{\prime0}\pi^0 K^0 )= 3\Gamma(\Xi_{bb}^{0}\to\Xi_{b}^{\prime0}K^0 \eta ),  \Gamma(\Xi_{bb}^{-}\to\Sigma_{b}^{+}\pi^- \pi^- )= 2\Gamma(\Omega_{bb}^{-}\to\Sigma_{b}^{+}\pi^- K^- ),\nonumber
\end{eqnarray}
\begin{eqnarray}
    \Gamma(\Xi_{bb}^{0}\to\Sigma_{b}^{0}\pi^0 \overline K^0 )= 3\Gamma(\Xi_{bb}^{0}\to\Sigma_{b}^{0}\overline K^0 \eta ),
    \Gamma(\Xi_{bb}^{0}\to\Xi_{b}^{\prime-}\pi^+ \pi^0 )= { }\Gamma(\Xi_{bb}^{-}\to\Xi_{b}^{\prime0}\pi^0 \pi^- ),\nonumber\\
    \Gamma(\Omega_{bb}^{-}\to\Omega_{b}^{-}\pi^0 \pi^0 )= \frac{1}{2}\Gamma(\Omega_{bb}^{-}\to\Omega_{b}^{-}\pi^+ \pi^- ),
    \Gamma(\Xi_{bb}^{-}\to\Sigma_{b}^{+}\pi^- K^- )= \frac{1}{2}\Gamma(\Omega_{bb}^{-}\to\Sigma_{b}^{+}K^- K^- ).\nonumber
\end{eqnarray}

\subsection{$\Xi_{bb}$ and $\Omega_{bb}$ decays into a bottom meson, a light decuplet baryon and a light meson}
\label{sec:relation_bb_baryon10_bm_lm_b2d}

The relations for decay widths are given as:
\begin{eqnarray}
     \Gamma(\Xi_{bb}^{0}\to\Delta^{-}\overline B^0\pi^+ )= 3\Gamma(\Xi_{bb}^{0}\to\Sigma^{\prime-}\overline B^0K^+ )= 3\Gamma(\Xi_{bb}^{0}\to\Xi^{\prime-}\overline B^0_sK^+ )= 3\Gamma(\Xi_{bb}^{0}\to\Sigma^{\prime-}\overline B^0_s\pi^+ ),\nonumber\\
        \Gamma(\Xi_{bb}^{-}\to\Delta^{-}B^-\pi^+ )= 3\Gamma(\Xi_{bb}^{-}\to\Sigma^{\prime-}B^-K^+ )= 3\Gamma(\Omega_{bb}^{-}\to\Xi^{\prime-}B^-K^+ )= 3\Gamma(\Omega_{bb}^{-}\to\Sigma^{\prime-}B^-\pi^+ ),\nonumber\\
            \Gamma(\Xi_{bb}^{0}\to\Delta^{0}B^-\pi^+ )= 2\Gamma(\Xi_{bb}^{0}\to\Sigma^{\prime0}B^-K^+ ).\nonumber
    \end{eqnarray}
\begin{eqnarray}
     \Gamma(\Xi_{bb}^{0}\to\Sigma^{\prime-}\overline B^0\pi^+ )= { }\Gamma(\Xi_{bb}^{0}\to\Xi^{\prime-}\overline B^0K^+ )= \frac{1}{3}\Gamma(\Xi_{bb}^{0}\to\Omega^-\overline B^0_sK^+ )= { }\Gamma(\Xi_{bb}^{0}\to\Xi^{\prime-}\overline B^0_s\pi^+ ),\nonumber\\
       \Gamma(\Xi_{bb}^{-}\to\Sigma^{\prime-}B^-\pi^+ )= { }\Gamma(\Xi_{bb}^{-}\to\Xi^{\prime-}B^-K^+ )= \frac{1}{3}\Gamma(\Omega_{bb}^{-}\to\Omega^-B^-K^+ )= { }=\Gamma(\Omega_{bb}^{-}\to\Xi^{\prime-}B^-\pi^+ ),\nonumber\\
       \Gamma(\Xi_{bb}^{0}\to\Sigma^{\prime0}B^-\pi^+ )= \frac{1}{2}\Gamma(\Xi_{bb}^{0}\to\Xi^{\prime0}B^-K^+ ).\nonumber
\end{eqnarray}

\end{appendix}

\end{document}